\documentclass[12pt]{mythesis}
\usepackage[utf8]{inputenc}
\usepackage{graphicx}
\usepackage{braket,amsfonts,amsmath,algorithmic,epstopdf,amsopn}
\usepackage{amssymb,amsbsy}
\usepackage{color}
\usepackage{xcolor}
\usepackage{setspace}
\usepackage{latexsym}
\usepackage{cite}
\graphicspath{ {images/} }
\usepackage{caption}
\usepackage{float}
\usepackage{longtable}
\usepackage{csquotes}
\usepackage{setspace}
\usepackage{mathrsfs}
\usepackage{amsfonts} 
\usepackage{amssymb} 
\usepackage{float} 
\usepackage{dcolumn}
\usepackage{bm}
\usepackage[font=small,justification=justified,format=plain]{caption}
\DeclareCaptionJustification{justified}{\justifying}
\usepackage[colorlinks,citecolor=blue,linkcolor=blue,urlcolor=blue,bookmarks=false,hypertexnames=true]{hyperref}

\usepackage{float} 
\usepackage{graphicx}
\usepackage[font=small,justification=justified,format=plain]{caption}
\usepackage{subcaption}

\usepackage{hyperref}
\hypersetup{
	colorlinks = true, 		   
	urlcolor    = cyan,  		
	linkcolor  = blue,  	  	
	citecolor  = {green!50!black},   
	urlcolor    = brown		 
}

\usepackage[width=160mm,top=32mm,bottom=25mm,bindingoffset=6mm]{geometry}

\theorembodyfont{\itshape}
\def\svhline{%
	\noalign{\ifnum0=`}\fi\hrule \@height2\arrayrulewidth \futurelet
	\reserved@a\@xhline}
\usepackage{array}
\newcolumntype{L}[1]{>{\raggedright\arraybackslash}p{#1}}
\newcolumntype{C}[1]{>{\centering\arraybackslash}p{#1}}
\newcolumntype{R}[1]{>{\raggedleft\arraybackslash}p{#1}}

\usepackage{fancyhdr}
\title{ThesisTitle}
\author{Author Name}
\date{Day Month Year}
\allowdisplaybreaks

\begin{document}
	\thispagestyle{empty}
	\begin{titlepage}
	\begin{center}
		\onehalfspacing
		\vspace*{0.2cm}
		\textbf{\Large Theoretical Studies on the Scanning Tunneling Microscope}
		
		\vspace{1cm}
		
		{\bf \large THESIS}\\
		\vspace{1.0cm}
		
		submitted in partial fulfillment\\
		of the requirements for the degree of\\
		\vspace{0.5cm}
		{\bf \large DOCTOR OF PHILOSOPHY}\\
		
		\vspace{0.8cm}
		by\\
		\vspace{0.8cm}
		
		\textbf{\large MALATI DESSAI}\\
		(\large 2013PHXF0101G)\\
		
		\vspace{1.2cm}
		Under the Supervision of\\
		{\bf \large Prof. Arun V. Kulkarni}\\
		
		\vspace{1.2cm}
		\includegraphics[scale=0.09]{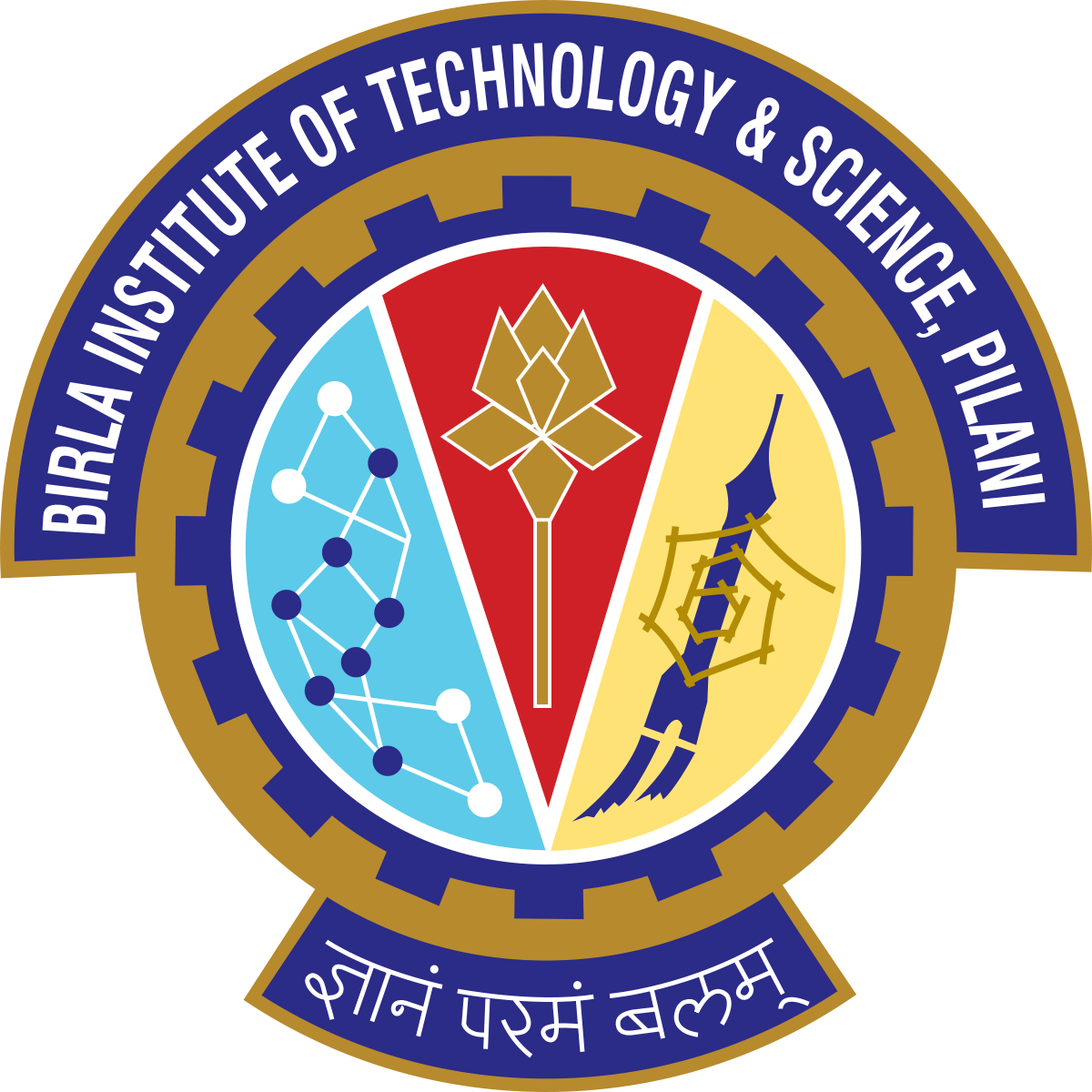}
		\vspace{1.0cm}
		
		{\bf  \normalsize  BIRLA INSTITUTE OF TECHNOLOGY \& SCIENCE, PILANI\\
		2023}
	\end{center}
	
\end{titlepage}

	
	\chapter*{\normalsize BIRLA INSTITUTE OF TECHNOLOGY AND SCIENCE, PILANI}
	\thispagestyle{empty}
\begin{center}
	\vspace{1.2cm}
	{\bf \large CERTIFICATE}
\end{center}
\vspace{1.5cm}
{\onehalfspacing
	\noindent
	This is to certify that the thesis entitled \textbf{``Theoretical Studies on the Scanning Tunneling Microscope"} and submitted by \textbf{Malati Dessai}, ID No \textbf{2013PHXF0101G} for the award of Doctor of Philosophy of the Institute embodies original work done by her under my supervision.

	\vspace{2.2cm}
	\begin{flushleft}
		Supervisor\\
		\vspace{0.25cm}
		Name: \textbf{Prof. Arun V. Kulkarni}\\	
		\vspace{0.25cm}
		Designation: \textbf{Professor}\\
		\;\;\;~~~~~~~~~~~~~~ Department of Physics\\
	    \;\;\;~~~~~~~~~~~~~~	BITS-Pilani, K.K. Birla Goa Campus\\
	    \vspace{0.25cm}
		Date: \today
	\end{flushleft}
}
	\pagenumbering{roman} 
	\onehalfspacing

	\chapter*{\normalsize BIRLA INSTITUTE OF TECHNOLOGY AND SCIENCE, PILANI}
	\thispagestyle{empty}
\begin{center}
	\vspace{1.2cm}
	{\bf \large DECLARATION}
\end{center}
\vspace{1.5cm}
{\onehalfspacing
	I hereby declare that the thesis entitled \textbf{``Theoretical Studies on the Scanning Tunneling Microscope
"} submitted by me for the award of the degree of \textit{Doctor of Philosophy} to BITS-Pilani University is a record of bonafide work carried out by me under the supervision of Prof. Arun V. Kulkarni.\\
	
	\noindent
	I further declare that the work in this thesis has not been submitted and will not be submitted, whether in part or whole, for the award of any other degree or diploma in this institute or any other institute or university.
	
	\vspace{2.2cm}
	\begin{flushleft}
		Name: \textbf{Malati Dessai}\\	
		\vspace{0.25cm}
		ID No.: 2013PHXF0101G\\
		\vspace{0.25cm}
		Date: \today
	\end{flushleft}
}

	\newpage
\thispagestyle{empty}
\vspace*{2.4in}
\begin{center}
{\Large \em Dedicated to}
\vspace{0.5cm} {\Large \em My Children}\\
 {\em \Large Sandhini and Ojas}
\end{center}

	\chapter*{Acknowledgment}
	After all these years of research, I have got quite a list of people who contributed in some way or the other to making this thesis possible. In this regard, I would like to express my gratitude to all those who gave me the possibility to complete this thesis.\\

\noindent
I take this opportunity to thank my research advisor, Prof. Arun V. Kulkarni, for his valuable guidance, consistent encouragement, patience, and support during my PhD. \vspace{0.3cm}

\noindent
I would like to thank my Doctoral committee members, Prof. Toby Joseph and Prof. Senthamarai Kannan, for their insightful comments on my research work. I would also like to express my sincere gratitude to Prof. Chandradew Sharma, Prof. Gaurav Dar for being an excellent instructor during my coursework. I am grateful to Prof. Radhika Vatsan, H.O.D. Department of Physics for her support, constructive comments and input on my research work. \vspace{0.3cm}

\noindent
I acknowledge BITS Pilani, K.K. Birla Goa campus for providing me with the personal computer and infrastructure to pursue my research. I want to thank all the faculty members of the Department of Physics for their help and support during my research. I would also like to acknowledge my gratitude to Directorate of Higher Education (DHE), Goa, the Management and Principal of Chowgule College of Arts and Science, Margao-Goa, for granting me the required study leave. \vspace{0.3cm}

\noindent
Thanks to my dear friends Dhavala, Mihir, Sharvari, Abhay, Meghaa, Selva, Chitira, Akhila, Sakhi, Sumit, Sharad, Akshay, Naresh, Tuhin, Asmita, Saorav, Megha, Saurish, Vidyunmati, Sarga, Mrunmay, Dattatray, Rishikesh for their friendship, and unyielding support. I am lucky to have met them here. I am deeply grateful to Mrs. Varsha Kulkarni, for her constant support and motivation. \vspace{0.3cm}

\noindent
I want to show my greatest appreciation to my whole family, especially my parents, brother and husband, for always believing in me and for their continuous love and support in all my decisions. \\

\noindent
And last but not least, I am thankful to those surrounding me on campus who treated me with love and respect, which I will always remember fondly.\\ 
\begin{flushright}
	{\bf Malati Dessai}
\end{flushright}

	\chapter*{Abstract}
	\addcontentsline{toc}{chapter}{\textbf{Abstract}}
Scanning Tunneling Microscopy (STM) have been known to provide a major tool of imaging surfaces so that various features of surfaces, and such as defects, growth of surface structures, etc. can be studied. Accurate calculation of the tunneling currents in a scanning tunneling microscope (STM) is needed for developing image processing algorithms that convert raw data of the STM into surface topographic images.

\bigskip

In this thesis, tunneling current densities are calculated for several pairs of planar conducting electrodes as a function of externally applied bias voltages, and separation between the two planes using only two properties viz. their Fermi energy and their work functions. Pauli blocking effects on both forward and reverse current densities are introduced. Airy function solutions are used for the trapezoidal barrier potential, and the results are compared with corresponding WKB results also found in this thesis. The planar model current densities are converted to currents for curved tips and flat samples by using an integration over a field line method. Thus, currents are determined for tip radii ranging from $R = $ 20 to 50 angstroms.

\bigskip

A new potential called the ‘Russell Potential’ is defined for each field line. It is slightly nonlinear as a function of the distance along the field line. To take care of this non-linearity as well as that introduced by the inclusion of the image force effects, a multi-slice method that uses the transfer matrix approach is developed, and is used to calculate tunneling currents for the Russell alone and for the trapezoid + image force potentials. The results for the trapezoid alone and Russell alone potentials for the shortest field line are almost identical. The Simmons image potential is due to a point sized charged electron in the barrier region, and with its introduction, tunneling currents are enhanced enormously which is considered unreasonable. Two models with distributed charge are constructed, and the image potentials in them are found to be negligible compared to the Trapezoidal potentials. Thus, image effects due to distributed charge models, which are considered reasonable, are shown to be negligible.

\bigskip

Tunneling currents are found to increase with bias voltage, and decrease exponentially with increasing tip-sample distance and increase with increasing radius of curvature of the tip.  The resolving power of the STM as defined in this thesis, is found to be degraded for blunter tips and for increasing bias voltage and for increasing tip-sample distances.

	{
		\hypersetup{linkcolor=black}
		\tableofcontents
		\newpage
		\addcontentsline{toc}{chapter}{\textbf{List of figures}}
		\listoffigures
		\newpage 
		\addcontentsline{toc}{chapter}{\textbf{List of tables}}
		\listoftables
	}

	\chapter{Introduction}\label{chap1}   
	\pagenumbering{arabic} 
	\setcounter{page}{1}
%
%
\section{Scanning Probe Microscopes}

All scanning probe microscopes involve a tip brought close to an almost plane surface. There is localized interaction between a sharp tip and almost plane sample which is placed at a distance of about a few nanometers to 100 nanometers. The nature of the interaction and the associated observable, define the type of microscope being used and the control variable used to generate different values of the associated observable for the same setting of the microscope. 

\bigskip

The key feature of the interaction is that it is localized which is why extremely sharp tips are used and further a very close proximity between tip and sample is maintained. The word scanning in all these microscopes is intended to describe the process in which the concerned observable is measured at various locations on the surface of the sample  and a surface distribution of the observable characteristic of the sample is obtained, keeping the control variable a constant. Realistic information concerning the surface profile and surface phenomena can be obtained from the profile of the observable by raster scanning the sample surface by the tip. The tip motion is closely controlled by piezoelectric crystals attached to the assembly so that currents to the crystals cause their motion and hence of the tip, and this can be done in a calibrated and controlled manner. 

\bigskip

The measured observable is an extremely sensitive function of tip sample distance. It is this feature of the interaction that is primarily responsible for the probe microscopes to be able to produce images with atomic size resolution. Since tip sample distances must be maintained constant to a great degree of precision, vibrational isolation is an extremely important requirement in all probe  microscopes. The source of these vibrations are usually be external such as vibrations of the building in which the microscope is placed. There are various ways in which one can achieve vibrational isolation. Binnig and Rohrer \cite{binnig5} used permanent magnets on a superconducting lead bowl to achieve magnetic levitation and hence vibrational isolation. However, use of spring coils for suspension of STM is more common \cite{Chen7}. It is also highly possible that the water molecules get adsorbed on the sample surface during the imaging process in ambient air and deviate from its original surface topography. Thus, it is best to use STM and AFM in high vacuum dry environments. Comparatively some samples can be studied in the ambient by STM.

\noindent The table below, lists the characteristics of various probe microscopes.

\begin{table}[htp]	
\begin{tabular}{|p{0.5cm}| p{3.4cm}|p{3.4cm}|p{2.4cm}|p{1.8cm}|p{2.0cm}|}\hline
	Sr. No. & Interaction & Observable & Microscope name & Abbrev. name & Control variable \\ \hline
	1.&	Quantum  tunneling of electrons& Tunneling current/tip positioning piezo current&	Scanning tunneling microscope&	STM &Tip-sample bias voltage \\ \hline
	2.& Attraction or repulsion of tip mounted on a cantilever&Cantilever deflection or shift in frequency of cantilever vibration&	Atomic Force Microscope&	AFM & Tip-Sample distance\\ \hline
	3.&Magnetic  Force&	Cantilever deflection caused by  Magnetic force \cite{Wiesendanger1}& Magnetic force microscopes&	MFM& Tip-sample distance\\ \hline
	4. &Tunneling of spin polarized electrons&Magnetized tip sensitive to magnetic moment  distribution of the sample&Spin polarized STM   &SP-STM & Tip-sample distance\\ \hline
\end{tabular}
\caption{\label{Table 1:}Various Probe Microscopes.}
\end{table}
In the above table several types of probe microscopes are listed. There is a lot of literature on the AFM (Atomic Force Microscopes) which is more versatile than the STM (Scanning Tunneling Microscope) because it admits non conducting samples. Some of the earliest papers in the subject are due to Binnig, Quate and Gerber \cite{binnig1982tunneling}. There are many papers on its application such as \cite{o1998atomic,horng2012vibration,ternes2008force,albrecht1991frequency}. The calculation of force between the atoms of the tip and the sample is well described by Israelachvili and Tabor \cite{israelachvili1973van,israelachvili2013intersection,zhang2021intermolecular,hou2022intermolecular,hubbard2011america}. The interactions in STM and AFM are described by Atkin et.al \cite{atkin2009afm}, Enevoldsen et.al.\cite{enevoldsen2008detailed}, Patil et al. \cite{patil}, Date et.al \cite{date}. Magnetic force microscopes are used to study surfaces with magnetic properties. The principles and applications of this microscope are well described in the following papers \cite{hartmann1999magnetic,kazakova2019frontiers,koblischka2003recent,wadas1989theoretical}. A scanning thermal microscope can also be constructed in which heat exchange between the tip and the sample is measured as described L.E. Ocola \cite{ocola}. The spin of the tunneling electron can be an important observable. Therefore one can have spin polarised STM's as described by Wulfhekel, Wulf and Kirschner, J\"{u}rgen \cite{doi:10.1146/annurev.matsci.37.052506.084342}.  Scanning Capacitance Microscopes are used for analyzing doping profiles in semiconductor-based structures. This microscope also has applications in the study of nanostructures such as quantum wells, nanowires and nanodots. A description of the theory and application of this microscope is described by Ruda and Shik \cite{shik2003theoretical,ruda2005scanning}. 

\bigskip

Most microscopes usually admit at least three modes of operation in both of which the control variable is kept constant. In the first mode of operation (constant observable mode) the measured observable is kept constant by raising or lowering of the tip during the scanning of the sample surface. Feedback circuits are used to track changes in the measured observable and appropriate currents are fed to z-spacing piezo crystal to regulate the tip sample distance so as to maintain a constant value for the measured observable. The z-spacing piezo currents are then the new measured observable, and this observable now contains the topographic information of the sample surface.
\vspace{4mm}\\
\indent In the second mode of operation (constant height mode) the tip – sample distance is kept constant while the observable is being measured. The value of the measured observable during scanning is  used to track the relevant surface property of the sample as a function of location on the sample surface. So the surface profile of the measured observable  contains information that involves both surface topographical  profile as well as the local interaction properties. One disadvantage of this mode is that, if the height is kept low, then during raster scanning of the sample there is always a danger if the tip crashing into a hill on the sample surface. This can potentially damage the tip or the sample or both.  
\vspace{4mm}\\
\indent In the third mode of operation (Spectroscopy) the control variable is varied and the corresponding value of the observable is measured. The slope of the curve of control variable versus measured observable is related to the Spectroscopic Factor (SF). Let the tip height and position with respect to the sample surface, be kept constant. Changing the control variable (such as the bias voltage in STM) would change the energy of the electrons and the observable value would also change. SF measures how the measured observable depends upon the energy which energy is varied by changing the control variable. Anything measured as a function of energy is said to provide spectroscopic information. Usually the local density of states of electrons in the sample can be inferred from such spectroscopic information. The measurement of SF is done at a single location on the sample surface rather than on several locations through raster scanning.  

\section{Introduction to Scanning Tunneling Microscopy(STM)}
\indent In 1986 Binnig and Rohrer\cite{binnig1,binnig3,binnig1987scanning,binnig1982tunneling,binnig5} got a Nobel Prize for their work on STM and AFM (Atomic Force Microscope) which showed that surface structures can be imaged using these microscopes in real time with almost atomic resolution. Since then both microscopes have become an indispensable tool for surface structure studies and for fabrication of surface nanostructures. The STM makes use of quantum mechanical tunneling  while the AFM uses force of attraction/repulsion due to atom-atom or intermolecular interactions to image the surface features of the sample. Since the tunnel current has to pass through both the tip and the sample, it is necessary that the tip and the sample in the STM must be conducting. Several authors describe the construction operation and the principle behind the STM \cite{wickramasinghe1990scanning,wickramasinghe2000progress,wickramasinghe1989scanned,abraham1988noise,wickramasinghe1989scanned,wiesendanger1992scanning,baumeister2013scanning,friedbacher1999classification,hansma1987scanning,chen2021introduction}. Another type of microscope ST-AFM (conductive AFM) can be constructed which uses the features of both the STM and the AFM in which a bias is applied between the tip and the sample and not only the tunneling current but also the tip-sample force is measured simultaneously. This has uses in the study os the surfaces of boron-doped diamond thin films \cite{holt2004scanning}. For such microscopes, the tip and the sample, both need to be conducting. In a pure AFM configuration this requirement can be relaxed. Therefore AFM finds greater use in studying non conducting samples such as biological samples \cite{weis}.

\subsection{ Description of STM Instrumentation}
\indent The components of an STM include scanning tip, piezoelectric controlled height and x, y scanner, coarse sample-to-tip control, vibration isolation system, and a computer. A probe tip usually made of W or Pt-Ir alloy, is attached to a piezodrive, which has x, y, z piezoelectric transducers. Applying a voltage to the piezoelectric transducer causes it to expands or contract. Now, using the coarse positioner and the z piezo the tip and the sample are brought to within a few nanometers distance from the sample surface which initiates the tunneling process. The tunneling current generated is converted to a voltage by the current amplifier, which is then compared with a reference value. The difference is amplified to drive the z piezo.
\vspace{4mm}\\
\begin{figure}[ht] 
	\begin{center}
		\includegraphics[width=3.5in,height=2.4in]{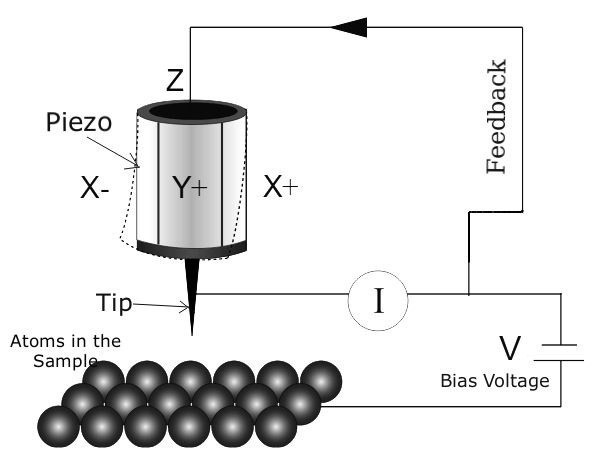}		
		\caption{A schematic picture of STM. Bias voltage $V$ is applied between the tip and the sample.}			
		\label{fig:piezo} 
	\end{center}  	
\end{figure}
During the topographical imaging process, the tunneling current is kept constant using the feedback adjustment. The output voltages from the feedback loop to the z-electrode of the piezo tube are used to deduce the vertical position of the tip as a function of its lateral position,(x,y). Effectively the vertical position is determined as a function of the lateral position for constant tunnneling current. This is called constant-current topographic imaging.
The most widely used mode of STM is the constant current imaging (CCI) mode. This mode of opertion is used for acquisition of profiles of 'work-function'(barrier height) and 'local density of states'(LDOS).
 The feedback in the CCI mode of operation of the STM, is very strong due to the strong dependence of the tunneling current on tip sample distance, the tip follows a contour generated by the sample topography. A topographic image of the surface is thus generated by recording the vertical position of the tip during the raster scanning.

\bigskip

In Constant Height Imaging (CHI) mode  of operation the scanner of STM moves the tip only in plane, so that current between the tip and the sample surface changes with the sample topography. The adjusting of the surface height is not needed in this mode, resulting in  higher scan speeds. CHI can only be applied if the sample surface is very nearly flat, because surface irregularities of higher than few  \r{A} may cause the tip to crash into the surface protrusions causing damage to either the tip or the sample or both.
 
\begin{figure}[hpt] 
	\begin{center}
		\includegraphics[width=3.5in,height=2.0in]{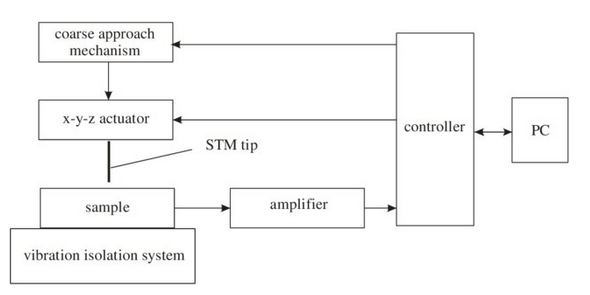}
		\caption{Block diagram of an STM system.}				
		\label{fig:blockdia} 
	\end{center}  	
\end{figure}

The topography of the surface is displayed as a grey scale image. The bright areas representing protrusions and the dark areas representing depressions \cite{Wiesendanger}. 

\subsection{ Theory of STM }
\indent The STM operation is based on the concept of quantum tunneling. In an STM the tip and the sample (both being conducting electrodes) are brought close to each other, and the narrow region between them ($5$ to $20$ \r{A}) is filled with either empty space or a dielectric. The region between the two electrodes behaves like a potential barrier. Electrons on either side of this region can tunnel across the barrier. The height and width of the barrier determines the probability of tunneling. In quantum mechanics, a particle will have a finite probability of tunneling through the potential barrier, even if its  value exceeds the particle's kinetic energy. This phenomenon, called tunneling has no analog in Classical Physics. The probability of tunnelling increases if the barrier height and width  are small. The electrons in both electrodes (tip and sample), occupy all energy levels right upto the Fermi level and beyond (depending upon the electrode temperature) in a manner that is consistent with the Fermi Dirac distribution. The Fermi levels of identical electrodes are displaced with respect to each other due to applied bias. The Fermi levels of dissimilar electrodes are displaced with respect to each other even in the absence of bias, and this is due to the contact potential. Applying a bias further changes the displacement of the Fermi levels of the dissimilar electrodes with respect to each other. The electrons in either electrode will have to overcome the potential barrier between the two electrodes to tunnel between them. These electrons preferentially tunnel in the direction determined by the direction of the externally applied bias voltage.  

\bigskip

Fowler-Nordheim (FN) \cite{FN} developed a comprehensive description of such tunneling in the vicinity of an electrode and used it to obtain expressions for field emission currents. Field emission is caused by electrons tunnelling through the surface barrier of a conducting metal surface. The barrier height and width are reduced by applying a strong electric field to the metal surface. The FN tunneling current formula uses WKB approximation to calculate the tunneling probability. 

\bigskip

Simmons, \cite{simmonsI}, Hartman and Chivian \cite{hartman} have used the FN formalism to develop an expression for the tunneling current densities in the planar model (electrode surfaces are treated as being parallel to each other and of infinite extent) for low and high bias voltages. This calculation uses WKB (Wentzel Krammer Brillouin) approximation to calculate the tunneling probability and image force corrections are also included. However Pauli blocking effects are ignored. These effects arise because an electron tunneling from one electrode can reach into the other electrode only if the corresponding energy level in the receptor electrode is vacant. If such a vacancy does not exist, then tunneling current corresponding to that energy is strongly inhibited. Such a phenomenon would occur also for reverse currents and this effect is due to the Pauli exclusion principle. 

\bigskip

In the simplest of models the barrier potential is linear and for this potential the Schrödinger equation for the electron wavefunction in the barrier region can be found analytically. The barrier in this case is said to be trapezoidal and the analytical wavefunctions can be written in terms of Airy functions\cite{dessai2022calculation}. A calculation using Bessel functions of order one-third which are related to Airy functions \cite{AS} has been reported by Shu et al. \cite{shu2002exactly}. Calculations based on Airy/Bessel function solutions are expected to be more accurate than those based on WKB approximation.

\bigskip

Chapter 3 describes the use of Airy function solutions \cite{AS} \cite{arfken} \cite{morse} for the Schrödinger Equation for the trapezoidal barrier in the tunneling region for calculating tunneling probability. This calculation includes temperature dependent Fermi Factors for each electrode and it involves integration over the electron energy. Thus the energy of the tunneling electron is not limited just to the Fermi energy, but ranges from zero to a maximum value given by the sum of the Fermi energy $\eta_1$ and the work function $\phi_1$ of the emitter electrode. However, the Fermi factor strongly damps out the contribution of energies greater than $\eta_1+20 \,\,kT$.   

\begin{figure}[hpt] 
	\begin{center}
		\includegraphics[width=3.0in,height=2.2in]{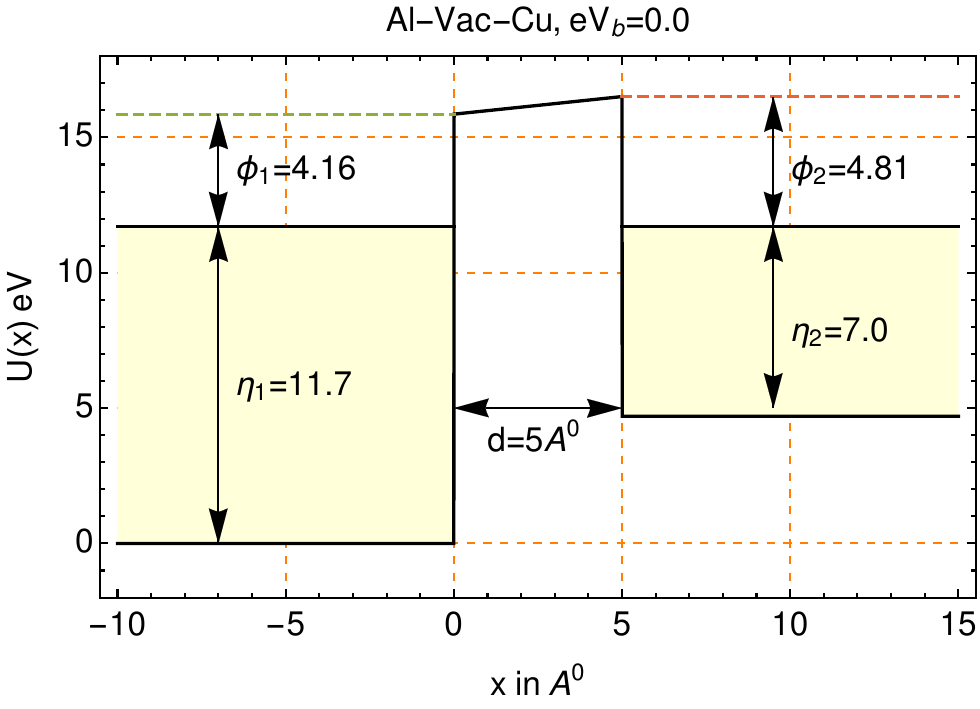}
		\caption{Energy diagram for dissimilar tunnel junction with zero bias. $\eta_1$ and $\eta_2$ are the fermi levels and $\phi_1$ and $\phi_2$ are the work functions of the two electrodes (the tip and the sample)}
		\label{fig:energydia} 
	\end{center}   	
\end{figure}

\bigskip

Pauli Blocking impacts the tunnelling transition rate and thereby has an effect on the tunelling current densities. Electron transition between two metals is possible only if the recipient metal has electron states vacant for the transiting electron to occupy. This is because electrons in the recipient metal obey Pauli's exclusion principle. If the receptor states are not vacant, the transition is forbidden and this effect is called Pauli Blocking. Many authors\cite{simmonsI,Shu,Ebeling} have implemented the general idea of Pauli blocking in a variety of contexts. Some calculations incorporate Simmon's treatment of Pauli blocking in tunneling microscopes. A modified approach to Pauli blocking is described in chapter 2, in which the Simmon's calculation is revisited and his approximations involving the integral in the exponent of the WKB amplitude is replaced by numerical integration. In this chapter tunneling current densities are calculated for several electrode pairs and for several bias voltages and tip sample distances. 

\bigskip

In calculations using Airy function, very low bias voltages are problematic due to the Airy functions Ai becomming very small, and Bi becomming  very large. These have the potential to cause underflow, overflow problems in the computer respectively. Therefore in this regime, regular methods are not computationally suitable. For these problems, Green functions for the $\nabla^2$ operator in the barrier region are used, with the linear term as the driving term. The solution is then found by iteration in powers of the driving term to construct the wavefunction. Chapter 3 describes this procedure and implements it along with other features such as Pauli Blocking to compute the tunneling current densities for the voltage range ($0.1-1$ V) used by Ting et.al. The tunneling currents found in chapter 4 for these voltages agrees well with Ting et.al.\cite{ting} results.

\bigskip

The Airy function calculation of the tunneling current densities has been carried out for tip and sample surfaces that are initially treated as plane surfaces with vacuum as the barrier. In this planar model, only tunneling current densities are calculated since tunneling currents would be infinite due to the assumed infinite extent of the electrode surfaces. However experimentally tunneling currents are measured, and these cannot be determined directly from calculated tunneling current densities in the Planar models. Comparison of theory with experiment is problematic in Planar Models of tunneling. Thus it is necessary to model actually used Tip and Sample shapes and convert calculated tunneling current densities into measuarable currents. Since tips are sharp non planar shapes must be introduced at least for the tip shape. In chapter 5, the tip and the sample are modeled as confocal surfaces in a prolate spheroidal coordinate system. This has the advantage that the Laplace equation becomes one dimensional and the entire calculation including the finding of the tunneling probabilities becomes one dimensional. By assuming that tunneling occurs principally along field lines, current densities are converted into currents, through a method described by Saenz and Garcia. Chapter 5 details the procedure for the calculation of currents, and report currents for several electrode pairs for different bias voltages and tip sample distances. 
	
\bigskip

A very frequently quoted model \cite{tersoff,garcia} is that of Bardeen \cite{bardeen}. In this model the electron wavefunctions in the tip and the sample are separately evaluated and the overlap between them is used to calculate the transfer matrix which determines the tunneling probability. This model was later simplified and exteded by Tersoff and Hamann \cite{tersoff} and Chen \cite{chen} to the planar model of the STM. These calculations show that the current density is proportional to the local electronic density of states (LDOS) of the surface. Thus the STM is believed to measure this LDOS. Since the tunneling probability depends upon the wavefunction overlap, and this overlap becomes very small at larger and larger tip-sample distances, it is necessary to maintain the tip sample distance to a small value ($~5$ \r{A}) to obtain sizable tunneling current. Thus Bardeen and other calculations based on Bardeen's model calculations may not work very well for larger tip sample seperations.  

\begin{figure}[ht] 
	\centering
		\includegraphics[width=3.0in,height=2.2in]{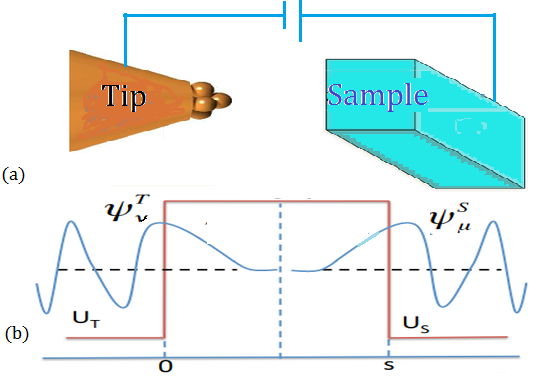}		
		\caption{
			(a) Schematic diagram of STM with the tip biased negative relative to the sample. (b) Figure shows the decaying of the electron wavefunction in the barrier region for similar metal electrodes.}		
		\label{fig:schedia}	  	
\end{figure}

\bigskip

The Schrödinger equation for the potential in the barrier region is coupled to the Poisson equation that determines the electrostatic potential in the barrier region. This it does by taking into account the charge density in the region, which in turn depends upon the wavefunction. Thus the effective Schrödinger equation becomes non linear and one way to solve it involves iteration to required order. Iteration of this equation imposes self consistency between the wavefunction and the charge density in the barrier region. This self consistency has been implemented by S. Banerjee and Peng Zhang \cite{SBPZ}. They proposed a model which characterizes the tunneling current in nano- and subnano-scale asymmetric MIM junctions, taking into account the effects of both space charge and exchange-correlation potential. They calculated the influence of electrode work functions on the forward and reverse tunneling. However their calculations still involve WKB approximation and is mostly modeled along the lines of the Simmon's \cite{simmonsI,simmonsII} calculation. Self consistent calculation with Airy function solutions are presented in chapter 5.

\bigskip

Ting et. al's \cite{ting} paper describes Spin Hall Effect in tunnel junctions in which a magnetic field perpendicular to the tunneling current is applied and the experimentally observed tunneling currents are studied as a function of the strength of the magnetic field and the bias voltage. They have also presented a benchmark data for zero applied magnetic field and this corresponds to the situation studied in this thesis. The bias voltages however are very low (10 to 70 mVolts).  In Chapter 3, Green functions are used to calculate tunneling currents for such low voltages and the calculated results (I-V curves) agree quite well with the experimental results of Ting et. al.

	\chapter{WKB Calculations of the STM Current Density}\label{chap2}
	%
%
%
%
\section{The WKB calculation}

In a Scanning Tunneling Microscope (STM) a sharp conducting tip is scanned across the sample’s surface. The region between the tip and the sample acts as a potential barrier. The tip-sample separation is kept very small ($\approx 1$ nm) so that appreciable quantum mechanical tunneling can occur. This separation is controlled very precisely with the help of piezo crystals attached to the tip. The tunneling current across the barrier in an STM is found to vary exponentially with tip-sample distance. This allows atomic sized resolution by an STM, which can therefore probe the atomic arrangements on conducting surfaces. 

\bigskip

The energy levels of the electrons in metallic conductors are grouped into conduction bands which are separated by forbidden bands. Within the conduction band, electrons of a metal are able to move almost freely inside the metal. Electrons are fermions; therefore they must obey Pauli Exclusion Principle, and because of this they cannot all congregate into a single energy state, but must distribute themselves into several energy states. The largest energy of the electrons at absolute zero temperature is nonzero, and has a finite value. This value is called the Fermi Energy of the metal. At finite non zero absolute temperatures, electron energies can exceed the Fermi energy and the distribution of electron energies is given by the Fermi-Dirac distribution function $f(E)$ which is given by 
\begin{equation}\label{FermiD}
f(E) = \big [\exp(\frac{(E-\mu)}{kT}) +1 \big]^{-1}
\end{equation}
where $\mu$ is the chemical potential. It satisfies $\mu = E_F$, at low enough temperatures, where $E_F$ is the Fermi energy, which is also the value of the chemical potential at absolute zero of temperature. The Boltzmann constant is  $k$, whose value is $8.617 \times 10^{-5}$ eV/K, and $T$ is the absolute temperature of the metal. If $T \rightarrow 0^+$ the Fermi-Dirac distribution function $f(E)$ becomes 
\begin{equation}\label{heavy}
  \lim_{T\rightarrow 0}f(E)=\Theta (E_F-E)\, = \,\left\{\begin{matrix}
1 \,\,\,\,\text{for}\,\, E\,<\,E_F\\ 0\,\,\,\,\text{for}\,\, E\,>\,E_F
\end{matrix}\right.
\end{equation}
where $\Theta(x)$ is the Heaviside function (also called the unit step function) whose value is 1 if $x >0$ and whose value is $0$ if $x \leq 0$. Different symbols such as  $\epsilon$, $\epsilon_F$, $E_F$ are used in literature to label the Fermi Energy. At large temperatures the chemical potential becomes a function of temperature and is given by a series expansion \cite{KH}.
$$ \mu(T) = \epsilon_F \Big[1 - \dfrac{\pi^2}{12}(\dfrac{kT}{\epsilon_F})^2 + \cdots \Big]$$ Thus the chemical potential differs appreciably from the Fermi energy only if the temperatures are comparable to $T_F = \epsilon_F/k$. This is called the Fermi temperature. Thus even at $T = 7000$K, $\mu$ differs from $\epsilon_F = 7$ eV by less than $1 \%$. Therefore for temperatures very less than the Fermi temperature, equation (\ref{FermiD}) can be used with $\mu$ replaced by the Fermi Energy $\epsilon_F$ with negligible error.
 In this thesis, the Fermi energy will be labelled by the symbol $\eta$.

\bigskip

 In the conduction band, the electrons can move freely in an unconstrained manner. They are however confined within the metal and cannot escape out of the metal across its free surface. This is due to cohesive forces acting on the electrons at the metal surface, and also image forces acting on an electron just outside the metal surface. Both these causes require that an electron has to be provided some extra energy in order to escape out of the metal across its surface. This extra energy is called the work function and is denoted by $\phi$ in this thesis. Therefore electrons whose energies exceed $E_\text{max} = (\eta +\phi)$ are unlikely to be confined within the metal and these electrons shall be emitted spontaneously, provided the temperature of the metal is greater than absolute zero. At absolute zero, spontaneous emission of electrons cannot occur unless an external electric field is applied.

\bigskip

When two metal (conducting) electrodes are brought close together, the intervening space between them may be either filled with an insulating dielectric or vacuum. The region between the two conducting surfaces behaves like a potential barrier for electrons confined to either metal electrode. Electrons from one metal can tunnel across to the other metal through this potential barrier. This tunneling can be either spontaneous or assisted by the application of a bias voltage across the two metallic conductors. Let $\eta_1$ and $\phi_1$ be the Fermi energy and the work function respectively of the metallic conductor which is placed to the left of the other conductor (also called the first conductor). Likewise let $\eta_2$ and $\phi_2$ be the Fermi energy and the work function respectively of the metallic conductor which is placed to the right of the first conductor, and this shall also be called the second conductor. When identical conductors, are brought close together keeping their plane surfaces parallel to each other, and separated by a thin (few angstroms width) gap of vacuum, their zero energy levels, their Fermi levels as well as their levels corresponding to $E_\text{max}$ will match. This is because in this case $\eta_1 = \eta_2$ and $\phi_1 = \phi_2$. The difference in the maximum energies $E_\text{max}$ of the two electrodes at zero bias, is called the contact potential. For identical electrodes the contact potential is zero. 
\begin{figure}[hpt]
	\centering
		\includegraphics[width=6.5in,height=6.0in]{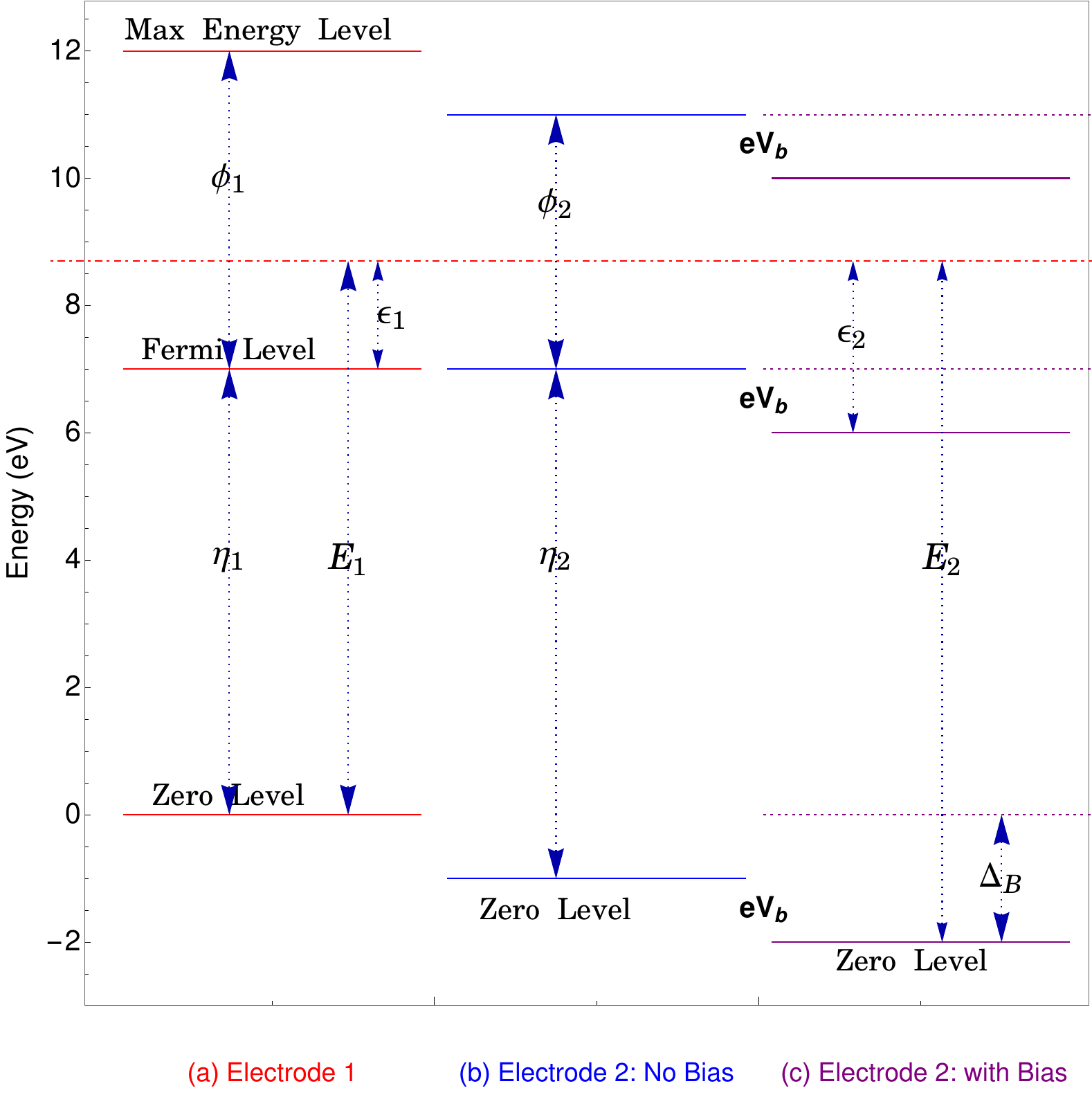}
		\caption{Energy level diagram of (a) electrode 1 (b) electrode 2 without bias and (c) electrode 2 with bias. $\Delta_B$ is the shift in the zero energy levels in the presence of bias.   ($\Delta_B=\eta_2-\eta_1+eV_b$).}
		\label{ED}	
\end{figure}

\bigskip

\par
\begin{figure}[hpt]
	 \begin{subfigure}{0.49\textwidth}
		\includegraphics[width=2.9in,height=2.6in]{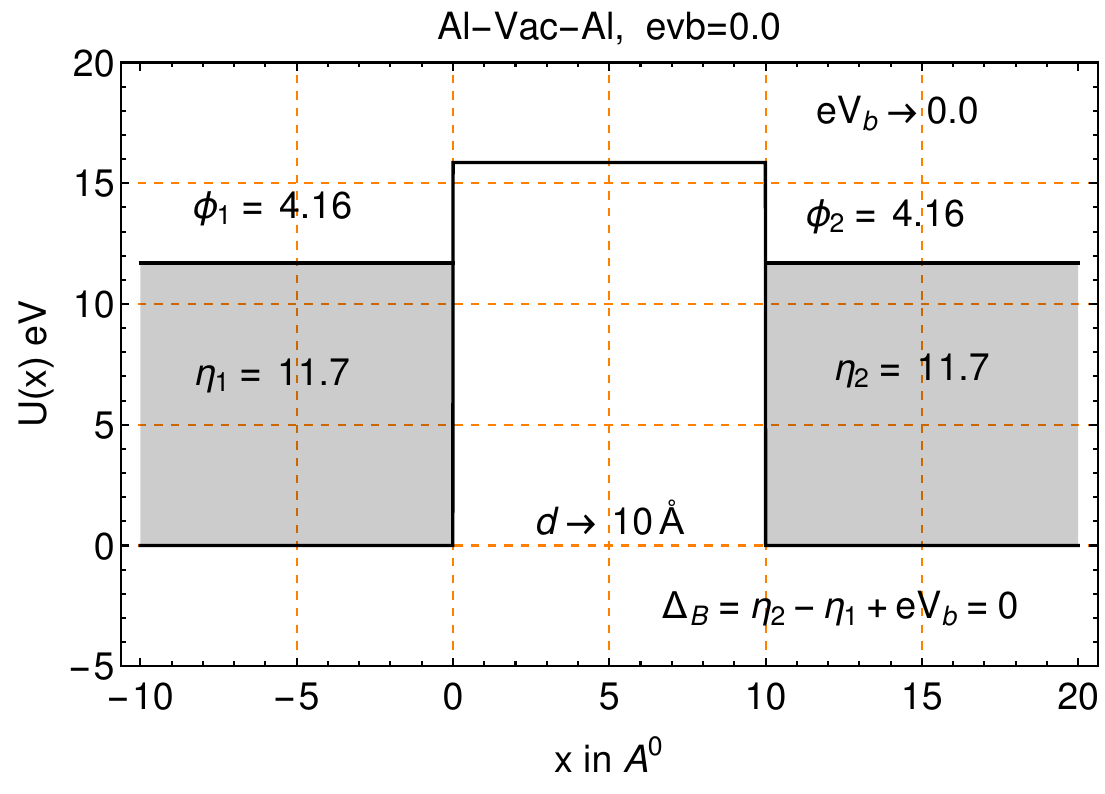}	\hspace{1cm}%
		\caption{Similar Electrodes at zero bias}	
		\label{fig:EDzeroa}
	 \end{subfigure}
	\hfill
  \begin{subfigure}{0.49\textwidth}
	\includegraphics[width=2.9in,height=2.6in]{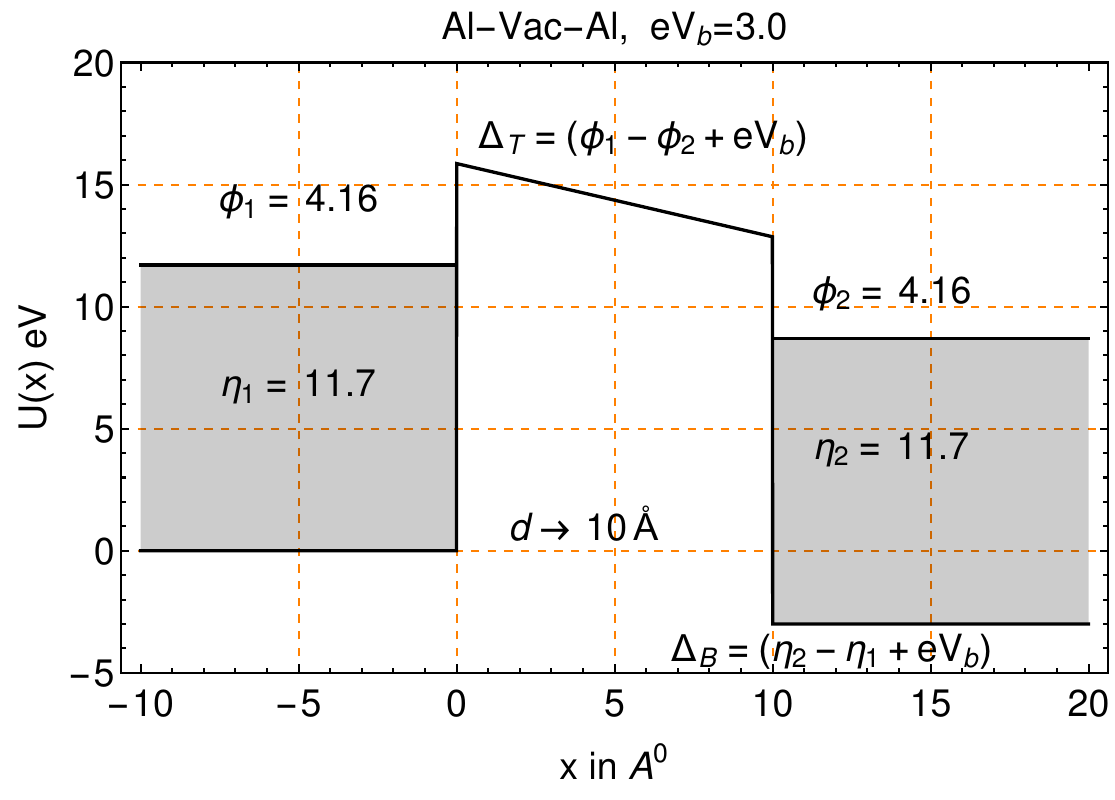}	
	 \caption{Similar Electrodes at $\text{Bias}\,=\,3\,V$}
	\label{fig:EDzerob}
\end{subfigure}
	\caption{Energy diagram for similar electrodes (Al-vac-Al) at (a) zero bias (b) external bias = 3 V applied to right electrode.}
	\label{AlAlV0V3}
\end{figure}

\begin{figure}[hpt]
	\begin{subfigure}{0.49\textwidth}
		\includegraphics[width=2.8in,height=2.5in]{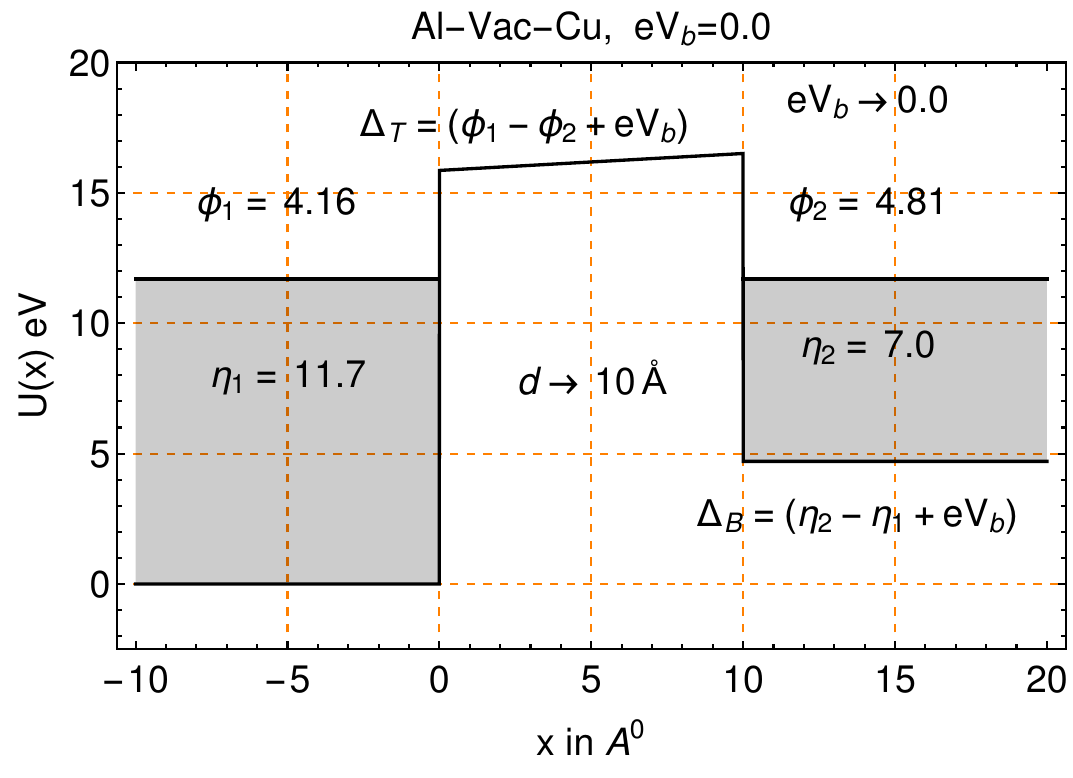}
		\caption{Dissimilar Electrodes at zero bias}	
		\label{fig:EDbias3a}
	\end{subfigure}
	\hfill
	\begin{subfigure}{0.49\textwidth}
		\includegraphics[width=2.8in,height=2.6in]{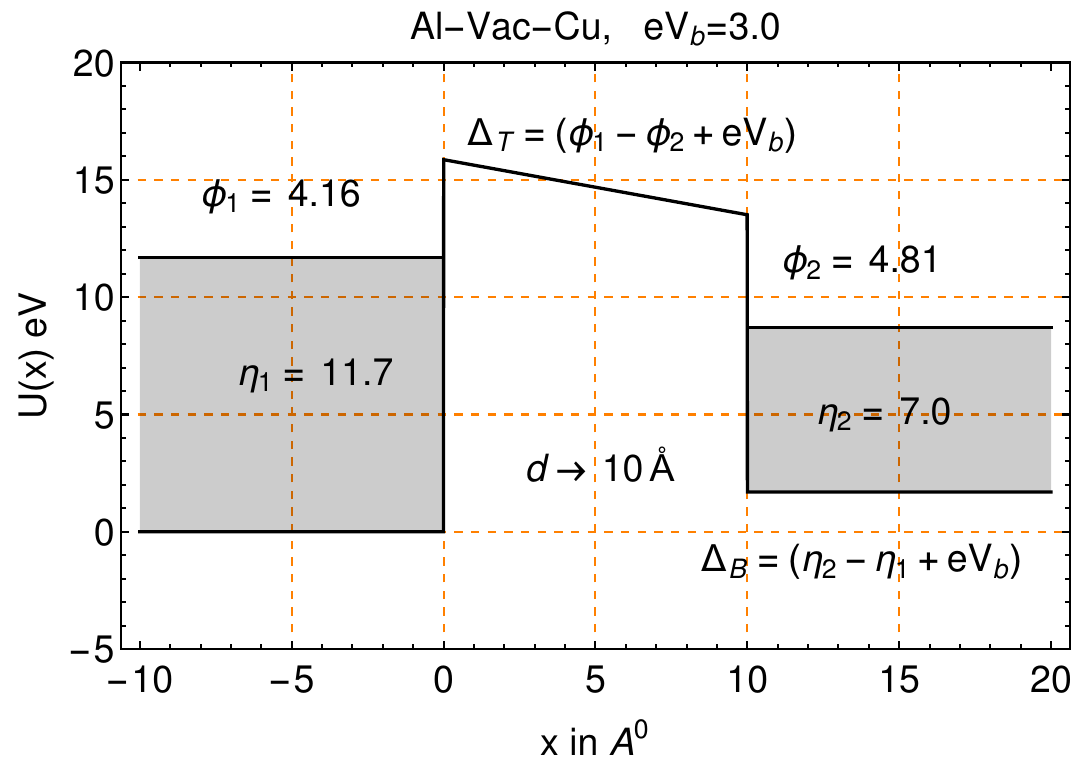}	
		\caption{Dissimilar Electrodes at $\text{Bias}\,=\,3\,V$}
		\label{fig:EDbias3b}
	\end{subfigure}
		\caption{Energy diagram for dissimilar electrodes (Al-vac-Cu) at (a) zero bias (b) external bias = 3 V applied to right electrode.}
	\label{fig:similarNdissim}
\end{figure}

 When non-identical conducting surfaces are brought close together, their $E_\text{max}$ values will initially match, but their Fermi levels will not match, since $\eta_1 \neq \eta_2$ and $\phi_1 \neq \phi_2$. In this situation, spontaneous tunnelling of electrons from one electrode to the other occurs, so that the net tunnel current forces the Fermi levels to match up, after which spontaneous tunneling stops. This leads to a mismatch in the $E_\text{max}$ values and also their zero energy values. In the absence of an external bias this difference $ \phi_1 - \phi_2$ is called the contact potential. When an external bias $V_b$ is applied then the Fermi levels of the two electrodes are again mis-matched, and this time by an amount equal to $e V_B$ where $e$ is the electron charge. All energies in this thesis are expressed in electron-volts (eV), and all distances in \r{A} $= 10^{-10}$ m. Fig.\ref{AlAlV0V3} and Fig.\ref{fig:similarNdissim} shows the relative positions (on the energy scale) of the zero energy levels, the Fermi levels and the $E_\text{max}$ levels for similar electrodes (Al-vac-Al) and for dissimilar (Al-vac-Cu) electrodes. Fig.\ref{fig:EDzerob} and Fig.\ref{fig:EDbias3b} shows how these levels are displaced upon applying an external bias.

\bigskip

\newpage
 Most calculations of the tunneling current start with the Planar model of the tunnel junction. In this model the two electrode surfaces are assumed to be of infinite extent and parallel to the $y-z$ plane and both surfaces being perpendicular to the $x - $ axis. The left (first) electrode surface is located at $x = 0$, and the right  (second) surface is located at $x = d$. The barrier region is then defined to exist within $0 < x < d$, where $d$ is called the tip - sample distance.
 Simmons \cite{simmonsI} evaluates the tunneling probability in the WKB approximation. The WKB approximation has been discussed by several authors. This approximation is most applicable to slowly varying potentials that do not change appreciably over a distance of the order of the De Broglie wavelength of the particle experiencing the potential. 
  The tunneling probability $D(E_x)$ for a particle with kinetic energy $E_x$ tunneling through a potential barrier $U(x)$ is given in the  WKB approximation as 
\begin{equation}\label{DWKB}
D(E_x)=\exp{\Bigg[-2\sqrt{\frac{2m}{\hbar^2}} \int_{x_1}^{x_2}dx\sqrt{U(x)-E_x} \Bigg ]}
\end{equation}

where the integral $I$ in the exponent has the limits $x_1$ and $x_2$ which are chosen appropriately. One could choose them to be the classical turning points of the potential $U(x)$. Alternatively, since tunneling occurs for electrons whose maximum energy at absolute zero is the Fermi energy $\eta_1$, Simmons \cite{simmonsI}  chooses the limits to be the roots of the function $(U(x) - \eta_1)$. Since the potential when image forces are included is extremely steep near the conducting surfaces, there isn't much difference between either of these two choices.
 
\bigskip

The calculation of tunneling current density from tunneling probability is well established and the earliest accounts of it are by Fowler and Nordheim \cite{FN} and Simmons \cite{simmonsI}. One of the principal assumptions made in these calculations is that the number of electrons that can occupy a unit volume in the 6 dimensional phase space spanned by the coordinate and momentum vectors ${\bf q}$ and ${\bf p}$, is $\dfrac{2}{h^3}$. The factor 2 arises from the two possible spin projections of a spin $\frac{1}{2}$ electron. Therefore the number of electrons in the infinitesimal volume $d^3q d^3 p$ in the 6 dimensional phase space is $$dN = \frac{2}{h^3}d^3 q d^3 p$$ The integral over $d^3 q$ gives the total volume $V$. Dividing both sides by $V$ gives $$\frac{dN}{V} =  \frac{2}{h^3} d^3 p$$ Since ${\bf p} = m{\bf v}$,  $d^3 p = m^3 d^3 v = m^3 dv_x dv_y dv_z$. Therefore the number of electrons per unit volume (of coordinate space) in the velocity space volume element $d^3 v = dv_x \,dv_y \,dv_z$ about the velocity vector ${\bf v}$ is given by
$$ \dfrac{dN}{V} = \frac{2m^3}{h^3} dv_x\, dv_y\,dv_z$$

\bigskip

The energy of these electrons is $E = E_x + E_y + E_z$, with $E_i = \frac{1}{2} m v_i^2$ for $i = (x,y,z)$. Then $E = E_x + E_r$ where $E_r = E_y + E_z = \frac{1}{2} m v_r^2$.  
The number density of electrons that are available for tunneling would be 
$$n_1 \, d^3 v = \dfrac{2 m^3}{h^3} f_1(E) \, d^3 v$$
where $f_1(E) $ is the Fermi Dirac Distribution function which can be regarded as the probability of occupation of the energy level $E$ in the electrode 1. It is given by 
$$f_1(E) = \big [\exp (\beta[E-\eta_1]) +1 \big ]^{-1}$$
where $\beta = 1/(kT)$. As per the Pauli Exclusion Principle, there ought to be a vacancy of the corresponding energy level in the receiving electrode if tunneling in the forward direction is to go through. However Simmons does not consider this feature in his calculations. This is tantamount to neglecting Pauli Blocking for the forward current. Thus neglecting Pauli Blocking, the number density of electrons that can tunnel through the barrier from electrode 1 to electrode 2, is 
\begin{equation}\label{numden}
	n_\text{For}(v) d^3 v =  D(E_x) n_1(v)\,d^3\, v   =   \dfrac{2 m^3}{h^3} D(E_x) f_1(E)  \,d^3 v.
\end{equation}
where $D(E_x)$ is the tunneling probability given by equation (\ref{DWKB}). 
The tunneling probability $D(E_x)$ is symmetric for tunneling in the forward (electrode 1 to electrode 2, $i.e.$ along the positive direction of the x-axis) as well for the reverse (electrode 2 to electrode 1, $i.e.$ along the negative direction of the x-axis) directions. The number density of electrons tunneling from electrode 2 to electrode 1 is calculated in similar way and is given by 
\begin{equation}
n_\text{Rev}(v) d^3 v = \dfrac{2 m^3}{h^3} D(E_x) f_2(E_2)  \,d^3 v 
\end{equation}
where  $$f_2(E_2) = \Big [1 + \exp(\beta[E_2-\eta_2])\Big ]^{-1}$$
All energies are measured with respect to the zero energy level of the first electrode. The argument of the Fermi-Dirac function $f_1(E)$  is simply the distance (on the energy scale) between the energy $E$ of the electron and the zero level of electrode 1 in which it is computed. If the first electrode is negatively biased with respect to the second electrode, then the Fermi level of the second electrode will be shifted negatively by an amount = $e V_b$. The zero level is further shifted due the difference $\Delta \eta =(\eta_2 - \eta_1)$ so that the net shift downward of the zero level of the second electrode with respect to that of the first electrode is $\Delta_B = eV_b + \Delta \eta$.  Of course  $\Delta \eta$ could be negative in which case the net shift can be either upward or downward. These shifts are shown in the Fig. \ref{fig:similarNdissim}. Here $\Delta_B$ stands for the shift of the zero energy level of the second electrode with respect to that of the first electrode. Note $\Delta_B$ is positive if the zero energy level of the second electrode lies below the zero of the energy level of the first electrode (see Fig. \ref{ED} and Fig. \ref{AlAlV0V3}). In Fig. \ref{fig:similarNdissim} the zero energy level of the second electrode is shifted up compared to the zero energy level of the first electrode. In this case $\Delta_B$ is negative. Thus the energy of the electron as measured with respect to the zero level of the electrode 2 will be $E_2 = E + \Delta_B$.  It can be seen that $E_2 - \eta_2 = E + eV_b -\eta_1$ and this is illustrated clearly in Fig. \ref{ED}. Therefore 
$$f_2(E_2) = f_1(E + eV_b) = f_1(E ^\prime)$$ 
where $E^\prime = E + eV_b$  

\bigskip

\par Thus the number density of electrons that will be found in the states of electrode 2, and which contribute to the reverse current are 
\begin{equation}\label{numdenR}
n_{Rev}(v) d^3 v = \dfrac{2 m^3}{h^3} D(E_x) f_1(E+eV_b) \,d^3 v 
\end{equation}
The forward current density is given by  $$J_\text{For} = e\int v_x\, n_\text{For}(v)\, d^3 v$$ 
Now, $E = E_x + E_r$. where $E_x = \frac{1}{2} m v_x^2 $ and $E_r = \frac{1}{2} m v_r ^2 $.
The $y, z $ velocity components $v_y$ and $v_z$ are expressed in terms of their polar components $(v_r, \theta)$ in the $y- z$ plane as 
$$ v_y = v_r \cos \theta \quad \text{and} \quad v_z = v_r  \sin \theta $$ Therefore
$$ v_x dv_x = (dE_x/m) \quad \text{and}  \quad dv_y dv_z = v_r dv_r d\theta = (dE_r/m) d \theta $$  
$$v_x  d^3v =(dE_x/m) \,(dE_r/m) \, d \theta $$ 
Substituting for $n_{For}$ and $v_x d^3v$ in equation for $J_{For}$ and integrating over $\theta$ gives
\begin{equation}
J_{\text{For}}=\frac{4\pi me}{h^3}\, \int_{0}^{E{max}} dE_x D(E_x)\int_{0}^\infty dE_r f_1(E_x+E_r) 
\end{equation}
Similarly the reverse current densities are then  
\begin{equation}
J_{\text{Rev}}=\frac{4\pi me}{h^3}\, \int_{0}^{E{max}} dE_x D(E_x)\int_{0}^\infty dE_r f_1(E_x+E_r + eV_b) 
\end{equation}

\bigskip

Simmons \cite{simmonsI} has carried out the tunelling current density calculation in which both electrodes are at $T = 0^o$K. As mentioned in equation (\ref{heavy}), at absolute zero temperature, the Fermi Distribution functions become Heaviside functions. Substituting and evaluating integral over $dE_r$ gives $$\int_{0}^{\infty}f_1(E)dE_r = \int_{0}^{\infty} \Theta( \eta_1 -E_x -E_r)dE_r = (\eta_1 - E_x) \Theta(\eta_1 -E_x) $$
For reverse currents $E_x$ must be replaced by $E_x+eV_b$ and the result of the integration over $dE_r$ for reverse currents case gives the factor $(\eta_1-E_x-eV_b)\Theta(\eta_1 -E_x-eV_b)$. 

The net current density $J_{Net} = J_\text{For} - J_\text{Rev \,}$ at $T=0 \,K$ is given by

\begin{equation}\label{jnetsimm}
J_{Net}=\frac{4\pi me}{h^3}\,\,\Bigg[ \int_{0}^{\eta_1}dE_x\, (\eta_1-E_x)\, \,D(E_x) - \,\int_{0}^{\eta_1-eV_b}dE_x(\eta_1-E_x-eV_b)\, D(E_x) \Bigg]
\end{equation}

\begin{equation}\label{jnetsim}
	J_{Net}=\frac{4\pi me}{h^3}\,\,\Bigg[ \int_{\eta_1-eV_b}^{\eta_1}dE_x\, (\eta_1-E_x)\, \,D(E_x) +eV_b\,\int_{0}^{\eta_1-eV_b}dE_x\, D(E_x) \Bigg]
\end{equation}

In the above expressions $D(E_x)$ refers to the tunneling probability. Simmons \cite{simmonsI} has developed approximate expressions for the tunneling probability for arbitrary potentials in which the WKB approximation plays a significan part. He defines a function $\Phi(x)$ as $U(x) = \eta_1 + \Phi(x)$. The tunneling probabilities obtained by the use of the WKB approximation will be called $D(E_x)_{WKB}$.  The integral in the exponent in the RHS of equation (\ref{DWKB}) is given by 
\begin{equation}\label{simspatI}
I = \int_{x_1}^{x_2} \sqrt {[U(x) - E_x]}\, dx
\end{equation}
where $x_1$ and $x_2$, are the zeroes of $\Phi(x)$. \\
\noindent Define 
\begin{equation}
D(E_x)_{WKB} = \exp[-A\times I]
\end{equation}
where $A=2\sqrt{\dfrac{2m}{\hbar^2}}$

\bigskip

\begin{figure}[h]
	\centering
		\includegraphics[width=5.0in,height=4.0in]{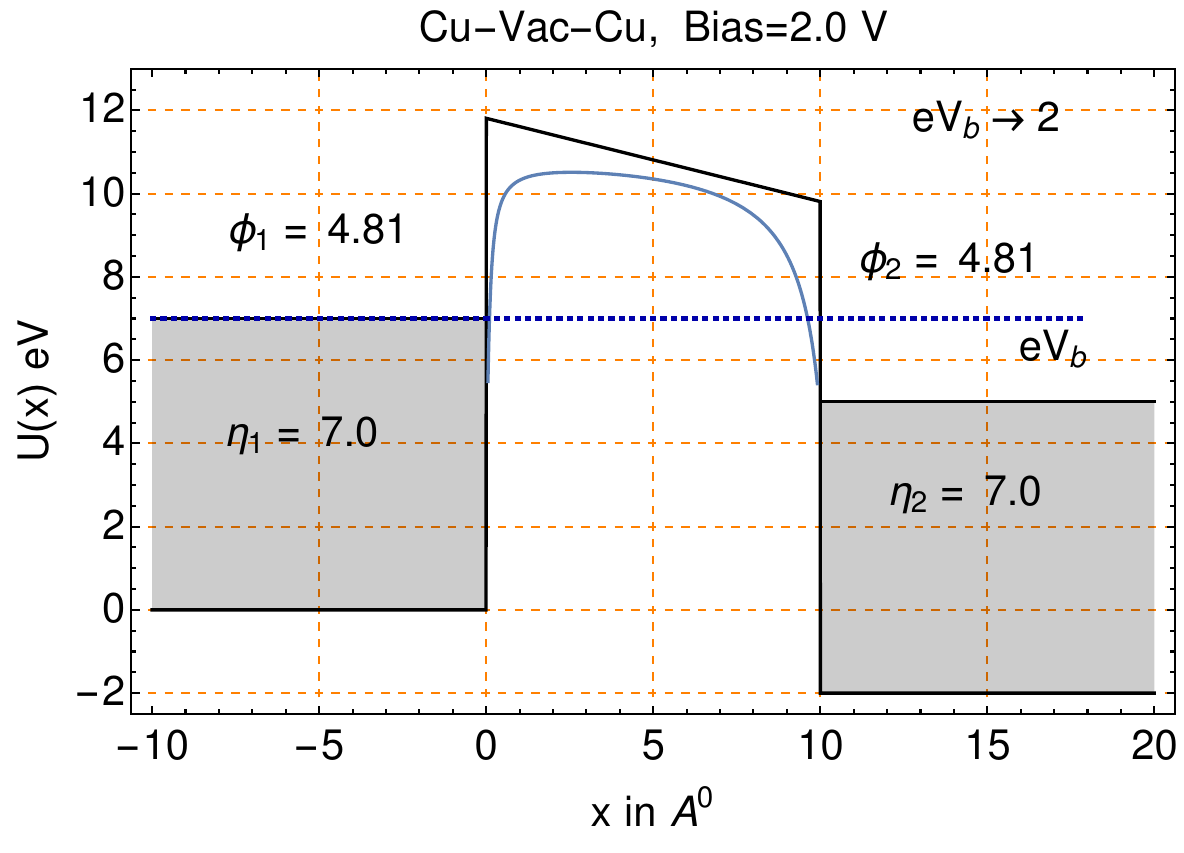}
	\caption{Energy lavel diagram for Cu-vac-Cu with Bias = $2$ V  }	
	\label{fig:simmdia}	
\end{figure}

Fig. \ref{fig:simmdia}	 shows the two electrodes, the trapezoidal potential between them and also the modified potential due to the addition of an image force term. The function $\Phi(x)$ is the blue curve that is underneath the trapezoid, but above the Fermi Level of the first electrode. Note that while $x_1$ is extremely close to $x = 0$, the other root $viz$, $x_2$ lies at $x <d$, thus indicating that image force terms have the effect of thinning the barier region and also shortening the effective height of the barrier.
This integral is approximated by Simmons \cite{simmonsI} as $$I \approx (x_2-x_1) \sqrt{[\eta_1 + \bar{\Phi} - E_x]}$$
where $\bar{\Phi}$ is the mean value of $\Phi(x)$ over the interval $x_1 \,< \, x \, < \, x_2$, 
and the justification for this approximation is described in the Appendix of the paper by Simmons \cite{simmonsI}. Thus, 
\begin{equation}\label{DExsim}
D(E_x) = \text{exp}[-A (x_2-x_1)(\eta_1+\bar{\Phi}-E_x)^{1/2}]
\end{equation}

\bigskip

At absolute zero temperature, ideally electrons with energies $E_x$ given by $\eta_1-eV_b \,\leqslant\, E_x\, \leqslant\, \eta_1$ contribute to tunneling in the forward direction, and reverse currents would be forbidden. The upper limit on energy integral would be $E_{\text{max}} = \eta_1$, which is the Fermi level of the first electrode because no electrons are available in the first electrode with energies greater than $\eta_1$ and the lower limit is $\eta_1 -eV_b$ because all levels below this minimum energy are occupied in the second electrode and Pauli Exclusion principle forbids tunneling for those energies. Further all energy levels in the first electrode for energies below the minimum value of $\eta_1 - eV_b$ (which is also the maximum energy of the electrons in the second electrode) are occupied. Therefore reverse tunneling currents ought to be forbidden according to the Pauli Exclusion Principle. Since Simmons \cite{simmonsI} does not explicitly include Pauli blocking effects and he obtains non zero tunneling current densities for energies that range from $0$ to $\eta_1$, This energy range is divided into two intervals, and the energy integral is split into two terms.
\begin{equation}
\int_{0}^{Emax} D(E_x)\, dE_x  = \int_{0}^{\eta_1\,-\,eV_b} D(E_x)\, dE_x + \int_{\eta_1\,-\,eV_b}^{\eta_1} \,D(E_x) \, dE_x
\end{equation}
Substituting the expression for $D(E_x)$ from equation (\ref{DWKB}) and evaluating the above integrals over $E_x$, and making some approximations for bias voltage small compared to $\eta_1$, in equation (\ref{jnetsimm}) at $T=0\,K$, gives

\begin{equation}
J_{Net}=\frac{e}{2\pi h\,d^2}\,\, \Big[-(\bar{\phi}+\frac{3}{2}eV_b)e^{-A\sqrt{(\bar{\phi}+eV_b)}}+\bar{\phi}e^{-A\sqrt{\bar{\phi}}}\Big]
\end{equation}
The corresponding equation in Simmon's paper \cite{simmonsI} (See [eqn 20] on page 1796) contains an error, it lacks the factor 3/2 contained in the above equation.

\bigskip

The barrier potential $U(x)$ shown by the solid black line in Fig.\ref{fig:simmdia}  is given by
\begin{equation}\label{Trappot}
U(x) = \eta_1+\phi_1-\frac{(\phi_1-\phi_2+eV_b)\,x}{d}
\end{equation}
This is usually referred to as the $linear$ or the $trapezoidal$ potential. For the trapezoidal potential given in equation (\ref{Trappot}), one can evaluate the spatial integral $I$ exactly without using Simmon's \cite{simmonsI} approximation mentioned above. Note also that the limits will be $x_1 = 0$ and $x_2 = d$ for low enough $eV_b$. Thus for the trapezoidal potential,  

$$I =  \,\left(\frac {2 d} {3 (\phi_ 1 - \phi_ 2 + eV_b)} \right) \Bigg[ \Big ( \eta_ 1 + \phi_ 1 - E_x \Big )^{3/2}-\Big (\eta_ 1 + \phi_ 2 - ev_b - E_x \Big )^{3/2} \Bigg ]$$

\begin{figure}[hpt]
	\centering
	\includegraphics[width=3.0in,height=2.5in]{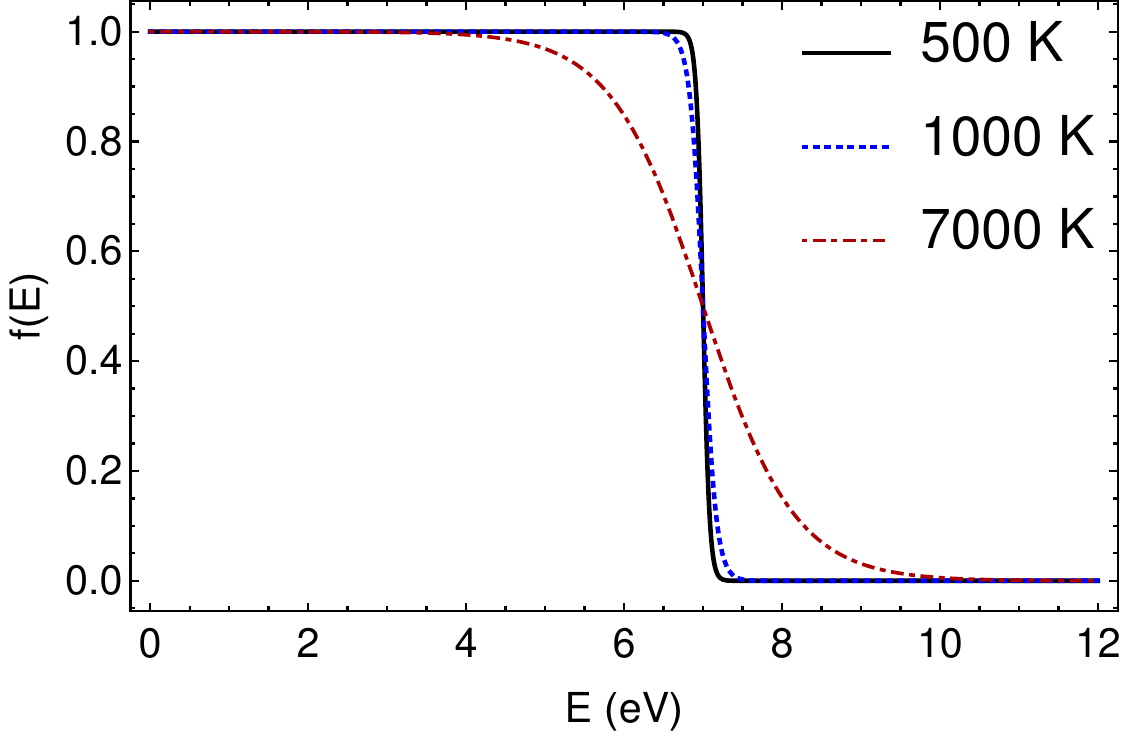}		\caption{Fermi Dirac distribution Function}	
		\label{fig:FDdisFun}
\end{figure}

At $T > 0$ K, electron energies will no longer be less than or equal to the Fermi energy. Some electrons can have energies that exceed the Fermi energy, these electrons shall cause vacancies in energy levels below the Fermi energies. Thus at $T > 0$ K, both electrodes will have some vacancies in all energy levels, and therefore tunneling can occur in both directions at all energies. Note however, the probability of occupation of an energy level and the probability of vacancy of the same energy level in the recipient electrode are determined by the respective Fermi Dirac Distribution functions. The behaviour of the Fermi Dirac Distribution as a function of the absolute temperature $T$ is shown in the Fig. \ref{fig:FDdisFun}.  Note that this figure clearly shows that at temperatures greater than absolute zero, some electrons would have energies greater than the Fermi Energy, and there is a finite  nonzero but small probability of their having large energies. As the temperature is increased more and more electrons will have energies greater than the Fermi Energy. However the probability of occupation of energies ( at low enough temperatures) far above the Fermi energy, decreases very rapidly as the energy $E-\eta$ becomes larger and larger. It is therefore convenient to cut off the energy spectrum at a reasonable value of $E_{max} = \eta_1+\phi_1$. 

\noindent Evaluating integral over $dE_r$ for forward currents,
gives 
\begin{equation}\label{F1Ex}
F_1(E_x) = \int_{0}^{\infty}f_1(E)dE_r = \frac{1}{\beta}\,\ln\Big[{1 + \exp{(-\beta[E_x-\eta_1])}}\Big ]
\end{equation}
and for reverse currents the $E_x$ in the above expression is replaced by $E_x + eV_b$.
The net current density $J_{Net} = J_\text{For} - J_\text{Rev}$ at $ T > 0 \,K$ is given by
\begin{equation}\label{jnetTgt0}
J_{Net}=\frac{4\pi me}{h^3\beta}\,\,\int_{0}^{E_{max}} D(E_x)\ln \left (\frac{1+ \exp[-\beta(E_x-\eta_1)]}{1 + \exp[-\beta(E_x+eV_b-\eta_1)]}\right)dE_x 
\end{equation}
This integral $J_{\text{Net}}$ for $T>0$ K is evaluted numerically. 

\bigskip 

According to Pauli's Exclusion principle an electron will be able to tunnel from electrode 1 to electrode 2 only if there is a vacancy in the energy level in electrode 2 corresponding to the energy of that electron. The probablity of vacancy in electrode 2 is given by $[1 - f_2(E_2)]$.
Let the probability that an electron in energy level $E_1$ is available in electrode 1 and the same energy level has a vacancy in the electrode 2 be $p_1$. The probability that an electron in energy level $E_1$ is available in electrode 2 and the corresponding level in electrode 1 is vacant be $p_2$. The energy dependent functions $p_1$ and $p_2$ may be called the forward and the reverse Pauli blocking factors respectively and are given by
\begin{equation}
p_1 = f_1(E_1)\big[1-f_2(E_2)\big] \quad \text{and} \quad p_2 = f_2(E_2)\big[1-f_1(E_1)\big]
\end{equation}
 The factor that enters the calculation of the net tunneling current is the difference of these Pauli blocking factors which is given by 
\begin{equation}
p_{12} = p_1-p_2
= f_1(E_1)\Big [1-f_2(E_2)\Big ] -\Big [1- f_1(E_1)\Big ]f_2(E_2) = f_1(E_1) - f_2(E_2) 
\end{equation}
Although Simmon's \cite{simmonsI} calculation has avoided explicitly introducing Pauli Blocking in the forward and reverse currents, it has nevertheless effectively incorporated Pauli Blocking. This is because the Pauli blocking effect on forward current exactly cancels that in the reverse current.
\begin{figure}[hpt]
	\begin{subfigure}{0.49\textwidth}
		\includegraphics[width=2.8in,height=2.5in]{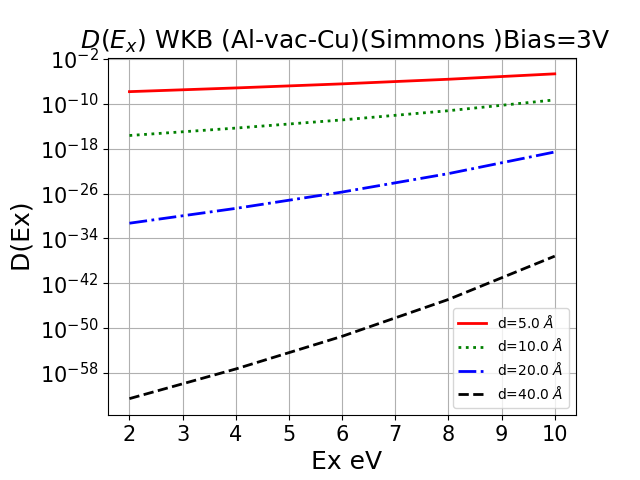}		
		\caption{}	
		\label{fig:TPWKB1}
	\end{subfigure}
	\hfill
	\begin{subfigure}{0.49\textwidth}
		\includegraphics[width=2.8in,height=2.5in]{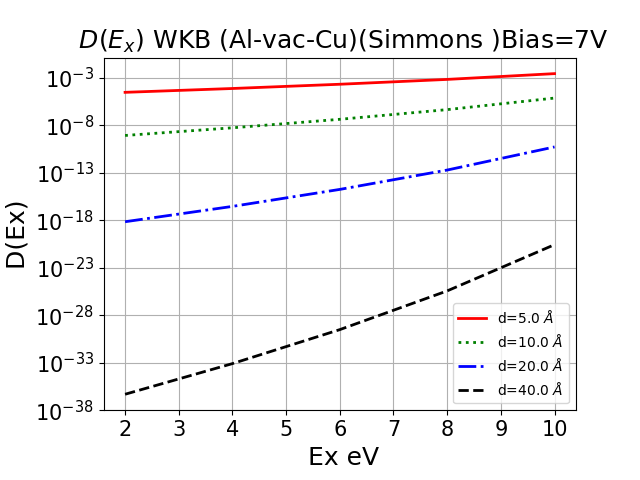}		
		\caption{}	
		\label{fig:TPWKB2}
	\end{subfigure}
	\caption{Plot of WKB tunneling probability (Al–vac–Cu) vs $E_x$ for d = 5, 10, 20, 40  \r{A} and fixed bias voltage of (a) 3 V and  (b) 7 V.}
	\label{fig:TPWKB}
\end{figure}

\bigskip

\begin{figure}[hpt]
	\begin{subfigure}{0.49\textwidth}
		\includegraphics[width=2.8in,height=2.35in]{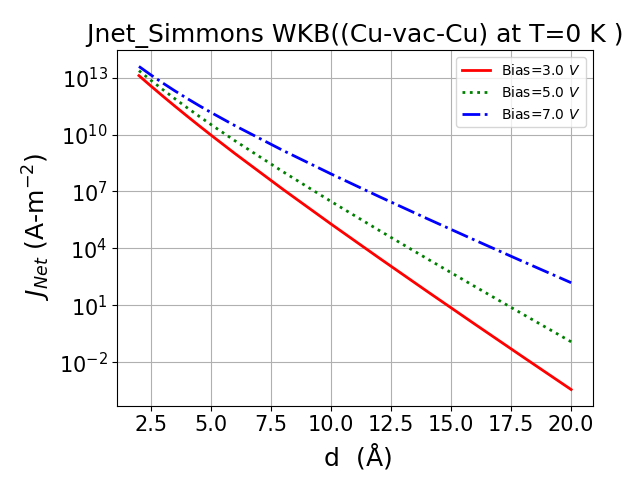}		
		\caption{ }	
		\label{fig:JTSD1}
	\end{subfigure}
	\hfill
	\begin{subfigure}{0.49\textwidth}
		\includegraphics[width=2.8in,height=2.35in]{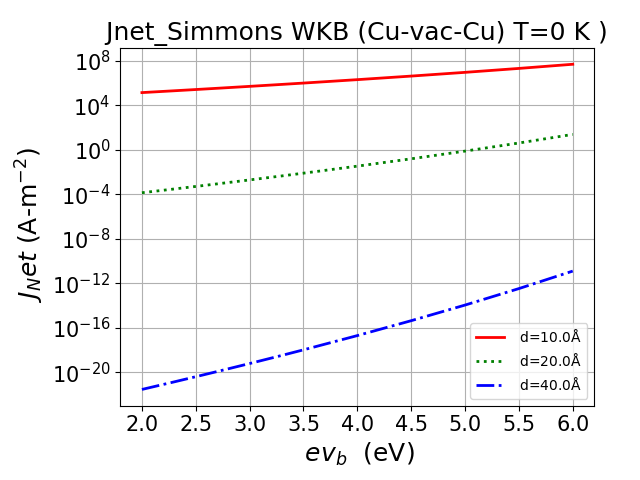}	
		\caption{ }
		\label{fig:JTSD2}
	\end{subfigure}
	\caption{Plot of (a) current vs the tip–sample distance for bias voltage = 3, 5,7 V (b) Plot of current vs bias voltage for d = 10, 20, 40 \r{A} at $T=0$ K}
	\label{fig:JTSD}
\end{figure}

Figures \ref{fig:TPWKB1} and  \ref{fig:TPWKB2} show the plot of the calculated tunneling probability $D(E_x)$ as a function of $E_x$ for two different bias voltages for $T = 0$ K. The tunneling probability is found to increase with energy as expected. $D(E_x$) increases much faster at lower tip-sample distances than at larger tip sample distances. This function is therefore a critical determinant of the dependence of the tunneling current density as a function of the tip- sample distance. The net tunneling current density $J_\text{Net}$ is shown in Figure \ref{fig:JTSD1} as a function of tip-sample distance for different bias voltages  for $T = 0$ K. The current densities are found to exponentially decrease with increase in the tip-sample distance. Therefore a very small change in the tip-sample seperation causes a large change in the tunneling current. Tunneling current density $J_\text{Net}$ as a function of bias voltage for various tip-sample distaces is shown in Fig. \ref{fig:JTSD2}. Current densities are found to increase with increasing bias voltage at all distances.
\begin{figure}[h]
	\begin{subfigure}{0.49\textwidth}
		\includegraphics[width=2.7in,height=2.29in]{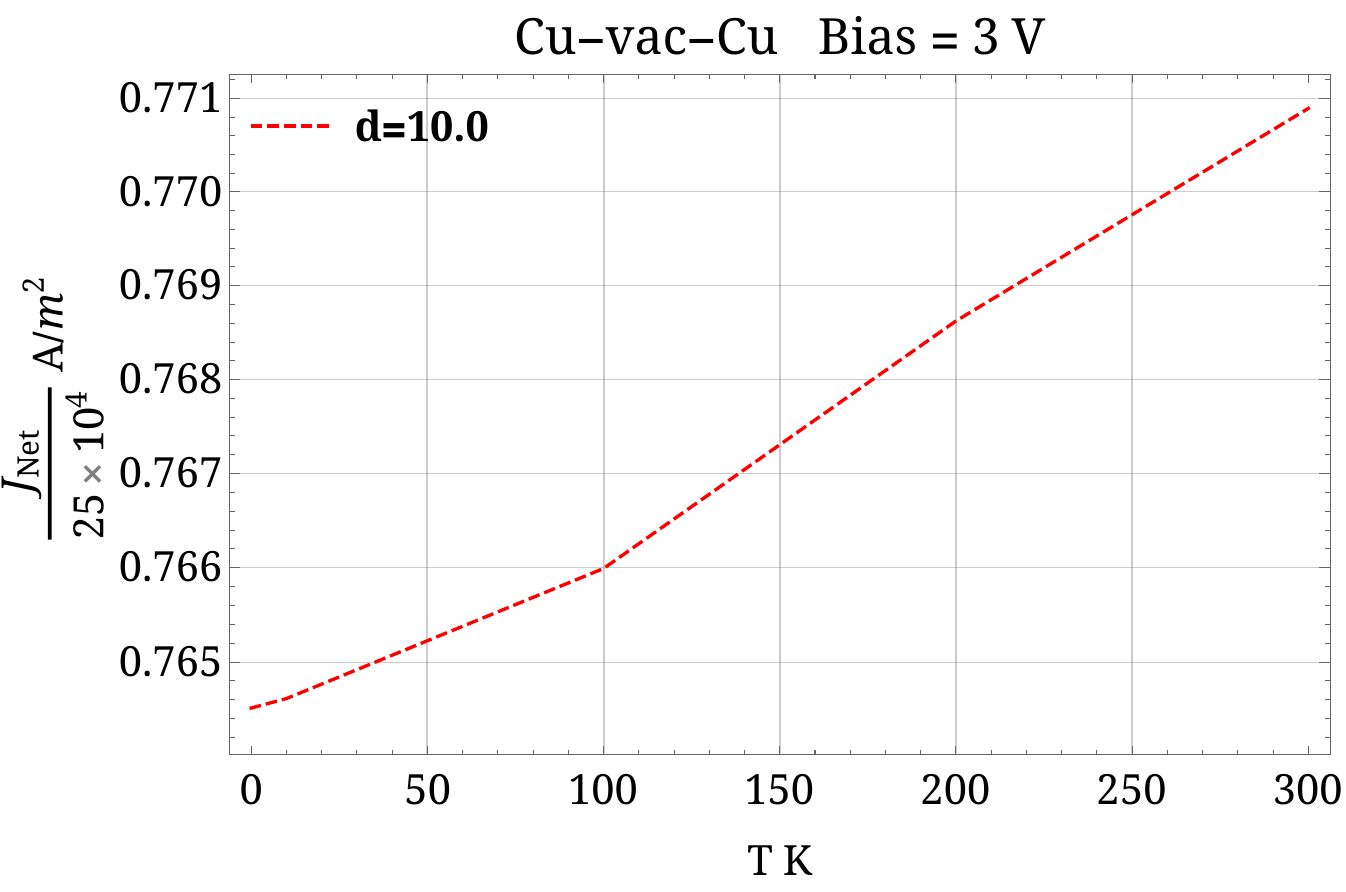}		
		\caption{ }	
		\label{fig:Temp1}
	\end{subfigure}
	\hfill
	\begin{subfigure}{0.49\textwidth}
		\includegraphics[width=3.0in,height=2.29in]{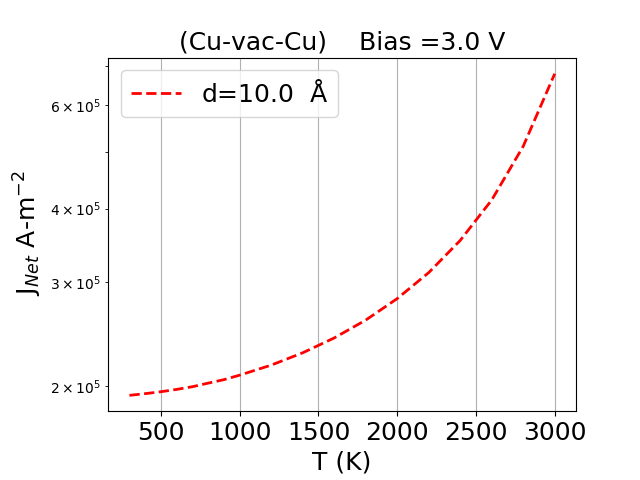}	
		\caption{ }
		\label{fig:Temp2}
	\end{subfigure}
	\caption{Plot of current vs temperature for fixed tip-sample diatance of $10$ \r{A} and bias voltage $3$ V. Temperatures varies from (a)  $0$ to $300$ K (b) $T = 300$ to $3000$ K }
	\label{fig:Temp}
\end{figure}
Figures \ref{fig:Temp1} and \ref{fig:Temp1} are plots of the net currrent densities for the Cu-Vac-Cu system as a function of the temperature for fixed distance of $10$ \r{A} and bias voltages of $3.0$ V and $7.0$ V. The Fig. \ref{fig:Temp1} shows that there is almost no change as the temperature rises from $0$ K to $300$ K. Fig. \ref{fig:Temp2} does show a slight increase at much larger temperatures from $500$ K to $3000$ K.

\bigskip

Approximate calculation of of tunneling probabilities using WKB ignore exact solution of the Schrödinger equation for the relevant potential. For arbitrary potentials where analytical solutions are not available, such approximations may be necessary, but for the trapezoidal potentials, exact analytical solutions are available in the form of Airy functions. It is therefore natural to attempt a more accurate calculation of the tunneling current densities using Airy functions to construct solutions to the Schrödinger equation. Chapter 3 describes such calculations in detail.  

%
%
%
%
%
%

	\chapter{Airy Function Calculation of the $J_{Net}$ for Linear Potentials}\label{chap3}
%
%
%
%
\section{Introduction}

The WKB approximation based calculations that were described in the previous chapter do not use exact solutions to Schr\"odinger Equation in the barrier region. More accurate calculation of the tunneling currents in an STM can only be found if the exact solution to the relevant Schr\"odinger equation is used. The consequent increase in accuracy of the tunneling current density can help in correctly converting experimentally determined data into realistic surface profiles. For trapezoidal potentials, exact analytically derived solutions are indeed available, and these can be written in terms of Airy Functions. In this chapter, Airy's functions are used to calculate the tunneling probabilities. The general method of calculating tunneling current densities from tunneling probabilities has been discussed in some detail in Chapter 2. This part of the calculations shall remain more or less intact. However tunneling probabilities that in Chapter 2 were calculated using the WKB approximations, will be replaced by those calculated using Airy Functions. 

\bigskip

The Airy function based wave functions are matched with wave functions within the respective electrodes which are at the barrier boundaries. In the interior of the left electrode the potential is zero, while it is a constant which is Max$\Big[0\,, \,(\eta_1-\eta_2-eV_b)\Big]$ inside the right electrode. The wave number of the electron in the interior of the left electrode is $k_1$ which is determined by the energy $E_x$ which is the kinetic energy of the electron for its motion along the $x -$ direction. The presence of a nonzero potential in the right electrode leads to a modification of the corresponding kinetic energy of the electron in the interior of the second electrode, which leads to the electron's wave number in this electrode being modified to $k_2$. Thus the wave function inside the electrodes will be that of plane waves with their corresponding wave numbers $k_1$ and $k_2$ respectively. The calculation is carried out at $T = 0$ K and at $T = 300$ K in which temperature dependent Fermi factors, and Pauli blocking are introduced for both forward and reverse currents. The chapter concludes with a discussion of the results of current density calculations and a comparison of these with the  corresponding WKB current densities.
\bigskip

\section{\label{sec:levelA}Wavefunctions in the Planar Model}

As mentioned in Chapter 2 the two planar electrodes are parallel to each other and to the y-z plane. Their surfaces are at $x=0$ and $x=d$  respectively, with the intervening space being occupied by vacuum. The entire extent of the x-axis can be divided up into three spatial regions (I, II, III) and the potential energy of an electron in these three regions is given by 
\begin{figure}[h]	
	\centering
	\includegraphics[width=4.5in,height=3.5in]{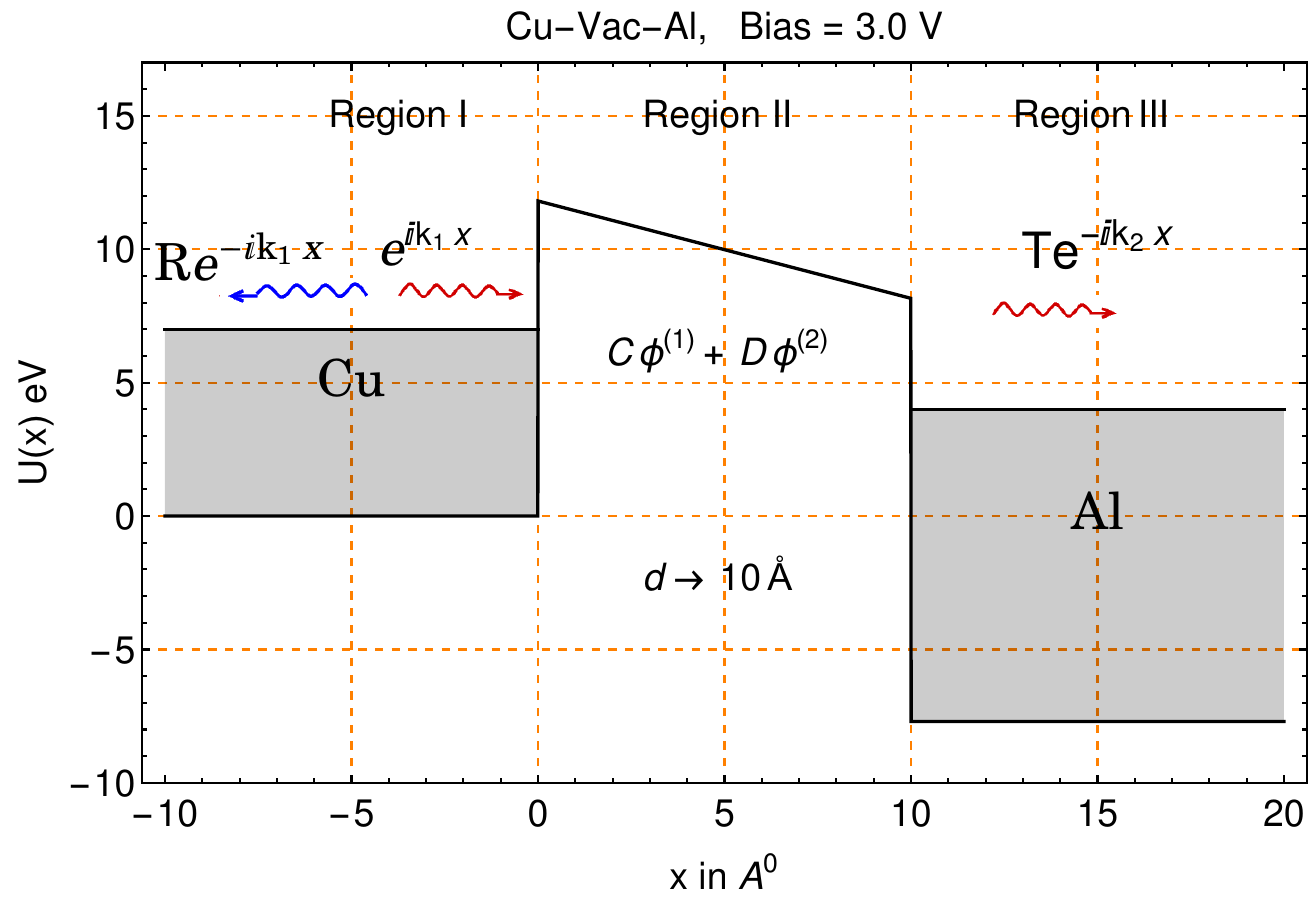}
	\caption{Dissimilar Electrodes with vacuum barrier at Bias $=3$ V, showing spatial regions I, II, and III.}			
	\label{fig:1a}
\end{figure}
\begin{equation}
U(x)=\left\{
\begin{array}{@{}lll@{}}
0 & 		x < 0\ &\quad \text{Region I} \\
U_{II}(x) & 		0 \leqslant x \leqslant d &\quad \text{Region II}\\
-\Delta_B&	 x > d &\quad \text{Region III} \\
\end{array}\right.
\end{equation} \\
where
$$\Delta_B=(\eta_2-\eta_1+eV_b) \quad \text{and} \quad U_{II}(x)=(\eta_1+\phi_1)-(\phi_1-\phi_2+eV_b) \dfrac{x}{d}$$
$U_{II}(x)$ is the trapezoidal (linear) potential described in equation (2.15) 
of Chapter 2. The wavefunctions in the two regions $(x < 0)$ and $(x > d)$ are given by   
\begin{equation}\label{psi1}
\Psi_1(x)=e^{ik_1x}+Re^{-ik_1x}, \hspace*{.2in} x < 0		
\end{equation}
\begin{equation}\label{psi3}
\Psi_3(x)=Te^{ik_2x}, \hspace*{.2in} x > d
\end{equation}
where, $R$, $T$ are the reflection and transmission amplitudes. Here $k_1$ and $k_2$ are the wave numbers of the electrons inside the respective electrodes. $k_1$ and $k_2$ are given by 
\begin{equation}\label{k1k2}
\hbar k_1=\sqrt{2mE_x}\,, \quad  \hbar k_2= \sqrt{2m(E_x+\Delta_B)} 
\end{equation}
The wavefunction $\Psi_b$ in the barrier region $0 \leqslant x \leqslant d$ is given by the  Schr\"odinger equation for the linear (trapezoidal) potential as
\begin{equation}\label{SETrap}
\frac{d^2 \Psi_b}{dx^2}-(A-Bx)\Psi_b= 0
\end{equation} 
where
\begin{equation}
A=\dfrac{2m}{\hbar^2} (\eta_1+\phi_1-E_x) 
\end{equation}	
\begin{equation}
B=\dfrac{2m}{\hbar^2 d} (\phi_1-\phi_2+eV_b) 
\end{equation} 
If both the electrodes were at absolute zero temperature, only electrons with energies less than or equal to their respective Fermi Energies will undergo tunneling. However at higher temperatures, electrons with a wider range of energies, that can exceed their Fermi energies will also be involved in tunneling. Note however that all energies are measured with respect to the zero energy level of the first electrode.  The energy range of electrons in the first electrode, can be divided into three Energy Stages. Fig.\ref{fig:1b} shows the three Energy Stages for the Al-vac-Cu system for which
$\eta_1=11.7$ eV, $\phi_1=4.16$ eV, $\eta_2=7.0$ eV, $\phi_2=4.81$ eV, $V_b = 3$ volts.\\
\begin{figure}[h]
	\centering
	\includegraphics[width=4.5in,height=3.5in]{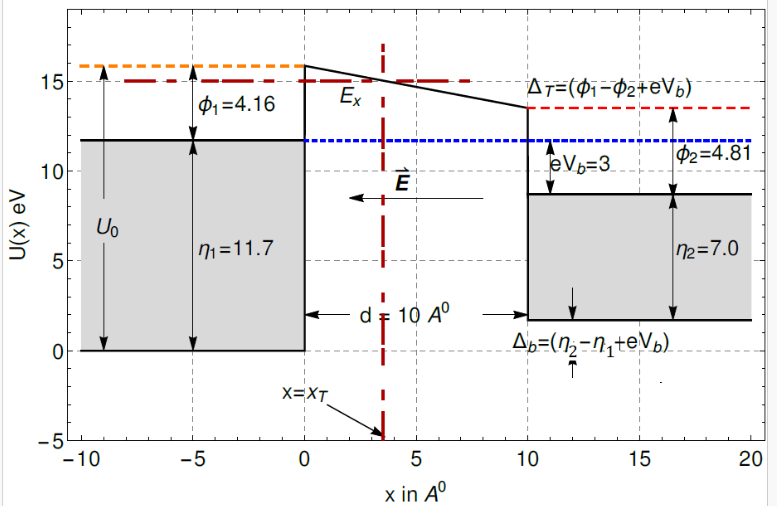}	
	\caption{Dissimilar Electrodes with vacuum barrier at Bias = 3.0 eV showing Energy Stages I ($0$ eV to $13.51$ eV), II ($13.51$ eV to $15.86$ eV), and III ($> 15.86$ eV) for Al-vac-Cu. The upper limits are indicated by red dotted line (Stage I), orange dotted line (Stage II).}
	\label{fig:1b}
\end{figure}

\noindent\textbf{Energy Stage I}: The barrier region in this stage does not contain a turning point. The energy in this region ranges from Max$\Big[0,(\eta_1-\eta_2-eV_b)\Big] < E_x < (\eta_1+\phi_2-eV_b)$. Therefore, the term $(A-Bx)\,\geqslant \, 0 , \quad \forall \, x \, \in \, [0,d]$. 

\noindent\textbf{Energy Stage II}: In this stage, any energy line $E = E_x$ will intersect the potential barrier in atleast one point, which is called the turning point.  The turning point inside the barrier region is at $x = x_T = \frac{A}{B}$. The energy range in this stage is $(\eta_1+\phi_2-eV_b)<E_x<(\eta_1+\phi_1)$ as shown in Fig.\ref{fig:1b}. 
Electron in this Energy Stage II, whose energies exceed $\eta_1+20\,kT$ will have negligible contribution to tunneling owing to the behaviour of Fermi Dirac distribution function. Therefore $(\eta_1 + \phi_2 - eV_b) < (\eta_1 + 20\,kT )$  or $ eV_b > (\phi_2 - 20\,kT)$. For bias voltages less than $(\phi_2 - 20\,kT)$ there will be negligible contribution to tunneling. \\


\noindent\textbf{Energy Stage III}: The energy of the electrons in this stage exceeds $(\eta_1+\phi_1)$ and these electrons should be emitted without having to tunnel through any barrier suggesting that the transmission probability would be very high. However for most values of $\phi_1$ the electron energy is so far above the Fermi energy $\eta_1$ that the Fermi factor (for room temperatures) is essentially zero and the contribution of electrons in this energy stage at these temperatures to the current density is vanishingly small. Therefore unless the electrode temperatures are very high (i.e. comparable to the Fermi temperature of the electrodes), the contribution of electrons in this Energy Stage will be neglected alltogether.\\
\par In Energy Stages I and II, the Schr\"odinger equation (\ref{SETrap}) can be reduced to the Airy differential equation as shown in the Appendix A-1.\\

\begin{equation}
\frac{d^2 \Psi_b}{dh^2} - h\Psi_b= 0
\end{equation} 
where $h=h(x)=\dfrac{A}{B^{2/3}}-B^{1/3}x$.
The general solution \cite{AS} can be written as 
\begin{equation}\label{phden18}
\Psi_b(x)= C \phi^{(1)}(x)+ D \phi^{(2)}(x)
\end{equation} 
where
\begin{equation}\label{phden19}
\phi^{(1)}(x)= \dfrac{Ai[h(x)]}{Ai[h(\bar{x})]} \, , \quad
\phi^{(2)}(x)= \dfrac{Bi[h(x)]}{Bi[h(\bar{x})]}
\end{equation}

\bigskip

\noindent are linearly independent solutions to the Airy Differential equation. The denominator of the functions $\phi^{(1)}$ and $\phi^{(2)}$ contain Airy functions evaluated at $\bar{x}$ (a fixed value of x near the centre of the barrier region). These denominators are sometimes required to avoid overflow and underflow errors in the calculation of the wavefunctions for different values of $x$ in the barrier. Since Wroskian of the Airy functions is $W[Ai(x),Bi(x)]=\frac{1}{\pi}$ therefore,
\begin{equation}
W[\phi^{(1)}(x),\phi^{(2)}(x)]=\dfrac{1}{\pi} \dfrac{1}{Ai[h(\bar{x})Bi[h(\bar{x})]}
\end{equation}

\section{The Fermi - Factor and $\bm{J_{Net}}$}

The tunneling probability $D(E_x)$ is the product of square modulus of tunneling amplitude $T(E_x)$ and the flux ratio  $\dfrac{J_{out}}{J_{inc}}$. The total current density $J_{Net}$ is obtained by integrating the product of this tunneling probability and a factor called the Fermi factor (which involves the Fermi-Dirac Distribution function) over the energy $E_x$ of the tunneling electrons for the translational motion along the x-direction. The range of the energies $E_x$ in the integration, includes those in Energy Stages I and II. The contribution to the tunneling currents of electrons with energies that exceed the maximum of Energy Stage II is considered zero.

\bigskip

The tunneling amplitude $T(E_x)$ is symmetric. That is, the amplitude for tunneling in the forward (electrode 1 to electrode 2; along the positive direction of the x-axis) is the same as for the reverse (electrode 2 to electrode 1; along the negative direction of the x-axis) direction, even though the barrier may not be symmetric. This has also been mentioned earlier in Chapter 2. The flux ratio $\dfrac{J_{out}}{J_{inc}}$ is the ratio of the probability currents in the emitter electrode and the receiver electrode. For forward currents, the $1^{st}$ (or left) electrode is considered as the emitter and the $2^{nd}$ (or right) electrode is considered as the receiver and vice versa for reverse currents.  If plane wave solutions are used within the electrodes then
\begin{equation}
\dfrac{J_{out}}{J_{inc}}=\left\{
\begin{array}{@{}ll@{}}
\dfrac{k_2}{k_1} & 	\text{for forward tunneling currents} \\ \\
\dfrac{k_1}{k_2} & 	\text{for reverse tunneling currents} \\	
\end{array}\right.
\end{equation} 
where $k_1$ and $k_2$ are given by equation (\ref{k1k2}).

\par The calculation of the number density of electrons tunneling from electrode 1 to 2 and vice versa has been discussed in equations (2.4, 2.5). 
The equations are modified due to explicit introduction of the Pauli blocking factors $p_1$ and $p_2$ and the corresponding flux ratios. Therefore the relevant electron number densities for forward and reversing tunneling are given by 
\begin{equation}
\begin{aligned}
n_\text{For} d^3 v= &\dfrac{2m^3}{h^3}D_{For}(E_x) p_1 d^3 v =
\dfrac{2m^3}{h^3}\frac{k_2}{k_1}\lvert T(E_x) \rvert^2 p_1 d^3 v \\
n_\text{Rev} d^3 v= &\dfrac{2m^3}{h^3}D_{Rev}(E_x) p_2 d^3 v =
\dfrac{2m^3}{h^3}\frac{k_1}{k_2}\lvert T(E_x) \rvert^2 p_2 d^3 v 
\end{aligned}
\end{equation}
where the factors $p_1$ and $p_2$ are defined in Chapter 2, and are given by 
$$ p_1(E,E^\prime) = f_1(E)\big[1 -f_1(E^\prime)\big]$$
$$p_2(E,E^\prime) = p_1(E^\prime,E)$$
where
$$f_1(E) = \Big[1 + \exp[{\beta(E-\eta_1)}] \Big]^{-1} \quad \text{and} \quad  f_1(E') = \Big[1 + \exp[{\beta(E + eV_b - \eta_1)}] \Big]^{-1}$$
The forward and reverse current densities are then given by 
\begin{equation}
J_\text{For} =\dfrac{4\pi me}{h^3} \int dE_x\, \dfrac{k_2}{k_1}\lvert T(E_x) \rvert^2\,\times \int\limits_0^\infty dE_r \,p_1(E,E^\prime)
\end{equation} 
\begin{equation}
J_\text{Rev} = \dfrac{4\pi me}{h^3} \int dE_x \dfrac{k_1}{k_2}\lvert T(E_x) \rvert^2\,\times\\ \int\limits_0^\infty dE_r \, p_2(E^\prime ,E)
\end{equation}

The net current density is given by $J_\text{Net} = J_\text{For} - J_\text{Rev}$, which becomes 
\begin{equation}
J_\text{Net} = \dfrac{4\pi me}{h^3} \int dE_x \lvert T(E_x) \rvert^2\,\times \mathcal{F}(E_x) 
\end{equation}
where the Fermi Factor $\mathcal{F}(E_x)$ is defined as 
\begin{equation}\label{np}
\mathcal{F}(E_x)=\int\limits_0^\infty dE_r \Bigg [f_1(E)\Big[1 - f_1(E')\Big] \,\dfrac{k_2}{k_1} - \Big[1 - f_1(E)\Big] f_1(E')\dfrac{k_1}{k_2} \Bigg] 
\end{equation}
Carrying out the integral over $E_r$ (shown in Appendix A-3)
\begin{multline}\label{paulib}
\mathcal{F}(E_x)= \dfrac{k_2}{k_1}F_1(E_x) - \dfrac{k_1}{k_2}F_1(E_x+eV_b)\, + \\ \frac{\Big [\dfrac{k_1}{k_2}-\dfrac{k_2}{k_1}\Big ]}{(1-e^{-\beta eV_b})}\Big [F_1(E_x+eV_b)-e^{-\beta eV_b}F_1(E_x)\Big] 
\end{multline}
where $$F_1(E_x)=\dfrac{1}{\beta}\text{Log}[1+e^{-\beta(E_x-\eta_1)}]$$ 

\begin{figure}[hpt]
	\centering
	\includegraphics[width=4.5in,height=4.0in]{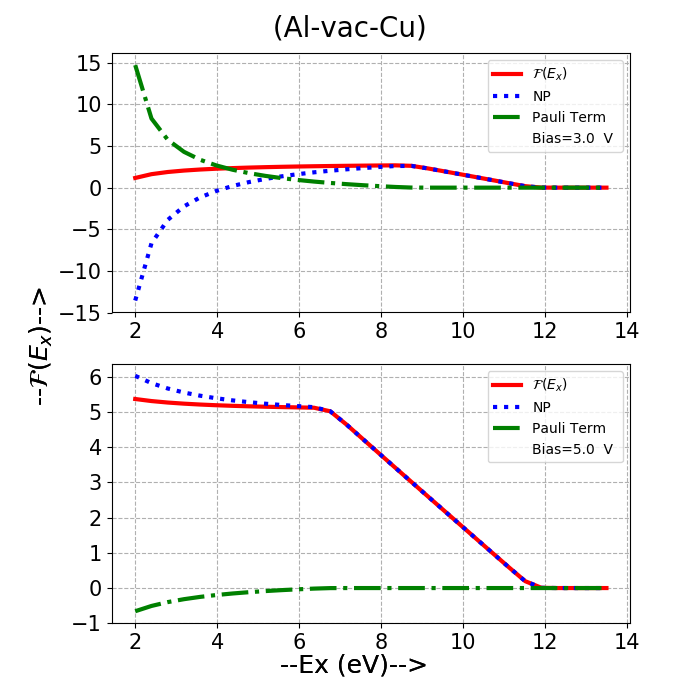}		
	\caption{Fermi factor for a bias voltage of 3 and 5 V for Al–vac–Cu. The solid red line refers to $F(E_x)$, the dotted line refers to the non-Pauli term, and the dotted–dashed line refers to the Pauli term }
	\label{fig:ThT3}
\end{figure}

\begin{figure}[hpt]
	\centering
	\includegraphics[width=4.5in,height=3.5in]{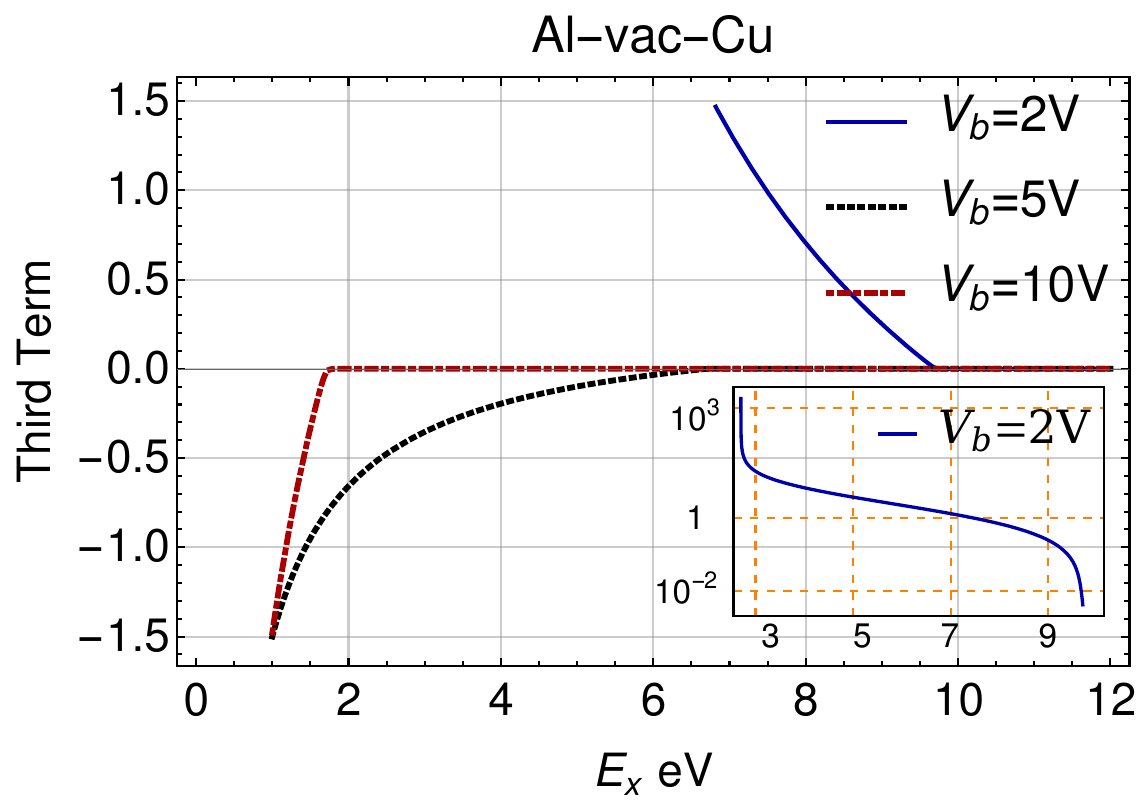}
	\caption{Third Term as a function of $E_x$ for bias voltages $2$ V, $5$  V,and $10$ V for Al-vac-Cu.}
	\label{fig:ThTerm1}
\end{figure}

\begin{figure}[hpt]
	\centering	
	\includegraphics[width=4.5in,height=3.5in]{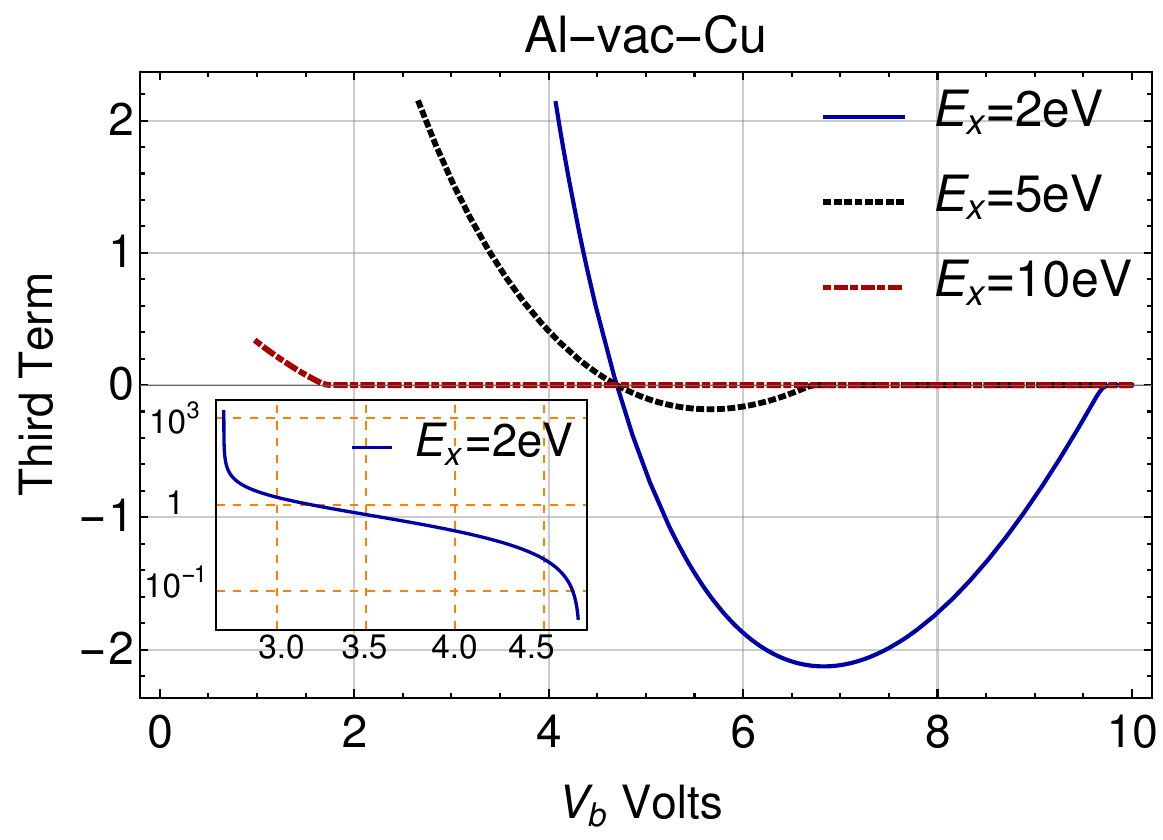}
	\caption{Third Term as a function of $V_b$ for electron energies of  $2\, eV, \,5 \, eV,\, \text{and} \,10\, eV$ for Al-vac-Cu.}
	\label{fig:ThTerm2}
\end{figure}

The first two terms in equation (\ref{paulib}) are the Non-Pauli contribution terms. i.e., these are the terms that one would get if Pauli blocking were
completely ignored. The third term in equation (\ref{paulib}) explicitely introduces Pauli Effects. The behaviour of these terms $viz$ Non-Pauli, (first two terms only) $\mathcal{F}(E_x)$, (all three terms) and the Pauli term (third term only), as a function of $E_x$ are displayed in Fig.\ref{fig:ThT3}. 
The Fermi Function $\mathcal{F}(E_x)$ (solid red line) is seen to peak at $ 8.7 \,\text{eV}$ $\approx \eta_1-eV_b$ for $eV_b=3 \,\text{eV}$; and it peaks at $ 6.7\, \text{eV}$ $\approx \eta_1-eV_b$ for $eV_b=5  \, \text{eV}$. Subsequently  $\mathcal{F}(E_x)$ is found to decrease rapidly to zero as $E_x$ exceeds $\eta_1$. This is found true for all bias voltages. The dotted line refers to the Non-Pauli (NP) term, which may have negative values for low bias voltages. For these voltages the reverse current exceeds the forward current. This is due to the different flux ratios for the forward and the reverse currents. The dot-dashed line refers to the third term of equation (\ref{paulib}). The third term is seen to go to zero at about $E_x=\eta_1-eV_b$. This term contains Pauli effects and it is studied in more detail in Figures \ref{fig:ThTerm1} and \ref{fig:ThTerm2}.

\bigskip

\par Fig. \ref{fig:ThTerm1} plots the third term for fixed bias voltages ($V_b=2\,V, \,V_b=5\, V, \,V_b=10\,V$) as a function of electron energies in the range ($1\, eV\,$ to $12\, eV$). The graph shows that Pauli Effects are significant for low energy electrons and especially for low bias voltages. For $E_x > \eta_1-eV_b$ the third term is essentially zero implying that there is negligible Pauli Effect. The inset in this figure shows the third term for $2$ V rising rapidly by at least $4$ orders of magnitude for very low values of $E_x$.

\bigskip

\par Figure \ref{fig:ThTerm2} plots the third term for fixed electron energies ($E_x=2 \, eV, \,E_x = 5\, eV,\, E_x = 10\, eV$) as a function of the bias voltage in the range ( $1$V to $10$V). The inset in this figure shows the third term for $E_x=2\,eV$ rising rapidly by at least $3$ orders of magnitude for very low values of bias voltage. All curves in Fig.\ref{fig:ThTerm2} pass through zero at $eV_b=\eta_1-\eta_2$ which in case of electrodes Al and Cu is given by $V_b = 4.7$ V. This is because $k_1 = k_2$ for this value of bias, and the zero level of the energy in the two electrodes become coincident. For bias voltages $eV_b < (\eta_1-\eta_2)$ the third term is positive and its effect is to increase the net tunneling current beyond the non-Pauli contribution. For bias voltages larger than this value the third term becomes negative and the Pauli Effect will be to reduce the net tunneling current from its Non-Pauli value. Thus, Pauli exclusion principle does not always lead to reduction of the net tunneling current density, but can cause enhancement in some cases. Thus the term Pauli blocking for the effect of the Pauli principle may be misleading and a more appropriate term \textit{Pauli effects} may be used instead.

\section{Calculation of $\bm {T(E_x)}$}
\subsection{Energy Stage I}
\hspace{0.2in} The energy $E_x$ of electrons in the Energy Stage I is given by  $E_x\, < \,U(x) \, \forall\, x \, \in [0,\,d] $. The spatial boundaries of the potential barrier are at $x=0$ and $x=d$. The wavefunction in this region is given by
\begin{equation}\label{ES_I}
\Psi_b(x)= C \phi_I^{(1)}(x)+ D \phi_I^{(2)}(x)
\end{equation} 
where the subscripts I on the functions $\phi^{(1)}$ and $\phi^{(2)}$ serve to indicate that they describe wavefunctios in Energy Stage I.
Define 
\begin{equation}
\phi^{(i)^{'}}(x) = \frac{d}{dx}\phi^{(i)}(x)=\frac{d\phi^{(i)}}{dh} \frac{dh}{dx} \,\,\, (i=1,2)
\end{equation}

\noindent The matching conditions on the wave function at $x=0$ are
\begin{equation}\label{reg1}
\begin{aligned}
1+R = & \,C \phi^{(1)}(0) +  D \phi^{(2)}(0)  \\
ik_1(1-R) =& C {\phi^{(1)}}^\prime(0) +  D {\phi^{(2)}}^\prime(0)
\end{aligned}
\end{equation}	
where $R$ is the reflection amplitude. 
From equations (\ref{reg1}) one can eliminate $R$ to get
\begin{equation}\label{at0}
C\Big[\phi^{(1)}(0) + \frac{\phi^{(1)^{'}}(0)}{ik_1} \Big]+ D\Big[\phi^{(2)}(0) + \frac{\phi^{(2)^{'}}(0)}{ik_1} \Big] =2
\end{equation}
Define
\begin{equation}
\begin{aligned}
\gamma_{10} = \Big (\phi^{(1)}(0) + \frac{\phi^{(1)^{'}}(0)}{ik_1} \Big),  \quad
\gamma_{20} = \Big (\phi^{(2)}(0) + \frac{\phi^{(2)^{'}}(0)}{ik_1} \Big) 
\end{aligned}
\end{equation}
Therefore equation (\ref{at0}) can be written as 
\begin{equation}\label{x_0}
\gamma_{10} \,C + \gamma_{20} \,D = 2
\end{equation}
The matching conditions on the wave function at $x=d$ gives,
\begin{equation}\label{reg3}
\begin{aligned}
C \phi^{(1)}(d)+ D \phi^{(2)}(d) = & T_I(E_x)e^{ik_2d}\\
C \phi^{(1)^{'}}(d)+ D \phi^{(2)^{'}}(d) = & ik_2 T_I(E_x) e^{ik_2d}
\end{aligned}
\end{equation}
where $T_I(E_x)$ is the transmission (tunneling) amplitude in the Energy Stage I.

Divide the second of equations (\ref{reg3}) by the first of equations (\ref{reg3})  to get 
$$
\frac{C \phi^{(1)^{'}}(d)+D \phi^{(2)^{'}}(d)}{C \phi^{(1)}(d) + D \phi^{(2)}(d)} = ik_2
$$
Rearranging terms in the above equation gives
\begin{equation}\label{atd}
C\Big[\phi^{(1)}(d) - \frac{\phi^{(1)^{'}}(d)}{ik_2} \Big]+ D\Big[\phi^{(2)}(d) - \frac{\phi^{(2)^{'}}(d)}{ik_2} \Big] = 0
\end{equation}In terms of
\begin{equation}
\begin{aligned}
\gamma_{1d} = \Big (\phi^{(1)}(d) - \frac{\phi^{(1)^{'}}(d)}{ik_2} \Big),  \quad
\gamma_{2d} = \Big (\phi^{(2)}(d) - \frac{\phi^{(2)^{'}}(d)}{ik_2} \Big)
\end{aligned}
\end{equation}
equation (\ref{atd}) can be written as 
\begin{equation}\label{x_d}
\gamma_{1d} \, C + \gamma_{2d} \,D = 0
\end{equation}
From equations (\ref{x_0}) and (\ref{x_d}) 

\begin{equation}\label{mat}
\begin{pmatrix}
\gamma_{10} & \gamma_{20}\\ \\ \gamma_{1d}
& \gamma_{2d}
\end{pmatrix}
\begin{pmatrix}
C \\ \\ D
\end{pmatrix}=\begin{pmatrix}
2 \\ \\  0 
\end{pmatrix}
\end{equation}
Inverting the Matrix in the above equation, 
\begin{equation}
\begin{pmatrix}
C \\  \\  D
\end{pmatrix}=\dfrac{1}{W_1}
\begin{pmatrix}
\gamma_{2d} & - \gamma_{20} \\ \\ - \gamma_{1d}
& \gamma_{10}
\end{pmatrix}
\begin{pmatrix}
2\\ \\ 0 
\end{pmatrix}
\end{equation}
where $W_1 = \gamma_{10} \gamma_{2d} - \gamma_{20} \gamma_{1d}\,$ which is the determinant of the square matrix in equation (\ref{mat}) is given by 

\begin{equation}
\begin{aligned}	
W_1 =\dfrac{1}{Ai[h(\bar{x})Bi[h(\bar{x})]}&\Bigg[ \frac{\sqrt[3]{3} Bi(d)-3^{5/6} Ai(d)}{3 \Gamma \left(\frac{2}{3}\right)} -\frac{i \left(3^{5/6} Ai'(d)-\sqrt[3]{3} Bi'(d)\right)}{3 k_2 \Gamma \left(\frac{2}{3}\right)} + \\ &\frac{i \left(3 \sqrt[6]{3} Ai(d)+3^{2/3} Bi(d)\right)}{3 k_1 \Gamma \left(\frac{1}{3}\right)} - \frac{\sqrt[6]{3} Ai'(d)+\frac{Bi'(d)}{\sqrt[3]{3}}}{k_1 k_2 \Gamma \left(\frac{1}{3}\right)} \Bigg] 
\end{aligned}
\end{equation}
where $\Gamma(x)$ is the Gamma function with values $\Gamma \left(\frac{1}{3}\right) = 2.67894$ and $\Gamma \left(\frac{2}{3}\right) =1.35412 \,\,$ \cite{AS}

\noindent Thus,  $$C=\frac{2\,\gamma_{2d}}{W_1}\quad \text{and} \quad   D= -\, \frac{2\gamma_{1d}}{W_1} $$ 
substituting in first of equation (\ref{reg3})
$$\frac{2\,\gamma_{2d}}{W_1} \phi_1(d)-\frac{2\,\gamma_{1d}}{W_1} \phi_2(d)=T_I(E_x) e^{ik_2d}$$

\begin{equation}\label{fig:TExStg1}
T_{I}(E_x) = \frac{2e^{-ik_2d}}{W_1}\Big [ \gamma_{2d} \, \phi_1(d) - \gamma_{1d} \, \phi_2(d)\Big]
\end{equation}
where $T_I(E_x)$ is the transmission amplitude for tunneling through the barrier. 

\subsection{Energy Stage II}
\hspace{0.2 in} In the Energy Stage II there is a turning point inside the barrier at $x=x_T=\frac{A_T}{B_T}$
where 
$$A_T =\frac{2m}{\hbar^2}(\eta_1+\phi_1 - E_x)$$
$$B_T =\frac{2m}{\hbar^2 d}(\phi_1 - \phi_2 + eV_b)$$					
In this stage, two separate wavefunctions $\Psi_{b1}$ and $\Psi_{b2}$ have to be determined for the two spatial regions $x < x_T$ and $x > x_T$ respectively.
The solution of the Airy differential equation in Energy Stage II for these regions can be written as 
\begin{equation}
\begin{aligned}
\Psi_{b1} = & C_1\,\,\phi_I^{(1)}(x) +  D_1\,\,\phi_I^{(2)}(x) \,\, \text{for}\,\,  x < x_T  \\ 
\Psi_{b2} = & C_2\,\,\phi_{II}^{(1)}(x) +  D_2\,\,\phi_{II}^{(2)}(x)  \,\, \text{for}\,\,  x > x_T
\end{aligned}
\end{equation} 
The constants $C_1$, $D_1$, $C_2$, and $D_2$ are determined from the matching of the wavefunctions and their derivative at $x=0$, $x=x_T$ and $x=d$. \\
The matching conditions on the wave function at $x=0$ give
\begin{equation}\label{reg1stg2}
\begin{aligned}
1+R = & \,C_1 \phi_I^{(1)}(0) +  D_1 \phi_I^{(2)}(0)  \\
ik_1(1-R) =&\Big[ C_1 \frac{d}{dx}\phi_I^{(1)}(x) +  D_1 \frac{d}{dx}\phi_I^{(2)}(x) \Big ]_{x=0}
\end{aligned}
\end{equation}	
From equation (\ref{reg1stg2}) one can eliminate $R$ to get an equation simillar to equation (\ref{reg1})
\begin{equation}\label{stg2x0}
C_1\Big[\phi_I^{(1)}(0) + \frac{\phi_I^{(1)^{'}}(0)}{ik_1} \Big]+ D_1\Big[\phi_I^{(2)}(0) + \frac{\phi_I^{(2)^{'}}(0)}{ik_1} \Big] = 2
\end{equation}
Define $$\tilde{\gamma}_{10}=\phi_I^{(1)}(0) + \frac{\phi_I^{(1)^{'}}(0)}{ik_1} , \quad \tilde{\gamma}_{20}=\phi_I^{(2)}(0) + \frac{\phi_I^{(2)^{'}}(0)}{ik_1} $$
Substituting in equation (\ref{stg2x0}) gives
\begin{equation}\label{at0stg2}
\tilde{\gamma}_{10} \, C_1 +  \tilde{\gamma}_{20} \, D_1 = 2
\end{equation}
The matching conditions on the wave function at $x=x_T$ give
\begin{equation}\nonumber 
C_1 \phi_I^{(1)}(x_T) +  D_1 \phi_{I}^{(2)}(x_T) = C_2 \phi_{II}^{(1)}(x_T) +  D_2 \phi_{II}^{(2)}(x_T) 
\end{equation}
\begin{equation}\nonumber
C_1 \phi_I^{(1)^{'}}(x_T) +  D_1 \phi_{I}^{(2)^{'}}(x_T) = C_2 \phi_{II}^{(1)^{'}}(x_T) +  D_2 \phi_{II}^{(2)^{'}}(x_T)
\end{equation}

\begin{equation}\label{mat2}
\begin{pmatrix}
C_2\\ D_2
\end{pmatrix}=\begin{bmatrix}
\phi_{II}^{(1)}(x_T) & \phi_{II}^{(2)}(x_T)\\ 
\phi_{II}^{(1)^{'}}(x_T) & \phi_{II}^{(2)^{'}}(x_T)
\end{bmatrix}^{-1}
\begin{bmatrix}
\phi_{I}^{(1)}(x_T) & \phi_{I}^{(2)}(x_T)\\ 
\phi_{I}^{(1)^{'}}(x_T) & \phi_{I}^{(2)^{'}}(x_T)
\end{bmatrix}
\begin{pmatrix}
C_1\\ D_1
\end{pmatrix}
=\bar{\bar{M}}
\begin{pmatrix}
C_1\\ D_1
\end{pmatrix}
\end{equation}

where
$$ \bar{\bar{M}}=\begin{bmatrix}
M_{11} & M_{12} \\
M_{21} & M_{22}
\end{bmatrix}
=
\begin{bmatrix}
\phi_{II}^{(1)}(x_T) & \phi_{II}^{(2)}(x_T)\\ 
\phi_{II}^{(1)^{'}}(x_T) & \phi_{II}^{(2)^{'}}(x_T)
\end{bmatrix}^{-1}
\begin{bmatrix}
\phi_{I}^{(1)}(x_T) & \phi_{I}^{(2)}(x_T)\\ 
\phi_{I}^{(1)^{'}}(x_T) & \phi_{I}^{(2)^{'}}(x_T)
\end{bmatrix}$$

In terms of the matrix elements of $ \bar{\bar{M}}$, equation (\ref{mat2}) can be written as 
\begin{equation} 
\begin{aligned}\label{Mmatrix}
C_2=&M_{11} C_1+M_{12} D_1\\
D_2=&M_{21} C_1+M_{22} D_1
\end{aligned}
\end{equation}
The matching conditions on the wavefunction at $x=d$ gives
\begin{equation}\label{atdstg2}
\begin{aligned}
C_2 \phi_{II}^{(1)}(d) +  D_2 \phi_{II}^{(2)}(d) = \,& T e^{ik_2d}\\
C_2 \phi_{II}^{(1)^{'}}(d) +  D_2 \phi_{II}^{(2)^{'}}(d) =\, & ik_2 T e^{ik_2d}
\end{aligned}
\end{equation}
Define $$\tilde{\gamma}_{1d}=\phi_{II}^{(1)}(d) - \dfrac{\phi_{II}^{(1)^{'}}(d)}{ik_2} , \quad \text{and} \quad \tilde{\gamma}_{2d}=\phi_{II}^{(2)}(d) - \dfrac{\phi_{II}^{(2)^{'}}(d)}{ik_2}$$

\begin{equation}\label{dcon}
\tilde{\gamma}_{1d}\, C_2 + \tilde{\gamma}_{2d} \, D_2 = 0 
\end{equation}
From equations (\ref{at0stg2}), (\ref{Mmatrix}),  (\ref{dcon}), we get 

\begin{equation}
D_1 = 2 \, \Big[ \tilde{\gamma}_2(0) - \mu \tilde{\gamma}_1(0)\Big]^{-1} \quad \text{and} \quad 
C_1 =  - D_1 \mu
\end{equation}
where 
$$\mu =  \frac{M_{22} \tilde{\gamma}_2(d) + M_{12} \tilde{\gamma}_1(d)}{M_{21} \tilde{\gamma}_2(d) + M_{11}\tilde{\gamma}_1(d)} $$
In terms of $\mu$ 
\begin{equation} 
C_2 = D_1 \, (M_{12} - M_{11} \mu) \quad \text{and} \quad D_2 = D_1 \, (M_{22} - M_{21} \mu)
\end{equation}
Substituting the coefficients $C_2$ and $D_2$ determined above, in the first of equations (\ref{atdstg2}) gives the tunneling amplitude $T_{II}(E_x)$ in the Energy Stage II to be
\begin{equation}
T_{II}(E_x) = e^{-ik_2d}\Big[ C_2\phi_{II}^{(1)}(d) + D_2\phi_{II}^{(2)}(d) \Big ]	
\end{equation}
\begin{figure}[hpt]
	\centering
	\includegraphics[width=4.3in,height=3.2in]{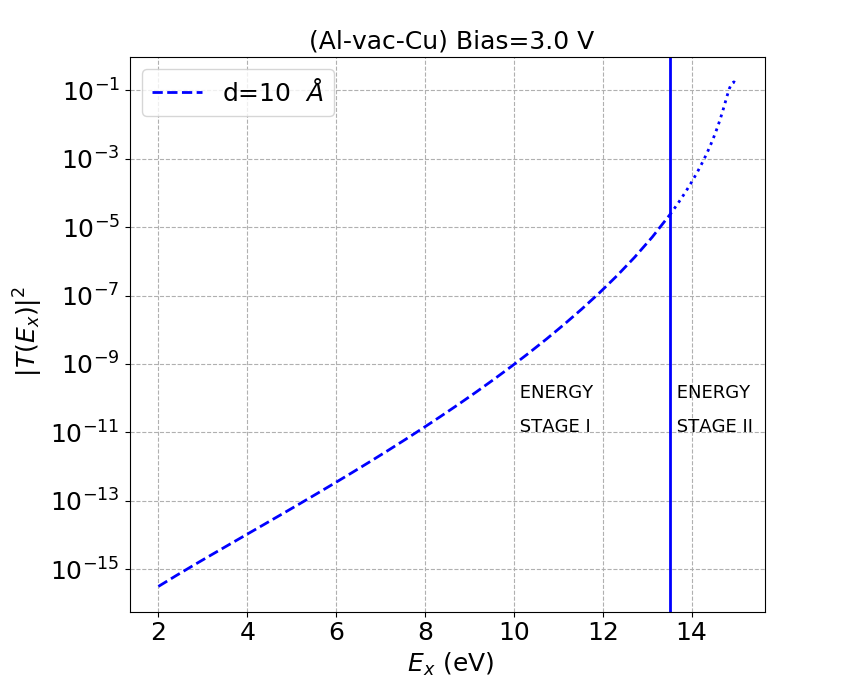}
	\caption{Plot of $\rvert T(E_x) \rvert^2$ vs $E_x$ for dissimilar electrodes with vacuum barrier at Bias = 3.0 V. Note both Energy Stages are included in the figure. }
	\label{fig:TEx}
\end{figure}
 The square modulus of the tunneling amplitude $\rvert T(E_x) \rvert^2$ as a function of $E_x$ for fixed tip-sample distance of $10$ \r{A} is shown in Fig. \ref{fig:TEx}. $\rvert T(E_x) \rvert^2$ increases with increase in energy $E_x$. Note that the energy $E_x$ in Fig. \ref{fig:TEx} runs through both the energy stages. 
\begin{figure}[hpt]
	\centering
	\includegraphics[width=4.0in,height=2.3in]{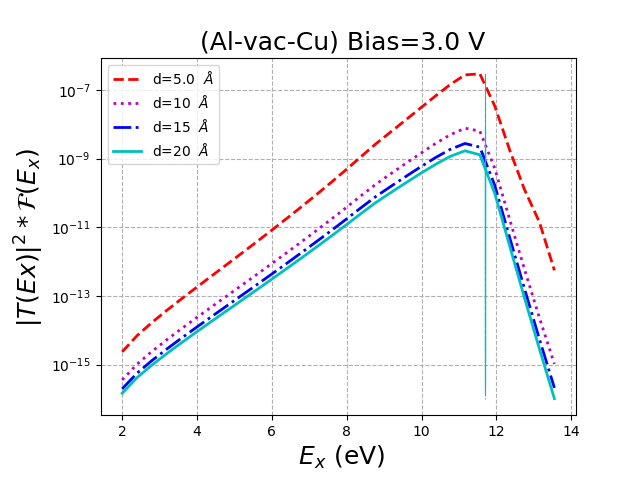}
	\caption{Plot of transmission probability vs $E_x$ for bias voltage $= 3\, V$ and $ d \,=\, 5, \,10,\, 15,\, \text{and}\, 20$ \r{A} (Al–vac–Cu)} 
	\label{fig:TBLow}	
\end{figure}

Fig. \ref{fig:TBLow} plots tunneling probability as a function of $E_x$ for fixed Bias of $3\,V$. Fig. \ref{fig:TBLow}, shows that the maximum contribution to the tunneling occurs for $E_x$ lying in a very narrow energy band whose upper limit is $\eta_1$. For $E_x \leqslant \eta_1$ the minimum value of $\frac{\hbar^2 \, \kappa^2}{2m}$ will be $\phi_1$. The condition that $\frac{\hbar^2}{2m} |f(x)|$ $\left (\text{where}\,f(x) = -\frac{2m}{\hbar^2} (\phi_1-\phi_2+eV_b)\dfrac{x}{d}\right)$ be smaller than this minimum value of $\frac{\hbar^2 \, \kappa^2}{2m}$ is $eV_b < \phi_2$. Hence for low bias voltages, $|f(x)|< \kappa$ is true, and perturbative solutions are valid. For $eV_b > \phi_2$ the Airy function methods will do just fine. 

\bigskip

\subsection{\label{sec:level4} Very Low Bias}
\par In the low bias regime (0.1V to 1V) the Airy Function $Ai$ becomes very small causing underflow problems and the Airy function $Bi$ becomes very large causing overflow problems in the calculation. Even introduction of denominators as in equation (\ref{phden19}), doesn't seem to help. Instead it is more useful to convert the Schr\"odinger equation to the following inhomogeneous equation
\begin{equation}\label{Greq}
\mathscr{L}\Psi(x)=f(x)\Psi(x)
\end{equation}  
where 
\begin{equation}\label{op}
\mathscr{L}=\dfrac{d^2}{dx^2}-\kappa^2
\end{equation} 
\begin{equation}\label{kap}
\dfrac{\hbar^2\, \kappa^2}{2m}=\eta_1+\phi_1-E_x
\end{equation}
\begin{equation}\label{fx}
\dfrac{\hbar^2}{2m}f(x) = -(\phi_1-\phi_2+eV_b)\dfrac{x}{d}
\end{equation}
The term $f(x)$ in the equation (\ref{Greq}), is usually very small for low bias voltages and low contact potentials, permitting a perturbative solution of equation (\ref{Greq}). This will be shown in the next subsection where the Fermi Factor is discussed in detail.

\bigskip

\par Define the Green function for the operator $\mathscr{L}$ of equation (\ref{op}) as 
\begin{equation}\label{Gr}
\mathscr{L}G(x,x')=\delta(x-x')
\end{equation}
Let the wavefunction in the barrier region be
\begin{equation}\label{21}
\Psi(x) = \phi_0(x) + \phi(x) 
\end{equation}
\noindent where $\phi_0(x)$ is the solution to the homogeneous equation  
\begin{equation}\label{hom}
\mathscr{L}\phi_0(x)=0
\end{equation}
From equations (\ref{21}) and (\ref{hom}), $\phi(x)$ will satisfy 
\begin{equation}\label{23}\nonumber
\mathscr{L}\phi(x)=f(x)\,[\phi_0(x)+\phi(x)]
\end{equation}
\begin{equation}\label{ept}
\phi(x') = \int_{0}^{d}G(x,x')f(x)\Big[\phi_0(x)+\phi(x)\Big]\,dx
 +\, \Bigg[ \phi (x) \frac{d G(x,x')}{dx} - G(x,x')\frac{d \phi (x)}{dx} \Bigg]_0^d
\end{equation}

\bigskip

The end point terms ($viz$ at $x=0$ and $x=d$) can be made to vanish if the Green function and the function $\phi$ are both made to vanish at  $x=0$ and $x=d$. Equation (\ref{ept}) can be iterated to obtain terms containing higher and higher powers of $f(x)$.  Here $f(x)$ will be regarded as small enough to neglect higher than first order terms. Keeping only the first order term, gives 
\begin{equation}
\phi(x')=\int_{0}^{d}G(x,x')f(x)\,\phi_0(x)dx 
\end{equation}
and therefore 
\begin{equation}\label{psi}
\Psi(x)=\phi_0(x)+\int_{0}^{d}G(x,x')f(x')\,\phi_0(x')dx'
\end{equation}
\noindent The tunneling amplitude $T^{Gr}$ is calculated using the wavefunction in equation (\ref{psi}) and joining it smoothly with the wavefunctions given in equations (\ref{psi1}) and (\ref{psi3}). 
\begin{figure}[hpt]
	\begin{subfigure}{0.49\textwidth}
		\includegraphics[width=3.2in,height=2.9in]{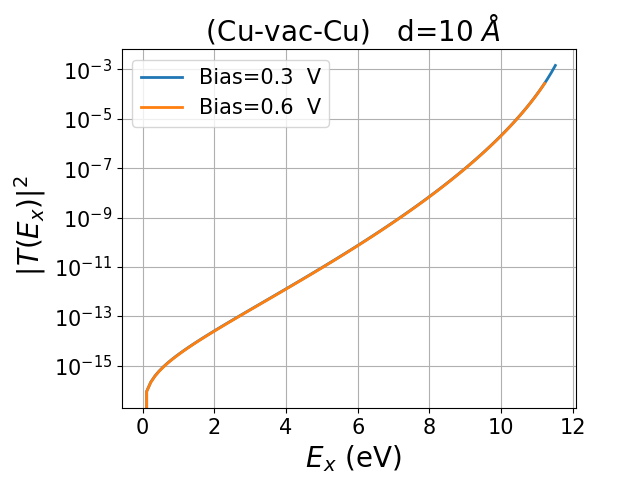}		
		\caption{}
		\label{fig:LowBiasSimmilar}
	\end{subfigure}
	\hfill
	\begin{subfigure}{0.49\textwidth}
		\includegraphics[width=3.2in,height=2.9in]{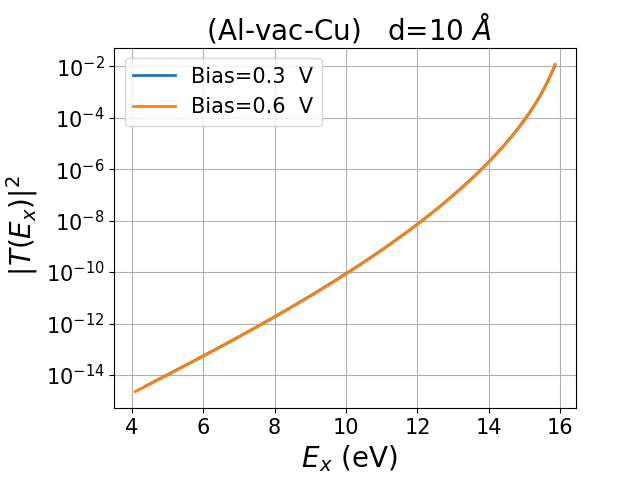}	
		\caption{}
		\label{fig:LowBiasdisSimmilar}
	\end{subfigure}
	\caption{Plot of tunneling probability vs $E_x$ with tip-sample distance = 10 \r{A} and Bias = $0.3,\,0.6\, V$ for (a) similar electrodes (Cu-vac-Cu) (b) dissimilar electrodes (Al-vac-Cu).}
	\label{LowBias}
\end{figure}
\begin{figure}[hpt]
	\begin{subfigure}{0.49\textwidth}
		\includegraphics[width=3.2in,height=2.9in]{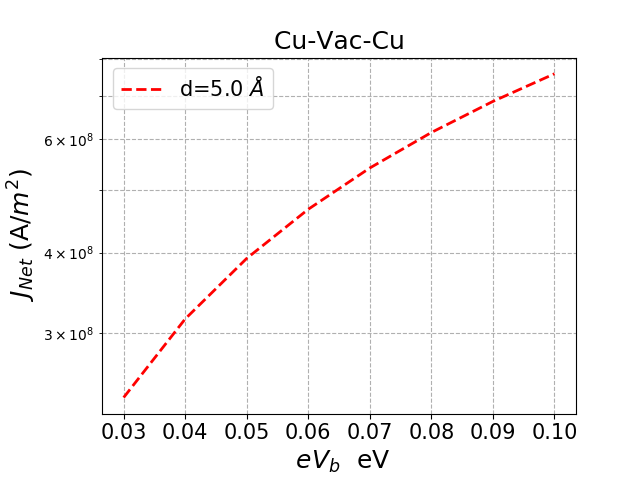}		
		\caption{}
		\label{fig:JLowBiasSimmilar}
	\end{subfigure}
	\hfill
	\begin{subfigure}{0.49\textwidth}
		\includegraphics[width=3.2in,height=2.9in]{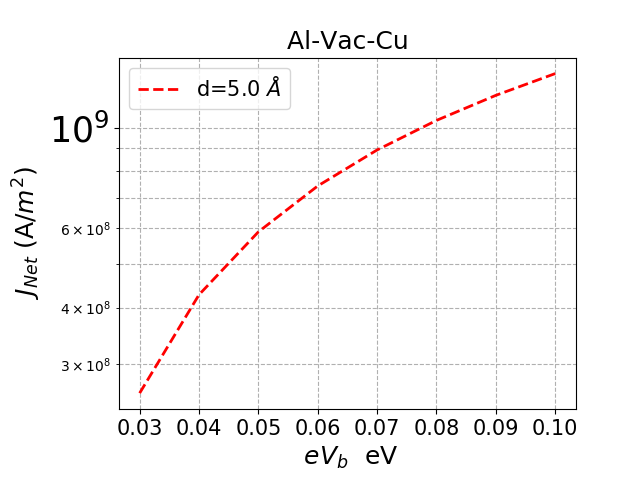}	
		\caption{}
		\label{fig:JLowBiasdisSimmilar}
	\end{subfigure}
	\caption{Plot of current density vs $eV_b$ with tip-sample distance = 5 \r{A} for (a) similar electrodes (Cu-vac-Cu) (b) dissimilar electrodes (Al-vac-Cu).}
	\label{JLowBias}
\end{figure}

This procedure is well discussed in the discussion leading to equation (\ref{fig:TExStg1}). For very low bias, the energy of the electrons in the Energy Stage II is well above the Fermi level of the first electrode and the Fermi Factor (to be discussed in the next section) is extremely small for these energies. Therefore these electrons will make a negligible contribution to tunneling and hence the Green function method outlined above is not extended to the second stage. Instead, the second stage contribution for very low bias voltages is zeroed out.  Fig. \ref{LowBias} shows a plots of tunneling probability as a function of $E_x$ for similar (Cu-vac-Cu) and dissimilar electrodes (Al-vac-Cu) for very low bias voltage. The tunneling probability is found to increase with $E_x$ in both the cases. For the dissimilar electrodes case  Fig. \ref{fig:LowBiasdisSimmilar} the slope of the potential is positive, and the energy diagram is not very different from that shown in Fig. [3 a]. However the Fermi level of the second electrode lies below that of the first electrode and it would appear that even in this case the forward tunelling will be more favored than the reverse tunneling as shown in the plots in Fig. \ref{JLowBias}.

\section{Results and Comparisons}
\begin{figure}[h]
	\centering
	\includegraphics[width=5.8in,height=4.2in]{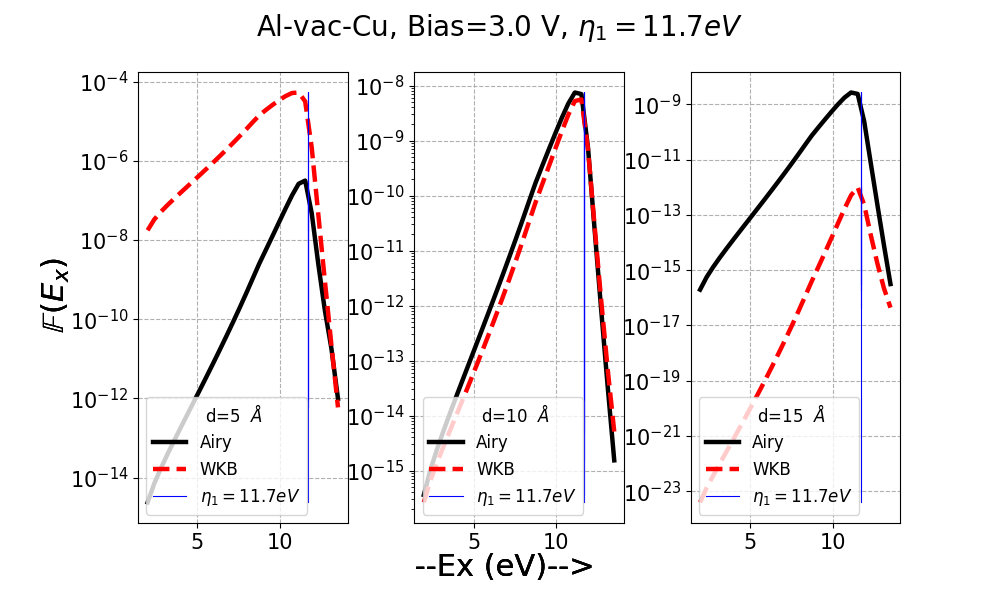}	
	\caption{Plot of transmission probability (Airy compared with WKB) vs $E_x$ for bias voltage $= 3$ V and d = 5, 10, and 15 \r{A} (Al–vac–Cu).}
	\label{fig:TPF}
\end{figure}

Fig. \ref{fig:TPF} shows a plot (Black solid line) of  ($\mathbb{F}_{Airy}(E_x)$) for dissimilar metal electrodes Al (tip) and Cu (sample) and contrasts them with plots (red dashed line) of the corresponding function $\mathbb{F}_\text{WKB}(E_x)$ for a bias voltage of $3$ V and tip-sample distances of $\,5 \,$, $\,10 \,$ and $\,15$ \r{A}. where
$$\mathbb{F}_{Airy}(E_x) =  \rvert T(E_x) \rvert^2  \mathcal{F}(E_x)$$
and 
$$\mathbb{F}_\text{WKB}(E_x) = D(E_x)_{WKB} \big ( F_1(E_x) -F_2(E_x) \big)$$

$D(E_x)_{WKB}$ is described in equation (12) of Chapter 2.
For a bias voltage of $3$ V, $\,\mathbb{F}_\text{WKB}(E_x)$ is found to exceed the $\mathbb{F}_\text{Airy}(E_x)$  for low tip-sample distances such as $d = 5$ \r{A}. The two are nearly equal for $d = 10$ \r{A}, however $\mathbb{F}_\text{Airy}(E_x) > \mathbb{F}_\text{WKB}(E_x)$ for larger tip sample distances such as $d = 15$ \r{A}. Thus the WKB approximation does not provide an accurate  dependence of the tunnel current densities on the tip-sample distances. This feature suggests that determination of surface topographies of samples would be inaccurate, if the image processing algorithm uses WKB approximation.

\bigskip

\begin{figure}[hpt]
	\begin{subfigure}{0.49\textwidth}
		\includegraphics[width=2.9in,height=2.6in]{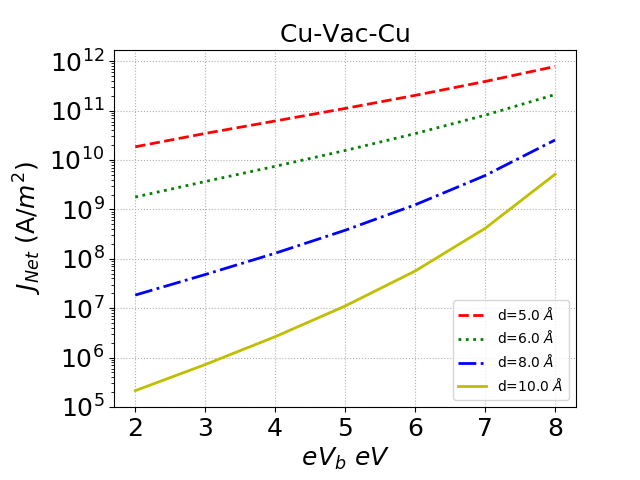}		
		\caption{}
		\label{fig:Jvsbias}
	\end{subfigure}
\hfill
	\begin{subfigure}{0.49\textwidth}
		\includegraphics[width=2.9in,height=2.6in]{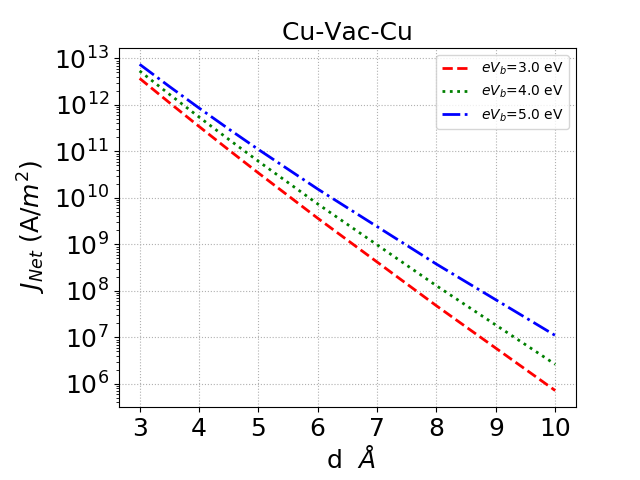}	
		\caption{}
		\label{fig:Jvsdist}
		\end{subfigure}
	\caption{(a) Plot of $J_{Net}$ vs Bias for tip-sample distance = 5, 6, 8, 10 \r{A}.(b) Plot of $J_{Net}$ vs tip-sample distance for bias = 3, 4, 5 V.}
	\end{figure}

The plots of the net current density $J_{Net}$ as a function of bias voltage for identical electrodes (Cu–vac–Cu) is shown in Fig. \ref{fig:Jvsbias}. The current density increases rapidly with the bias voltage at all distances. 
Fig. \ref{fig:Jvsdist} shows the plot of tunneling currents density $J_{Net}$ vs tip–sample distance d in similar (Cu–vac–Cu) pair of electrodes for bias voltages of 3, 4, and 5 V. $J_{Net}$ is found to decrease exponentially with increasing tip-sample distance for all bias voltages. This behavior is also qualitatively reproduced by almost all calculations of tunneling current density, including those that use the WKB approximation. The current density calculated using Simmon's approximation in Chapter 2 (Fig. 2.7 b) for tip-sample distance $d=10$ \r{A} agrees qualitatively with the current densities calculated using (Fig. \ref{fig:Jvsdist}) Airy functions in the calculation of tunneling probability. 

\begin{figure}[hpt]
	\begin{subfigure}{0.49\textwidth}
		\includegraphics[width=2.9in,height=2.6in]{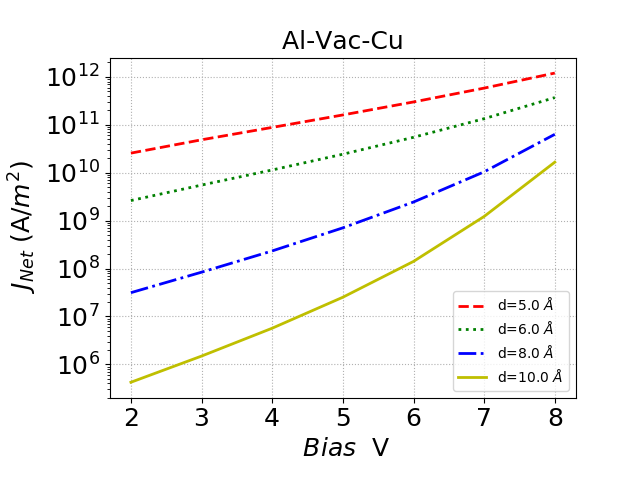}		
		\caption{}
		\label{fig:JFor}
	\end{subfigure}
	\hfill
	\begin{subfigure}{0.49\textwidth}
		\includegraphics[width=2.9in,height=2.6in]{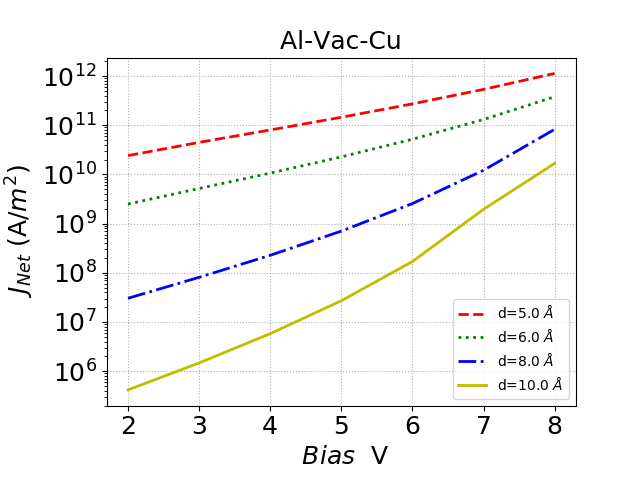}	
		\caption{}
		\label{fig:JRev}
	\end{subfigure}
	\caption{(a) Plot of $J_{Net}$ vs Bias for tip-sample distance = 5, 6, 8, 10 \r{A} for forward current.(b) Plot of $J_{Net}$ vs tip-sample distance for bias = 3, 4, 5 V for reverse current.}
	\label{JForRev}
\end{figure}

For dissimilar metal electrodes, reversing the bias is equivalent to interchanging the electrodes and maintaining positive bias. Results of such switching are presented for the case of Al-vac-Cu where $\vert J_{Net} \vert$ is plotted as a function of $\vert eV_b \vert$ for $2 < eV_b < 8 V$ and several tip-sample distances (refer Fig. \ref{JForRev}). The magnitude of the net current densities are found to be nearly the same for both  Al-vac-Cu and for the Cu-vac-Al systems. The current density for either of these two systems is only a function of $\vert eV_b \vert$ despite the fact that the electrode parameters of the two electrodes are considerably different from each other. Thus the plot of $J_\text{net}$ versus $V_b$ is nearly symmetric for positive and negative values of the bias.

\begin{figure}[h]	
	\begin{subfigure}{0.49\textwidth}
		\includegraphics[width=3.0in,height=2.55in]{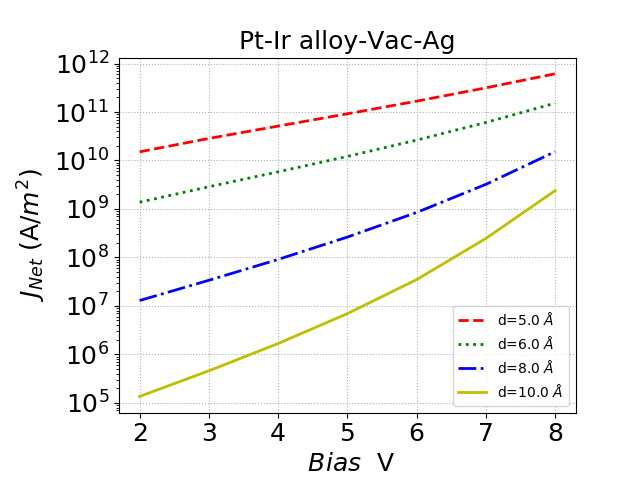}	
		\caption{}
		\label{fig:JWAu}
	\end{subfigure}
	\hfill
	\begin{subfigure}{0.49\textwidth}
		\includegraphics[width=3.0in,height=2.55in]{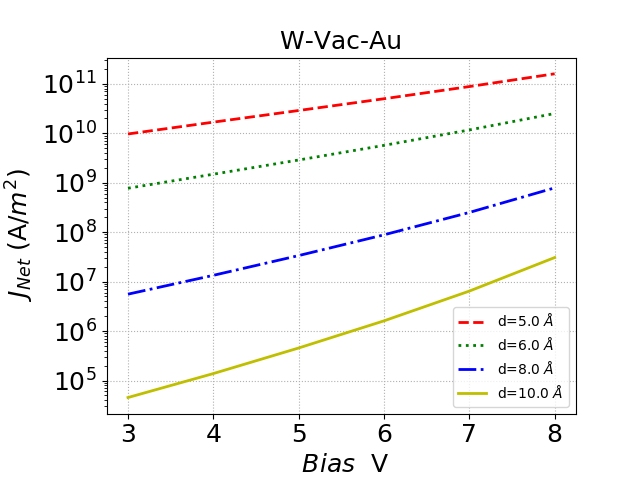}	
		\caption{}
		\label{fig:JPtIrAg}
	\end{subfigure}
	\caption{Plot of $J_{Net}$ vs Bias for tip-sample distance = 5, 6, 8, 10 \r{A} for (a) Pt-Ir alloy-vac-Ag.(b) W-vac-Au.}
	\label{JWAuPtIrAg}
\end{figure}

\bigskip
STM tips are usually made either of Pt-Ir alloy or tungsten \cite{RScIn}. The Pt-Ir alloy is suitable for the use under ambient conditions because Pt is relatively inert to oxidation and the addition of a small fraction of Ir increases the hardness of the tip and hence its durability.  The composition of the alloy that is used is usually $90 \%$ Pt, and $10 \%$ Ir.  The Fermi energy and work function for this composition is approximately estimated to be $\eta_{tip} = 0.9\, \eta (\text{Pt}) + 0.1 \, \eta (\text{Ir})$ and $ \phi_{tip} = 0.9\, \phi (\text{Pt}) + 0.1\, \phi (\text{Ir})$. The calibrating sample is usually silver (Ag) 111 face whose $\eta_{sample}(\text{Ag}\, (111))= 5.51$ eV and $\phi_{sample}(\text{Ag}\, (111))= 4.46$ eV.
Tungsten tips are also widely used because of their low cost. Tungsten is inert, hard and can be easily produced by electrochemical etching. Fig. \ref{JWAuPtIrAg} shows the plot of current density vs bias voltage for  Pt-Ir alloy electrode (tip) and Ag electrode (sample) and also for tungsten (first electrode) and gold (second electrode) W-vac-Au. Both plots show that the current density increases with bias for all distances, as expected.

\bigskip

In this Chapter a Planar Model of the Electrode-Vacuum-Electrode configuration for STM is used to calculate tunneling current densities for both low and high bias voltages. Airy function solutions to the Schrödinger equation in the barrier region are used to compute tunneling probabilities, from which current densities are obtained. The contribution of electrons whose energies range from $0$ to $(\eta_1 + \phi_1)$ are summed to determine the current density. Pauli effects are explicitly introduced in the calculation of current density. The calculation of current densities for the trapezoidal potential is reported by Malati and Kulkarni \cite{dessai2022calculation}

\bigskip

%
%
%
%


	\chapter{Currents From Current Densities}\label{chap4}
%
%
%
%
\section{Introduction}

\par Chapters 2 and 3 discusses the planar model calculation of current densities. Planar models cannot directly give currents. In order to obtain currents, which are measured in a Scanning Tunneling Microscope (STM),  one has to depart from the planar model, and work with realistic shapes for the tip and the sample surfaces. At least the tip will be far from planar in shape.  A direct calculation of currents for curved surfaces in a 3-dimensional configuration would be very complicated. Such a calculation would be computationally expensive, without providing a commensurate gain in physical understanding of the tunneling process inside of an STM. A reasonably simple planar model calculation does indeed provide most of the insight necessary to understand the working of an STM and the means to derive some of its characteristics. Therefore instead of discarding the current densities obtained in the planar model, in favour of a full 3-dimensional curved surfaces current calculation, a method to utilise the planar model current densities, and obtain currents by incorporating the tip and sample shapes, needs to be found.

\bigskip

This method was first devised  and implemented by Saenz and Garcia \cite{SG} in year 1994. Their idea consisted of first identifying the field lines in the curved-surfaces geometry, and to assume that all tunneling would occur principally along the field lines. One could think of a realistic STM being made up of 
infinitesimally thin STM's, with infinitesimal areas $dS$. Each infinitesimally thin STM would consist of two infinitesimally-small plane-area conducting surfaces, and the region between them would consist of an infinitesimally thin tube containing the field line (line of force) that joins a point on the tip to the corresponding point on the sample. This infinitesimal planar STM would have a tip-sample distance equal to the length of the field line that joins the point $P$ on the tip surface and the corresponding point Q on the sample surface. The current density $j_P$ for this infinitesimal STM is calculated using the planar model techniques described in Chapter 2. Let $dS_P$ be the infinitesimal area on the tip surface that is the cross section of the tube surrounding the field line. The current $dI$ carried by this tube is then  given by $dI=j_P\,dS_P$. These currents are summed over all elements of area of all tubes contained in the original realistic STM. Since each such thin tube encloses a field line, the sum over tubes can be regarded as a {\it sum over field lines}.  

\bigskip

This task is made much easier if the tip shape and the sample shape are coordinate surfaces of the same orthogonal curvilinear coordinate system. This would ensure that one set of surfaces could model the surfaces of the tip and the sample, and the field lines would all lie on the orthogonal coordinate surfaces. If the Lapalce equation is seperable in this coordinate system, then it could be reduced to a one dimensional equation in the coordinate that specifies the equipotential surfaces.  A very convenient coordinate system which describes the tip shape very well would be the prolate spheroidal coordinate system (PSC System). In this system, both the tip and the sample surface would be confocal hyperboloids and the field lines would lie on the prolate spheroid. In this coordinate system the  hyperboloids are equipotential surfaces and the field lines would lie on the prolate spheroids. The Laplace equation for the electrostatic potential in this coordinate system becomes one dimensional and can be analytically solved. This coordinate system and some of its properties are described in the next section.
 
\section{Tip \& Sample Surfaces In The PSC System}
\begin{figure}[h]
	\centering
	\includegraphics[width=3.5in,height=3.0in]{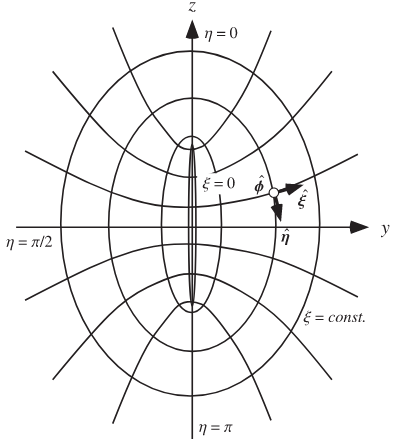}
	\caption{Prolate Spheroidal coordinates}
	\label{fig:PSC}
\end{figure}

The prolate spheroidal coordinate system \cite{SG}\cite{moon}\cite{morse}\cite{russel} is a three-dimensional orthogonal coordinate system $(\eta, \xi, \phi)$ that results from rotating the two-dimensional elliptic coordinate system about the focal axis of the ellipse, i.e., the symmetry axis on which the foci are located as shown in Fig. \ref{fig:PSC}. 
The transformation equations between the prolate spheroidal coordinate system and the cartesian coordinate system are given as follows 
\begin{equation}
x=\rho \cos\phi, \quad y=\rho \sin\phi, \quad  z=a\xi \eta
\end{equation}
where $\rho=a \sqrt{\xi^2-1} \sqrt{1-\eta^2}$,  and $a$ is a constant.
The ranges of the coordinate values are given by
\begin{equation}\label{range}
1\leqslant \xi \leqslant \infty, \quad -1\leqslant \eta \leqslant 1, \quad  0\leqslant \phi \leqslant 2\pi
\end{equation}
\noindent The distance element in this coordinate system is $ds^2 = d\rho^2+\rho^2d\phi^2+dz^2$ is evaluated to 
\begin{equation}\label{len}
ds^2 = a^2 \Big(\dfrac{\xi^2-\eta^2}{\xi^2-1} \Big)d\xi^2 + a^2 \Big(\dfrac{\xi^2-\eta^2}{1-\eta^2} \Big)d\eta^2 + \rho d\phi^2
\end{equation}
\begin{equation}\label{lensca}
ds^2 = h_\xi^2 d\xi^2 + h_\eta^2 d\eta^2 + h_\phi^2 d\phi^2 
\end{equation}
where,  $h_\xi,  h_\eta, h_\phi $ are the scale factors given by
\begin{equation}\label{scafac}
h_\xi = a \sqrt{ \Big(\dfrac{\xi^2-\eta^2}{\xi^2-1} \Big) }, \quad
h_\eta = a \sqrt{ \Big(\dfrac{\xi^2-\eta^2}{1-\eta^2} \Big) }, \quad
h_\phi = a \sqrt{(\xi^2-1)(1-\eta^2)}
\end{equation}
The volume element in this coordinate system is then
\begin{equation}\label{volele}
 dV = a^3 (\xi^2-\eta^2) \, d \xi \,d\eta \, d\phi \, = 2 \pi a^3(\xi^2-\eta^2) \, d\xi \, d\eta
\end{equation}
and the surface element on the hyperboloid sheet is
\begin{equation}\label{surele}
dS =  a^2 \sqrt{(\xi^2 - \eta^2)(1-\eta^2)}\,d\xi \, d \phi
\end{equation}
The length of a field line characterized by $\xi$ is 
\begin{equation}\label{dfl}
d_{fl}(\xi) = a \int_{\eta_{sample}}^{\eta_{tip}}\frac{\sqrt{\xi^2-\eta^2}}{\sqrt{1-\eta^2}} d\eta
\end{equation}
This is independent of the azimuthal angle $\phi$, due to the rotational symmetry about the $z - $ axis. The smallest value of $d_{fl}$ would correspond to $\xi = 1$ and for this it would be 
\begin{equation}\label{dfl_1}
d = d_{fl}(1) = a(\eta_{tip}- \eta_{sample})
\end{equation}
$d$ is usually called the tip-sample distance. If the sample surface is the $x - y$ plane then $\eta_{sample} = 0$ and the equation (\ref{dfl_1}) becomes $ d_{fl}(1) = a \,\eta_{tip}$

\begin{figure}[h]
	\centering	
	\includegraphics[width=3.2in,height=3.7in]{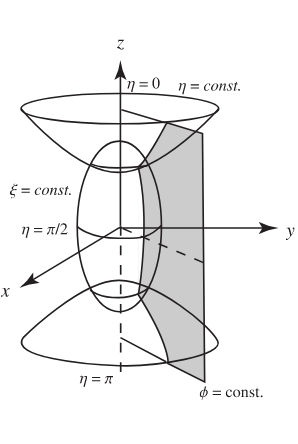}
	\caption{Coordinate Surfaces for constant $\eta$, $\xi$ and $\phi$ in Prolate Spheroidal coordinate system}
	\label{fig:2}
\end{figure} 

\bigskip

Fig.\ref{fig:2} describes the coordinate surfaces of this coordinate system. The two sheeted confocal hyperboloids shown in this figure, are a one parameter family of surfaces, whose equation in cartesian coordinates is  
\begin{equation}\label{hyp}
\frac{z^2}{a^2\eta^2} - \frac{(x^2+y^2)}{a^2(1-\eta^2)}=1
\end{equation}
The above equation represents two hyberboloid sheets characterised by positive and negative values of $\eta$. Note that these sheets intersect the $z - $ axis at $z = \pm a \eta$. Only positive values of $\eta$ are considered here. Note $\eta=0$ represents the x-y plane and can be chosen to represent the sample surface if it is flat. Note very small local surface features such as hills and valleys which are small departures from the plane shape in the sample surface cannot be described by this coordinate system. 

\bigskip

The positive part of the z-axis is characterised by $\eta=1$. Therefore an extremely sharp tip like surface can be generated for a value of $\eta < 1$ but very close to it. Typical values used are $\eta_{tip} \in [0.8 \,\,\text{to}\, \,0.95]$. It is desired to characterize the sharpness of the tip represented by the hypoboloid sheet by means of the radius of curvature of the surface measured at its lowest point $viz$ at $z = a\, \eta_{tip}$, assuming that the sample surface is the $x - y$ plane, so that $\eta_{sample} = 0$. Consider the curve formed by the intersection of this hyperboloid sheet with the $z - x$ plane. Actually any plane containing the $z - $ axis will do. The equation of this curve will be 
\begin{equation}\label{eq5}
\frac{z^2}{a^2\eta_{tip}^2} - \frac{x^2}{a^2(1-\eta_{tip}^2)}=1
\end{equation}
In terms of $z^\prime = \dfrac{dz}{dx}$ and $z^{\prime \prime} = \dfrac{dz^\prime}{dx}$, the radius of curvature is given by 
$R = \dfrac{(z^{\prime 2}+1)^{3/2}}{z^{\prime \prime}}$ \\
\noindent For $x = 0$, $z^\prime(0) = 0 \quad \text{and} \quad z^{\prime \prime}(0) = \dfrac{\eta_{tip}}{a(1-\eta_{tip}^2)} $ using equation (\ref{eq5}) for $x =0$ gives\\  

\begin{equation}\label{etatip}
R =  \dfrac{a(1-\eta_{tip}^2)}{\eta_{tip}}, \quad \eta_{tip}=\sqrt{\dfrac{d}{d+R}} \quad \text{and} \quad a = \sqrt{d(d+R)}
\end{equation}
The radius of curvature of the tip apex $R$, and the tip sample distance $d$, determine the value  of $\eta_{tip}$. 
\bigskip

The prolate spheroids in this coordinate system are generated by rotating an ellipse about the $z - $ axis. Each prolate spheroid is characterized by the value of the coordinate $\xi$, The value $\xi = 1$ represents a segment of length $2 a$ on the $z - $ axis. The equation of this surface in Cartesian coordinates is given by  
\begin{equation}
\frac{\rho^2}{a^2(\xi^2-1)}+\frac{z^2}{a^2\xi^2}=1
\end{equation} 
These are surfaces of constant $\xi$ and they are every where orthogonal to the surfaces of constant $\eta$. If the surfaces of constant $\eta$ are chosen to be equipotential surfaces, then the lines of constant $\xi$ and $\phi$ represent electric field lines or the lines of force of the electrostatic potential. These lines will lie on the surfaces of confocal prolate spheroids.

\section{The Russell Potential}

Since the geometry of the tip - sample system is sought to be described by confocal surfaces of the Prolate Spheroidal Coordinate system, it should be natural to use the electrostatic potential, that is a solution to the Laplace equation in this coordinate system. The  Laplacian operator $\nabla^2$ in the Prolate Spheroidal Coordinate System is given by 
$$\nabla^2 \Phi= \frac{1}{a^2 \left(\xi ^2-\eta ^2\right)}\left[\frac{\partial }{\partial \eta }\left \{(1-\eta ^2)\frac{\partial \Phi }{\partial \eta }\right \}+\frac{\partial }{\partial \xi }\left \{(\xi ^2-1)\frac{\partial \Phi }{\partial \xi }\right\}\right]+\frac{1}{\rho^2}\frac{\partial ^2\Phi }{\partial \phi ^2}$$
Due to the rotational symmetry in the system about the $z- $ axis, the potential must be independent of the azimuthal angle $\phi$. If boundary conditions are specified confocal hyperboloids, which are equipotentials, then the potential $\Phi$ is only a function of $\eta$, and is independent of $\xi$. The Laplace Equation for $\Phi$ becomes 
$$\frac{\partial }{\partial \eta }\left[(1-\eta ^2)\frac{\partial \Phi }{\partial \eta }\right]=0 ; \quad \eta \in [\eta_s,\, \eta_{tip}]$$
whose solution is 
\begin{equation}\label{eqPhi}
\Phi(\eta)=\frac{A}{2}  \log \left(\frac{1+\eta}{1-\eta }\right)+B
\end{equation}
where $A$ and $B$ are constants, to be determined by the boundary conditions imposed on $\Phi$ at the surfaces $\eta = \eta_{s}$ which represents the sample surface, and $\eta = \eta_{tip}$ which represents the tip surface. 
The potential energy of an electron in the potential $\Phi(\eta)$ in equation (\ref{eqPhi}) is $U(\eta) = -e \Phi(\eta)$ where $e$ is the magnitude of the charge of the electron. The Boundary conditions on $U(\eta)$ are 
\begin{equation}\label{15}
U(\eta = \eta_{tip}) = \eta_1 + \phi_1\quad \text{and} \quad U(\eta = 0) = \eta_1+\phi_2 - eV_b
\end{equation} 
where the $\eta_s = 0$ is imposed so as to describe a flat sample surface. Here $\eta_1, \phi_1$ are the Fermi energy and the work function of the first electrode, $\eta_2$ is the Fermi energy of the second electrode and $V_b$ is the bias voltage, defined in such a way that the tip (first electrode) is negatively biased with respect to the sample (second electrode).
Define the function 
\begin{equation}\label{16}
\lambda (\eta) = \log (\dfrac{1+\eta}{1-\eta})
\end{equation}
and $\lambda_t = \lambda(\eta_{tip})$, and $\lambda_s = \lambda(\eta_s) = 0$,
\begin{figure}[h]
	\centering	
	\includegraphics[width=4.3in,height=3.8in]{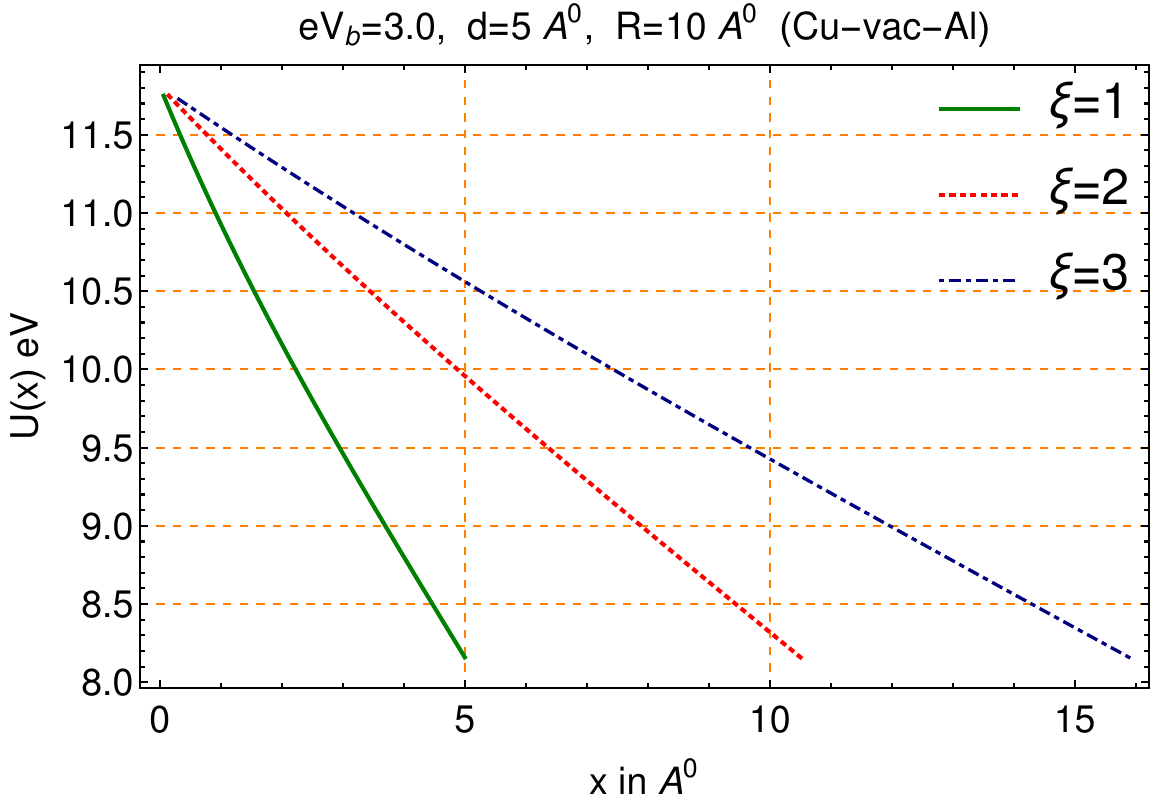}
	\caption{Plot of $U(x)$ vs $x$ for Bias = $3 \,\,V$, d $= 5$ \r{A}, R = $10$\r{A} }
	\label{fig:3}
\end{figure} 

\bigskip

With the Boundary conditions on $U$ specified in equation (\ref{15}), the potential energy becomes 
\begin{equation}\label{17}
U(\eta) = (\eta_1+\phi_2-eV_b) + (\phi_1-\phi_2+eV_b)\, \dfrac{\lambda(\eta)}{\lambda_{\eta_{tip}}}
\end{equation}
 For a given field line which is described by $\,\xi\,$ the length along the field line can be calculated as 
$$ x_{fl}(\eta, \xi) = a \int\limits_\eta^{\eta_{tip}} d \eta \,\sqrt{\Big (\dfrac{\xi^2 - \eta^2}{1-\eta^2} \Big)} $$
It should be possible to invert this equation to find $\eta = \eta(\xi, x_{fl})$. This value of $\eta$ can be substituted in equation (\ref{17}) to get $U[x_{fl}(\xi)] = U[\eta(\xi,x_{fl})]$. This function expressed as a function of a linear distance $x_{fl}$ may be called the Russell Potential. Fig. \ref{fig:3} shows the plots of the Russell Potential as a function of $x$  where $x$ runs from $0$ to $ d_{fl}(\xi)$ for different values of $\xi$. It is seen that the Russell potential shows slight concave curvature, but doesn't depart very much from a straight line joining the points $[x = 0,U = (\eta_1+\phi_1)]$ and $[x = d_{fl}(\xi), U = U(\eta = 0)]$. The equation of the straight line joining these two points is 
\begin{equation}\label{18}
U^L(x(\xi)) = (\eta_1+\phi_1) - (\phi_1-\phi_2+eV_b)\, \dfrac{x}{d_{fl}(\xi)}
\end{equation}
The above equation represents the linearised version of the Russell potential. Note that this potential is trapezoidal. However it is different for different field lines. 

\section{Integration Over Field Lines}

\begin{figure}[h]	
		\centering
		\includegraphics[width=5.0in,height=3.5in]{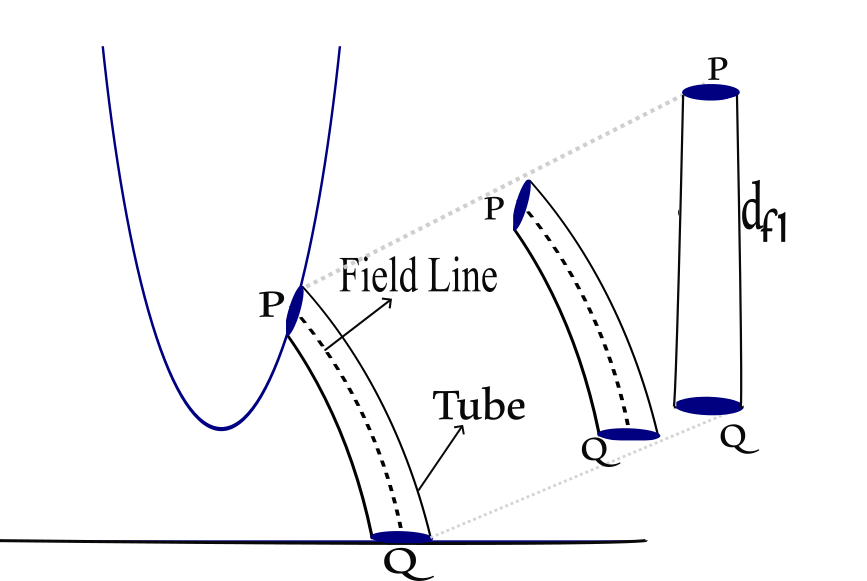}
	\caption{Field line plus tube in original STM, Pull out of the infinitesimal STM, infinitesimal STM straightend out.} 
	\label{FLineTube}
\end{figure}
Fig. \ref{FLineTube} shows a tip sample configuration in which the tip is hyperboloid sheet with $\eta_{tip} > 0$ and $\eta_{tip}$ is very close to 1 and in which the sample surface is flat and is described by $\eta = 0$. Consider a point P on the tip surface.  Its coordinates on the PSC system would be $(\eta_{tip},\,\xi_P,\, \phi_P)$. The field line that originates from P touches the sample surface at a point Q. The length of the field line PQ is $d_{fl}(\xi_P)$ where the function $d_{fl}$ is given in equation (\ref{dfl}) with $\eta = \eta_{tip}$. Consider a small element of area $dS_p$ at the point P on the tip surface, bounded by the field lines through $\xi_P \pm d \xi/2 $ and the azimuthal angles $\phi_P \pm d \phi/2$. Construct a thin tube that contains $dS_P$ and encloses the field line through P. The element of area $dS_P$ is given by equation (\ref{surele}) above with $\eta=\eta_{tip}$. Fig. \ref{FLineTube} shows a pull out of this combination of conducting surfaces with infinitesimally small area $dS_P$ separated by the distance $d_{fl}(\xi_P)$. Two basic assumptions will now be made. These may be stated as follows\\

\noindent\underline{Asumption 1.}  The STM with curved tip and flat sample is made up of infinitesimally thin planar STM's each of which consist of two infinitesimally small area conducting surfaces separated by a thin tube of length $= d_{fl}(\xi)$ which encloses the line of force which passes through the point $\xi$.\\
\underline{Asumption 2.} The potential energy of the electron in this infinitesimally thin planar STM is approximated by a linear trapezoidal potential $U^L(x)$ given by equation (\ref{18}).

\bigskip

\begin{figure}[h]
	\centering
	\includegraphics[width=2.0in,height=2.1in]{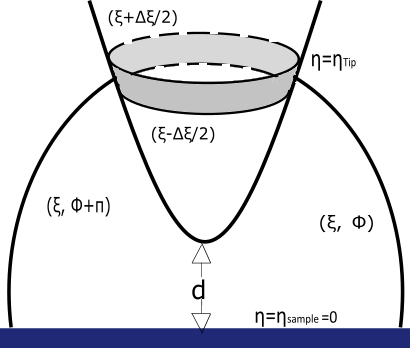}
	\caption{Hyperboloid Tip and Spheroidal shaped field line.} 
	\label{fig:tip}	
\end{figure}

The method of calculating the current density $J_{Net}$ due to the linear potential of equation(\ref{18}) will be almost identical to that discussed in Chapter 3.  The tunnel amplitudes will be given by equations 33 and 44 of Chapter 3 in which the tip sample distance $d$ has to be replaced by $d_{fl}(\xi)$ for each field line defined by the value of $\xi$. Thus the Tunnel amplitudes are not only a function of the energy $E_x$ but also of the value of $\xi$. They can then be referred to as $T_I(E_x, \xi)$ and $T_{II}(E_x, \xi)$ respectively for the first and second energy stages. The current density can be calculated from these amplitudes using equation (2.18) in which $d$ is replaced by $d_{fl}(\xi_P)$. The current densities so obtained may be labelled as $j_{Net}(\xi)$. The current through this tube is given by 
\begin{equation}
dI(\xi_P) = J_{Net}(\xi_P)\,a^2\,\sqrt{(\xi_P^2-\eta_{tip}^2)(1-\eta_{tip}^2)}\,d\xi_P\,d\phi
\end{equation}
The above can be integrated over $\phi \, \in \,[0,2 \pi]$ and $\xi_P \, \in\,[1,\xi_{max}]$ to find the net current due to the curved tip and the flat sample. This integration is called {\it integration over the field lines}. Since there is rotational symmetry about the z-axis, the $\phi$ integration yields a factor of $2\, \pi$  and therefore
 \begin{equation}
 I = 2\, \pi \, a^2 \sqrt{(1-\eta_{tip}^2)} \,\int_{1}^{\xi_{max}} J_{Net}(\xi)\,\sqrt{\xi^2-\eta_{tip}^2}\,d\xi
 \end{equation}
Due to the symmetry under rotation about the $z -$ axis, integrating over $\phi$ makes the elemental area $dS_P$ of the tube to assume the shape of a ribbon that wraps around the tip. This is shown in Fig. \ref{fig:tip}. The two field lines shown in this figure correspond to angles $\phi$ and $\pi + \phi$. 
It is necessary to introduce an upper cutoff at $\xi_{max}$ for computational convenience. It has been shown in Chapter 2 that the current density in the planar model decreases exponentially with increasing tip-sample distance. In the above calculation the variable that plays the role of the tip-sample distance is $d_{fl}(\xi)$ which increases linearly with $\xi$. Thus for large enough $\xi$, $J_{Net}$ will be negligible and therefore a $\xi_{max}$ can be chosen such that any contribution to the integral for $\xi > \xi_{max}$ will be negligibly small compared to that for $1\, \leqslant \, \xi  \,\leqslant \, \xi_{max}$. It is found that any value for $\xi_{max}$ between $3$ and $5$ will suffice. The net current density $J_{Net}(\xi)$ is obtained by integrating the product of the tunneling probability $|T|^2$ and the Fermi Factor over the energy $E_x$ of the tunneling electrons for the translational motion along the x-direction as shown in equation (3.16). 

While the discussion of the integration over the field lines above may leave the impression that this method is primarily valid only for the prolate spheroidal coordinate system. But this is not true. Assumption 1 will be valid for any pair of curved conducting surfaces on which Boundary Conditions on either the electrostatic potential or the potential energy can be specified. However for a given pair of surfaces it is necessary to construct a large number of field lines. It is not often possible to develop analytical expressions describing these field lines, much less being able to integrate over them easily. Much of this work will be entirely computational. Two interesting examples would that of a sphere - plane combination or a cone - plane combination. In both these instances the potential function is given by an infinite series and it is nearly impossible to find analytical expressions for the equations of the field lines for these geometries. It is a fortunate coincidence that the analytical expressions are possible for the hyperboloid surfaces in the prolate spheroidal Coordinate system. Hence the calculation of nonplanar geometries is limited to the confocal surfaces of this coordinate system only.
	
\bigskip

The tip - sample configuration of an STM with a sharp tip and a flat sample is well described by confocal hyperboloids in the prolate spheroidal coordinate system. The relations between the radius of curvature at the tip apex of the tip and the tip - sample distance $d$ and the parameter $a$ of the coordinate system are developed. The STM is now assumed to be made up of infinitesimally thin cylindrically shaped tube like STM's with infinitesimal surface area dS and length equal to that of the particular field line it encloses. The potential within this tube is assumed to be a linearised form of the Russell potential, the latter being the solution of the Laplace equation in this coordinate system. Current densities within each tube is calculated exactly as in chapter 3 with the tip - sample distance being replaced by $d_{fl}$; the length of the field line. The current densities in the tubes so obtained are integrated over their surface areas to obtain a current. This method called "Integration over field lines" is described and developed in the prolate spheroidal coordinate system. While the same method could be applied for other pair of curved surfaces, the paucity of analytical expressions for the 'field lines' makes the exercise entirely numerically computational.  

	\chapter{Multislice Calculation of $T(E_x)$ For Nonlinear Potentials}	\label{chap5}
%
%
%
\section{Introduction}
The calculation of the tunnel amplitude for the planar model in Chapter 3 was helped by the fact that the potential throughout the barrier region was trapezoidal and the same exact solution to the  Schr\"odinger equation was valid throughout this region. However it may so happen that for various reasons such as the introduction of the Russell potential or the image potentials, the net potential may not be trapezoidal in the barrier region. In such a case the barrier region is split up into slices and the potential is assumed to be piecewise linear within the slices. Linear combinations of the linearly independent Airy functions $Ai$ and $Bi$ viz $C_j Ai + D_j Bi$ are determined for $j^{th}$ slice and the solutions are matched at the common boundaries of neighbouring slices. This gives matrix relations between the constants $C_j$, $D_j$ and $C_{j-1}$, $D_{j-1}$ $viz$ between the constants belonging to neighbouring slices. The solutions in the first slice are matched to the plane wave solutions in the first electrode and that in the last slice with the plane wave solution in the second electrode. The Tunneling amplitude $T$  is found by eliminating the constants $C_j$ and $D_j$ of all the slices in the barrier region. A calculation of the tunnel amplitude for non trapezoidal potentials using this method which may be called the `Transfer Matrix Method' is described in this Chapter.

\section{Multislice Method for $T_I(E_x)$ in the Planar Model}
Let there be an arbitrary potential described by some function given by 
\begin{equation}
U(x)=\left\{
\begin{array}{@{}lll@{}}
0 & 		x < 0\ &\quad \text{Region I} \\
U_{II}(x) & 		0 \leqslant x \leqslant d &\quad \text{Region II}\\
-\Delta_B&	 x > d &\quad \text{Region III} \\
\end{array}\right.
\end{equation} \\
where $\Delta_B=(\eta_2-\eta_1+eV_b)$. Note $U_{II}(x)$ need not be linear in $x$.
\begin{figure}[h]
	\centering
	\includegraphics[width=5.5in,height=3.0in]{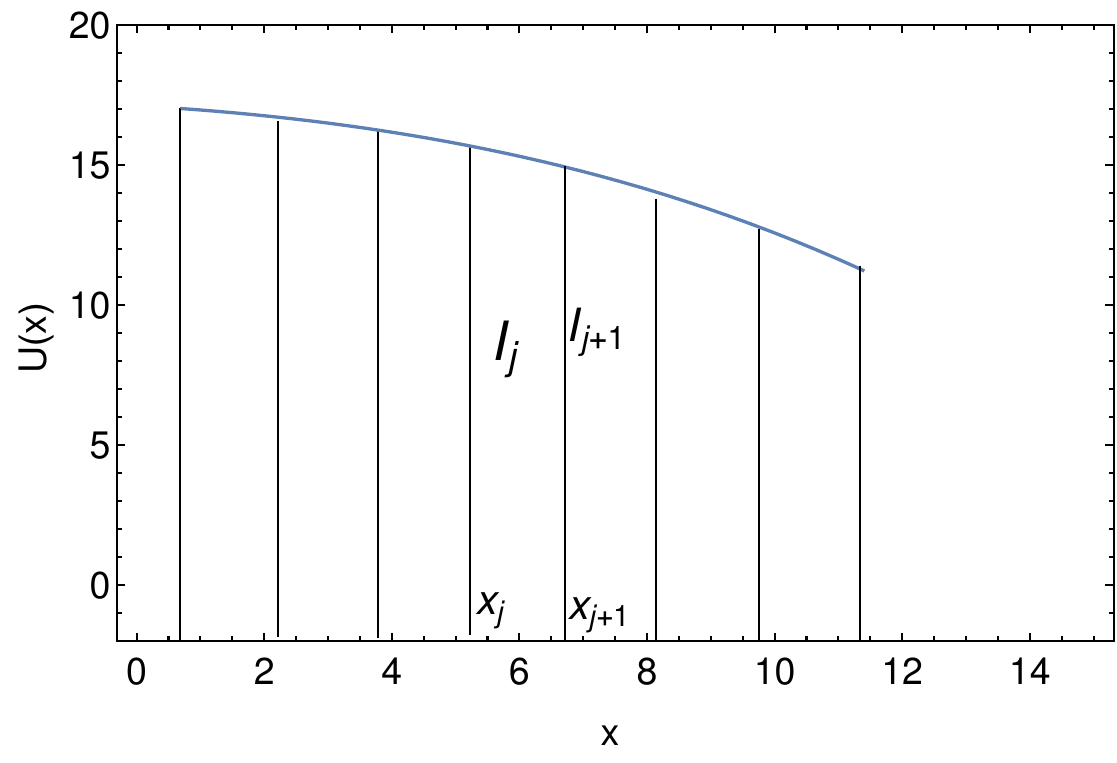}
	\caption{Multiple slices in the barrier region showing two node points at $x_j$ and $x_{j+1}$}
	\label{fig:1}
\end{figure}  
Let the entire region between $[0,d]$ is divided up into $N$ intervals, each of thickness $\Delta x = \frac{d}{N}$. Each interval (slice) is labelled as $I_j$. The left and right boundaries of $I_j$ are at $x_j = j \, \Delta x$  and at $ x = x_j + \Delta x = x_{j+1} $, where $j=0,1,2,3,\cdots N-1$. The boundary points $x_j$ of the intervals are called the nodal points. The value of the potential function $U(x)$ at the node point $x=x_j$ may be denoted by $U(x_j) = U_j$. The linear approximation for the potential $U(x)$ in the interval $I_j$, may be constructed  by 
\begin{equation}
U^L_j(x) =  U_j+\gamma_j (x-x_j)
\end{equation}
where $\gamma_j = \dfrac{U_{j+1}-U_j}{\Delta x}$ \quad for $\,\,x_j < x < x_{j+1}.\,$  Note that the superscript $L$ in $U_j^L(x)$ refers to linearized potential function. Define a remainder function $R_j(x)$ in the interval $I_j$ as 
$$ R_j(x) = U(x) - U^L_j(x) = U(x) - \big[ U_j+\gamma_j (x-x_j) \big] $$
The Schr\"odinger equation for the potential $U(x)$ in the interval $I_j$ is given by 
\begin{equation}\label{Sch}
\Big [\frac{- \hbar^2}{2m} \frac{d^2}{dx^2} + U^L_j(x)+R_j(x) \Big ] \Psi_j(x) = E_x \Psi_j(x)
\end{equation}
substituting for $U^L_j(x)$ and rearranging equation (\ref{Sch})
\begin{equation}
\Big [ \frac{d^2}{dx^2} -\frac{2m}{\hbar^2} (U_j-\gamma_j x_j-E_x +\gamma_j x) \Big ] \Psi_j(x) = \frac{2m}{\hbar^2}R_j(x) \Psi_j(x)
\end{equation}
Define 
\begin{equation}
A_j=\dfrac{2m}{\hbar^2} (U_j-\gamma_j x_j-E_x) \,, \quad 
B_j = - \dfrac{2m}{\hbar^2} \gamma_j 
\end{equation}	 
The Schr\"odinger equation reduces to a following differential equation
\begin{equation}\label{SE_ArbPot}
\frac{d^2 \Psi_j}{dx^2}-(A_j-B_j x)\Psi_j = \dfrac{2m}{\hbar^2} R_j(x)\Psi_j(x)
\end{equation} 
If $R_j(x)$ is small, a piecewise linear approximation to the potential $U(x)$ is obtained by the  $U^L_j(x)$ $\forall \,x \,\in I_j$.  For $R_j(x)$ to be small, the function $U(x)$ must be slowly varying and the $j^{th}$ slice must be thin enough, so  that there isn't appreciable curvature of $U(x)$ within the slice. The slices should therefore be numerous enough to meet the above conditions. This approximation ensures that the solution of the Schr\"odinger equation with the linearized potential $U^L_j(x)$ in each interval $I_j$ can be transformed to the Airy differential equation as shown in Appendix A-1 \cite{AS}. The general solution to this equation can be written as a linear combination of Airy functions $Ai[h_j(x)]$ and  $Bi[h_j(x)]$, where x is transformed to the function $h_j(x)$;  one for each interval. The function $h_j(x)$ may be written as   
$$\,\,h_j = \dfrac{A_j}{B_j^{2/3}}-B_j^{1/3}x$$
Neglecting the remainder function in the R.H.S of equation (\ref{SE_ArbPot}), the transformed Airy differential equation \cite{AS} becomes 
\begin{equation}
\frac{d^2 \Psi_j}{dh^2} - h\Psi_j= 0
\end{equation} 

 The general solution to the above Airy differential equation in the interval $I_j$ is given by 
 \begin{equation}
 \Psi_j[(h_j(x)] = C_j Ai[(h_j(x)]+D_j Bi[(h_j(x))]
  \end{equation}
  where $C_j$ and $D_j$ are constants.
The conditions of continuity of the wavefunction and its derivative at the common boundary $x_j$ of interval $I_j$ and $I_{j-1}$ is 
 $$\Psi_j(x)\vert_{x_j}=\Psi_{j-1}(x)\vert_{x_j} \quad \text{and} \quad \dfrac{d\Psi_j(x)}{dx} \Big \vert_{x_j} = \dfrac{d\Psi_{j-1}(x)}{dx} \Big \vert_{x_{j}}$$
 where $\Psi_j(x)$ and $\Psi_{j-1}(x)$ are the wavefunctions in the $j^{th}$ and $(j-1)^{th}$ intervals. Expressing the above continuity conditions in matrix form  and rearranging gives
 \begin{equation}
 \begin{bmatrix}
 C_j\\ D_j
 \end{bmatrix}=\dfrac{1}{W}\begin{bmatrix}
 Bi^\prime[h_j(x_j)]& -Bi[h_j(x_j)]\\ 
 -Ai^\prime[(h_j(x_j)]& Ai[(h_j(x_j)] 
 \end{bmatrix}
 \begin{bmatrix}
 Ai[h_{j-1}(x_j)] & Bi[h_{j-1}(x_j)]\\ 
 Ai^\prime[h_{j-1}(x_j)]&  Bi^\prime[(h_{j-1}(x_j)]
 \end{bmatrix}
 \begin{bmatrix}
 C_{j-1}\\ D_{j-1}
 \end{bmatrix}
 \end{equation}
 where the prime denote derivative with respect to $x\,$ and $\,W$ is a Wronskian of Airy functions and is defined as \cite{AS} 
 $$W = \Big [ Ai[(h_j(x_j)] Bi{^\prime}[(h_j(x_j)]-Ai^\prime[(h_j(x_j)] Bi[h_j(x_j)] \Big] = \dfrac{1}{\pi}$$  and
$$Ai^\prime[h_j(x_j)]= - B_j^{1/3}\dfrac{d}{dh_j} Ai[(h_j(x)] \Big \vert_{x_j} , \,\, Ai^\prime[h_{j-1}(x_j)]= - B_{j-1}^{1/3}\dfrac{d}{dh_{j-1}} Ai[(h_{j-1}(x)] \Big \vert_{x_j}$$
 $$Bi^\prime[h_j(x_j)]= - B_j^{1/3}\dfrac{d}{dh_j} Bi[(h_j(x)] \Big \vert_{x_j}, \,\,  Bi^\prime[h_{j-1}(x_j)]= - B_{j-1}^{1/3}\dfrac{d}{dh_{j-1}} Bi[(h_{j-1}(x)] \Big \vert_{x_j}$$\\ 
\noindent Define  $\bar{\bar{M_j}}=
 \pi \begin{bmatrix}
 Bi^\prime[h_j(x_j)]& -Bi[h_j(x_j)]\\ 
 -Ai^\prime[h_j(x_j)]& Ai[h_j(x_j)] 
 \end{bmatrix}
 \begin{bmatrix}
 Ai[(h_{j-1}(x_j)] & Bi[(h_{j-1}(x_j)]\\ 
 Ai^\prime[(h_{j-1}(x_j)]&  Bi^\prime[(h_{j-1}(x_j)]
 \end{bmatrix}$
 as the $j^{th}$ transfer matrix. In terms of $\bar{\bar{M_j}}$, the relation between coefficient is 
 \begin{equation}\label{Mjmat}
 \begin{bmatrix}
 C_j\\ D_j
 \end{bmatrix}=\bar{\bar{M_j}}
 \begin{bmatrix}
 C_{j-1}\\ D_{j-1}
 \end{bmatrix}
  \end{equation}
Rewriting the above equation by replacing $j$ by $j-1$ gives  
\begin{equation}
\begin{bmatrix}
C_{j-1}\\ D_{j-1}
\end{bmatrix}=\bar{\bar{M_{j-1}}}
\begin{bmatrix}
C_{j-2}\\ D_{j-2}
\end{bmatrix}
\end{equation}
substituting in equation (\ref{Mjmat}) we have
\begin{equation}
\begin{bmatrix}
C_j\\ D_j
\end{bmatrix}=\bar{\bar{M_j}}\bar{\bar{M_{j-1}}}
\begin{bmatrix}
C_{j-2}\\ D_{j-2}
\end{bmatrix}
\end{equation}
Using similar approach one can find the transfer matrices for all slices in the barrier region in reverse order
 \begin{equation}\label{M_mat}
 \begin{bmatrix}
 C_{N-1}\\ D_{N-1}
 \end{bmatrix}=\bar{\bar{M}}_{N-1}\cdot\bar{\bar{M}}_{N-2}\cdots\bar{\bar{M}}_{1}
 \begin{bmatrix}
 C_{0}\\ D_{0}
 \end{bmatrix} \equiv \bar{\bar{M}}  \begin{bmatrix}
 C_{0}\\ D_{0}
 \end{bmatrix}
 \end{equation}
 where $\bar{\bar{M}}$ is the inner products of transfer matrices, and it is given by 
 \begin{equation}
 \bar{\bar{M}}=\bar{\bar{M}}_{N-1}\cdot\bar{\bar{M}}_{N-2}\cdot \cdots\bar{\bar{M}}_{1}
 \end{equation}
 The wavefunctions in the left electrode $(x < 0)$ is given by equation (3.3). Imposing continuity of wavefunctions and its derivative at $ x=0 $ gives 
\begin{equation}\label{x0}
 1+R = C_{0}\, Ai[h_0(0)] + D_{0}\, Bi[h_0(0)]
\end{equation}
\begin{equation}\label{xd}
 1-R = \dfrac{C_{0}}{i k_1}\, Ai^{\prime}[h_0(0)] + \dfrac{D_{0}}{i k_1}\, Bi^{\prime}[h_0(0)]
\end{equation}
 Adding equations (\ref{x0}) and (\ref{xd}) gives
\begin{equation}\label{add_x0}
 2 =  C_{0} \Bigg[ Ai[h_0(0)] +   \dfrac{Ai^{\prime}[(h_0(0)]}{i k_1} \Bigg]+ D_{0} \Bigg[ Bi[h_0(0)] +   \dfrac{Bi^{\prime}[h_0(0)]}{i k_1}   \Bigg]
\end{equation}
 Define 
 $$
 U_{11} = \dfrac{1}{2}\Bigg[ Ai[h_0(0)] +   \dfrac{Ai^{\prime}[h_0(0)]}{i k_1} \Bigg] \quad \text{and} \quad U_{12} =  \dfrac{1}{2}\Bigg[ Bi[h_0(0)] +   \dfrac{Bi^{\prime}[h_0(0)]}{i k_1}   \Bigg]
 $$
In terms of $U_{11}$ and $U_{12}$ equation (\ref{add_x0}) becomes
\begin{equation}\label{U0}
 1 = C_{0} U_{11} + D_{0} U_{12}
\end{equation} 
 Imposing continuity conditions at $ x=d $ and using the wavefunction inside second electrode $(x > d)$ given by equation (3.4) of Chapter 3 gives 
\begin{equation}\label{T_d}
		T_{I} \, e^{ik_2d} =  C_{N-1}  Ai[h_{N-1}(d)] +  D_{N-1} Bi[h_{N-1}(d)] \\
\end{equation}

\begin{equation}\label{T_primed}
 T_{I} \, e^{ik_2d} = \dfrac{C_{N-1}}{ik_2} Ai^\prime[h_{N-1}(d)] +  \dfrac{D_{N-1}}{ik_2} Bi^\prime[h_{N-1}(d)] 
\end{equation} 
where $T_I$ is the transmission amplitude for tunneling in Energy Stage I. 
Writing equations (\ref{T_d}) and (\ref{T_primed}) in matrix form,
 $$
 T_{I} \begin{bmatrix}
 1\\\\1 	
 \end{bmatrix}= e^{-ik_2d}\begin{bmatrix}
 Ai[h_{N-1}(d)]& Bi[h_{N-1}(d)]\\ \\
 \dfrac{1}{ik_2}Ai^\prime[h_{N-1}(d)] & \dfrac{1}{ik_2}Bi^\prime[h_{N-1}(d)]
 \end{bmatrix}\begin{bmatrix}
 C_{N-1}\\\\  D_{N-1}	
 \end{bmatrix} = \bar{\bar{V}} e^{-ik_2d} \begin{bmatrix}
 C_{N-1}\\\\  D_{N-1}	
 \end{bmatrix}
 $$
where  $\bar{\bar{V}}=\begin{bmatrix}
 Ai[h_{N-1}(d)]& Bi[h_{N-1}(d)]\\ \\ 
 \dfrac{1}{ik_2}Ai^\prime[h_{N-1}(d)] & \dfrac{1}{ik_2}Bi^\prime[h_{N-1}(d)]
 \end{bmatrix}
 $ 
 
 \bigskip
 
\noindent substituting for $ \begin{bmatrix}
  C_{N-1}\\ D_{N-1}	
  \end{bmatrix}$  from equation (\ref{M_mat}) gives 
 $$ T_{I} \begin{bmatrix}
 1\\\\1 	
 \end{bmatrix}= \bar{\bar{V}}\, \bar{\bar{M}}\,\,e^{-ik_2d} \begin{bmatrix}
 C_{0}\\\\  D_{0}	
 \end{bmatrix} = \bar{\bar{S}}\,\,e^{-ik_2d} \begin{bmatrix}
 C_{0}\\\\  D_{0}	
 \end{bmatrix} $$
where $\bar{\bar{S}} = \bar{\bar{V}} \cdot \bar{\bar{M}}$. In terms of the matrix elements of $\bar{\bar{S}}$ $viz.$ $S_{ij} \,\,(i,j=1,2)$ the tunnel amplitude in the Energy Stage I may be obtained as   
 \begin{equation}\label{TunAmp}
 T_{I} = S_{11}\,C_{0} + S_{12}\,D_{0} = S_{21}\,C_{0} + S_{22}\,D_{0}
 \end{equation}
 Solving for $C_{0}$ and $D_{0}$ using equations (\ref{U0}) and (\ref{TunAmp}) gives
 $$C_{0} = \dfrac{S_{22}-S_{12}}{U_{12}(S_{11}-S_{21})+U_{11}(S_{22}-S_{12})}
 \quad \text{and} \quad D_{0} = \dfrac{1-C_{0} U_{11}}{U_{12}}$$ 
 Substitute in the first of equations (\ref{TunAmp}) to get
 \begin{equation} 
 T_I =  \dfrac{ e^{ik_2d}(S_{11}S_{22}-S_{12}S_{21})} {U_{12}(S_{11}-S_{21})+U_{11}(S_{22}-S_{12})}
 \end{equation}
 Thus the tunneling amplitude for the Energy Stage I has been found in the multislice (transfer matrix) approach.

\section{Multislice Method for $T_{II}(E_x)$ in the Planar Model}
In Energy Stage II there is a turning point at $x=x_T$ inside the barrier region. At $x=x_T$, the value of the potential function is $U(x_T)=E_x$, where $E_x$ is the energy of the electron associated with its motion along the $x - $ direction. The energy in Energy Stage II ranges from $(\eta_1+\phi_2-eV_b) < E_x < (\eta_1+\phi_1)$. In the transfer matrix approach the barrier region defined by $x \in [0,d]$, is divided into two spatial regions from $0 < x < x_T$ (Region 1) in which $U(x) \geqslant E_x$ and $x_T < x < d $ (Region 2) in which $U(x) \leqslant E_x$. The point $x=x_T$ is chosen as one of the nodal points common to both the regions 1 and 2. Furthermore regions 1 and 2 are divided each into several intervals so that the potential in each interval can be approximated by an appropriate linear function. Let the number of intervals in the regions 1 and 2 be $N_1$ and $N_2$ respectively. 

\bigskip

\noindent Consider the Schrödinger equation in the two regions divided by the turning point $x = x_T$.\\
1) Spatial Region 1 $(0 < x < x_T)$ : In each slice of this region, the transformed Schrödinger equation with a linear approximation of the potential $U(x)$ becomes
\begin{equation}\label{Airy_l_xT}
\frac{d^2 \Psi^{(1)}_j}{dh_j^{(1)^2}}- h_j^{(1)}\Psi^{(1)}_j= 0; \quad h_j^{(1)} \, \geqslant \, 0
\end{equation} 
where $\Psi^{(1)}_j$ is the wavefunction in the $j^{th}$ slice in region 1. $j=0,1,2,3\cdots (N_1-1)$. These wavefunctions  are composed of a linear combination of Airy Functions which are monotonic in this region.\\
2) Region 2 $(x_T < x < d)$ : Similarly, in this region, the transformed Schrödinger equation is given by   
\begin{equation}\label{Airy_g_xT}
 \frac{d^2 \Psi^{(2)}_j}{dh_j^{(2)^2}} - h_j^{(2)}\Psi^{(2)}_j= 0;\quad h_j^{(2)} \, \leqslant \, 0
\end{equation} 
where $\Psi^{(2)}_j$ is the wavefunction in the $j^{th}$ slice in region 2. $j=0,1,2,3\cdots (N_2-1)$. These wavefunctions  are composed of a linear combination of Airy Functions which are oscillatory in this region. Using the linear approximation for the potential in the $j^{th}$ interval, (Note $j$ runs from $0$ to $N_1-1$ in region 1 and from $0$ to $N_2-1$ in region 2) let the general solutions to the equations (\ref{Airy_l_xT}) and (\ref{Airy_g_xT}) be given by
\begin{equation}
\Psi^{(k)}_j[h_j^{(k)}(x)] = C^{(k)}_j Ai[h^{(k)}_j(x)]+D^{(k)}_j Bi[h_j^{(k)}(x)]\,\, ; k=1,2
\end{equation}
where $C^{(k)}_j$, $D^{(k)}_j$, are constants.
Imposing the relevant matching conditions across the common bouundary at $x = x_j$ for the intervals $I_j$ and $I_{j-1}$ gives in matrix form 
\begin{equation}\label{Mk_Matrix}
\begin{bmatrix}
C_j^{(k)}\\ D_j^{(k)}
\end{bmatrix} = \bar{\bar{M_j}}^{(k)}
\begin{bmatrix}
C_{j-1}^{(k)}\\ D_{j-1}^{(k)}
\end{bmatrix}
\end{equation}
where\\
$\bar{\bar{M_j}}^{(k)} =
\dfrac{1}{W}\begin{bmatrix}
{Bi^\prime}[h^{(k)}_j(x_j)]& -Bi[h^{(k)}_j(x_j)]\\ \\
-{Ai^\prime}[h^{(k)}_j(x_j)]& Ai[h^{(k)}_j(x_j)] 
\end{bmatrix}
\begin{bmatrix}
Ai[h^{(k)}_{j-1}(x_j)] & Bi[h^{(k)}_{j-1}(x_j)]\\ \\
{Ai^\prime}[h^{(k)}_{j-1}(x_j)]&  {Bi^\prime}[h^{(k)}_{j-1}(x_j)]
\end{bmatrix}$\\
and $W$ is the Wronskian. $W = [Ai\, , \,Bi] = \dfrac{1}{\pi}$\\\\
Repeating equation (\ref{Mk_Matrix}) for $j=1,2,\cdots N_k-1$ $\,\,(k=1,2)$ gives 
\begin{equation}\label{M_Reg1_Planar}
\begin{bmatrix}
C_{N_k-1}^{(k)}\\ D_{N_k-1}^{(k)}
\end{bmatrix}=\bar{\bar{M}}^{(k)}_{N_k-1} \cdot \bar{\bar{M}}^{(k)}_{N_k-2}\cdots\bar{\bar{M}}^{(k)}_{1}
\begin{bmatrix}
C_{0}^{(k)}\\ D_{0}^{(k)}
\end{bmatrix}  \equiv \bar{\bar{M}}^{(k)}  \begin{bmatrix}
C_{0}^{(k)}\\ D_{0}^{(k)}
\end{bmatrix}
\end{equation}
where $\bar{\bar{M}}^{(k)}=\bar{\bar{M}}^{(k)}_{N_k-1}  \cdot \bar{\bar{M}}^{(k)}_{N_k-2}\cdots\bar{\bar{M}}^{(k)}_{1}$ is the inner product of transfer matrices $\bar{\bar{M}}^{(k)}$ $j=1,2,\cdots N_k-1$, taken in reverse order.

\bigskip

At the turning point $x=x_T$, the transfer matrix is given by 
\begin{equation}
\bar{\bar{M_T}} = \dfrac{1}{W}\begin{bmatrix}
Bi^\prime[h_0(x_T)]& -Bi[h_0(x_T)]\\ 
-Ai^\prime[h_0(x_T)]& Ai[h_0(x_T)] 
\end{bmatrix}
\begin{bmatrix}
Ai[h_{N_1-1}(x_T)] & Bi[h_{N_1-1}(x_T)]\\ 
Ai^\prime[h_{N_1-1}(x_T)]&  Bi^\prime[h_{N_1-1}(x_T)]
\end{bmatrix} 
\end{equation} 
The matrix elements of $\bar{\bar{M_T}}$ are 
$$(M_T)_{11} = \dfrac{1}{W} \Big [ Bi^\prime[h_0(x_T)] Ai[h_{N_1-1}(x_T)] - Bi[h_0(x_T)]Ai^\prime[h_{N_1-1}(x_T)] \Big ]$$
$$(M_T)_{12} = \dfrac{1}{W} \Big [ Bi^\prime[h_0(x_T)] Bi[h_{N_1-1}(x_T)] - Bi[h_0(x_T)]Bi^\prime[h_{N_1-1}(x_T)] \Big ]$$
$$(M_T)_{21} =  \dfrac{1}{W} \Big [ Ai[h_0(x_T)]Ai^\prime[h_{N_1-1}(x_T)]-Ai^\prime[h_0(x_T)]Ai[h_{N_1-1}(x_T)] \Big ]$$
$$(M_T)_{22} =  \dfrac{1}{W} \Big [ Ai[h_0(x_T)]Bi^\prime[h_{N_1-1}(x_T)]- Ai^\prime[h_0(x_T)]Bi[h_{N_1-1}(x_T)] \Big ]$$
the argument of all Airy's functions become zero at $x=x_T$. Thus elements of the $\bar{\bar{M_T}}$ become
$$(M_T)_{11} =  \dfrac{1}{W} \Big [Bi^\prime[0] Ai[0] - Bi[0]Ai^\prime[0] \,\Big ]= \,1 $$
$$(M_T)_{12} =  \dfrac{1}{W} \Big [ Bi^\prime[0] Bi[0] - Bi[0]Bi^\prime[0] \,\Big]= \,0 $$
$$(M_T)_{21} =  \dfrac{1}{W} \Big [Ai[0]Ai^\prime[0]-Ai^\prime[0]Ai[0]    \,\Big ]= \,0 $$
$$(M_T)_{22} =  \dfrac{1}{W} \Big [ Ai[0]Bi^\prime[0]- Ai^\prime[0]Bi[0]   \,\Big ]= \,1 $$
%
 \begin{equation}
 \bar{\bar{M_T}} =   \begin{bmatrix}
 1&0\\0&1
 \end{bmatrix}
 \end{equation}
  
Thus
\begin{equation}\label{atXT}
\begin{bmatrix}
C_{0}^{(2)}\\ D_{0}^{(2)}
\end{bmatrix}=\begin{bmatrix}1&0\\0&1
\end{bmatrix}
\begin{bmatrix}
C_{N_1-1}^{(1)}\\ D_{N_1-1}^{(1)}
\end{bmatrix}
\end{equation}
Thus the coefficients in the $I^{(1)}_{N_1-1}$ interval of region 1 are equal to the  coefficients in the $I_0^{(2)}$ interval of region 2.   Substituting for $\begin{bmatrix}
C_{N_1-1}^{(1)}\\ D_{N_1-1}^{(1)}
\end{bmatrix}$ from equation (\ref{M_Reg1_Planar}) we have 
$$\begin{bmatrix}
C_{0}^{(2)}\\ D_{0}^{(2)}
\end{bmatrix}=  \bar{\bar{M}}^{(1)}  \begin{bmatrix}
C_{0}^{(1)}\\ D_{0}^{(1)}
\end{bmatrix} $$
From equations (\ref{M_Reg1_Planar}) and (\ref{atXT}) we have 
\begin{equation}\label{M_Reg12_Planar}
\begin{bmatrix}
C_{N_2-1}^{(2)}\\ D_{N_2-1}^{(2)}
\end{bmatrix}=\bar{\bar{M}}^{(2)} \cdot \bar{\bar{M}}^{(1)}
\begin{bmatrix}
C_{0}^{(1)}\\ D_{0}^{(1)}
\end{bmatrix}
\end{equation}
As before we impose the matching conditions at $x=0$. The point $x=0$ is a common boundary of the $0^{th}$ slice of region 1 and the interior of the $1^{st}$ electrode. This gives the relation 
\begin{equation}\label{U_reg1_x0}
1 = C_{0}^{(1)} \tilde{U}_{11} + D_{0}^{(1)} \tilde{U}_{12}
\end{equation}
where $\tilde{U}_{11}$ and $\tilde{U}_{12}$ are the matrix elements of the transfer matrix that connect coefficients relevant for the common boundary at $x=0$ and are given as 
$$
\tilde{U}_{11} = \dfrac{1}{2}\Bigg[ Ai(h_0^{(1)}(0)) +   \dfrac{Ai^{\prime}(h_0^{(1)}(0))}{i k_1} \Bigg] \quad \text{and} \quad \tilde{U}_{12} =  \dfrac{1}{2}\Bigg[ Bi(h_0^{(1)}(0)) +   \dfrac{{Bi^{\prime}}(h_0^{(1)}(0))}{i k_1}   \Bigg]
$$
The point $x=d$ is a common boundary of the $(N_2-1)^{th}$ slice of the region 2 and the interior of the $2^{nd}$ electrode. The transfer matrix for this common boundary is $\bar{\bar{\tilde{V}}}$.
$$
\bar{\bar{\tilde{V}}} = \begin{bmatrix}
Ai(h_{N_2-1}^{(2)}(d))& Bi(h_{N_2-1}^{(2)}(d))\\ \\
\dfrac{1}{ik_2}{Ai^\prime}(h_{N_2-1}^{(2)}(d)) & \dfrac{1}{ik_2}{Bi^\prime}(h_{N_2-1}^{(2)}(d))
\end{bmatrix}
$$
The combined transfer matrix is 
$$\bar{\bar{\tilde{S}}} = \bar{\bar{\tilde{V}}}\,\cdot \bar{\bar{M}}^{(2)}\, \cdot \bar{\bar{M}}^{(1)}$$
where $\bar{\bar{M}}^{(1)}$ is the product of matrices in region 1 and $\bar{\bar{M}}^{(2)}$ is the product of matrices in region 2.
The tunnel amplitude $T_{II}$ is related to the matrix elements of $\bar{\bar{\tilde{S}}}$ as follows.
\begin{equation}\label{T_II}
T_{II} = \tilde{S}_{11}\,C_{0}^{(1)} +  \tilde{S}_{12}\,D_{0}^{(1)} =  \tilde{S}_{21}\,C_{0}^{(1)} +  \tilde{S}_{22}\,D_{0}^{(1)}
\end{equation}
solve for $C_{0}^{(1)}$ and $D_{0}^{(1)}$ using equations (\ref{U_reg1_x0}) and (\ref{T_II})
$$C_{0}^{(1)} = \dfrac{\tilde{S}_{22}-\tilde{S}_{12}}{\tilde{S}_{11}U_{12} - \tilde{S}_{12}\tilde{U}_{11}-\tilde{S}_{21}\tilde{U}_{12}+\tilde{S}_{22}\tilde{U}_{11}} \quad \text{and} \quad D_{0}^{(1)} = \dfrac{1-C_{0}^{(1)} \tilde{U}_{11}}{\tilde{U}_{12}}$$ 
Thus the tunneling amplitude for the Energy Stage II, using transfer matrix approach is given by 
\begin{equation}
T_{II} =  \dfrac{ e^{-ik_2d}(\tilde{S}_{11}\tilde{S}_{22}-\tilde{S}_{12}\tilde{S}_{21})}{\tilde{S}_{11}\tilde{U}_{12} - \tilde{S}_{12}\tilde{U}_{11}-\tilde{S}_{21}\tilde{U}_{12}+\tilde{S}_{22}\tilde{U}_{11}}
\end{equation}

\section{A Trivial Demonstration (Trapezoidal Potential)}
The multislice method described in the previous section ought to work even if the potential is pure trapezoid. The transfer matrix for each interval in a multislice calculation becomes a unit matrix, and this is demonstrated in this section. Consider the Energy Stage I.
The wavefunction for the trapezoidal potential in $j^{th}$ interval in this Energy Stage can be written as  
\begin{equation}
\Psi_j(h_j) = C_j Ai(h_j)+D_j Bi(h_j)
\end{equation}
Imposing the continuity of wavefunctions and its derivatives at $x=x_j$ gives the same set of equations as in equation (9)
\begin{equation}
\begin{bmatrix}
Ai[(h_j(x_j)] & Bi[h_j(x_j)]\\ 
Ai^\prime[(h_j(x_j)]& Bi^\prime[h_j(x_j)]
\end{bmatrix}
\begin{bmatrix}
C_j\\ D_j
\end{bmatrix}=
\begin{bmatrix}
Ai[h_{j-1}(x_j)] & Bi[h_{j-1}(x_j)]\\ 
Ai^\prime[h_{j-1}(x_j)]&  Bi^\prime[(h_{j-1}(x_j)]
\end{bmatrix}
\begin{bmatrix}
C_{j-1}\\ D_{j-1}
\end{bmatrix}
\end{equation} 

\noindent For pure trapezoidal potential $B_j$ which involves the slope of the potential in the $j^{th}$ slice, is independent of $j$ $\,\,(i.e., B_{j} = B_{j-1})$ and $h_j(x_j) = h_{j-1}(x_j)$. This implies that the square Matrices on both sides of the above equation become identical and

\begin{equation}
\begin{bmatrix}
	C_j\\ D_j
\end{bmatrix}=
\begin{bmatrix}
	C_{j-1}\\ D_{j-1}
\end{bmatrix}
\end{equation}
therefore the transfer matrix for trapezoidal potential is a unit matrix as expected. Hence the calculation of $T_I(E_x)$ in the multislice method reduces to the single slice calculation described in equation (3.31) of Chapter 3.
In the Energy Stage II the spatial region $x \in [0,d]$ is divided into two regions on each side of the turning point $x = x_T$. The Transfer Matrix $\bar{\bar {M}}^{(1)}_j$ relating the constants $C_j$ and $D_j$ for the $j^{th}$ slice to those for the $(j-1)^{th}$ slice, both lying in the segment that is to the left of $x_T$ and  the analogous Transfer Matrix $\bar{\bar {M}}^{(2)}_j$ for consecutive slices that lie in the  right side of the turning point $x_T$ need to be constructed. For reasons identical to that for the case of the Energy Stage I, both these Transfer Matrices for the case of a pure trapezoidal potential become the identity matrix. Hence the  multislice calculation for Energy Stage II to find $T_{II}(E_x)$ will also be identical to the single slice calculation described in equation (3.42) of Chapter 3.

\section{Multi-Slice Calculation for the Russell Potential}

The Russell potential for each field line, can be constructed from equation (4.17) of Chapter 4.
\begin{equation}\label{russs}
U(\eta) = (\eta_1+\phi_2-eV_b) + (\phi_1-\phi_2+eV_b)\, \dfrac{\lambda(\eta)}{\lambda_{\eta_{tip}}}
\end{equation}
For a given field line which is described by $\,\xi\,$ the length along the field line can be calculated as 
\begin{equation}\label{xfl} 
x_{fl}(\eta, \xi) = a \int\limits_\eta^{\eta_{tip}} d \eta \,\sqrt{\Big (\dfrac{\xi^2 - \eta^2}{1-\eta^2} \Big)} 
\end{equation}
The linear distance used will be $x_{fl}$ instead of $x$. Inverting the above equation \ref{xfl} to get $\eta = \eta[x_{fl}(\xi),\xi]$ it should be possible to construct $\lambda(\eta)$ as an explicit function of $x_{fl}(\xi)$. Substituting this in the equation \ref{russs} gives $U = U[x_{fl}(\xi)]$.  This expression is called the Russell Potential for the field line '$\xi$'.  

\bigskip 
 
For the infinitesimal STM, the distance variable $x_{fl}$ runs from $0$ to $d_{fl}(\xi)$ for a field line characterized `$\xi$'. For this infinitesimal STM, the interval $[0,\, d_{fl}]$ is split into multislices indexed by $j$ which runs from $0,1,2,\cdots (N-1)$. The transfer matrices  are obtained for consecutive slices exactly as described before for Energy Stage I. For Energy Stage II a turning point $x_T$ is identified and both the left and the right side intervals are themselves split into several slices, and corresponding transfer matrices obtained for consecutive slices in both regions 1 and 2 which lie to the left and to the right of the turning point respectively.  The subsequent calculations are almost identical to those described in Sections 3 and 4 above, except that $d_{fl}$ replaces $d$ everywhere, and the tunnel amplitudes  $T_I(E_x,\xi)$ and $T_{II}(E_x,\xi)$ are determined by equations similar to equations (19) and (31) respectively, where the matrices in these expressions are defined appropriately.

\section{$T(E_x)$ Plots and Comparisons}

\begin{figure}[h]
	\begin{subfigure}{0.49\textwidth}
			\centering
				\includegraphics[width=3.0in,height=2.3in]{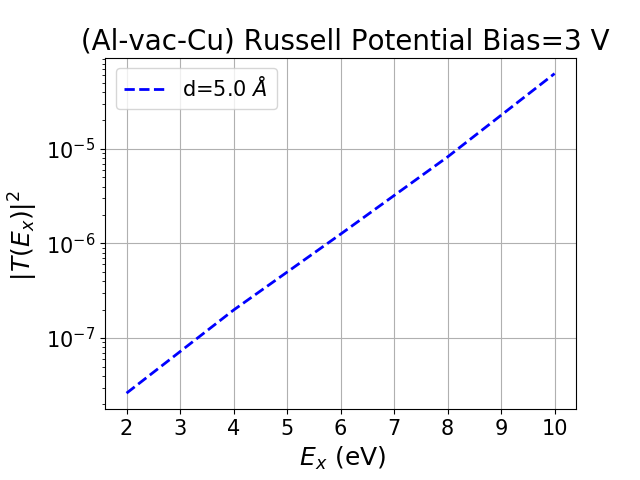}	
					\caption{} 
					\label{fig:TEx11}
				\end{subfigure}	
				\hfill
				\begin{subfigure}{0.49\textwidth}
					\centering
						\includegraphics[width=3.0in,height=2.3in]{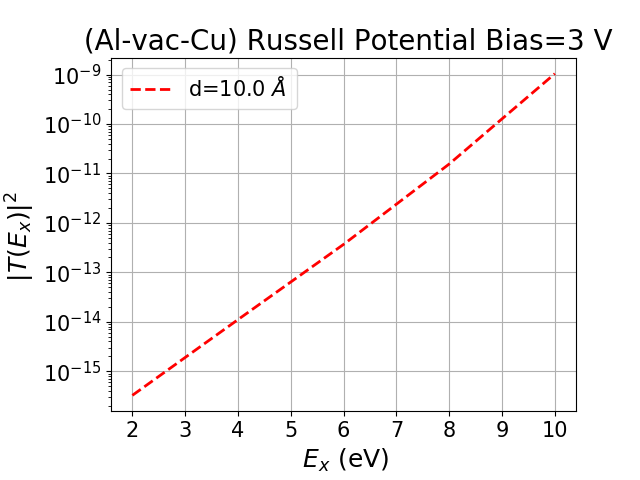}
					\caption{} 
					\label{fig:TEx12}	
				\end{subfigure}	
	\caption{Plot of $\vert T(E_x) \vert^2$ vs $E_x$ for bias voltage of $3\, V$ (a) $d = 5 $ \r{A} (b) $10 $ \r{A} using Airy functions}
	\label{fig:TExd5N10}
\end{figure} 
\begin{figure}[hpt]		
		\centering
	\begin{subfigure}{0.49\textwidth}
		\includegraphics[width=3.0in,height=2.0in]{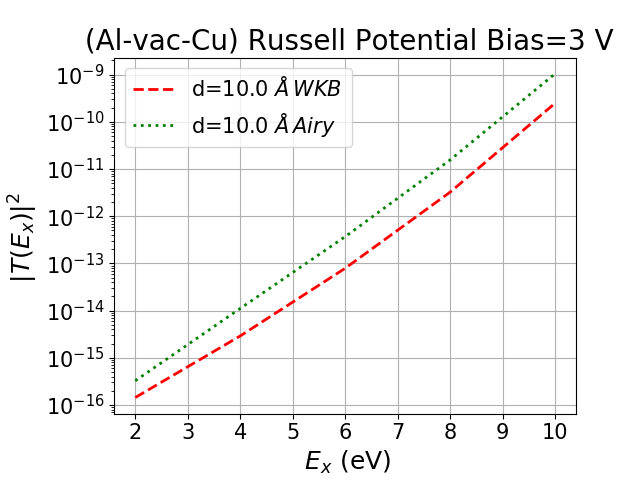}
		\caption{} 
		\label{fig:TEx13}
	\end{subfigure} 
\hfill
	\begin{subfigure}{0.49\textwidth}	
		\centering
		\includegraphics[width=3.0in,height=2.0in]{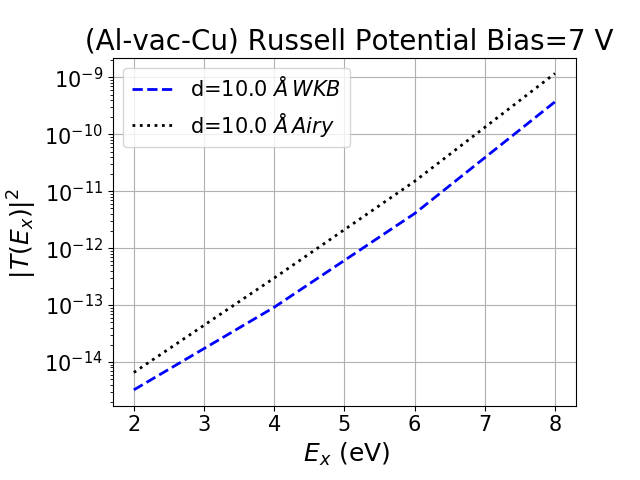}
		\caption{} 
		\label{fig:TEx7}
	\end{subfigure}
	\caption{Plot of $\vert T(E_x) \vert^2$ vs $E_x$ for bias voltage of $7\, V$ and  $d = 10 $ \r{A} using Airy functions and WKB approximation} 
	\label{fig:TEx37} 
\end{figure} 
Figs. \ref{fig:TEx11} and \ref{fig:TEx12} show plots of the tunneling probability  $\vert T(E_x) \vert^2$ as a function of energy of the tunneling electron for bias voltage of $3 \,V$ and tip sample distance $= 5 $ \r{A} and $ 10 $ \r{A} for the  Russell potential. These plots show that for both tip-sample distances, the $\vert T(E_x) \vert^2$ increases with $E_x$, which is to be expected. These plots also show that the tunneling probability decreases sharply with increasing tip-sample distance. Fig. \ref{fig:TEx13} and \ref{fig:TEx7} compares the tunneling probability $\vert T(E_x) \vert^2$ as a function of $(E_x)$ for Russell potential calculated using Airy functions and using the WKB approximation. For these bias voltages of $3 \, V \text{and} 7 \,V$, the tunneling amplitude is found to exceed for Airy dependent calculation, although the general behaviour is that both probabilities increase with increasing $E_x$ as expected. 

\bigskip
 
In this chapter, the calculation of the tunneling probability for non-trapezoidal potential using the 'Transfer Matrix Method' is explained. This method involves approximating the potential as linear within each slice and determining transfer matrices by imposing the continuity of the wavefunction and its derivative at each nodal point.
A calculation of the tunnel amplitude for Russell potential (non trapezoidal) using this method is described in this Chapter. Chapter 6 includes the calculation of tunnel amplitude for non trapezoidal potential with images.

	\chapter{The Simmon's Image Potential}\label{Chap6}
	%
%
%
%
%
\section{Introduction}
 
Chapter 3 describes the calculation of tunneling amplitude for trapezoidal potential barrier. In this calculation the effect of tunneling electron inside the barrier was not considered. The potential in the barrier region may be modified due to field of the tunneling electron, and the net potential will not be trapezoidal. The method of images can be used to calculate the modified potential. In this method, the problem of a point charge placed in front of grounded conducting plane can be transformed into a simpler problem of  the source point charge and a mirror point charge of opposite sign, placed on the opposite side of the plane, and placed equidistantly from it. The electrostatic potential due to both charges at all points on the plane vanishes, due to reflection symmetry about the plane. The vanishing of the potential on the grounded conducting plane is required by the Boundary condition imposed on the plane. However in the problem at hand, the source point charge (viz the electron) is between two conducting surfaces, one of which is grounded and the other is maintained at a potential $V_b$. 

\bigskip

The net potential is a linear superposition of the trapezoidal potential applied externally + the potential due to image charges that appear on the surfaces of the two conducting planes that are grounded. The latter can be calculated by first constructing an image charge $q_{L1}$ due to one grounded conducting plane (on the left) of the source charge $q$. Next construct images $q_{R1}$ of the source charge $q$, and also the image $q_{R2}$ of the image charge $q_{L1}$ due to the other the other grounded conducting plane. More images can be generated by constructing images of images through either of the two planes. Note these images will be infinite in number and will be formed further and further away from the two conducting plane surfaces. Once a sufficient number of images (of charges with alternating signs) have been located, the sum of the electrostatic potential evaluated at the location of the source point charge is the image potential $U_{im}$

\bigskip

\begin{figure}[hpt]
	\centering
	\begin{subfigure}{0.49\textwidth}
		\includegraphics[width=3.0in,height=2.5in]{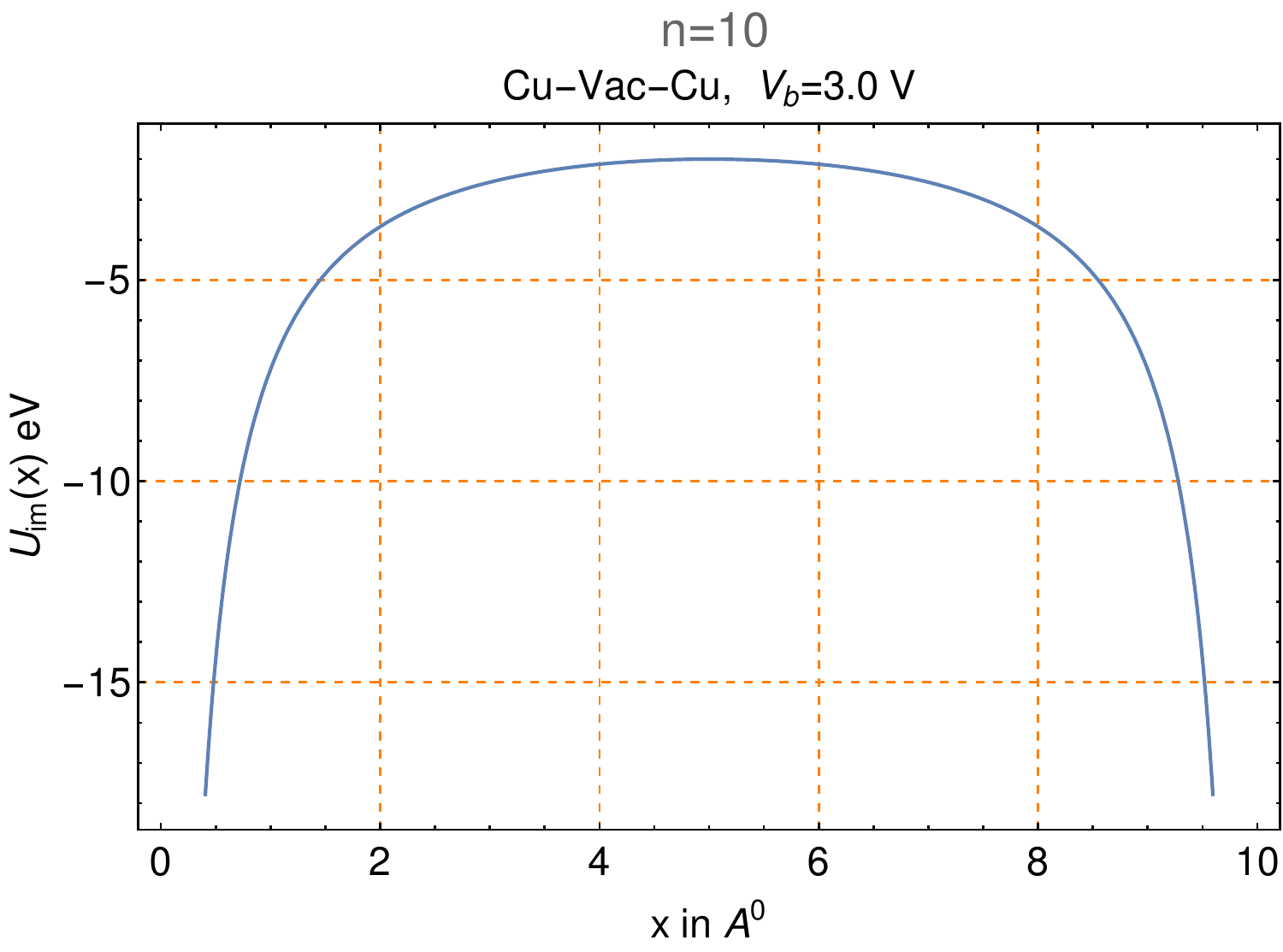}
		\caption{} 
		\label{fig:img10}
	\end{subfigure}
		\hfill
	\begin{subfigure}{0.49\textwidth}	
		\includegraphics[width=3.0in,height=2.5in]{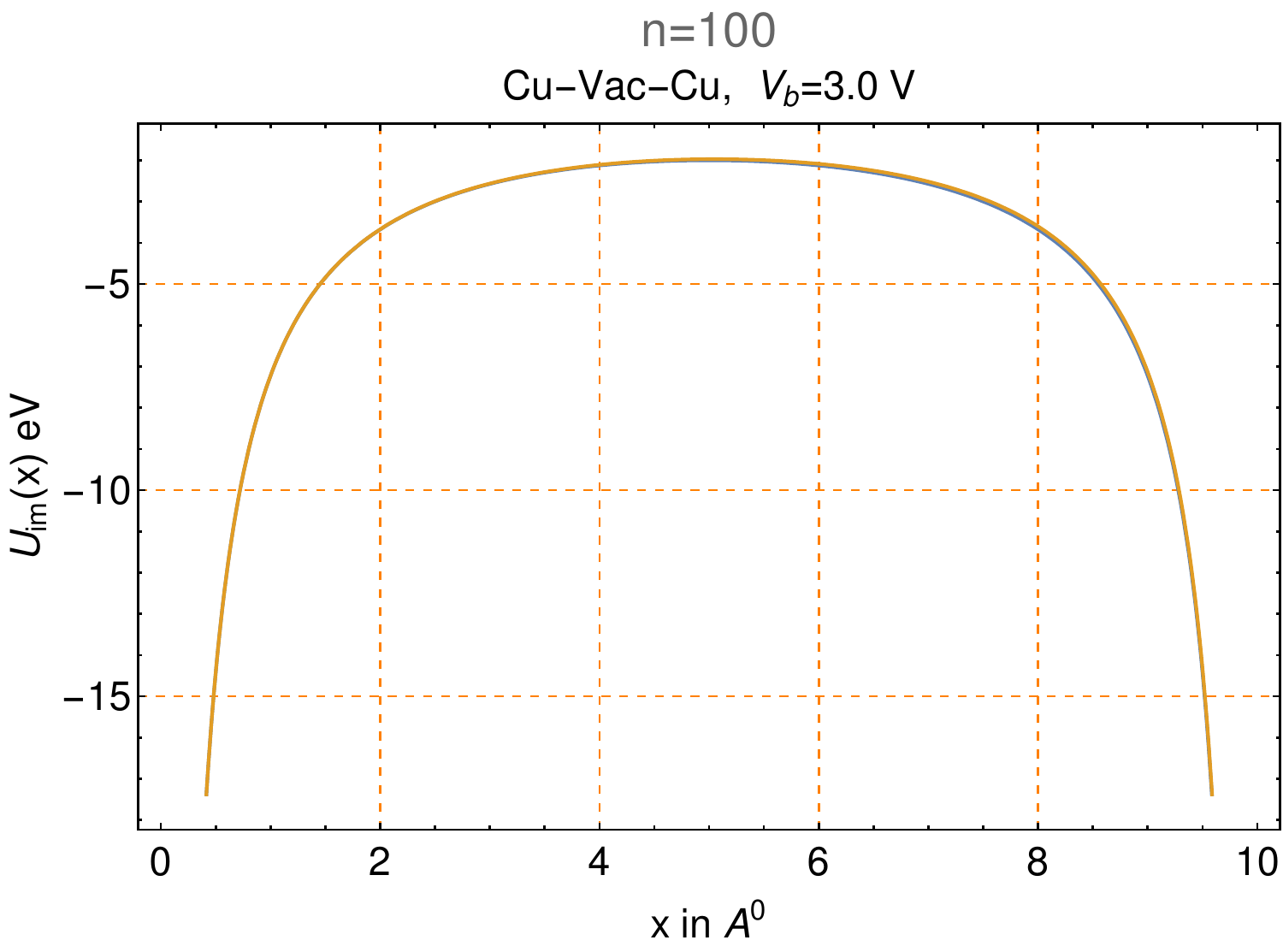}
		\caption{} 
		\label{fig:img100}
	\end{subfigure}
	\caption{Plot of image potential vs $x$ for (a) n = 10 and (b) for n = 100} 
	\label{fig:img}
\end{figure}

The potential due to these infinite set of images has been found \cite{CRC75} to be 
\begin{equation}\label{Uim_mulImg}
-\frac{q_e^2}{4 \pi  \epsilon_0 }\left(\frac{1}{2 x}+ \sum_{n=1}^{n_{max}} \left[\frac{n\,d}{(n\,d)^2 - x^2}-\frac{1}{n\,d}\right] \right)
\end{equation}
where $x$ is the distance of the electron from electrode 1, and $n_{max} = \infty$.  Fig. \ref{fig:img} show plots of the image potential calculated from equation  (\ref{Uim_mulImg}) in which $n_{max} = 10$, and  $n_{max} = 100$.  It is seen that the sum converges very fast since the two plots in Fig.\ref{fig:img}  are almost identical. Simmons \cite{simmonsI} has approximated the RHS of equation (\ref{Uim_mulImg}) by the following expression (called the Simmon's Image Potential).
\begin{equation}\label{Uim_approx}
U^S_{im}(x) = -\dfrac{\alpha d^2}{x(d-x)} 
\end{equation}
where $\alpha = 1.15 \lambda, \, \,$ and $\, \lambda = \dfrac{q_e^2 \log[2]}{8\pi \epsilon_0 d}$.
\begin{figure}[hpt]
	\centering
	\includegraphics[width=3.0in,height=2.5in]{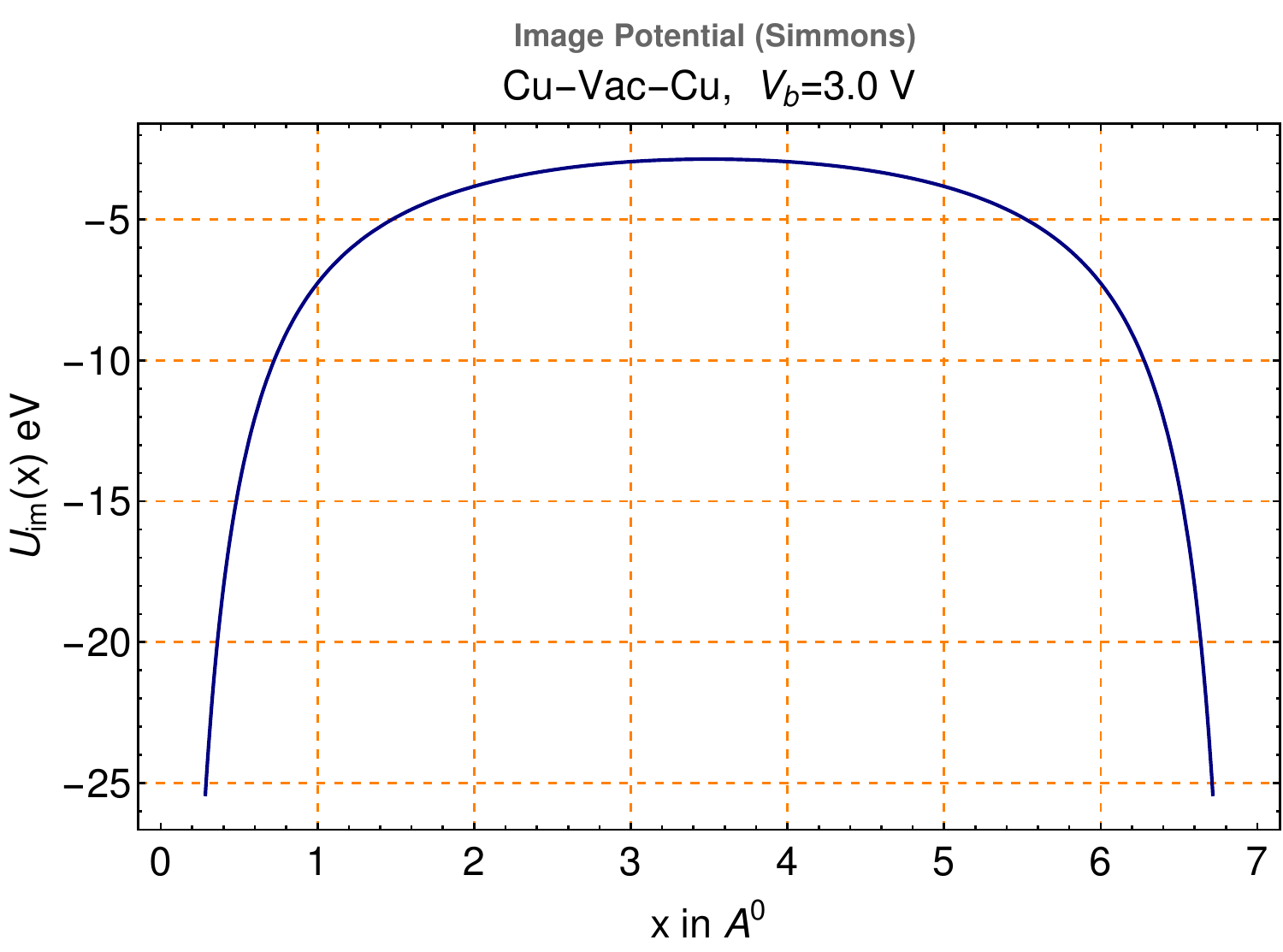}
	\caption{Simmons approximation for image potential} 
	\label{fig:img_simmapprox}
\end{figure}
Fig. \ref{fig:img_simmapprox} shows the plot of Simmon's approximation of the image potentials due to a point charge between two grounded conducting planes. It is seen that Simmon's approximation works extremely well when compared to the exact result for the image potential $U_{im}(x)$. 

\section{Multi - slice method for the Trapezoid + the Simmons Image Potential}
The sum of the Trapezoidal potential and the Simmon's Image potential is given by 
\begin{equation}\label{UTNim}
U_{Tim} = (\eta_1 + \phi_1) - (\phi_1-\phi_2 + eV_b) \dfrac{x}{d} -\dfrac{\alpha d^2}{x(d-x)} 
\end{equation}
The net potential spread over the three spatial regions $x < 0$, $0 < x < d$ and $x > d$ is given by 
\begin{equation}
U(x)=\left\{
\begin{array}{@{}lll@{}}
0 & 		x < 0\ &\quad \text{Region I} \\
U_{Tim}(x) & 		0 \leqslant x \leqslant d &\quad \text{Region II}\\
-\Delta_B&	 x > d &\quad \text{Region III} \\
\end{array}\right.
\end{equation} \\
where $\Delta_B=(\eta_2-\eta_1+eV_b)$. \\

\begin{figure}[h]
	\begin{subfigure}{0.49\textwidth}
		\centering
		\includegraphics[width=3.0in,height=3.0in]{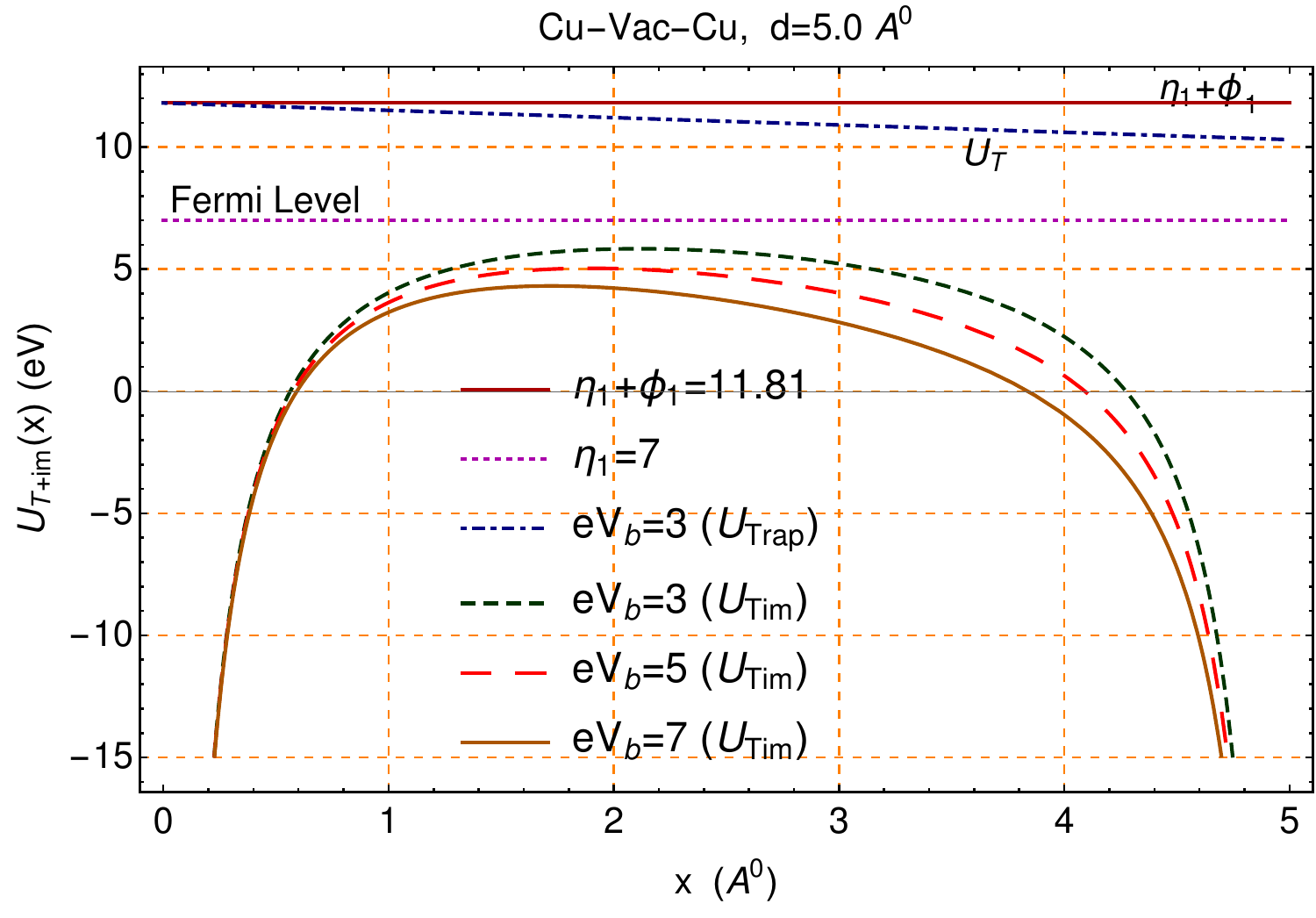}	
		\caption{} 
		\label{fig:UTimd5}
	\end{subfigure}	
	\hfill
	\begin{subfigure}{0.49\textwidth}
		\centering
		\includegraphics[width=3.0in,height=3.0in]{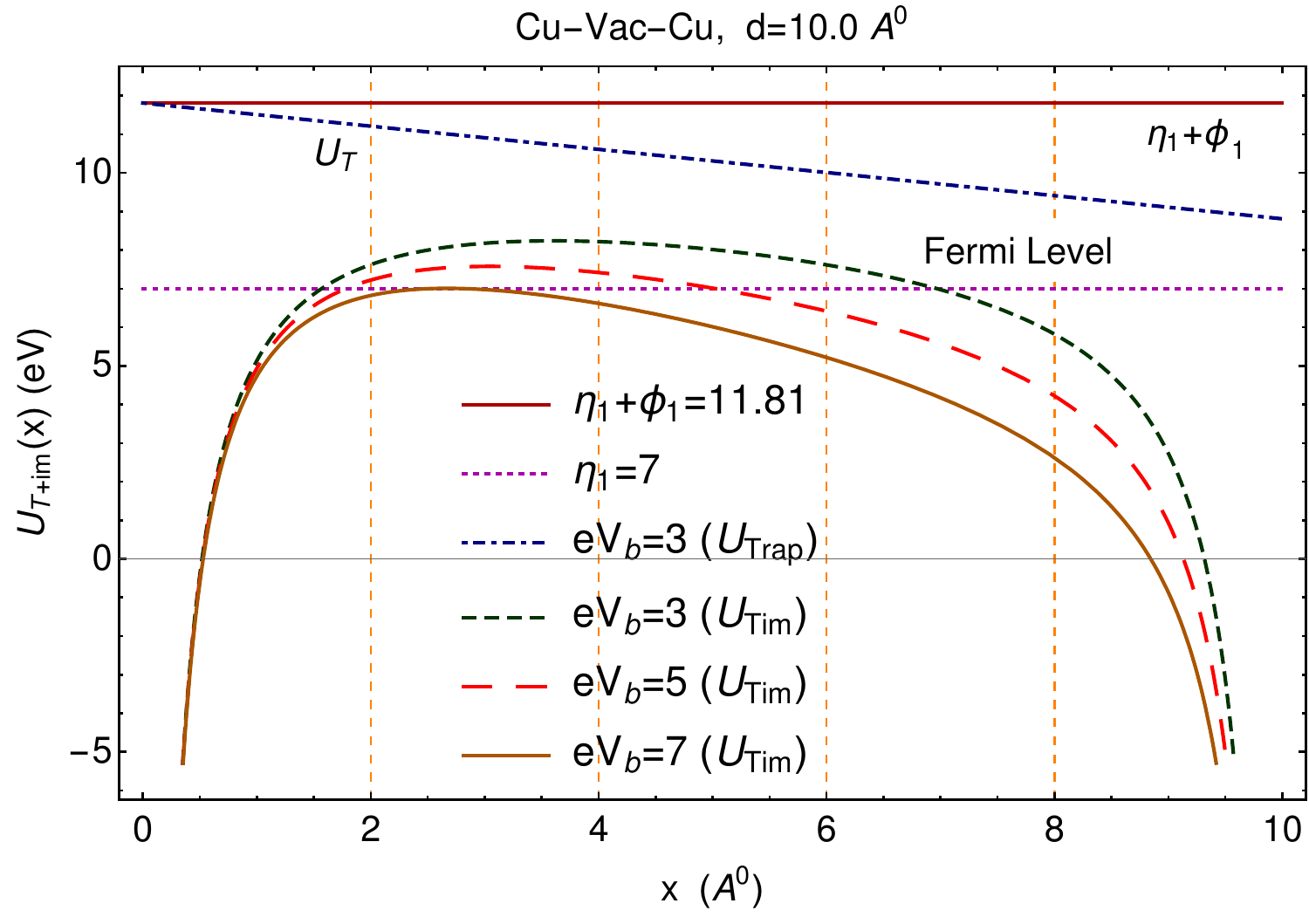}
		\caption{} 
		\label{fig:UTimd10}	
	\end{subfigure}	
	\caption{Plot of Trapezoid + Image potential vs $x$ for  Bias voltage of $V_b= 3,\,5,\,7\, V$. The dotted magenta line shows the Fermi level of the first electrode. $U_T$ is the trapezoidal potential for Bias $V_b= 3 \,V$.  The red solid line shows the plot of $\eta_1+\phi_1$.}
	\label{fig:UTimd510}
\end{figure}
Fig. \ref{fig:UTimd5} and \ref{fig:UTimd10} shows the effect of the image potential in the barrier region for $d = 5,\, 10 $ \r{A} and for bias voltage of $3, \, 5,\,$ and $7 \, V$. The height and the width of the barrier is seen to reduce for both tip-sample distances and all bias voltages of $3 \,,5,\, 7 \,V$ compared to the pure trapezoidal potential (blue dot-dashed line). This is likely to  increase the tunneling probability and perhaps also the tunneling current. 
The dotted line (Magenta) shows the Fermi level of the first electrode $(\eta_1 = 7\, V)$.  For all bias voltages the barrier potential shifts \underline{below} the Fermi level. Fig. \ref{fig:UTimd5} shows that for all bias voltage of $3,\, 5,\, 7 \, V$ the barrier potential is seen to lie below the Fermi level of the first electrode. This implies that the field emission becomes dominant over tunneling for electron energies above the Fermi energy.  Conversely, for electron energies below $\eta_1$ there is net tunneling of electrons from electrode 1 to electrode 2. Fig. \ref{fig:UTimd10} shows the plot of Trapezoid+Image potential for $d = 10 $ \r{A}. The height of the potential barrier increases with increasing $d$, suggesting  that the current decreases with increasing $d$. However, the tunneling probability with images is expected to be greater than that for the pure trapezoidal potential. 

\bigskip

The electron tunneling through the barrier encounters two classical turning points within the potential barrier for every energy $E_x$. The location of these points at zero energy are labelled as $s_{10}$ and $s_{20}$ which are the roots of the cubic equation given by equation (\ref{UTNim}).   At non zero energy $E_x > 0 $, the classical turning points occur at $x_{R1}$ and $x_{R2}$ within the potential barrier region, which are the roots of the cubic equation 
\begin{equation}\label{xR12}
U_{Tim} - E_x = 0
\end{equation}
The roots of the cubic equations  (\ref{UTNim}), (\ref{xR12}) are found analytically, using the formula for the roots of a cubic. The third root in both the cases lies outside the barrier region and is ignored. 

\bigskip

\begin{figure}[hpt]
	\centering
	\includegraphics[width=5.2in,height=2.7in]{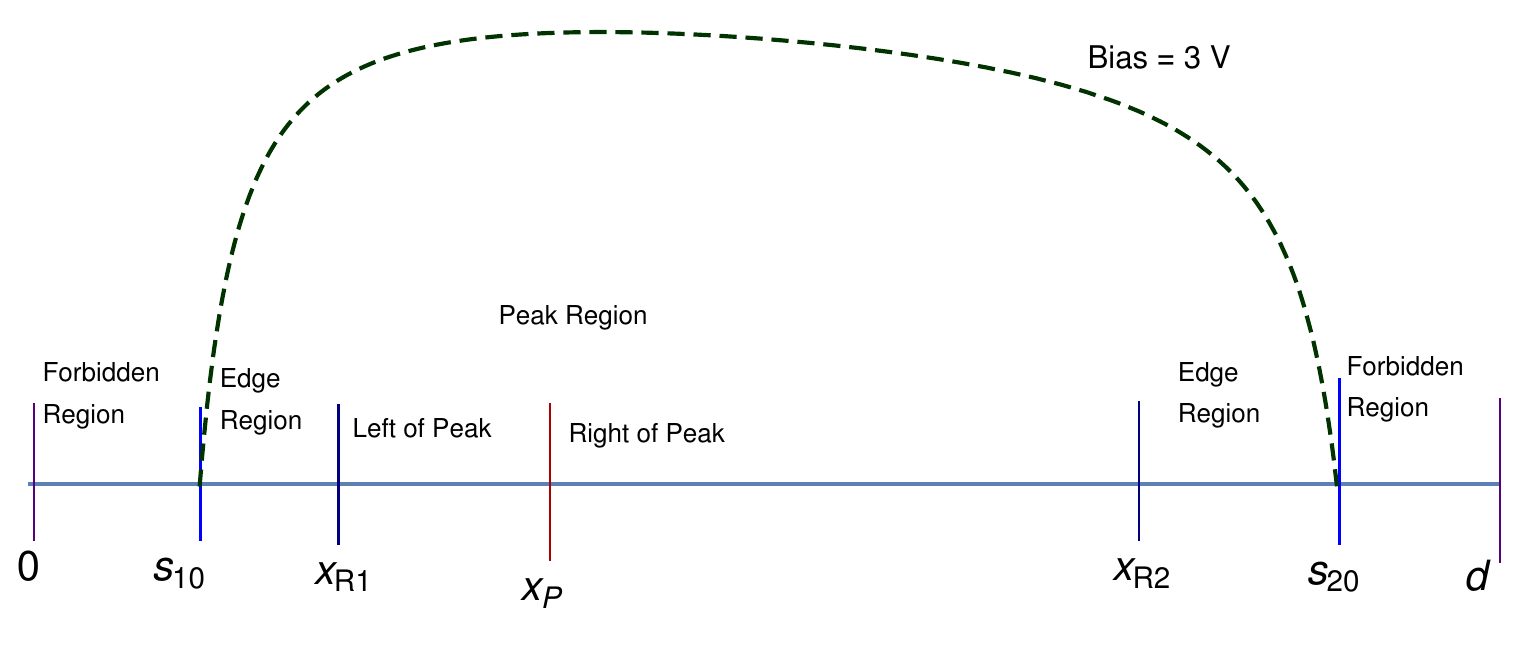}	
	\caption{Spatial Region between $[0,\,d]$ is divided into Forbidden regions, Edge regions, and Peak region.} 
	\label{fig:Reg}
\end{figure}

The location of the maxima of the potential function $U_{Tim}$ is calculated by solving the equation  \begin{equation}\label{Xpeak}
U^\prime_{Tim}(x) = -\dfrac{\phi_1-\phi_2+eV_b}{d}-\dfrac{\text{$\alpha $} d^2 \left(2 x-d\right)} {\left[x \left(d-x\right)\right]^2} = 0
\end{equation} 
This roots of this quartic equation is found numerically by employing the Newton-Raphson method. Based on the locations of the roots of equations (\ref{UTNim}), (\ref{xR12}), and (\ref{Xpeak}) the spatial region between $[0, \,d]$ is divided into  the following spatial regions (refer Fig. \ref{fig:Reg})\\
\begin{enumerate}
\item[1] \underline{The Forbidden Regions} The region defined by $0 < x < s_{10}$ is the left forbidden region and  $s_{20} < x < d$ is the right forbidden region. These regions are introduced because the potential goes to negative infinity which can lead to bound states of the electron. However electrons that are undergoing tunneling will have positive energies and therefore cannot be in bound states. It is best to move the electrodes closer together so that their positions may be effectively assumed to be at $x = s_{10}$ and at $x = s_{20}$. Thus the effective potential barrier width due to the image potential is reduced to $2d - (s_{10} + s_{20})$.

\item[2]\underline{ Left Edge Region} The Left Edge region lies between $s_{10} \,\,\text{and}\,\, x_{R1}$ and a single slice is introduced with slice width $ = x_{R1} - s_{10}$.     

\item[3] \underline{The Right Edge Region} The Right Edge is between $x_{R2}\,\,\text{and}\,\,  s_{20}$.The region from $x_{R2} < x < s_{20}$ is divided into $N_3$ slices with slice width $ \Delta = \dfrac{s_{20}-x_{R2}}{N_3}$ with nodes at $x_j= (j-1) x + \Delta$. where $j = 1, \cdots, N_3$.

\item[4] \underline{Left of Peak Region} The region defined by $(x_{R1} < x < x_{P})$ is called the left of peak region.  The left of peak region is divided into $N_1$ intervals of width $\Delta x_1 = \dfrac{x_{R1}-s_{10}}{N_1}$ with nodes at $x_j= (j-1) x + \Delta x_1$. where $j = 1, \cdots, N_1$.  

\item[5] \underline{Right of Peak Region} The region defined by $(x_{P} < x < x_{R2})$ is called the right of peak region. This region is divided into $N_2$ intervals of width $\Delta x_2 =\dfrac{s_{20}-x_{R2}}{N_2}$ with nodes at $x_j = (j-1) x + \Delta x_2$. where $j=1, \cdots, N_2$.

\end{enumerate}

Using a linear approximation for $U_{Tim}$ in each of the slices, and using the conditions of continuity of the wavefunction and its derivative at the common boundaries $x_j$ of interval $I_j$ and $I_{j-1}$, the corresponding transfer matrices can be obtained. From these transfer matrices, the tunnel amplitude can be calculated as described in Chapter 5 section 4.

The roots $s_{10}$, $s_{20}$, $x_{R1}$, $x_{R2}$, and the location of peak $x_P$ are chosen to be the nodal points in multislice method used in the calculation of tunneling probability. The roots $s_{k0}\,, \quad (k=1,2)$ are given by the standard trigonometric methods \cite{CRC75} to analytically solve cubic equations such as that in equation (\ref{UTNim})    
$$s_{k0}=2 \sqrt{\dfrac{p}{3}}\cos\ \Big[{\dfrac{1}{3}\cos^{-1}\Big(\dfrac{3q}{2p}\sqrt{\dfrac{3}{p}}\Big) - \dfrac{2(3-k)\pi}{3}} \Big] - \dfrac{b}{3};\quad k = 1,2$$
where 
$$
p = \dfrac{b^2}{3} -c \,;\quad
q = - \dfrac{2b^3}{27}  + \dfrac{bc}{3} - e
$$
$$
b = - \dfrac{d(\eta_1+\phi_1+\Delta_1)}{\Delta_1};\quad
c =  \dfrac{d^2(\eta_1+\phi_1)}{\Delta_1};\quad
e = - \dfrac{\alpha d^3}{\Delta_1}
$$
and for non zero energy, the cubic equation is that of equation (\ref{xR12}) whose roots are given by the same methods to be  
$$x_{R_k}=2 \sqrt{\dfrac{\tilde{p}}{3}}\cos\ \Big[{\dfrac{1}{3}\cos^{-1}\Big(\dfrac{3\tilde{q}}{2\tilde{p}}\sqrt{\dfrac{3}{\tilde{p}}}\Big) - \dfrac{2(3-k)\pi}{3}} \Big] - \dfrac{\tilde{b}}{3};\,\, k = 1,2$$
where 
$$
\tilde{p} = \dfrac{\tilde{b}^2}{3} -\tilde{c};\quad
\tilde{q} = - \dfrac{2\tilde{b}^3}{27}  + \dfrac{\tilde{b}\tilde{c}}{3} - \tilde{e}
$$

$$
\tilde{b} = - \dfrac{d(\eta_1+\phi_1-E_x+\Delta_1)}{\Delta_1};\quad
\tilde{c} =  \dfrac{d^2(\eta_1+\phi_1-E_x)}{\Delta_1};\quad
\tilde{e} = - \dfrac{\alpha d^3}{\Delta_1}
$$

\bigskip

In the left forbidden region the wave function is assumed to be the same as that given inside the first electrode. The transfer matrix at $s_{10}$ connects the Left Forbidden Region with the single slice which constitutes the left edge. The conditions of continuity of the wave function and its derivative at $s_{10}$ gives the following equations 
\begin{equation}\label{s10}
e^{ik_1 s_{10}} + R e^{-ik_1 s_{10}} = C_L \phi_L^{(1)}(s_{10}) + D_L \phi_L^{(2)}(s_{10}) 
\end{equation}
\begin{equation}\label{s10prime}
i k_1 e^{ik_1 s_{10}} - i k_1 R e^{-ik_1 s_{10}} = C_L \phi_L^{(1)^\prime}(s_{10}) + D_L \phi_L^{(2)^\prime}(s_{10})
\end{equation}
where 
\begin{equation}
\phi_L^{(1)}(s_{10})= \dfrac{Ai[h(s_{10})]}{Ai[h(\bar{x})]} \, , \quad
\phi_L^{(2)}(s_{10})= \dfrac{Bi[h(s_{10})]}{Bi[h(\bar{x})]}
\end{equation}
and where $C_L$ and $D_L$ are the coefficients of the independent Airy's functions which make up the wave function in the Left Edge Region. Note the primes represent derivatives with respect to $x$. $\bar{x}$ is usually the midpoint of the slice in question. This notation shall be consistently followed throughout the discussion in this Chapter. The transfer matrix $M_{FL}$ at $s_{10}$ is then given by 
$$M_{FL} = \begin{bmatrix}
1 &e^{-ik_1 s_{10}} \\ \\
ik_1 e^{ik_1 s_{10}}& -ik_1 e^{-ik_1 s_{10}}
\end{bmatrix}^{-1}
\begin{bmatrix}
\phi_L^{(1)}(s_{10})&\phi_L^{(2)}(s_{10}) \\ \\
\phi_L^{(1)^\prime}(s_{10})& \phi_L^{(2)^\prime}(s_{10})
\end{bmatrix}$$

Define
$$M_C =  \phi_L^{(1)}(s_{10}) +\dfrac{\phi_L^{(1)^\prime}(s_{10})}{ik_1} \quad \text{and} \quad M_D =  \phi_L^{(2)}(s_{10}) +\dfrac{\phi_L^{(2)^\prime}(s_{10})}{ik_1}$$
Adding equations (\ref{s10}) and (\ref{s10prime}) gives
\begin{equation}\label{addAts10}
2e^{ik_1 s_{10}} = C_L M_C + D_L M_D
\end{equation}
\begin{figure}[hpt]
	\centering
	\includegraphics[width=5.0in,height=3.0in]{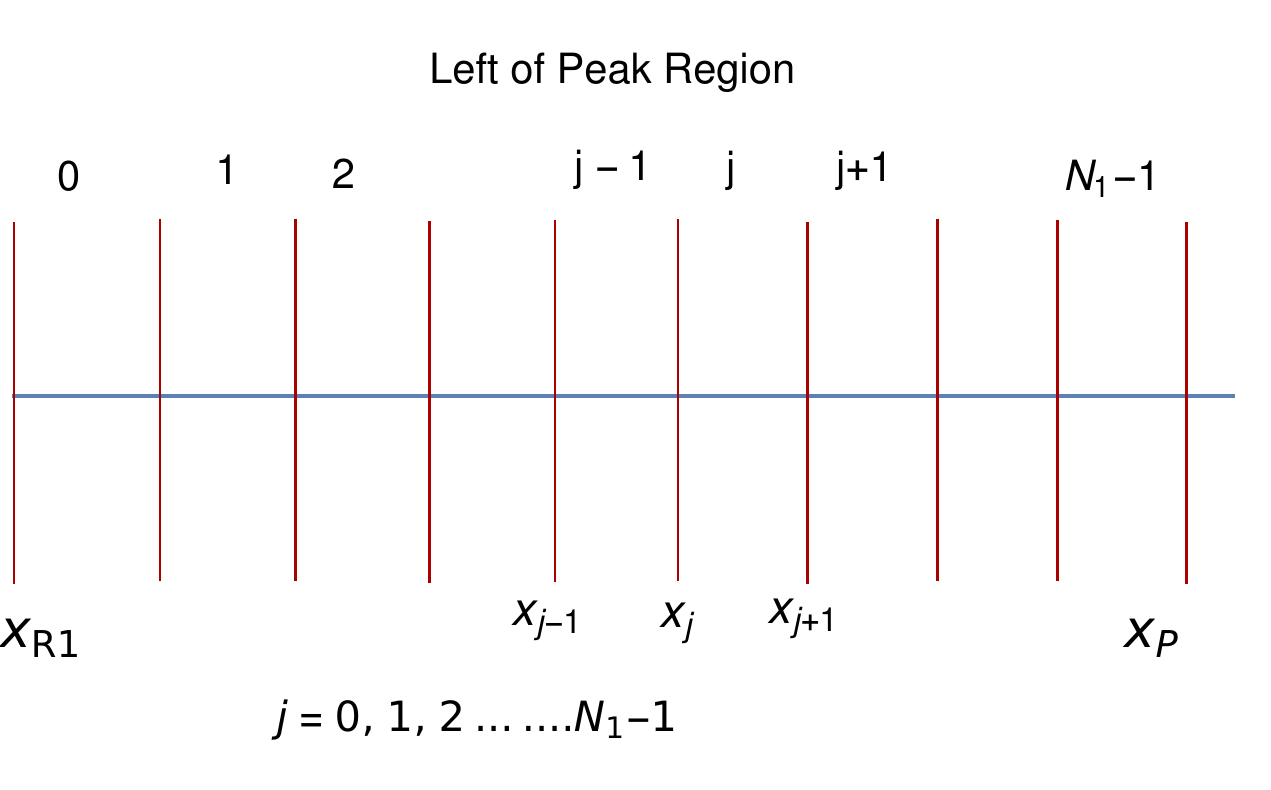}
	\caption{Figure shows the $N_1$ slices in spatial region between $[x_{R1},\,x_P]$ for Left of Peak Region.} 
	\label{fig:LPR}
\end{figure}
The transfer matrix $\bar{\bar{M}}_{x_{R1}}$ at $x_{R1}$ relates the coefficients $(C^{LP}_0$ and $D^{LP}_0$) in the $0^{th}$ slice of Left of Peak Region  to the coefficients ($C_L$ and $D_L$) in Left Edge Region.
$$ \begin{bmatrix}
\phi_{0LP}^{(1)}(x_{R1}) & \phi_{0LP}^{(2)}(x_{R1})\\ \\
\phi_{0LP}^{(1)^\prime}(x_{R1})&\phi_{0LP}^{(2)^\prime}(x_{R1})
\end{bmatrix}
\begin{bmatrix}
C^{LP}_0\\\\D^{LP}_0 
\end{bmatrix}=
\begin{bmatrix}
\phi_{0L}^{(1)}(x_{R1}) & \phi_{0L}^{(2)}(x_{R1})\\ \\
\phi_{0L}^{(1)^\prime}(x_{R1})&\phi_{0L}^{(2)^\prime}(x_{R1})
\end{bmatrix}
\begin{bmatrix}
C_L\\\\D_L 
\end{bmatrix} 
$$
$$
\begin{bmatrix}
C^{LP}_0\\\\D^{LP}_0 
\end{bmatrix}= \bar{\bar{M}}_{x_{R1}} \begin{bmatrix}
C_L\\\\D_L 
\end{bmatrix}$$
where $\bar{\bar{M}}_{x_{R1}} = \begin{bmatrix}
\phi_{0LP}^{(1)}(x_{R1}) & \phi_{0LP}^{(2)}(x_{R1})\\ \\
\phi_{0LP}^{(1)^\prime}(x_{R1})&\phi_{0LP}^{(2)^\prime}(x_{R1})
\end{bmatrix}^{-1}
\begin{bmatrix}
	\phi_{0L}^{(1)}(x_{R1}) & \phi_{0L}^{(2)}(x_{R1})\\ \\
	\phi_{0L}^{(1)^\prime}(x_{R1})&\phi_{0L}^{(2)^\prime}(x_{R1})
\end{bmatrix}$\\

The Left of Peak Region is divided into $N_1$ slices and the $j^{th}$ slice in this region is bounded by $x_j$ and $x_{j+1},$ so that the transfer matrix $M^{LP}_j$ at $x_j$ relates the coefficients in the interval $I_{j-1}$ to those in the interval $I_j$, [See Fig. \ref{fig:LPR}]. The relation is given by 
$$ \begin{bmatrix}
\phi_{j,LP}^{(1)}(x_j) & \phi_{j,LP}^{(2)}(x_j)\\ \\
\phi_{j,LP}^{(1)^\prime}(x_j)&\phi_{j,LP}^{(2)^\prime}(x_j)
\end{bmatrix}
\begin{bmatrix}
C^{LP}_j\\\\D^{LP}_j 
\end{bmatrix}=
\begin{bmatrix}
\phi_{j-1,LP}^{(1)}(x_j) & \phi_{j-1,LP}^{(2)}(x_j)\\ \\
\phi_{j-1,LP}^{(1)^\prime}(x_j)&\phi_{j-1,LP}^{(2)^\prime}(x_j)
\end{bmatrix}
\begin{bmatrix}
C^{LP}_{j-1}\\\\D^{LP}_{j-1} 
\end{bmatrix} 
$$
$$ 
\begin{bmatrix}
C^{LP}_j\\\\D^{LP}_j 
\end{bmatrix} = \begin{bmatrix}
\phi_{j,LP}^{(1)}(x_j) & \phi_{j,LP}^{(2)}(x_j)\\ \\
\phi_{j,LP}^{(1)^\prime}(x_j)&\phi_{j,LP}^{(2)^\prime}(x_j)
\end{bmatrix}^{-1}
\begin{bmatrix}
\phi_{j-1,LP}^{(1)}(x_j) & \phi_{j-1,LP}^{(2)}(x_j)\\ \\
\phi_{j-1,LP}^{(1)^\prime}(x_j)&\phi_{j-1,LP}^{(2)^\prime}(x_j)
\end{bmatrix}
\begin{bmatrix}
C^{LP}_{j-1}\\\\D^{LP}_{j-1} 
\end{bmatrix} 
\equiv \bar{\bar{M}}^{LP}_j
\begin{bmatrix}
	C^{LP}_{j-1}\\\\D^{LP}_{j-1} 
\end{bmatrix} 
$$\\
Using a similar approach one can find the transfer matrices for all slices in the Left of Peak Region. The relation between the coefficients for the $(N_1-1)^{th}$ slice and the coefficients for the $0^{th}$ slice in the Left of Peak Region is given by the product of all the transfer matrices for the intermediate nodes, taken in reverse order. 
\begin{equation}\label{M_mat}
\begin{bmatrix}
C^{LP}_{N_1-1}\\ D^{LP}_{N_1-1}
\end{bmatrix}=\bar{\bar{M}}^{LP}_{N_1-1}\cdot\bar{\bar{M}}^{LP}_{N_1-2}\cdots\bar{\bar{M}}^{LP}_{1}
\begin{bmatrix}
C^{LP}_{0}\\ D^{LP}_{0}
\end{bmatrix} \equiv \bar{\bar{M}}^{LP}  \begin{bmatrix}
C^{LP}_{0}\\ D^{LP}_{0}
\end{bmatrix}
\end{equation}
where $C^{LP}_{N_1-1}$ and $D^{LP}_{N_1-1}$ are the coefficients for ${(N_1-1)}^{th}$ slice and $C^{LP}_{0}$ and $D^{LP}_{0}$ are the coefficients for $0^{th}$ slice in Left of Peak Region. The transfer matrix for the whole of the Left of Peak Region is $\bar{\bar{M}}^{LP}$. It is given by the inner product of all the intermediate transfer matrices taken in reverse order.
\begin{equation}
\bar{\bar{M}}^{LP}=\bar{\bar{M}}^{LP}_{N_1-1}\cdot\bar{\bar{M}}^{LP}_{N_1-2}\cdot \cdots\bar{\bar{M}}^{LP}_{1}
\end{equation}

\begin{figure}[hpt]
	\centering
	\includegraphics[width=5.0in,height=3.0in]{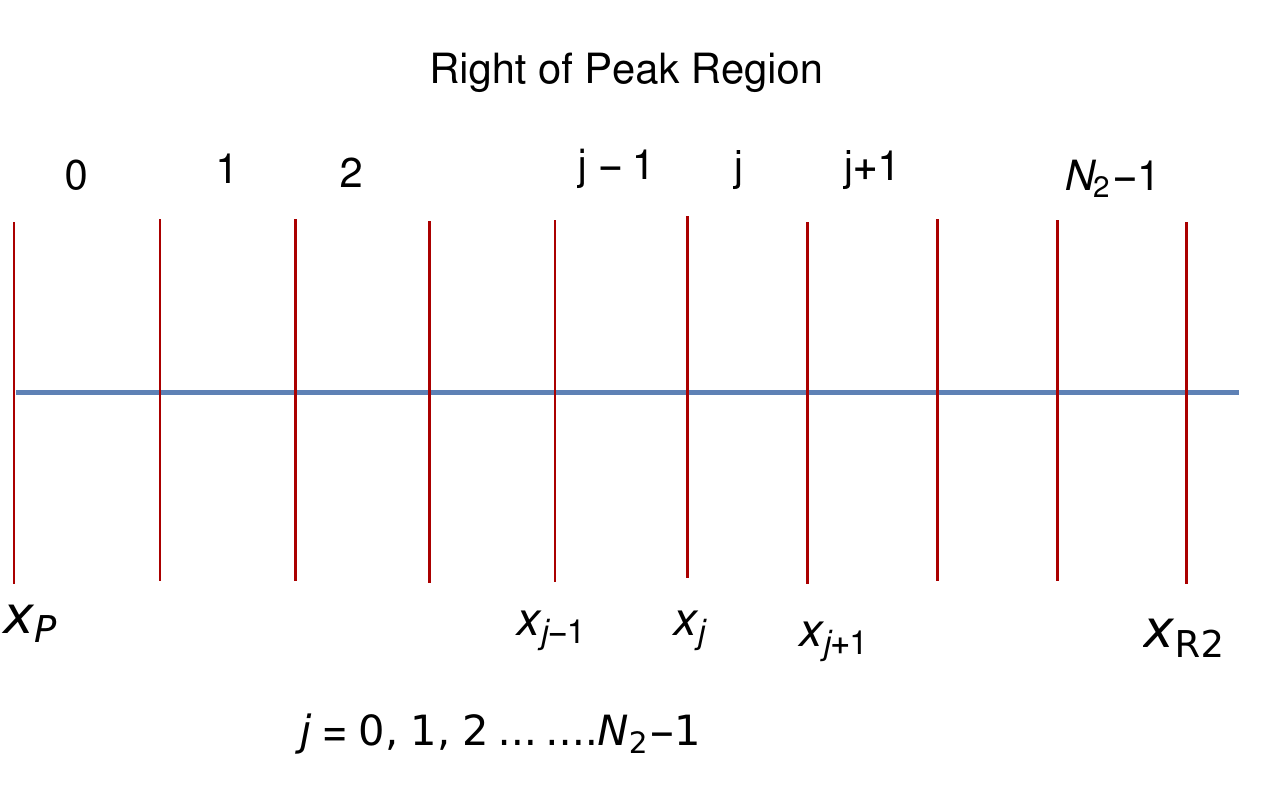}
	\caption{Figure shows $N_2$ slices in spatial region between $[x_P,\,x_{R2}]$ for Right of Peak Region.} 
	\label{figRPR}
\end{figure}
Next, the Right of Peak Region is divided into $N_2$ slices. Define a transfer matrix $\bar{\bar{M}}^{xP}$ at $x_{P}$ which relates the coefficients for the $0^{th}$ slice in Right of Peak Region ($C^{RP}_0$ and $D^{RP}_0$)  to the coefficients for $(N_1-1)^{th}$ slice in  Left of Peak Region ($C^{LP}_{N_1-1}$ and $D^{LP}_{N_1-1}$).
$$ 
\begin{bmatrix}
C^{RP}_0\\\\D^{RP}_0 
\end{bmatrix} = \begin{bmatrix}
\phi_{0,RP}^{(1)}(x_P) & \phi_{0,RP}^{(2)}(x_P)\\ \\
\phi_{0,RP}^{(1)^\prime}(x_P)&\phi_{0,RP}^{(2)^\prime}(x_P)
\end{bmatrix}^{-1}
\begin{bmatrix}
\phi_{N_1-1,LP}^{(1)}(x_P) & \phi_{N_1-1,LP}^{(2)}(x_P)\\ \\
\phi_{N_1-1,LP}^{(1)^\prime}(x_P)&\phi_{N_1-1,LP}^{(2)^\prime}(x_P)
\end{bmatrix}
\begin{bmatrix}
C^{LP}_{N_1-1}\\\\D^{LP}_{N_1-1} 
\end{bmatrix}
$$ 
$$
\begin{bmatrix}
C^{RP}_0\\\\D^{RP}_0 
\end{bmatrix} =
 \bar{\bar{M}}^{xP}
\begin{bmatrix}
C^{LP}_{N_1-1}\\\\D^{LP}_{N_1-1} 
\end{bmatrix} 
$$
where  $\bar{\bar{M}}^{xP}=\begin{bmatrix}
\phi_{0,RP}^{(1)}(x_P) & \phi_{0,RP}^{(2)}(x_P)\\ \\
\phi_{0,RP}^{(1)^\prime}(x_P)&\phi_{0,RP}^{(2)^\prime}(x_P)
\end{bmatrix}^{-1}
\begin{bmatrix}
\phi_{N_1-1,LP}^{(1)}(x_P) & \phi_{N_1-1,LP}^{(2)}(x_P)\\ \\
\phi_{N_1-1,LP}^{(1)^\prime}(x_P)&\phi_{N_1-1,LP}^{(2)^\prime}(x_P)
\end{bmatrix}$\\\\
The Right of Peak Region between $x_P$ to $x_{R2}$ is divided into $N_2$ slices labelled by $j = 0,1,2, \cdots (N_2 -1)$. Using the calculation similar to that undertaken for the Left of Peak Region, one can find the transfer matrices for all slices in the Right of Peak Region. The relation between the coefficients for the $(N_2-1)^{th}$ slice and that for the $0^{th}$ slice, both within the Right of Peak Region, is the product of all the transfer matrices for the intermediate nodes in this region taken in reverse order. 
\begin{equation}\label{RP}
\begin{bmatrix}
C^{RP}_{N_2-1}\\ D^{RP}_{N_2-1}
\end{bmatrix}=\bar{\bar{M}}^{RP}_{N_2-1}\cdot\bar{\bar{M}}^{RP}_{N_2-2}\cdots\bar{\bar{M}}^{RP}_{1}
\begin{bmatrix}
	C^{RP}_0\\ D^{RP}_0
\end{bmatrix} \equiv \bar{\bar{M}}^{RP}\begin{bmatrix}
C^{RP}_0\\ D^{RP}_0
\end{bmatrix}
\end{equation}
where $\bar{\bar{M}}^{RP}$ is given by the inner products of intermediate transfer matrices taken in reverse order.  
\begin{equation}
\bar{\bar{M}}^{RP}=\bar{\bar{M}}^{RP}_{N_2-1}\cdot\bar{\bar{M}}^{RP}_{N_2-2}\cdot \cdots\bar{\bar{M}}^{RP}_{1}
\end{equation}
\begin{figure}[hpt]
	\centering
	\includegraphics[width=5.0in,height=3.0in]{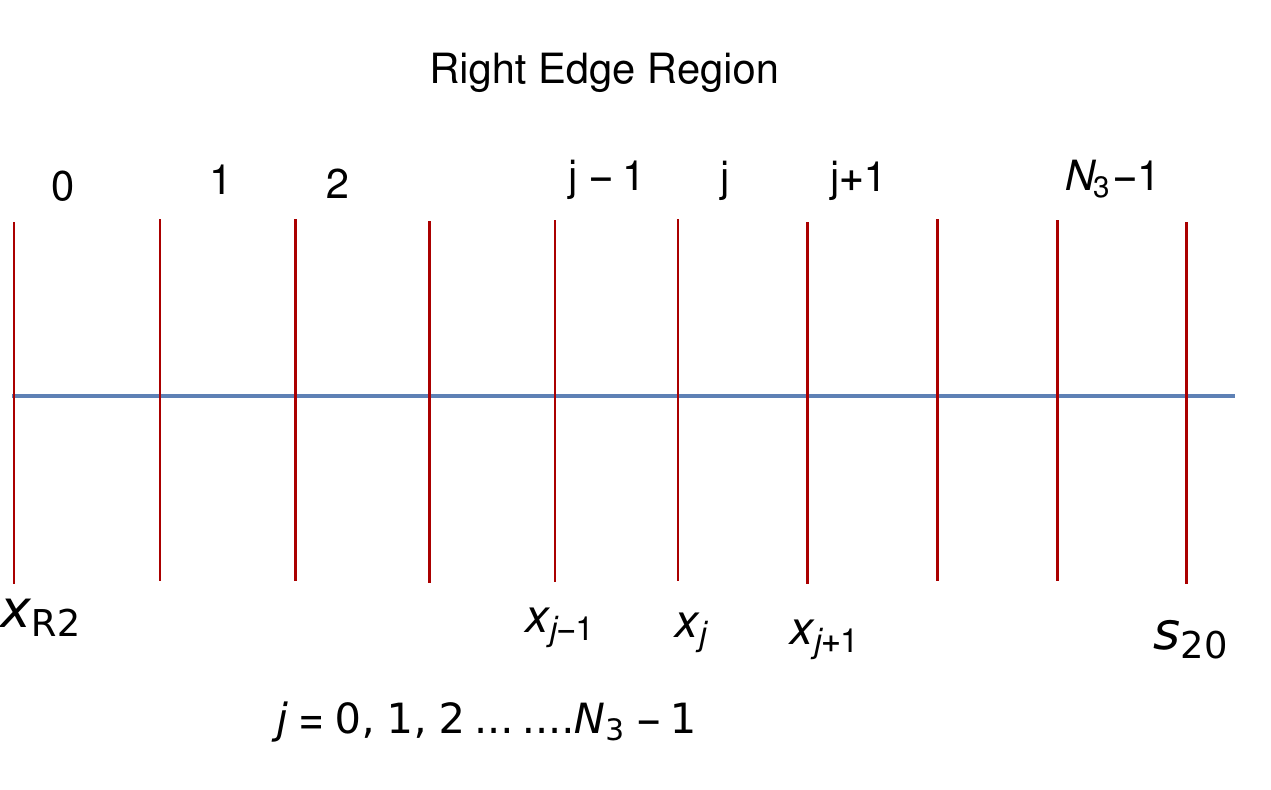}
	\caption{Figure shows $N_3$ slices in spatial region between $[x_{R2},\,s_{20}]$.} 
	\label{figRE}
\end{figure}
Next the region between $x_{R2}$ to $s_{20}$ is divided into $N_3$ slices. The transfer matrix at $x_{R2}$ relates the coefficients for the $0^{th}$ slice in Right Edge $(C^R_0$ and $D^R_0$) to  the coefficients for $(N_2-1)^{th}$ slice in Right of Peak Region ($C^{RP}_{N_2-1}$ and $D^{RP}_{N_2-1}$).
$$ \begin{bmatrix}
C^R_0\\\\D^R_0 
\end{bmatrix} = \bar{\bar{M}}_{x_{R2}} 
\begin{bmatrix}
C^{RP}_{N_2-1}\\\\ D^{RP}_{N_2-1}
\end{bmatrix} $$
where $\bar{\bar{M}}_{x_{R2}} =\begin{bmatrix}
	\phi_{0,R}^{(1)}(x_{R2}) & \phi_{0,R}^{(2)}(x_{R2})\\ \\
	\phi_{0,R}^{(1)^\prime}(x_{R2})&\phi_{0,R}^{(2)^\prime}(x_{R2})
\end{bmatrix}^{-1}
\begin{bmatrix}
	\phi_{N_2-1,RP}^{(1)}(x_{R2}) & \phi_{N_2-1,RP}^{(2)}(x_{R2})\\ \\
	\phi_{N_2-1,RP}^{(1)^\prime}(x_{R2})&\phi_{N_2-1,RP}^{(2)^\prime}(x_{R2})
\end{bmatrix}$\\\\ 
Now for $N_3$ slices in Right Edge, the relation between the coefficients for the $(N_3-1)^{th}$ slice and that for the $0^{th}$ slice, both within the Right Edge Region, is given by the product of all the transfer matrices for the intermediate nodes in this region taken in reverse order. 
\begin{equation}\label{R}
\begin{bmatrix}
C^{R}_{N_3-1}\\ D^{R}_{N_3-1}
\end{bmatrix}=\bar{\bar{M}}^{R}_{N_3-1}\cdot\bar{\bar{M}}^{R}_{N_3-2}\cdots\bar{\bar{M}}^{R}_{1}
\begin{bmatrix}
C^{R}_0\\ D^{R}_0
\end{bmatrix} \equiv \bar{\bar{M}}^{R}\begin{bmatrix}
C^{R}_0\\ D^{R}_0
\end{bmatrix}
\end{equation}
$$
\bar{\bar{M}}^R = \prod_{k=1}^{N_2-1}M^R_{N_2-k}
$$
In the Right Forbidden Region, the wave function is assumed to be the same as that given inside the second electrode. The transfer matrix at $s_{20}$ connects the $(N_3-1)^{th}$ slice in Right Edge Region, with a single slice which constitutes the Right Forbidden Region. The conditions of continuity of the wave function and its derivative at $s_{20}$ lead to the following equations 
$$
Te^{ik_2s_{20}} = C^R_{N_3-1}\phi_{N_3-1,R}^{(1)}(s_{20})+ D^R_{N_3-1}\phi_{N_3-1,R}^{(2)}(s_{20})
$$
$$
ik_2Te^{ik_2s_{20}} = C^R_{N_3-1}\phi_{N_3-1,R}^{(1)^\prime}(s_{20})+ D^R_{N_3-1}\phi_{N_3-1,R}^{(2)^\prime}(s_{20})
$$
The above equations can be written in matrix form as 
$$T\begin{bmatrix}
	1\\\\1 
\end{bmatrix} = e^{-ik_2s_{20}} 
\begin{bmatrix}
	\phi_{N_3-1,R}^{(1)}(s_{20})& \phi_{N_3-1,R}^{(2)}(s_{20})\\ \\
	\dfrac{1}{ik_2}\phi_{N_3-1,R}^{(1)^\prime}(s_{20})&\dfrac{1}{ik_2}\phi_{N_3-1,R}^{(2)^\prime}(s_{20})
\end{bmatrix}
\begin{bmatrix}
	C^R_{N_3-1}\\ \\
	D^R_{N_3-1}
\end{bmatrix}$$
\begin{equation}\label{T_atd}
T\begin{bmatrix}
1\\\\1 
\end{bmatrix} = e^{-ik_2s_{20}} 
\bar{\bar{M}}_{s_{20}} 
\begin{bmatrix}
C^R_{N_3-1}\\ \\
D^R_{N_3-1}
\end{bmatrix}
\end{equation}
where
$$\bar{\bar{M}}_{s_{20}} = \begin{bmatrix}
\phi_{N_3-1,R}^{(1)}(s_{20})& \phi_{N_3-1,R}^{(2)}(s_{20})\\ \\
\dfrac{1}{ik_2}\phi_{N_3-1,R}^{(1)^\prime}(s_{20})&\dfrac{1}{ik_2}\phi_{N_3-1,R}^{(2)^\prime}(s_{20})
\end{bmatrix}
$$
Define $\bar{\bar{V}}$ as
$$ \bar{\bar{V}} = M_{s_{20}} \cdot M_R \cdot M_{x_{R2}} \cdot M_{RP} \cdot M_{xP} \cdot M_{LP} \cdot M_{x_{R1}}$$ 
the equation (\ref{T_atd}) becomes 
$$T\begin{bmatrix}
	1\\\\1 
\end{bmatrix} = e^{-ik_2s_{20}} 
\bar{\bar{V}}
\begin{bmatrix}
	C_L\\ \\
	D_L
\end{bmatrix}
$$
the tunnel amplitude $T$ in terms of matrix elements of $\bar{\bar{V}}$ is given by
\begin{equation}\label{ats20}
 T = e^{-ik_2s_{20}}\big[ V_{11} C_L + V_{12} D_L \big] = e^{-ik_2s_{20}}\big[ V_{21} C_L + V_{22} D_L \big]
\end{equation}
solving equations (\ref{addAts10}) and (\ref{ats20}) for the coefficients $C_L$ and $D_L$  gives  \\

$ C_L = \dfrac{2 e^{ik_1 s_{10}} (V_{22} - V_{12})}{M_D(V_{11} - V_{21}) + M_C(V_{22} - V_{12})}  \quad \text{and} \quad  D_L = \dfrac{2e^{i k_1 s_{10}} - M_C C_L}{M_D} $\\\\
Substitute for these coefficients in the first of equations (\ref{ats20}) to obtain the tunneling amplitude $T(E_x)$ for the Trapezoid + Simmons Image Potential. 
\begin{equation}
T(E_x) = e^{-ik_2s_{20}}\Bigg \{ V_{11}
\Bigg[ \dfrac{2 e^{ik_1 s_{10}} (V_{22} - V_{12})}{M_D(V_{11} - V_{21}) + M_C(V_{22} - V_{12})} \Bigg]
 + V_{12} \Bigg [ \dfrac{2e^{i k_1 s_{10}} - M_C C_L}{M_D} \Bigg] \Bigg \}
\end{equation}
\begin{figure}[hpt]
	\begin{subfigure}{0.49\textwidth}
		\centering	
		\includegraphics[width=3.0in,height=2.3in]{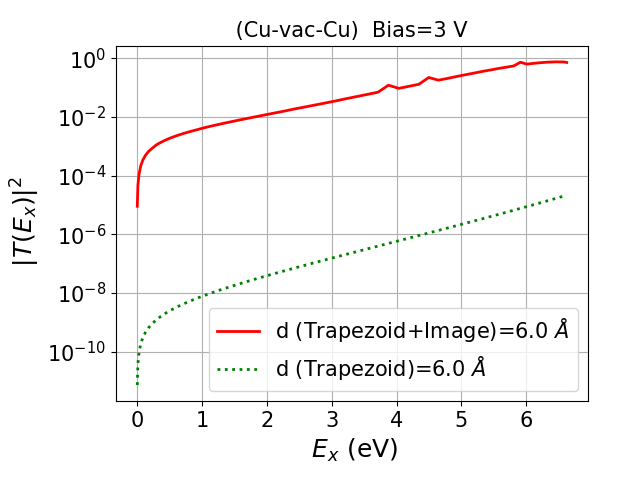}	
		\caption{} 
		\label{fig:TEx1}
	\end{subfigure}	
	\hfill
	\begin{subfigure}{0.49\textwidth}
		\centering
		\includegraphics[width=3.0in,height=2.3in]{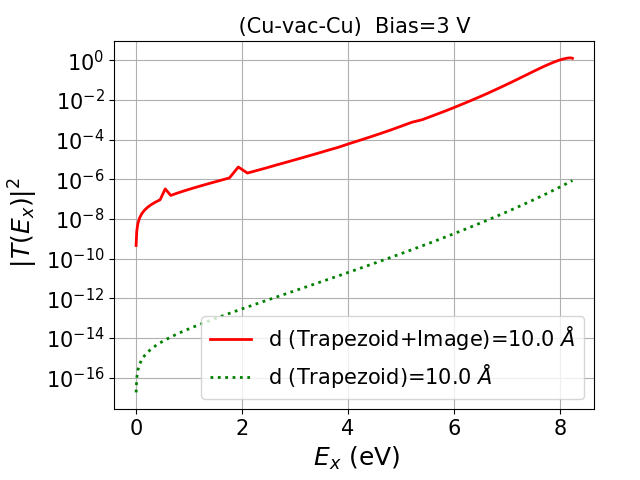}
		\caption{} 
		\label{fig:TEx2}	
	\end{subfigure}	
	\caption{Plot of $\vert T(E_x) \vert^2$ vs $E_x$ comparing Trapezoid and Trapezoid+Image potential for bias voltage of $3\, V$ and tip-sample distance (a) $d = 6$ \r{A} (b) $d = 10 $ \r{A}}
	\label{fig:TExd5N6}
\end{figure} 

\bigskip

Figs. \ref{fig:TEx1} and \ref{fig:TEx2} plot and compare the tunneling probability for Trapezoid and Trapezoid+Simmons Image potential for fixed bias voltage of $3\, V$ and tip sample distances of  $6$ \r{A} and $ 10 $ \r{A} respectively. With the inclusion of image potential, the tunneling probability increases by almost three orders of magnitude in comparison to pure Trapezoid potential. This increase can be traced back to how the image term reduces both the width and height of the barrier potential (refer figure \ref{fig:UTimd510}).  But such a drastic change in tunneling probability due to the image potential alone seems unlikely. 
\begin{figure}[hpt]
	\begin{subfigure}{0.49\textwidth}
		\centering
		\includegraphics[width=3.0in,height=2.3in]{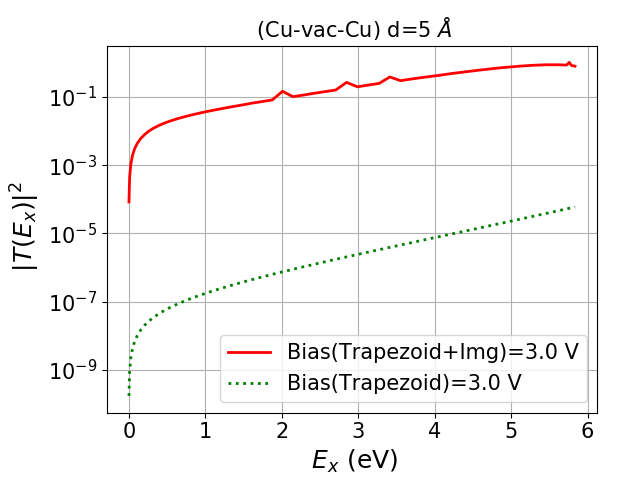}	
		\caption{} 
		\label{fig:TEx3}
	\end{subfigure}	
	\hfill
	\begin{subfigure}{0.49\textwidth}
		\centering
		\includegraphics[width=3.0in,height=2.3in]{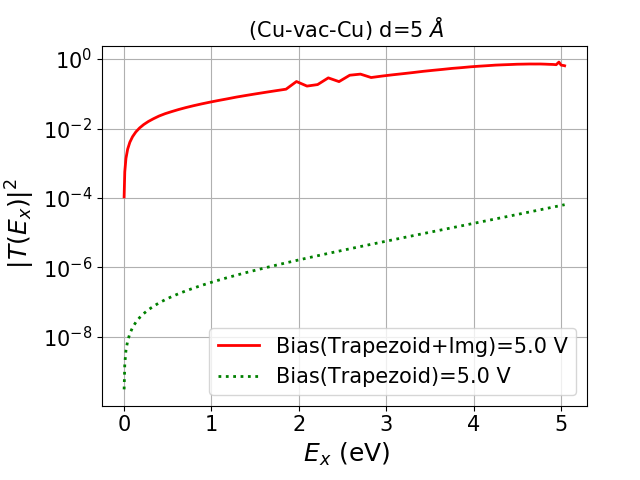}
		\caption{} 
		\label{fig:TEx4}	
	\end{subfigure}	
	\caption{Plot of $\vert T(E_x) \vert^2$ vs $E_x$ comparing Trapezoid and Trapezoid+Image potential for tip-sample distance $d = 5 $ \r{A} and bias voltage of (a) $3\, V$ and (b) $5\, V$}
	\label{fig:TExBias5N7}
\end{figure} 
Figs. \ref{fig:TEx3} and \ref{fig:TEx4} plot and compare the tunneling probability for Trapezoid and Trapezoid+Simmons Image potential for fixed tip-sample distance of $ 5$ \r{A} and bias voltages of $3 \,V$, and $5 \,V$. These plots also show that for both bias voltages $\vert T(E_x) \vert^2$ increases with $E_x$. The wiggles in the plot of image potential is purely an artifact of numerical calculation and increase in the number of slices tends to smoothen most of the wiggles. However increasing the number of slices only adds to the computational expense without any significant change in the average behavior. The tunneling probability as a function of $E_x$ shows a monotonic increasing trend for results with and without image contribution. This trend is along expected lines. 

\section{Multi - slice method for the Russell + the Simmons Image Potential}
The Russell potential for each field line is given in equation (5.38) of Chapter 5. The inclusion of Simmon's image term to the Russell potential modifies the equation to 
 \begin{equation}\label{russ}
  U(\eta) = (\eta_1+\phi_2-eV_b) + (\phi_1-\phi_2+eV_b)\, \dfrac{\lambda(\eta)}{\lambda_{\eta_{tip}}} - \dfrac{\alpha d^2}{x_{fl}(d_{fl}-x_{fl})}
\end{equation}   
\begin{figure}[hpt]
	\centering
	\includegraphics[width=5.0in,height=3.0in]{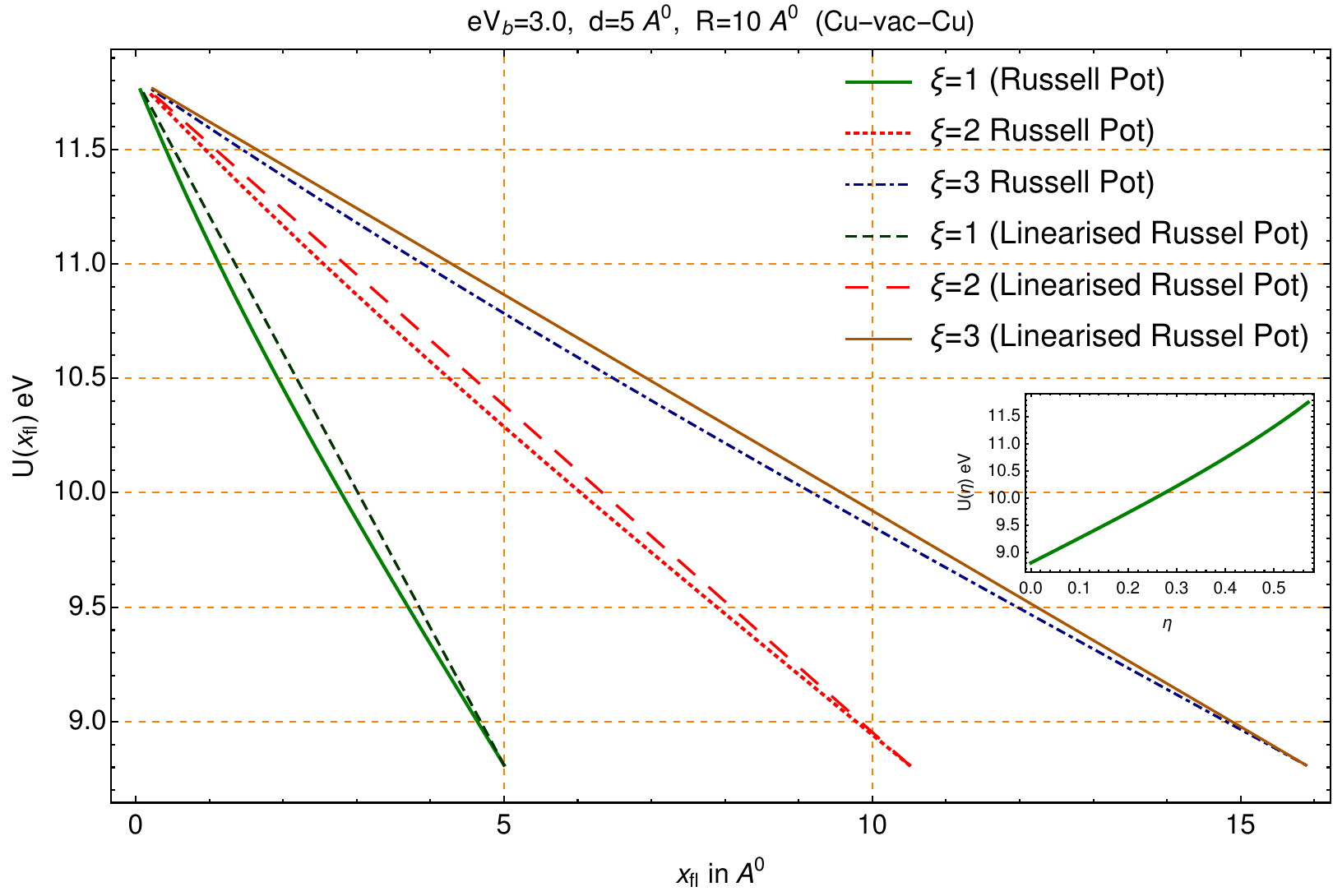}
	\caption{Plot of Russell Potential and its linearised version as a function of $x_{fl}$ for $\xi = 1,2,3$. The inset shows the behaviour of $U$ as a function of $\eta$. Note $\eta=0$ represents the sample surface while higher values of $\eta$ signify approaching closer to tip surface.} 
	\label{fig:RussellNlinear}
\end{figure}
where $x_{fl}$ is the linear distance used instead of $x$. From the graph shown in Fig.\ref{fig:RussellNlinear}, it can be seen that the Russel potential is almost linear as a function of the linear distance $x_{fl}$ along the field line for all $\xi$. Thus, instead of trapezoidal potential in the planar model that is defined as a linear function of $x \,\in \, [0,\,d]$, a linearised Russel potential is defined as a function of $x_{fl} \in [0,d_{fl}]$ for each $\xi$.  With this approximation, equation (\ref{russ}) for the Russell potential on the field line $\xi$ can be written as
\begin{equation}\label{RussLinsed}
U^L(x_{fl}(\xi)) = (\eta_1+\phi_1) - (\phi_1-\phi_2+eV_b)\, \dfrac{x_{fl}}{d_{fl}(\xi)}- \dfrac{\alpha d_{fl}^2}{x_{fl}(d_{fl}-x_{fl})}
\end{equation}
The inclusion of image potential modifies the barrier to a non-trapezoidal shape. The electron tunneling through the modified barrier encounters two classical turning points within the potential barrier for evergy energy $E_x$. 
The location of the turning points at $E_x = 0$ are the roots of the cubic equation given by equation (\ref{RussLinsed}). At non zero energy $E_x \, > \,0$, the location of the classical turning points are given by equation 
\begin{equation}\label{RussXR1R2} 
 U^L(x_{fl}(\xi)) - E_x = 0
\end{equation}
The third root in both cases lies outside the barrier region. The location of the maxima of the potential function $U_{Tim}$ is calculated by solving the equation  
\begin{equation}\label{RussXP}
U^\prime(x) = -\dfrac{\phi_1-\phi_2+eV_b}{d_{fl}}-\dfrac{\text{$\alpha $} d_{fl}^2 \left(2 x_{fl}-d_{fl}\right)}    {\left[x_{fl} \left(d_{fl}-x_{fl}\right)\right]{}^2} = 0
\end{equation}
The roots of the quartic equation are found numerically by using Newton-Raphson method. Bsaed on the location of the roots of equations  (\ref{RussLinsed}),  (\ref{RussXR1R2}), and (\ref{RussXP}) the spatial region $[0,\,d_{fl}]$ is divided into the Forbidden, Left Edge, Right Edge, Left of peak, and Right of peak regions. These regions are further subdivided into an appropriate number of slices analogous to the method described in section 2. Using linear approximation for $U$ in each of the slices, and imposing the  continuity of wavefunction and its derivative at the common boundaries, the corresponding transfer matrices are obtained. From these transfer matrices tunnel amplitude can be calculated as described in section 4 of Chapter 5. 

\begin{figure}[hpt]
	\centering
	\begin{subfigure}{0.49\textwidth}
	\includegraphics[width=3.0in,height=2.3in]{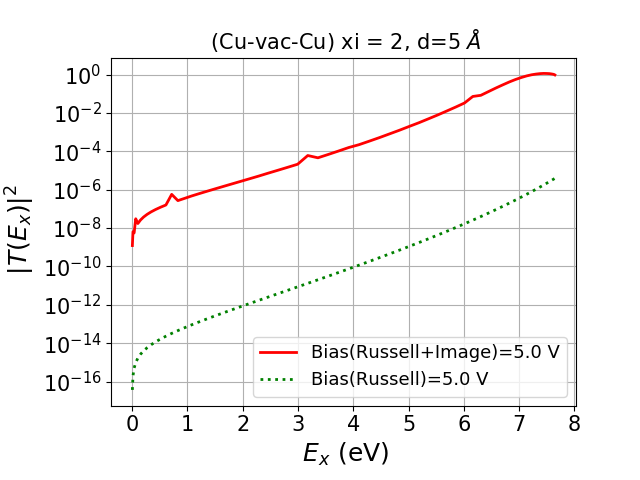}	
	\caption{}
	\label{fig:TExRd5Bias3}
	\end{subfigure}
\hfill
	\begin{subfigure}{0.49\textwidth}
	\centering
	\includegraphics[width=3.0in,height=2.3in]{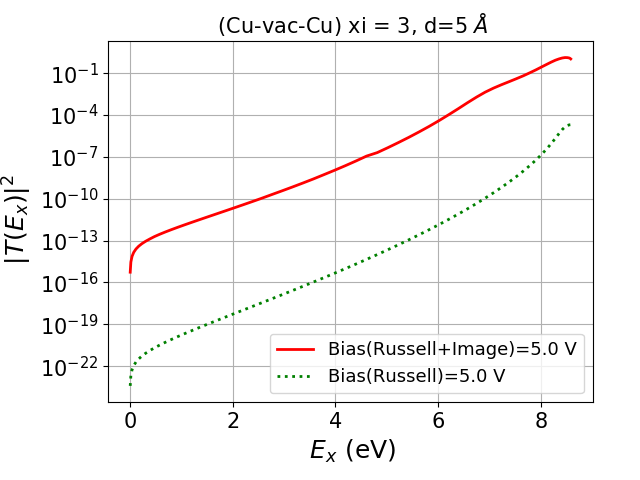}
	\caption{}
	\label{fig:TExRd5Bias5}
	\end{subfigure}	
	\caption{Plot of $\vert T(E_x) \vert^2$ vs $E_x$ comparing Russell and Russell+Image potential for tip-sample distance $d = 5 $ \r{A}  and bias voltage of (a) $3\, V$ and (b) $5\, V$}
	\label{fig:TExRd5Bias35}
\end{figure}

\bigskip

Fig. \ref{fig:TExRd5Bias3} and \ref{fig:TExRd5Bias5} plot and compare the tunneling probability for Russell and Russell+Simmons Image potential for $\xi = 2,\,3$ with fixed tip-sample distance of $ d = 5 $ \r{A} and bias voltage of $V_b = 5 \,V$. These plots again show that the tunneling probability $\vert T(E_x) \vert^2$ with images increases  by three orders of magnitude in comparison to that for the pure Russell potential (without images contribution). This huge increase due to the image contribution is seen for both the Trapezoid as well as for the Russell potential cases. This huge increase caused by the image potential alone, appears unphysical. The increase of the tunneling probability with increasing $E_x$ for pure Russell and for pure Russell+Image is similar to that seen for Trapezoid and Trapezoid + Image potential and both behaviours are expected.   

\bigskip

\section{Comparison of Current Densities $J_{Net}$ With and Without Image Potential}
The net current density calculated from tunneling probabilites (using the method described in Chapter 3) is plotted as a function of tip-sample distance and bias voltage. 
 \begin{figure}[hpt]
	\begin{subfigure}{0.49\textwidth}
		\centering
		\includegraphics[width=3.0in,height=2.3in]{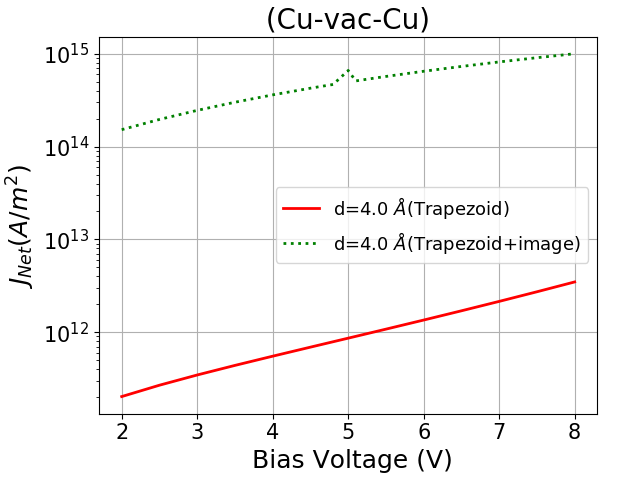}
		\caption{} 
		\label{fig:JnetTimd5}
	\end{subfigure}	
	\hfill
	\begin{subfigure}{0.49\textwidth}
		\centering
		\includegraphics[width=3.0in,height=2.3in]{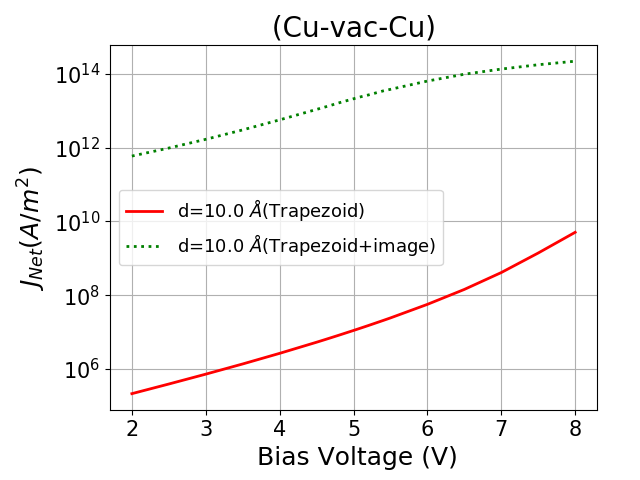}
		\caption{} 
		\label{fig:JnetTimd10}	
	\end{subfigure}	
	\caption{Plot of current density $J_{Net}$ vs Bias voltage comparing Trapezoid and Trapezoid+Image potential for tip-sample distance (a) $d = 4 $ \r{A} (b) $d = 10 $ \r{A} }
	\label{fig:JnetTimd510}
\end{figure} 
Fig \ref{fig:JnetTimd5} and \ref{fig:JnetTimd10} compares the current density $J_{Net}$ for Trapezoid and Trapezoid+Simmon's Image potential as a function of bias voltage for fixed tip-sample distance of $4 $ \r{A} and $10 $ \r{A}.   The increase in current density with images by nearly three orders of magnitude in comparison to pure trapezoid is expected because the tunneling probability increases by three orders of magnitude with images as seen in Figures \ref{fig:TExd5N6}, \ref{fig:TExBias5N7}, \ref{fig:TExRd5Bias35}.  In the Fig. \ref{fig:JnetTimd510}, the current density increases with bias for both Trapezoid and Trapezoid+Simmon's Image potential which behaviour is along expected lines.
 \begin{figure}[hpt]
 	\begin{subfigure}{0.49\textwidth}
 		\centering
 		\includegraphics[width=3.0in,height=2.3in]{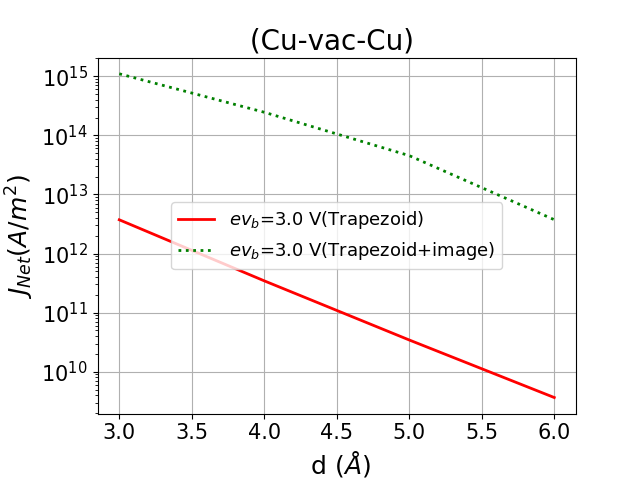}	
 		\caption{} 
 		\label{fig:JnetTimV3}
 	\end{subfigure}	
 	\hfill
 	\begin{subfigure}{0.49\textwidth}
 		\centering
 		\includegraphics[width=3.0in,height=2.3in]{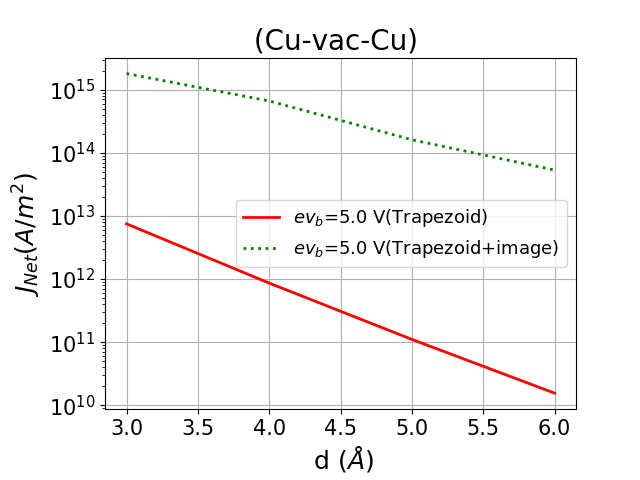}
 		\caption{} 
 		\label{fig:JnetTimV5}	
 	\end{subfigure}	
 	\caption{Plot of $J_{Net}$ vs  $d$ comparing Trapezoid and Trapezoid+Image potential for  Bias voltage (a) $V_b= 3\, V$ and (b) $V_b= 5\, V$ }
 	\label{fig:Jnet35}
 \end{figure}  
Fig. \ref{fig:JnetTimV3} and \ref{fig:JnetTimV5} compare the current density  $J_{Net}$ for pure Trapezoidal potential and Trapezoid + Simmons Image Potential as a function of the tip-sample distance for bias voltages of $3 \,V$ and  $5 \,V$. In both cases, the current density decreases precipitously as the tip-sample distance ($d$) increases. 
\begin{figure}[hpt]
	\begin{subfigure}{0.49\textwidth}
		\centering
		\includegraphics[width=3.0in,height=2.3in]{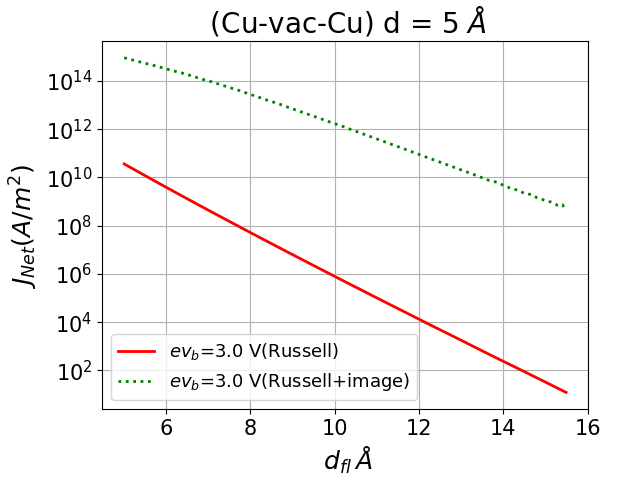}
		\caption{} 
		\label{fig:JnetRimV3}
	\end{subfigure}	
	\hfill
	\begin{subfigure}{0.49\textwidth}
		\centering
		\includegraphics[width=3.0in,height=2.3in]{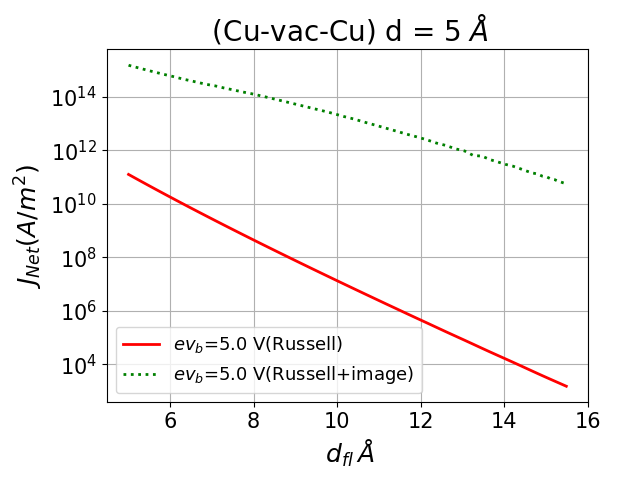}
		\caption{} 
		\label{fig:JnetRimV5}	
	\end{subfigure}	
	\caption{Plot of $J_{Net}$ vs  $d$ comparing Russell and Russell+Image potential for  Bias voltage (a) $V_b= 3\, V$ and (b) $V_b= 5\, V$}
	\label{fig:JnetRim35}
\end{figure}  
Fig. \ref{fig:JnetRimV3} and \ref{fig:JnetRimV5} compares current density $J_{Net}$ for Russell potential and Russell + Simmons Image Potential  as a function of $d_{fl}(\xi)$ which is the length along the field line for bias voltage of $3 \,V$ and $5 \,V$. The current density with images is again found to increase by three orders of magnitude as compared to Russell potential alone. In both cases, the current density decreases as ($d_{fl}$) increases.

\bigskip

With the inclusion of image potential, the boost in tunneling probability and tunneling current density in comparison to pure Trapezoid / Russell potential is too drastic. In the calculation of Simmon's image potential,  the fundamental idea was that the tunneling electron inside the barrier is a point particle. This is probably not well founded, because the electron in the barrier region, is described by a wavefunction which does not describe a freely moving point particle. Instead it represents a charge that is distributed throughout the barrier region, and the image potential produced by this distributed charge may NOT cause such strong enhancement as seen for that due to a point charge model. It is therefore necessary to explore the response of the system to the image potential generated by a distributed charge in the barrier region. This is done in the next Chapter, where models of the charge distribution in the barrier region, are constructed, and the image effects of these on tunneling probabilities, and current densities, are studied.  

%
%
%
%

	\chapter{Distributed Charge Models for the Image Potentials}\label{Chap7}
%
%

\section{Introduction}

In Chapter 6 the image potential was assumed to be due to the image charges produced on the surfaces of the electrodes which were maintained at constant specified potentials. These potentials could be $V_\text{tip}$ for the tip (first electrode) and $V_\text{sample}$ for the sample (second electrode). These image charges would be due to the electrostatic field produced by a point electron tunneling through the barrier region. However this assumption is not justified, owing to the fact that the electron in the barrier region is \underline{not} described by a travelling sinusoidal wave which is characteristic of a freely moving point particle, but is instead described by a wave function which is monotonic in the regions where the Kinetic Energy of the electron is negative, and by an oscillatory function, not necessarily sinusoidal in the regions where its kinetic energy is positive. Therefore the image potential must be calculated through a the solution of a Poisson equation in which the charge density $\rho(\mathbf{r})$ in the barrier region is supplied by the solution to the Schroedinger Equation for the net potential $V_\text{net}$ which is the sum of the applied potential and the image potential. Thus the problem of finding the image potential involves  solving of the coupled Poisson and Schroedinger equations which are given below.
\begin{equation}\label{eqn7.1}
[-\dfrac{\hbar^2}{2 m} \bm \nabla^2  + (-e) V_\text{net}] \psi(\mathbf{r}) = E_x \psi(\mathbf{r}) 
\end{equation}
\begin{equation}\label{eqn7.2}
\bm \nabla^2 V_\text{net} = -(-e)\dfrac{1}{\epsilon_0} |\psi(\mathbf{r}) |^2  \, \, ; \quad V(x= 0) = V_\text{tip} \,\, \text{and} \,\, V( x= d) = V_\text{sample} 
\end{equation}
Because these two equations are coupled, the potential $V_\text{net}$ is consistent with the wave function and vice a versa. The self consistent expression for the image potential is then defined by 
\begin{equation}\label{eqn7.3}
V_\text{image} = V_\text{net} - V_\text{ext}
\end{equation}
where the same boundary conditions are satisfied by $V_\text{ext}$ as by $V_\text{net}$ except that it satisfies the Laplace Equation.

\bigskip

The solution of the coupled equations (\ref{eqn7.1}\, \& \, \ref{eqn7.2}), can be carried out through iterations. First  equation (\ref{eqn7.1}) can be solved with $V_{ext}$ in place of $V_{net}$, and the (zeroth iterated) wave function so obtained is substituted in equation (\ref{eqn7.2}), and it is solved to find the first iterate for the potential $V^{(1)}_{net}$, and this is now again substituted in equation (\ref{eqn7.1}), and the whole process repeated again and again. Thus one can in principle obtain a sequence of functions $V^{(1)}_{net}$, $V^{(2)}_{net}\, \cdot \, V^{(n)}_{net}$, which hopefully meet some appropriate convergence criteria. The zeroth iterate for the potential is $V_{ext}$ and for planar models is the trapezidal potential or the Russell Potential for non planar models described in the prolate spheroidal coordinate system. In this Chapter, the planar models are not considered. Further the two 3 - dimensional equations (\ref{eqn7.1}\, \& \,  \ref{eqn7.2}), are reduced to one dimensional equations in a suitable independent variable $\eta$. Two different models for the charge density $\rho$ involving the zeroth iterate for the wave function are constructed, and for each of them the Poisson equation in Prolate Spheroidal Coordinate System, is modified to look like a one dimensional equation in $\eta$, with an average source function $\bar{f}(\eta)$. This equation (in each model) is solved by direct integration to find the first iterate for the potential. This is compared with the zeroth iterate for the potential ($viz$ the Russell Potential) and the effects of the difference ( which is the image potential in the first iteration) on the tunneling probability $|T(E_x, \xi) |^2$, and the current density $j_{Net}(\xi)$ are studied. Note only results for the first order iterative calculations are reported, since the effects are found small enough to neglect higher order corrections. This Chapter gives a detailed description of the procedures involved and discusses the results of the calculations.
 
\bigskip

\section{The function $\bar{f}(\eta)$ and the Image Potential}

The exact solution of the coupled equations (\ref{eqn7.1}\,  \& \,\ref{eqn7.2}) in 3 dimensions is beyond the scope of this thesis. Instead, an approximate approach is followed. First the solution of the Schroedinger equation with only the external potential is constructed. Note there is no image potential in this equation. This makes the equation 
\begin{equation}\label{eqn7.4}
[-\dfrac{\hbar^2}{2 m} \bm \nabla^2  + (-e) V_\text{ext}] \psi_0(\mathbf{r}) = E_x \psi_0(\mathbf{r}) 
\end{equation}
uncoupled.  Although the above equation is 3 - dimensional, an appropriate choice of a coordinate system along with symmetry considerations, can reduce this to a one dimensional equation. In the Planar models this reduces to an equation in which the independent variable is $x$, and the barrier region is defined by the interval $ 0 \, \leqslant \, x \, \leqslant \, d$. In the sharp - tip \&  plane sample configuration of the STM, described in the prolate spheroidal coordinate system is employed and in this system the independent variable is $\eta$ and the barrier region is defined by the interval $0 \, \leqslant \, \eta \, \leqslant  \, \eta_\text{tip}$.  

\bigskip

Some very plausible models for the charge density can be constructed. A model dependent charge density function can be constructed using the wavefunction of equation(\ref{eqn7.4}), and the charge density so obtained is now substituted into the following Poisson equation 
\begin{equation}\label{eqn7.5}
\bm \nabla^2 V_\text{net} = -(-e)\dfrac{1}{\epsilon_0} \rho [|\psi_0(\mathbf{r}) |^2] \, \,; \quad V(x= 0) = V_\text{tip} \,\, \text{and} \,\, V( x= d) = V_\text{sample} 
\end{equation}
Note that the above equation is uncoupled, and the inhomogeneous term $\rho$ is a one dimensional function. It would be desirable to ensure that the Poisson equation also could also be rendered one dimensional. This is trivial for planar models, but because of the infinite extent of the electrodes, there will be infinite volumes for the barrier region. Any finite charge in this region would give zero charge density. So it is relatively difficult to carry out a self consistent solution in the planar models. However for the sharp tip and plane sample configuration of the STM, described in the prolate spheroidal coordinate system, the region can be made finite by the limits $0 \, \leqslant \, \eta \, \leqslant  \, \eta_\text{tip}$, $1 \, \leqslant \, \xi \, \leqslant  \, \xi_\text{max}$ and $0 \, \leqslant \, \phi \, \leqslant  \, 2 \pi$. Hence self consistent models shall be constructed in this coordinate system alone.

\bigskip

In this coordinate system, the STM is assumed to be an assembly of infinitesimally thin STM's (as discussed in Chapter 4) with each infinitesimally thin STM covering a field line characterized by the value of the $\xi$ coordinate. The distances along this field line are measured by $x_{fl}$ which in addition to being dependant on $\xi$ will also depend upon $\eta \, ,  i.e.\,\, x_{fl} =  x_{fl}(\xi \, , \, \eta)$. Thus the wave function which solves the Schroedinger equation (\ref{eqn7.4}) which is one dimensional in $x_{fl}$ will be a function of $x_{fl}$ only, and therefore will be a function of  two coordinates $\xi$ and $\eta$. Thus the charge density function $\rho[|\psi_0|^2]$ must also be a function of two coordinates $\xi$ and $\eta     \, ,  i.e.\,\, V_{net} =  V_{net}(\xi \, , \, \eta)$. The Poisson equation (\ref{eqn7.5}) must be rewritten in the prolate spheroidal coordinate system as 
\begin{equation}\label{eqn7.6}
\dfrac{1}{a^2(\xi^2 - \eta^2)}\Big[\dfrac{\partial}{\partial \eta}\big \lbrace (1 - \eta^2) \dfrac{\partial}{\partial \eta}\big \rbrace + \dfrac{\partial}{\partial \xi}\big \lbrace (\xi^2 - 1) \dfrac{\partial}{\partial \xi} \big \rbrace \Big] V_{net}(\xi,\, \eta) = -\dfrac{1}{\epsilon_0} \rho (\xi,\, \eta)
\end{equation}
with boundary conditions given by
\begin{equation}\label{eqn7.7}
 V(\eta= \eta_{tip}) = V_\text{tip} \,\, \text{and} \,\, V( \eta= 0) = V_\text{sample} 
\end{equation}
It is much more convenient to deal with the potential energy of the electron than with the electrostatic potential. After all the two are related by a constant factor $(-e)$ which is the charge of the electron. Thus the potential energy function $U(\xi,\, \eta)$ will be used everywhere. It will satisfy the equation 
\begin{equation}\label{eqn7.6b}
\dfrac{1}{a^2(\xi^2 - \eta^2)}\Big[\dfrac{\partial}{\partial \eta}\big \lbrace (1 - \eta^2) \dfrac{\partial}{\partial \eta}\big \rbrace + \dfrac{\partial}{\partial \xi}\big \lbrace (\xi^2 - 1) \dfrac{\partial}{\partial \xi} \big \rbrace \Big] U(\xi,\, \eta) = \dfrac{e}{\epsilon_0} \rho (\xi,\, \eta)
\end{equation}
Ideally this equation would be solvable by means of a Green function for the problem. The determination of the Green function is indeed carried out by L. H. Pan et. al. \cite{cutler}. However the calculations involve the functions $P^m_{-1/2 + i \tau}(\mu)$, which functions are very difficult to compute numerically, and integrals of products of such functions over the variable $\tau$ are even more difficult to compute. Thus the exact solution of the two dimensional Poisson-like equation (\ref{eqn7.6b}) in the variables $\xi$  and  $\eta$ will not be attempted. It would be much easier if this equation could somehow be reduced to an equation in one variable only  viz. $\eta$. Let us assume that there does exist such a function viz. $U(\eta) = (-e) V_{net}(\eta)$ which is independent of $\xi$, and  which satisfies the equation 
\begin{equation}\label{eqn7.8}
\Big[\dfrac{\partial}{\partial \eta}\big \lbrace (1 - \eta^2) \dfrac{\partial}{\partial \eta}\big \rbrace \Big] U(\eta) = \bar{f}(\eta)\,  \quad  0 \, \leqslant \, \eta \, \leqslant  \, \eta_\text{tip}
\end{equation}
where $U$ is the potential energy of the electron in the electrostatic potential $V_{net}$. It satisfies the boundary conditions 
\begin{equation}\label{eqn7.9}
U(\eta= \eta_{tip}) = U_\text{tip} \,\, \text{and} \,\, U( \eta= 0) = U_0 
\end{equation}
and $\bar{f}(\eta)$ is the average of the function 
\begin{equation}\label{eqn7.10}
f(\xi, \eta) =  \Big[- (-e)(\dfrac{1}{\epsilon_0})a^2 (\xi^2 - \eta^2) \rho(\xi,\, \eta) \Big] 
 \end{equation}
over the surface of the hyperboloid $\eta$ bounded by the values of $\xi$, given by $1 \, \leqslant \, \xi \, \leqslant  \, \xi_\text{max}$ .  This is given by
\begin{equation}\label{eqn7.11}
\bar{f}(\eta) = \dfrac{1}{S(\eta)}\int\limits_1^{\xi_{max}} dS(\xi, \eta) f(\xi, \eta)
\end{equation}
where $dS(\xi, \eta)$ is the surface element described in equation (7) of Chapter 4. Here $S(\eta)$ is the area of the hyperboloid surface bounded by a circle which is generated by the intersection of the  prolate spheroid described by $\xi = \xi_{max}$ with the hyperboloid described by $\eta$. The area $S(\eta)$ can be calculated exactly to be 
\begin{equation}\label{eqn7.12}
\begin{aligned}
S(\eta) = \pi a^2\sqrt{(1 - \eta^2)}\Bigg[ \Big \lbrace \xi_{max}\sqrt{(\xi_{max}^2 - \eta^2)} - \sqrt{(1 - \eta^2)}\Big \rbrace \\
- \eta^2 \Big \lbrace \cosh^{-1}(\xi_{max}/\eta) - \cosh^{-1}(1/\eta) \Big \rbrace \Bigg ] 
\end{aligned}
\end{equation}
Thus the exact solution $U(\xi,\, \eta)$ of the equation (\ref{eqn7.6b}) will be replaced by the solution $U(\eta)$ of the equation (\ref{eqn7.8}) as an approximation. The validity of this approximation rests on the physical assumption that charge is distributed in the barrier region in such a way that equipotential surfaces and field lines are not altered very much, and any change in them may be ignored in the first order. Integrate equation (\ref{eqn7.8}) directly to find 
$$ (1 - \eta^2) \dfrac{\partial}{\partial \eta} U(\eta) = \int\limits_0^{\eta} \bar{f}(\eta) \,d \eta + C_1
$$ 
where $C_1$ is an integration constant. Define a function $\bar{G}(\eta)$ such that $\bar{G}(\eta) = \int\limits_0^{\eta} \bar{f}(\eta^\prime ) \,d \eta^\prime$.
Divide both sides by $(1 - \eta^2)$ and integrate again to get 
$$
U(\eta) = \int\limits_0^\eta \dfrac{\bar{G}(\eta^\prime )}{(1 - {{\eta}^\prime}^2)} d \eta^\prime + C_1 \int\limits_0^{\eta^\prime} \dfrac{d \eta^\prime}{(1 - {{\eta}^\prime}^2)} + C_2
$$ 
where $C_2$ is another integration constant.  Define another function $\bar{H}(\eta) = \int\limits_0^\eta \dfrac{\bar{G}(\eta^\prime )}{(1 - {{\eta}^\prime}^2)} d \eta^\prime $ so that 
$$U(\eta) = \bar{H}(\eta) + C_1 \lambda(\eta) + C_2$$
where $\lambda(\eta)$ has been described in equation (16) of Chapter 4. The constants $C_1$ and $C_2$ can be found by imposing the boundary conditions given in equation (\ref{eqn7.9}). This gives 
\begin{equation}\label{eqn7.13}
U(\eta) = \Big[U_0 + \dfrac{\lambda(\eta)}{\lambda(\eta_{tip})} [U_{tip} -U_0]\Big] + \Big[\bar{H}(\eta) - \dfrac{\lambda(\eta)}{\lambda(\eta_{tip})} \bar{H}(\eta_{tip})\Big]
\end{equation}
The first term in the RHS of equation (\ref{eqn7.13}) is clearly the Russell Potential which is the zeroth iterate for the potential energy and is the same as $U_{ext}$, the potential energy due to the externally applied bias potential. The second term in the RHS must clearly be the image potential term. Thus in this calculation [$\lambda(\eta_{tip}) = \lambda_{tip }$ and $\bar{H}(\eta_{tip}) =  \bar{H}_{tip}$] the image potential term is given by 
\begin{equation}\label{eqno 7.13}
U_\text{image}(\eta) =  \Big[\bar{H}(\eta) - \dfrac{\lambda(\eta)}{\lambda_{tip}} \bar{H}_{tip}\Big]
\end{equation}
As expected this image potential vanishes at $\eta = 0$ and at $\eta = \eta_{tip}$.

\bigskip

\section{The Iterative Calculation}
\bigskip

At any level of iteration the several quatities discussed below pertaining to the previous iteration, are assumed to be available or have been calculated. In the $(n-1)^\text{th}$  iteration, 
\begin{enumerate} 
\item[1.] The potential $U^{(n-1)}(\eta)$ is available and is considered known. The image potential in this order of iteration will be $U^{(n-1)}_\text{image}(\eta) = U^{(n-1)}(\eta) - U_\text{Russell}(\eta)$. 
\item[2.] The Tunneling amplitude $T^{(n-1)}(E_x, \xi)$ is calculated for each energy $E_x$, and for each field line $\xi$ using the Multislice method described in Chapter 5.  
\item[3.] Back calculating from the tunneling amplitude, and using the transfer matrices, all the coefficients  $C_i$ and $D_i$ for each slice in the barrier region can be calculated.
\item[4.] Using these coefficients the wave function in each slice can be calculated and therefore the wave function in the entire barrier region is known. This may be called the $(n-1)^\text{th}$ iterate of the wave function, and be labelled by $\psi^{(n-1)}(E_x, \xi)$.
\item[5.] In this order of iteration the calculation of the current density $j^{(n-1)}_{net}(\xi)$, and the net current $I^{(n-1)}_\text{net}$ can be obtained.
\item[6.] Two models will be constructed. In Model 1 the charge density is assumed to be uniform and the net charge in the barrier region is assumed to be equal to the electron charge. In Model 2 the  charge density in the barrier region is obtained from the tunneling probability  $|T^{(n-1)}(E_x, \xi)|^2$. The details of these two models shall be discussed in the next section.
\item[7.] From this charge density, calculate the function $\bar{f}(\eta)$, which will be useful for the next iteration.
\end{enumerate}

\bigskip

In the next $viz \, \,   n^\text{th}$ iteration the function $\bar{f}(\eta)$ calculated in step 7 of the above list can be used to to calculate the potential energy  $U^n(\eta)$ by solving the inhomogeneous equation (\ref{eqn7.8}) by the process of direct integration, and the process indicated in the steps 1 to 7 in the list above can be repeated with the new potential energy function $U^n(\eta)$ to obtain results in the $n^\text{th}$ iteration. In this thesis only the zeroth and the 1st iterations will be calculated. The results in the zeroth order involves the potential that is purely the Russell Potential, and the results for these have been calculated in Chapter 6. In this thesis the results of Chapter 6 for the Russell Potential without image potential, shall constitute results of the zeroth order irteration. These zeroth order results can be used to obtain the first order correction to the tunneling amplitudes, the current densities and the net current. The iteration procedure shall stop at the first order only, as the results will warrant the neglect of higher order contributions. The potentials in the zeroth and the first order interations shall be plotted to show the extent of the effect of the presence of the image potentials. \\

\bigskip

\section{Models for $\bm \rho(\xi, \, \eta, E_x)$ \&  the associated function  $\bm \bar{f}(\eta)$}

\bigskip

The main impetus in constructing these models is to replace a point sized electron in the barrier region with an extended charge distribution, that occupies the whole of barrier region. If the effective barrier region is infinite in extent, then any negative definite charge distribution will lead to there being an infinite amount of charge in the barrier region. And likewise a finite total charge entails a zero charge density in the region. It is difficult to work with both these results. Hence planar models will have difficulty incorporating distributed charge in the barrier region. However in the sharp-tip \& plane sample configuration well described by confocal hyperboloidal surfaces in the Prolate Coordinate system, it has been found that only a finite extent of these surfaces are relevent, since for $\xi > \xi_{max}$, there is negligible contribution to the tunneling current. Hence one can consider a finite effective volume in this system, which is bounded by the following coordinate values $viz \, $, $1 \, \leqslant \, \xi \, \leqslant \, \xi_{max}$, $0 \, \leqslant \, \eta \, \leqslant \, \eta_{tip}$ and finally $0 \, \leqslant \, \phi \, \leqslant \, 2 \pi$. The volume of this region can be found by integrating the expression for the volume element given in equation (6) of Chapter 4. The volume of this region can be calculated to be 
\begin{eqnarray}\label{eqn Vmax}
V_{max} &=& 2 \pi a^3 \int\limits_0^{\eta_{tip}} d\eta \int\limits_1^{\xi_{max}} d \xi \,(\xi^2 - \eta^2) \\
 &=& \dfrac{2 \pi a^3}{3}\eta_{tip}(\xi_{max} - 1)\big[(\xi^2_{max} - \eta_{tip}^2) + (\xi_{max} +1) \big]
\end{eqnarray}

\bigskip

\bigskip

\subsection{Model I. Uniform Charge Density}

The simplest model would be that of uniform charge density distribution throughout the allowed volume of the barrier region and the associated function $\bar{f}(\eta)$ shall be computed for this model. This model is likely to give a good order of magnitude estimate of the image effect. This model requires that the electron charge to be spread uniformly over the available volume $V_{max}$, so that $\rho = \dfrac{-e}{V_{max}}$. The function $f(\xi, \eta)$ becomes [See equation(\ref{eqn7.10})] 
$$f(\xi, \eta) = -a^2\dfrac{e^2}{\epsilon_0 V_{max}}(\xi^2 - \eta^2)$$
and the function $\bar{f}(\eta)$ becomes 
\begin{eqnarray}\label{fbarmodel1}
\bar{f}(\eta) &=&  -e^2\dfrac{a^2}{\epsilon_0 V_{max}}\int\limits_1^{\xi_{max}} (\xi^2 - \eta^2) \dfrac {dS(\xi,\, \eta)}{S(\eta)}\\
 &=& - \dfrac{2 \pi a^4}{\epsilon_0 V_{max} S(\eta)} \sqrt{1- \eta^2} \int\limits_1^{\xi_{max}} d \xi (\xi^2 - \eta^2)^{(3/2)}
 \end{eqnarray}\\

The functions $\bar{G}(\eta)$ and $\bar{H}(\eta)$ have to be calculated numerically using the equation (\ref{fbarmodel1}). Thus $U(\eta)$ and $U_\text{image}(\eta)$ can also be determined, and the multislice method of Chapter 5 can be used to determine the tunnel amplitudes and from these the current densities and the currents. Note that this model is not in the self consistent category of models, since the charge density is fixed without reference to the wave function or the tunneling amplitudes. Note also that in this model the charge density will be energy independent, which means that the resulting image potential will be independent of the energy of the tunneling electron. Further, this model does not require multiple iterations. The first iteration alone will give the final results.

\bigskip

\subsection{Model II. Self Consistent Charge Density}

The expressions for the number of electrons per unit volume in coordinate space per unit volume in the velocity space travelling in the forward and in the reverse directions have been found in Chapter 2 as well as in Chapter 3 ref equation (13) and is given by 
\begin{equation}\label{ndcubedv}
\begin{bmatrix}
n_\text{For}({\bf v})&\\
n_\text{Rev}({\bf v})&
\end{bmatrix} d^3 v\, = \, \dfrac{2 m^3}{h^3}|T(E_x)|^2 \times \begin{bmatrix}
\dfrac{k_2}{k_1} f_1(E) [1 - f_1(E +eV_b)]&\\
\dfrac{k_1}{k_2} f_1(E+ eV_b) [1 - f_1(E )]
\end{bmatrix} d^3 v
\end{equation}
Adding the two rows together gives the total number of electrons (per unit coordinate space volume per unit velocity space volume) present in the barrier region regardless of whether they belong to the forward or the reverse tunneling current. 

\noindent Thus 
\begin{equation}\label{ntotd3v}
\begin{aligned}
n_\text{tot}(v) \, d^3 v =& \dfrac{2 m^3}{h^3}|T(E_x)|^2\Big[ \dfrac{k_2}{k_1} f_1(E) [1 - f_1(E +eV_b)] \\+& \dfrac{k_1}{k_2} f_1(E+ eV_b) [1 - f_1(E )] \Big]\,d^3v
\end{aligned}
\end{equation}
where $V_b$ is the bias potential. Note that in calculating current densities the difference between the two rows is relevant and not the sum as has been done in the above equation. The total charge density in the barrier region is then $\rho_\text{total} = \int n_\text{tot}(v) d^3v$. Putting $E_x = \frac{1}{2} mv_x^2$, gives $dE_x = \sqrt{2mE_x} dv_x$  and $v_x = \sqrt{\dfrac{2 E_x}{m}}$ so that $dv_x = \dfrac{dE_x}{\sqrt{2mE_x}}$. \\

\bigskip

Further put $E_r = \frac{1}{2} m v_r^2 =  \frac{1}{2} m (v_y^2 + v_z^2) $ and $E = E_x + E_r$. Introducing polar coordinates for the integration over the $y$ and $z$ components of the velocity one finds $dv_y \, dv_z = v_r \, dv_r\,  d \theta$ where $v_r$ ranges from $0$ to $\infty$ and $\theta$ ranges from $0$ to $2 \pi$. Put $v_r dv_r = ( dE_r/m) $ and integrate over $dE_r d\theta$ to get the total charge density in the barrier region to be 
\begin{equation}\label{rhotot}
\begin{aligned}
\rho_\text{tot} = -\dfrac{\sqrt{2} m^{3/2}e}{h^3}\int \dfrac{dE_x}{\sqrt{E_x}}|T(E_x)|^2 (2 \pi) \int dE_r \Big[ &\dfrac{k_2}{k_1} f_1(E) [1 - f_1(E +eV_b)] \\+& \dfrac{k_1}{k_2} f_1(E+ eV_b) [1 - f_1(E )] \Big]
\end{aligned}
\end{equation} 
Define the charge density Fermi Factor $\zeta(E_x)$ as 
$$\zeta(E_x) = \int dE_r \Big[ \dfrac{k_2}{k_1} f_1(E) [1 - f_1(E +eV_b)] + \dfrac{k_1}{k_2} f_1(E+ eV_b) [1 - f_1(E )] \Big]$$
Note that the $\zeta$ function differs from the Fermi factor $\mathcal{F}$. Thus some aspect of Pauli effects is included in the expression for the charge density. 
The integral over $E_r$ can be carrier out analytically as shown in Appendix A-3, so that 
\begin{equation}\label{zeta}
\begin{aligned}
\zeta(E_x) =& \dfrac{1}{\beta} \Big[ \dfrac{k_2}{k_1}F_1(E_x) + \dfrac{k_1}{k_2}F_1(E_x + eV_b)\Big] \\-& \dfrac{\frac{k_1}{k_2}+\frac{k_2}{k_1}}{\beta (1 - e^{-\beta eV_b})} \Big[F_1(E_x+eV_b) - e^{\beta eV_b} F_1(E_x) \Big]
\end{aligned}
\end{equation}
Therefore the charge density used in this model is given by 
\begin{equation}\label{Model2rho}
\rho_\text{tot} = -\Big(\dfrac{2m}{\hbar^2}\Big)^{3/2}e\int \dfrac{dE_x}{\sqrt{E_x}} |T(E_x)|^2 \zeta(E_x)
\end{equation}
The model may be considered semiclassical, because the charge density is obtained by multiplying the number density with the charge of the electron. It is true that the number density is obtained using the quantum mechanically determined tunneling amplitudes. The finding of the function $\bar{f}(\eta)$ and from it the net potential $U(\eta)$ etc. will all have to be done numerically. 

\bigskip

\section{Plots and Comparisons}

\bigskip

\begin{figure}[h]
	\centering
	\includegraphics[width=4.2in,height=2.8in]{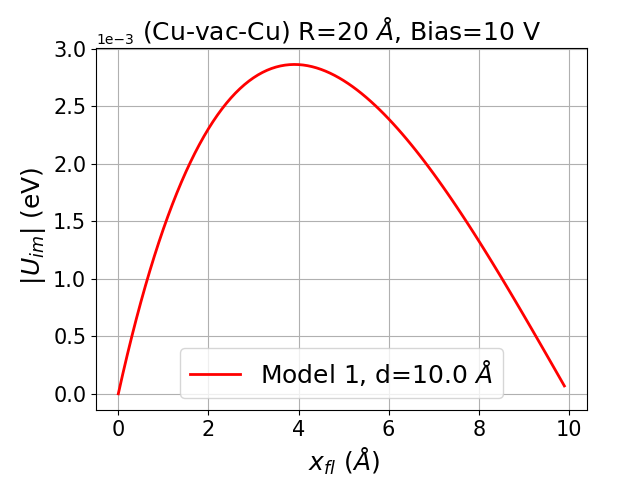}
	\caption{Plot of image potential vs $x_{fl}$ for uniform charge density model with $d = 10 $ \r{A} and Bias $= 10$ V}
	\label{fig:Uim_Model1}
\end{figure}

The image potential can be calculated using the equations (18) and (19) and the equation (14). Fig. \ref{fig:Uim_Model1} shows the plot of $-U_{im}$ as a function of $x_{fl}$. It can be see that the image potential does indeed vanish at the end points $x_{fl} = 0 $ \r{A} and $x_{fl}= 10 $ \r{A}. 
\begin{figure}[hpt]
	\centering
	\begin{subfigure}{0.49\textwidth}
		\includegraphics[width=3.05in,height=2.5in]{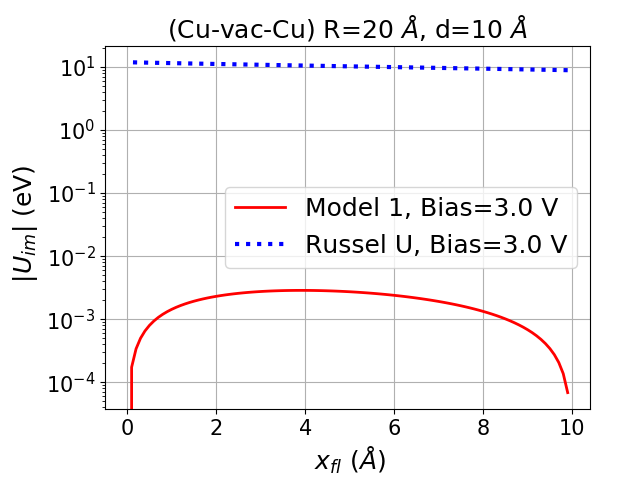}	
		\caption{} 
		\label{fig:M1RussV3}
	\end{subfigure}
	\hfill
	\begin{subfigure}{0.49\textwidth}	
		\includegraphics[width=3.05in,height=2.5in]{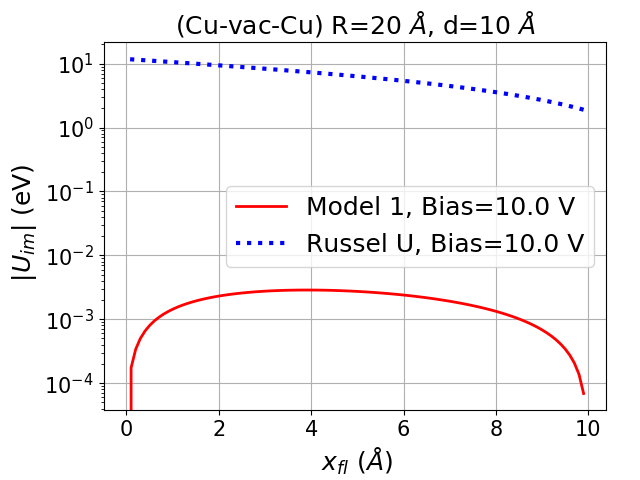}
		\caption{} 
		\label{fig:M1RussV10}
	\end{subfigure}
	\caption{Plot of potential vs $x_{fl}$ showing the comparison of image potential and Russel potential for $d=10\, $ \r{A} and  Bias (a) $= 3$ V and (b) $= 10$ V}
	\label{fig:Uim_Model1_Russ}
\end{figure}

Fig. \ref{fig:Uim_Model1_Russ} shows a comparison of the magnitudes of the image potential with the Russell Potential, in which it can be seen that the image potentiasl magnitudes of Model-1 are as much as 3 to 4 orders of magnitude smaller than the Russell Potential. Clearly such a small change in the net potential is insignificant and will cause negligible change in the tunneling amplitudes, the tunneling current densities and finally the currents. Figures \ref{fig:M1RussV3} and \ref{fig:M1RussV10} also show that this feature is independent of the bias voltage.  Therefore the changed values of the  current densities and currents are not calculated and only magnitudes $|U_{im}$ in eV are calculated and plotted against the distance $x_{fl}$ along the field line, for a fixed value of $\xi$. Experimental results on measurements of tunneling currents must decide whether the Simmons type of models in which the electron in the barrier region behaves as a point charge or whether it bahves as a distributed charge as in Models 1 and 2. The paper of \cite{Huang} calculates the effect of the image potential on the $j-V$ characteristics in the WKB approximation, for a semi-conductor-dielectric-conductor tunnel junction, and comes to the conclusion that the effect of the image potential is too small to be detectable through experiment. While this calculation deos differ from ours in several ways, their conclusions agrres with the assumption of this thesis that the charge in the barrier region is distributed.

\bigskip

\begin{figure}[hpt]
	\centering
	\begin{subfigure}{0.49\textwidth}
		\includegraphics[width=3.05in,height=2.3in]{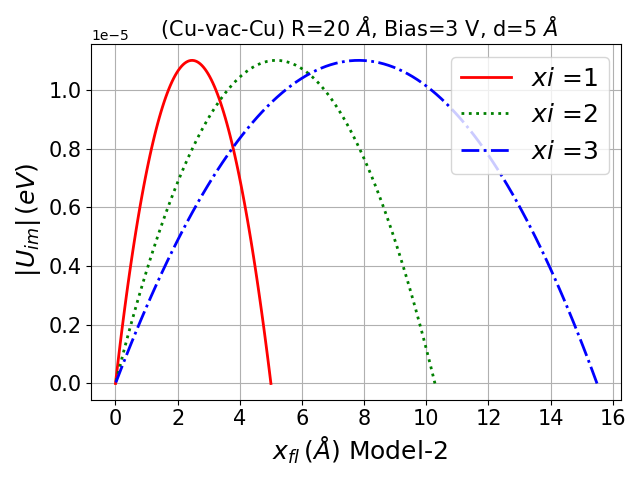}	
		\caption{} 
		\label{fig:CuCUM2V3d5}
	\end{subfigure}
	\hfill
	\begin{subfigure}{0.49\textwidth}	
		\includegraphics[width=3.05in,height=2.3in]{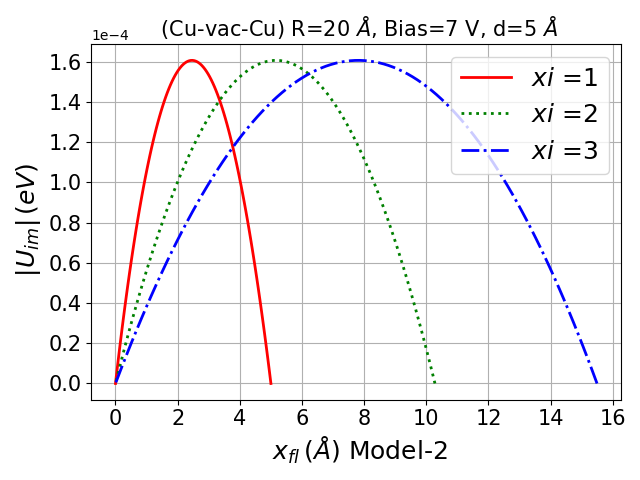}
		\caption{} 
		\label{fig:CuCUM2V7d5}
	\end{subfigure}
	\caption{Plot of image potential $|U_{im}|$ (Model-2) vs $x_{fl}$ for  $d = 5 $ \r{A} and Bias $= 3,\,7$ V for similar electrodes (Cu-vac-Cu)}	
	\label{fig:CuCuM2V37d5}
\end{figure}

\begin{figure}[hpt]
	\centering
	\begin{subfigure}{0.49\textwidth}
		\includegraphics[width=3.05in,height=2.3in]{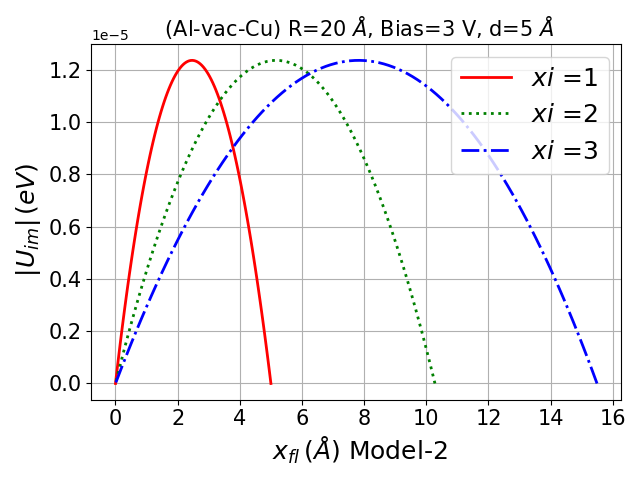}	
		\caption{} 
		\label{fig:AlCuM2V3d5}
	\end{subfigure}
	\hfill
	\begin{subfigure}{0.49\textwidth}	
		\includegraphics[width=3.05in,height=2.3in]{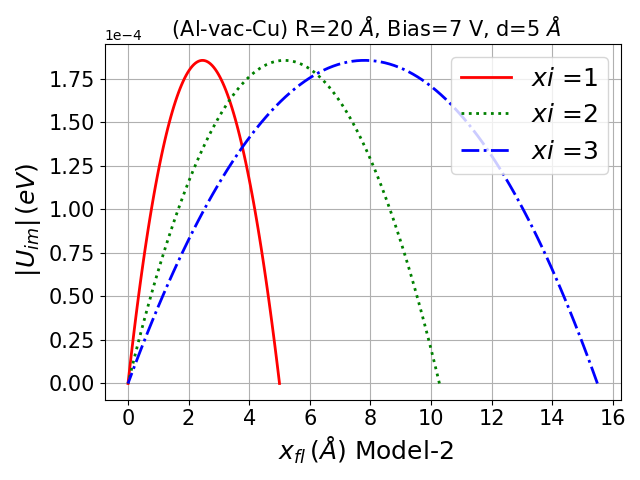}
		\caption{} 
		\label{fig:AlCuM2V7d5}
	\end{subfigure}
	\caption{Plot of image potential $|U_{im}|$ (Model-2) vs $x_{fl}$ for  $d = 5 $ \r{A} and Bias $= 3,\,7$ V for dissimilar electrodes (Al-vac-Cu)}
	\label{fig:AlCuM2V37d5}
\end{figure}

\begin{figure}[hpt]
	\centering
	\begin{subfigure}{0.49\textwidth}
		\includegraphics[width=3.05in,height=2.3in]{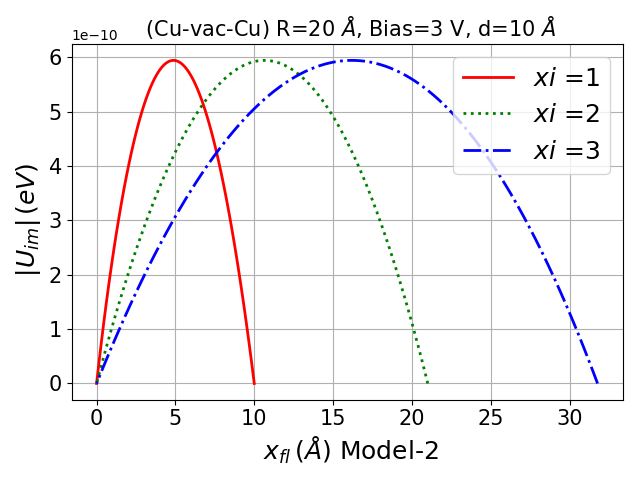}	
		\caption{} 
		\label{fig:CuCUM2V3d10}
	\end{subfigure}
	\hfill
	\begin{subfigure}{0.49\textwidth}	
		\includegraphics[width=3.05in,height=2.3in]{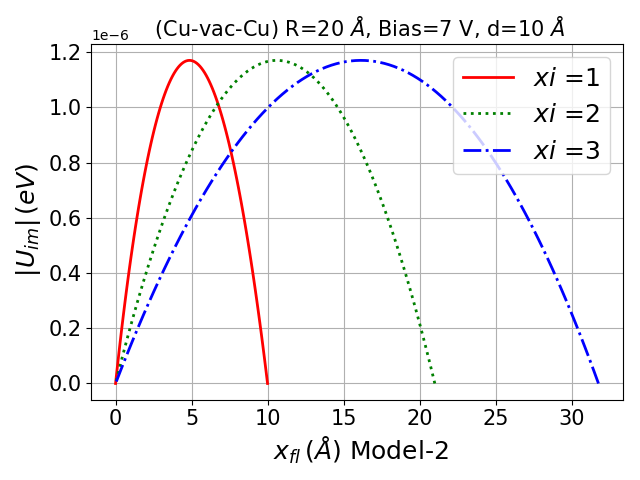}
		\caption{} 
		\label{fig:CuCUM2V7d10}
	\end{subfigure}
	\caption{Plot of image potential $|U_{im}|$ (Model-2) vs $x_{fl}$ for  $d = 5 $ \r{A} and Bias $= 3,\,7$ V for similar electrodes (Cu-vac-Cu)}	
	\label{fig:CuCuM2V37d10}
\end{figure}

\begin{figure}[hpt]
	\centering
	\begin{subfigure}{0.49\textwidth}
		\includegraphics[width=3.05in,height=2.3in]{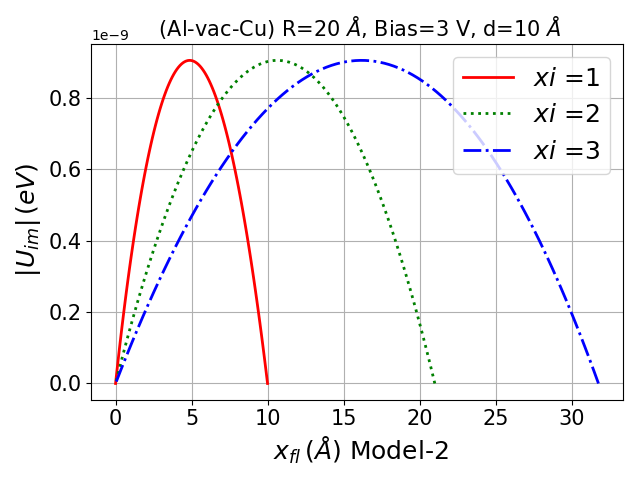}	
		\caption{} 
		\label{fig:AlCuM2V3d10}
	\end{subfigure}
	\hfill
	\begin{subfigure}{0.49\textwidth}	
		\includegraphics[width=3.05in,height=2.3in]{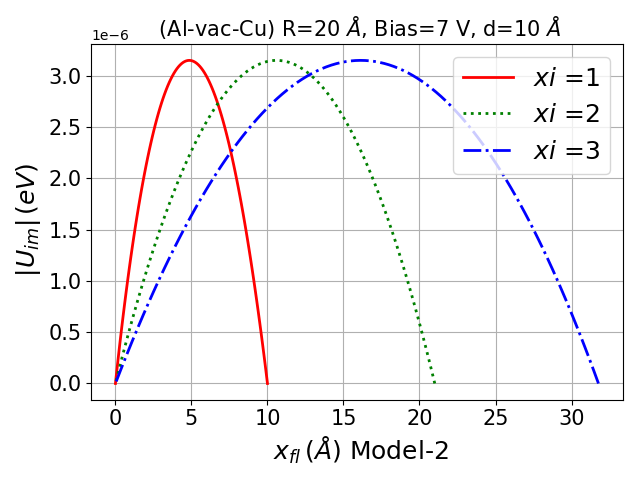}
		\caption{} 
		\label{fig:AlCuM2V7d10}
	\end{subfigure}
	\caption{Plot of image potential $|U_{im}|$ (Model-2) vs $x_{fl}$ for  $d = 5 $ \r{A} and Bias $= 3,\,7$ V for dissimilar electrodes (Al-vac-Cu)}
	\label{fig:AlCuM2V37d10}
\end{figure}

\begin{figure}[hpt]
	\centering
	\includegraphics[width=4.0in,height=3.3in]{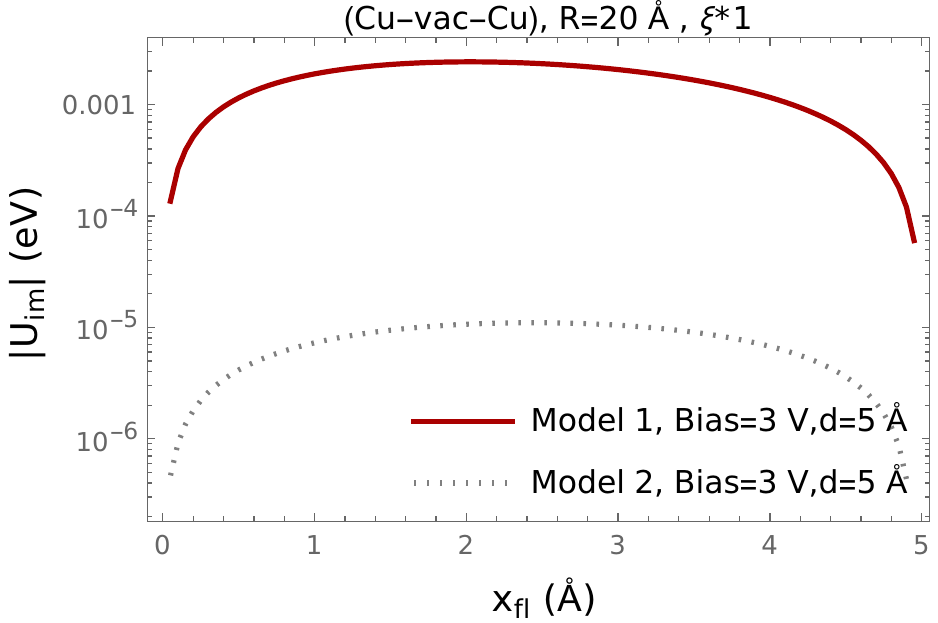}
	\caption{Plot of image potential $U_{im}$ vs $x_{fl}$ for Model-1 and Model-2, with $d = 5 $ \r{A} and Bias $= 3$ V}
	\label{fig:Uim_Model1N2Comparison}
\end{figure}

Fig. \ref{fig:CuCuM2V37d5}, \ref{fig:AlCuM2V37d5}, \ref{fig:CuCuM2V37d10}, and \ref{fig:AlCuM2V37d10} show  plots of $|U_{im}|$ obtained in Model 2 as a function of $x_{fl}$ different bias voltages and for simillar (Cu-vac-Cu) and dissimillar  (Al-vac-Cu) electrodes. It can be seen that the image potential magnitudes are  very small and such a small change in the net potential is insignificant and will cause negligible change in the tunneling amplitudes, the tunneling current densities and finally the currents. All these figures show that the 
image potential peaks near about the midpoint of the field line. This is found true for all values of $\xi$. The behaviour of the Image potential in this model as $\xi$ increases from $1$ to $3$ resembles the rescaling of the abcissa, with no change in the basic shape of the curve. Note also that that the image potential does indeed vanish at the end points, unlike in the Simmons case where the image potential is infinity at the end points. 
Further a comparison of the results of models 1 and 2, [See Fig. \ref{fig:Uim_Model1N2Comparison}] shows that the Model 2 image potentials are about two or more orders of magnitude smaller than those of Model 1. Note that in Model 1 (Uniform Charge Density) it was assumed that the net charge in the barrier region is equal to the electron charge. In Model 2 however the net charge is decided by the tunnel amplitudes, and it appears that the net charge is perhaps 2 orders of magnitude smaller than that in Model 1.


\bigskip

A plot of the Simmons image potential + Trapezoidal potential is shown in Fig. \ref{fig:UTimd510} (figures 6.3 a \& b) of Chapter 6. It can be seen that the Simmons image potential is comparable to the Trapezoidal potential and the net effect reduces the height of the barrier potential and also reduces its width. Thus the Simmons Image potential has an astronomical effect on the tunnel amplitudes [Chapter 6 Figures 6.8 (a\& b)]  and thereby also on the current densities [Chapter 6 Figures 6.12 (a \& b)]. In contrast the image potentials of the distributed charge models are at least three orders of magnitude smaller than the Trapezoid or the Russell potentials. Therefore the effect of these potentials is expected to be negligible. So in displaying the results of the current density and current calculations, image-contributions due to the distributed charge models will not be included. 

\bigskip

In Figure \ref{fig:Uim_Model1N2Comparison} the image potentials of Model 1 and Model 2  are compared. The Model 2 image potential is about 3 to 4 orders of magnitude smaller than the Model 1 image potential. One can understand this by comparing the total charge (called the source charge) contained in the barrier region in the two models. The source charge, is reponsible for producing image charges on the boundary surfaces. These  induced charges on the surfaces produce the image  potential. The magnitude of the total source charge $|Q_1|$ in Model 1 is $|e|$, independent of input parameters such as the bias voltage ot the tip-sample distance. 

\begin{figure}
	\centering
	\begin{subfigure}{0.49\textwidth}	
		\includegraphics[width=2.9in,height=2.3in]{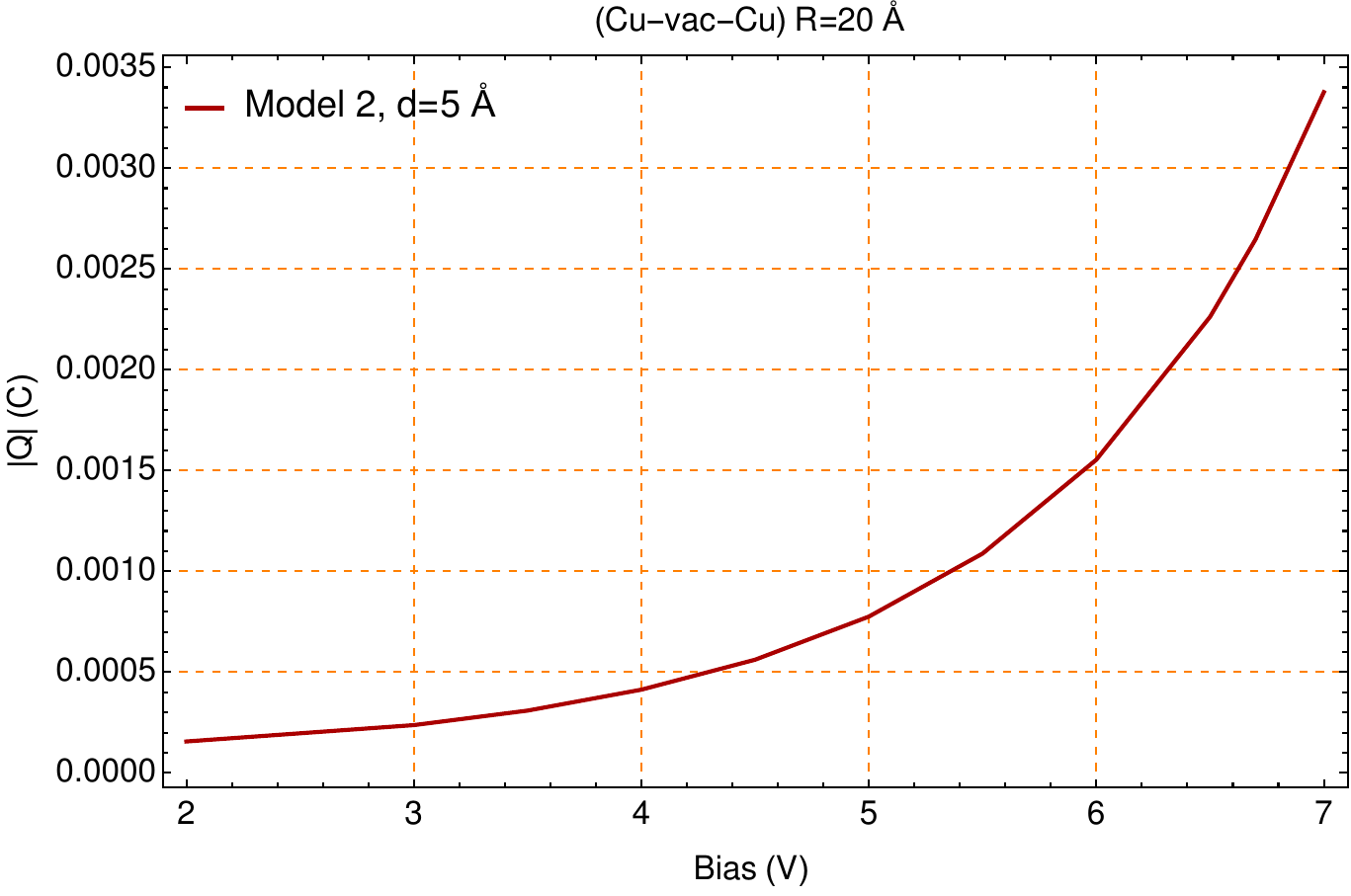}
		\caption{}
		\label{fig:Qcharged5}	
	\end{subfigure}
	\hfill
	\begin{subfigure}{0.49\textwidth}	
		\includegraphics[width=2.9in,height=2.3in]{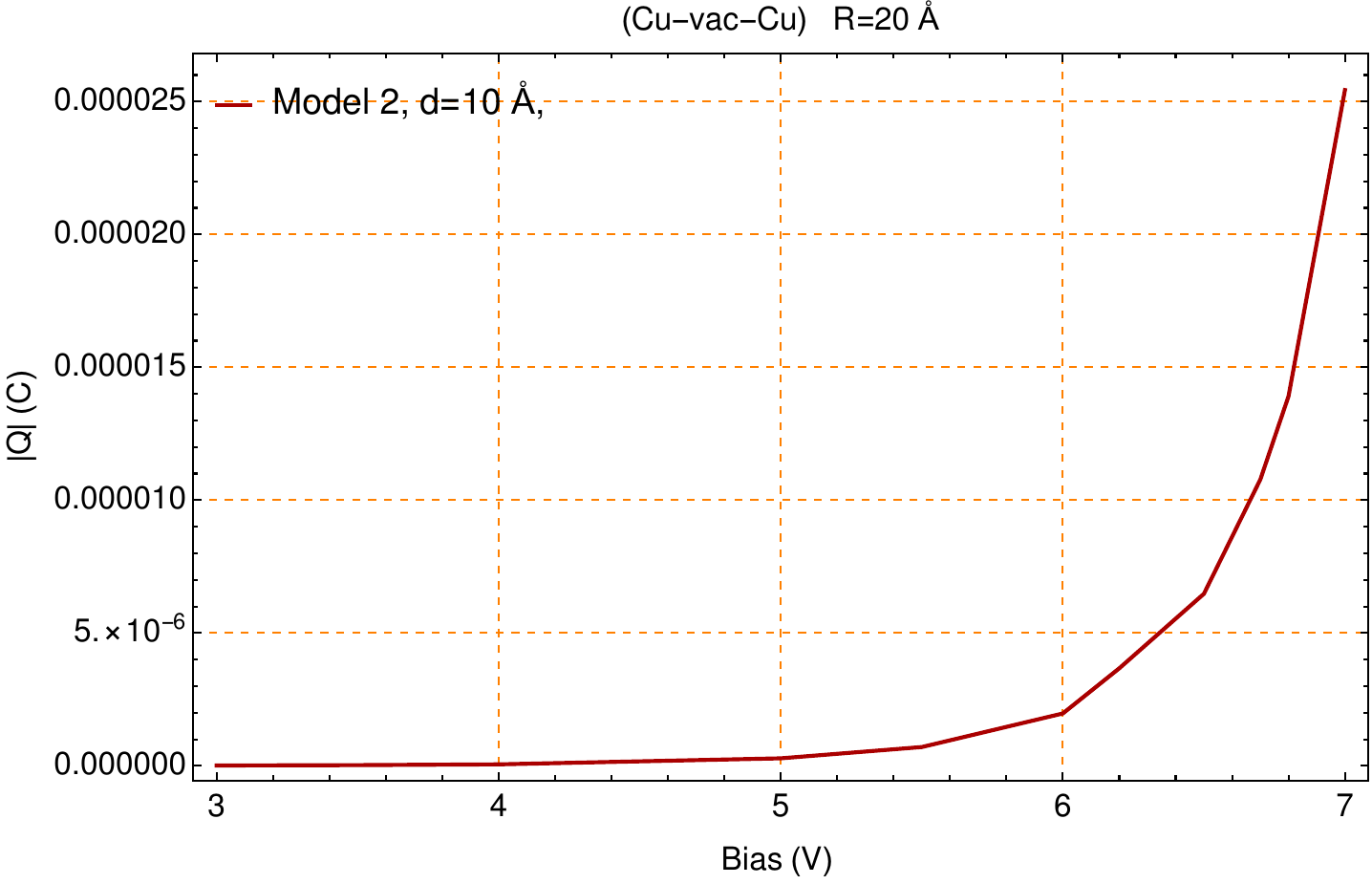}
		\caption{}
		\label{fig:Qcharged10}
	\end{subfigure}
	\caption{Plot of effective charge $Q$ vs Bias for (a) $d = 5 $ \r{A} and (b) $d = 10 $ \r{A}}
	\label{fig:Uim_Qcharged510}
\end{figure}

\bigskip

The source charge in Model 2 depends upon the input parameters, and have to be calculated separately for each combination of these parameters. Figures \ref{fig:Qcharged5} and \ref{fig:Qcharged10} show the scource charge $|Q_2|$ in Model 2 in units of the electron charge $|e|$ as a function of the bias voltage for a fixed tip-sample distance of $(5)$ \r{A} and $(10)$ \r{A} respectively. It is found that the charge $|Q_2|$ is larger at shorter distances. The volume of the barrier region increases with distance, and this suggests that the charge $|Q_2|$ ought to increase with distance. However the charge density in the region depends upon the tunneling probability, which is much much larger at shorter distances. Thus at shorter distances, the increase in the charge density more than compensates for the reduction in the volume of the barrier region, thereby explaining the relative maginitudes of $|Q_2|$ at the two distances $ 5$ \r{A} and $10 $ \r{A}.

\bigskip

In this thesis, the charge in the barrier region will be assumed to be distributed, for which image effects have been shown to be negligible. Thus an accurate experimental determination of the I-V characteristics of the electrode - vacuum - electrode tunnel junctions can provide an answer to a fundamental question regarding the nature of tunneling electrons. Do the tunneling electrons in the barrier region interact with conducting boundaries as if they were point particles, or as if they were distributed charges with magnitude that could be very much less than the electron charge. 

\bigskip

Chapter 8 discusses the general behavior of currents determined from planar model current densites, using the multislice method described in Chapter 5 and the prolate spheroid coordinate system based integration over field lines described in Chapter 4, as a function of various input parameters concerning the electrode pairs, the bias voltage and the distance between tip and sample. 





	
	\chapter{Results and Conclusions}\label{Chap8}
%
%
%
%
%
\section{Introduction}
The calculation of current densities (planar model) for the linear trapezoidal potential in the barrier region 
is described in Chapters 2 and 3. To calculate the currents from current densities, the planar model for the trapezoidal potential is modified to include the realistic geometry of the tip. The tip is considered as a sharp pointed surface and the sample is treated as a flat surface. The key assumption here is that tunneling from curved tips primarily occurs along the field lines. Note for a sharp tip, its radius of curvature at the point closest to the sample must be small. For such tips, the length of the electric field lines increases rapidly as one goes away from the centre. As the tunneling current densities fall off exponentially with increasing length of the field line, the currents are also expected to decrease with increasing length of the field line. The current densities along field lines when integrated over all field lines emanating from a finite region of the tip surface will lead to a finite value for the current.

\bigskip

For a linear trapezoidal potential, Airy function solutions to the Schr\"odinger equation in the barrier region are used to compute tunneling probabilities, from which current densities and currents are obtained. For the Russell potential and for potential with images, the potential energy is no longer linear in the distance and for these cases the transfer matrix method is used to compute tunneling probabilities, from which current densities are obtained. The contribution of electrons whose energies range from $0$ to $\eta_1+\phi_1$ are summed to determine the current density. Pauli effects are explicitely introduced and the current densities so obtained are converted to currents for hyperboloidal tips.  For energies beyond $\eta_1 + \phi_1$, Pauli dependent Fermi factors completely extinguish all contribution.

\bigskip

The behaviour of the tunnel currents as a function of input parameters such as tip-sample distance, bias voltage and tip radius of curvature is studied and the results are diplayed in next section. A measure of lateral resolution is also introduced and it is shown to be degraded for blunter tips, increasing bias voltages and increasing tip sample distances. The tunneling currents with and without image contributions are also included. Here the effect of the point charge image model with Simmons \cite{simmonsII} approximation is contrasted against the contribution of the distributed charge models. It is seen that unlike the Simmons approx. model, the distributed charge shows negligible effect. 
  
\section{\label{sec:level7}Behaviour of Tunneling Currents}
\begin{figure}[hpt]
	\centering
	\begin{subfigure}{0.49\textwidth}
		\includegraphics[width=3.2in,height=3.0in]{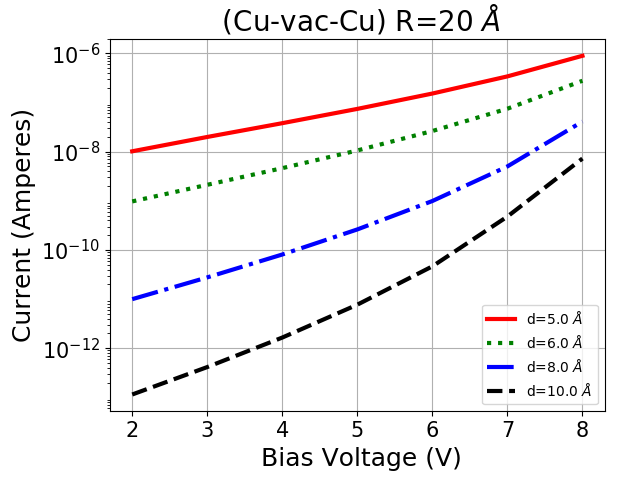}
		\caption{} 
		\label{fig:TrapCuCu}
	\end{subfigure}
	\hfill
	\begin{subfigure}{0.49\textwidth}	
		\includegraphics[width=3.2in,height=3.0in]{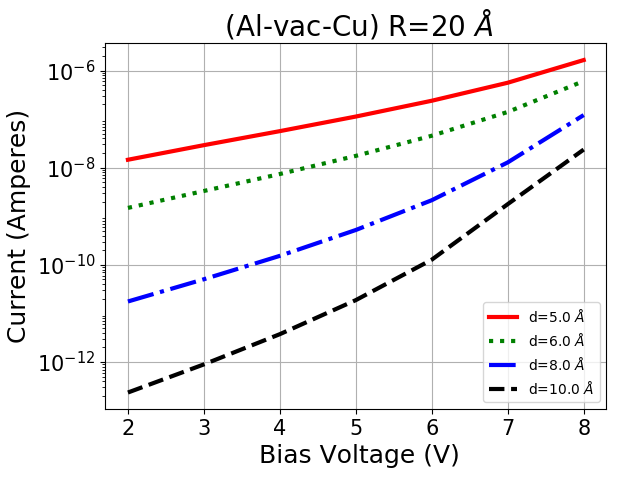}
		\caption{} 
		\label{fig:TrapAlCu}		
	\end{subfigure}
	\caption{Plot of Current\, versus\, Bias\, Potential for $R=20$ \r{A} and $d=5,\, \, 10\,  $ \r{A} for similar (Left) and dissimilar (Right) electrodes.}
	\label{fig:TrapAlCuCuCu}
\end{figure}

Fig.\ref{fig:TrapAlCuCuCu} shows calculated I-V characteristics for similar electrodes (Cu-vac-Cu), and for dissimilar electrodes (Al-vac-Cu),  for several tip-sample distances of $5 \, $, $6 \,$, $8 \, $ and $10 $ \r{A}. 

\begin{figure}[hpt]
	\centering
	\begin{subfigure}{0.49\textwidth}
		\includegraphics[width=3.2in,height=2.7in]{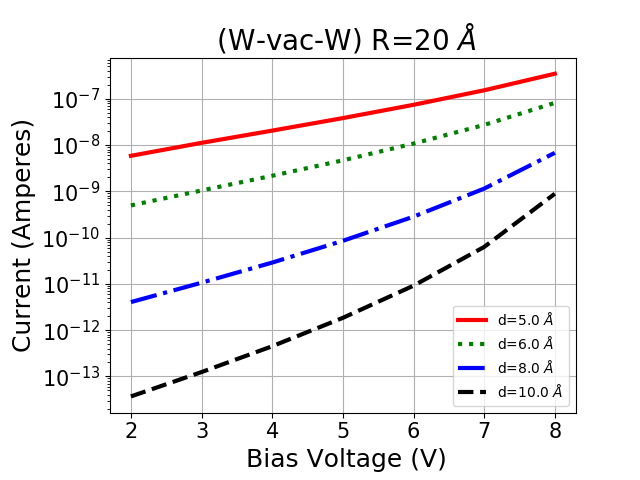}
		\caption{} 
		\label{fig:WW}
	\end{subfigure}
	\hfill
	\begin{subfigure}{0.49\textwidth}	
		\includegraphics[width=3.2in,height=2.7in]{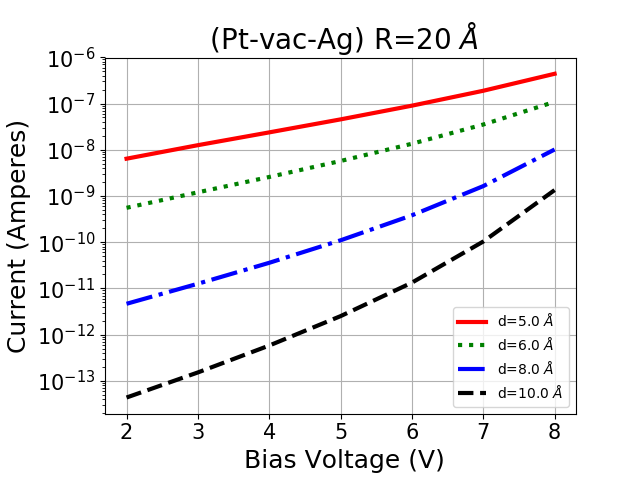}
		\caption{} 
		\label{fig:PtAg}
	\end{subfigure}
	\hfill	
	\begin{subfigure}{0.49\textwidth}	
	\includegraphics[width=3.2in,height=2.7in]{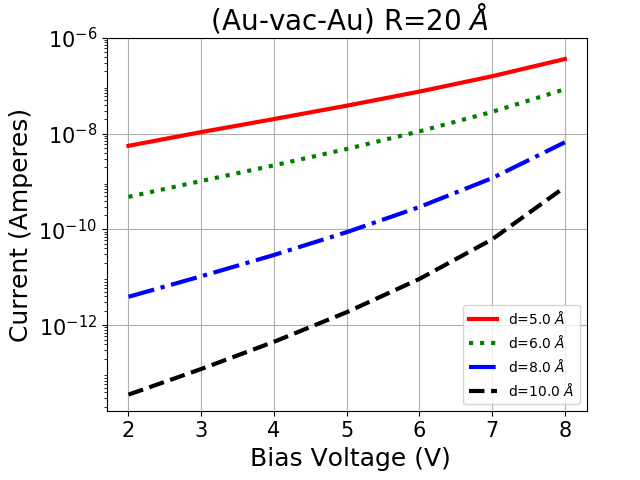}		
	\caption{} 
	\label{fig:AuAu}
	\end{subfigure}
	\hfill
	\begin{subfigure}{0.49\textwidth}	
		\includegraphics[width=3.2in,height=2.7in]{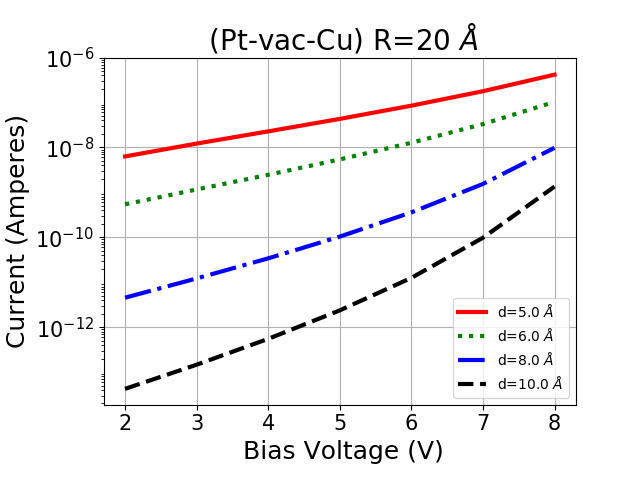}		
		\caption{} 
		\label{fig:PtCu}
	\end{subfigure}
	\caption{Plot of Current\, versus\, Bias\, Potential for $R=20$ \r{A} and $d=5,\, \, 10\,  $ \r{A} for similar (Left) and dissimilar (Right) electrodes.}
	\label{fig:WWPtAgAuAu}
\end{figure}

Fig. \ref{fig:WWPtAgAuAu} shows calculated I-V characteristics for similar electrodes (W-vac-W), (Au-vac-Au) and for dissimilar electrodes (Pt-vac-Ag), (Pt-vac-Cu) for several tip-sample distances of $5 \, $, $6 \,$, $8 \, $ and $10 $ \r{A}. The current is seen to increase rapidly with bias voltage for all pair of electrodes. This behaviour is similar for all tip sample distances.

\bigskip 

\bigskip

\begin{figure}[hpt]
	\centering
	\begin{subfigure}{0.49\textwidth}
		\includegraphics[width=3.2in,height=2.7in]{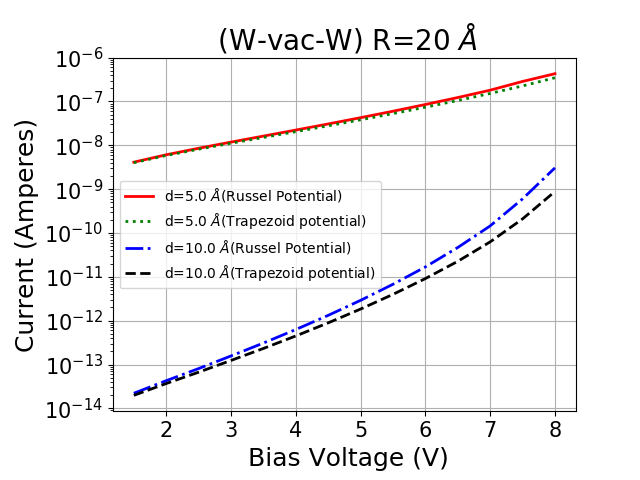}		
		\caption{} 
		\label{fig:RussTrapWW}
	\end{subfigure}
	\hfill
	\begin{subfigure}{0.49\textwidth}
		\includegraphics[width=3.2in,height=2.7in]{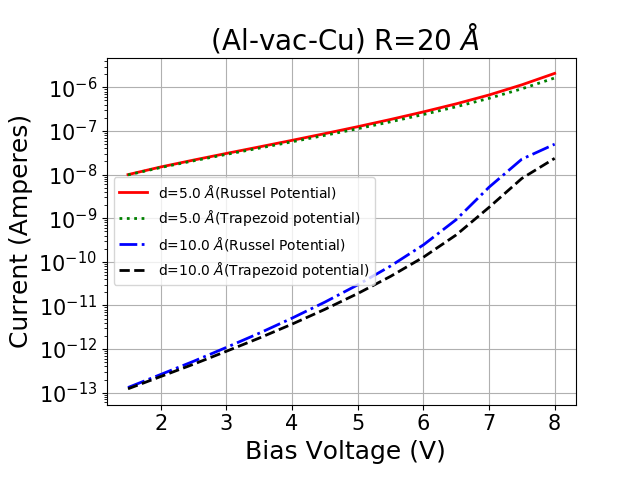}		
		\caption{} 
		\label{fig:RussTrapAlCu}
	\end{subfigure}
	\hfill
	\begin{subfigure}{0.49\textwidth}	
		\includegraphics[width=3.2in,height=2.7in]{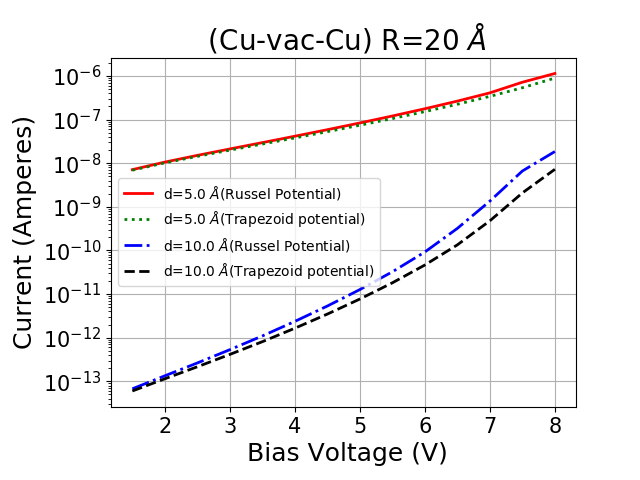}		
		\caption{} 
		\label{fig:RussTrapCuCu}
	\end{subfigure}
	\hfill
	\begin{subfigure}{0.49\textwidth}
		\includegraphics[width=3.2in,height=2.7in]{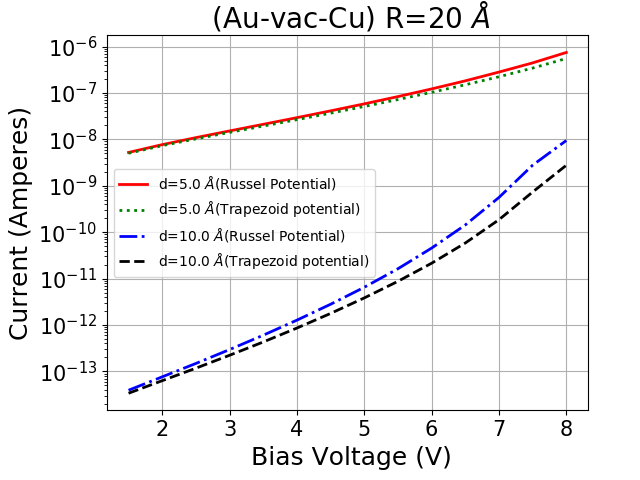}		
		\caption{} 
		\label{fig:RussTrapAuCu}
	\end{subfigure}
	\caption{Comparison of I-V characteristics fot Russell and Trapezoidal potential for $R=20$ \r{A} and $d=5,\, \, 10$ \r{A} for similar (Left) and dissimilar (Right) electrodes.}
	\label{fig:RussTrap1}
\end{figure}

Fig. \ref{fig:RussTrap1} compares the calculated I-V characteristics for Trapezoid and Russell Potential in case of similar (Cu-Vac-Cu), (W-Vac-W) and dissimilar  electrodes (Al-vac-Cu), for tip-sample distances of $\,5 $ \r{A} and $10 $ \r{A}.
The current is seen to increase rapidly with bias voltage for both Trapezoid and Russell Potential. The current decreases with increase in tip-sample distance as expected. Since the Russell potential does not depart very much from the trapezoidal potential (refer Chapter 4 Fig. 4.3), the currents for the Russell potential also do not vary from the currents for the trapezoidal potential. 

\begin{figure}[hpt]
	\centering
	\begin{subfigure}{0.49\textwidth}
		\includegraphics[width=3.2in,height=2.7in]{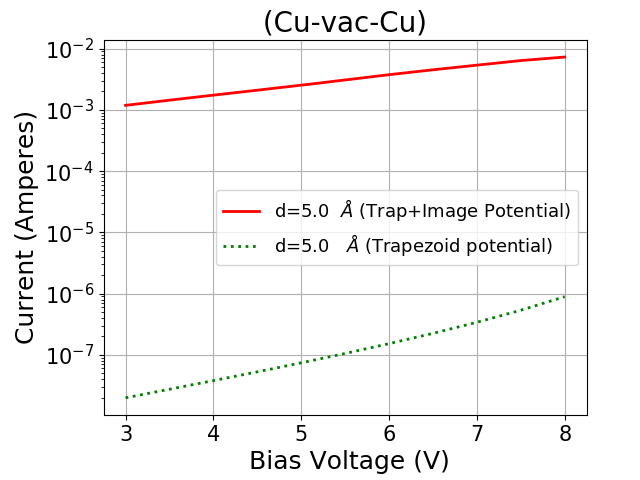}
		\caption{} 
		\label{fig:CuCu}
	\end{subfigure}
	\hfill
	\begin{subfigure}{0.49\textwidth}	
		\includegraphics[width=3.2in,height=2.7in]{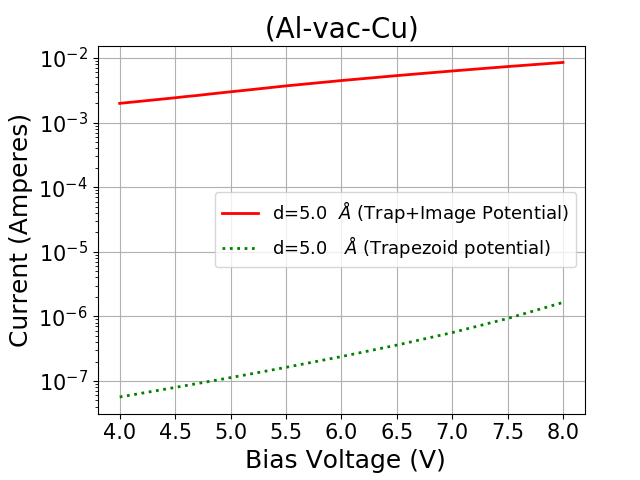}
		\caption{} 
		\label{fig:AlCu}
	\end{subfigure}
	\hfill
	\begin{subfigure}{0.49\textwidth}	
		\includegraphics[width=3.2in,height=2.7in]{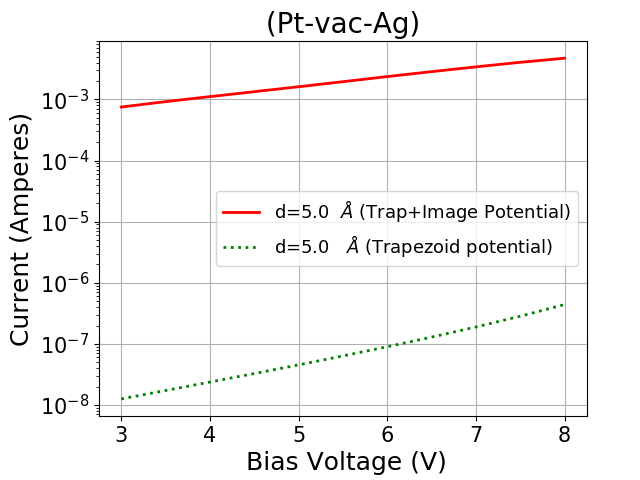}		
		\caption{} 
		\label{fig:PtAu}
	\end{subfigure}
	\hfill
	\begin{subfigure}{0.49\textwidth}	
		\includegraphics[width=3.2in,height=2.7in]{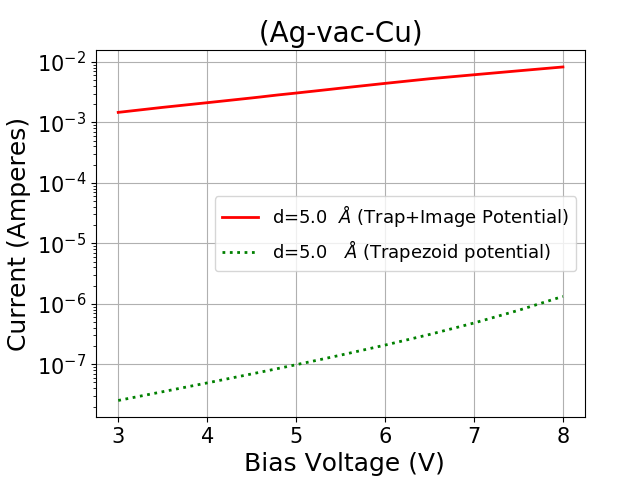}		
		\caption{} 
		\label{fig:AgCu}
	\end{subfigure}
	\caption{Comparison of I-V characteristics for Trapezoidal and Trapezoidal with image potential for for $R=20 $ \r{A} and $d=5 $ \r{A} for similar (Cu-vac-Cu) and dissimilar (Al-vac-Cu), (Pt-vac-Ag), (Ag-vac-Cu) electrodes.}
	\label{fig:AlAlPtAuAgCu}
\end{figure}

\begin{figure}[hpt]
	\centering
	\begin{subfigure}{0.49\textwidth}
			\includegraphics[width=3.3in,height=3.0in]{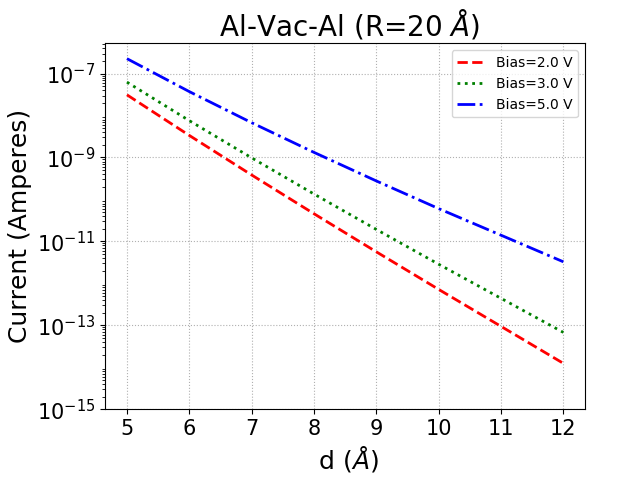}
		\caption{} 
		\label{fig:RussdAlAl}
	\end{subfigure}
	\hfill
	\begin{subfigure}{0.49\textwidth}	
		\includegraphics[width=3.3in,height=3.0in]{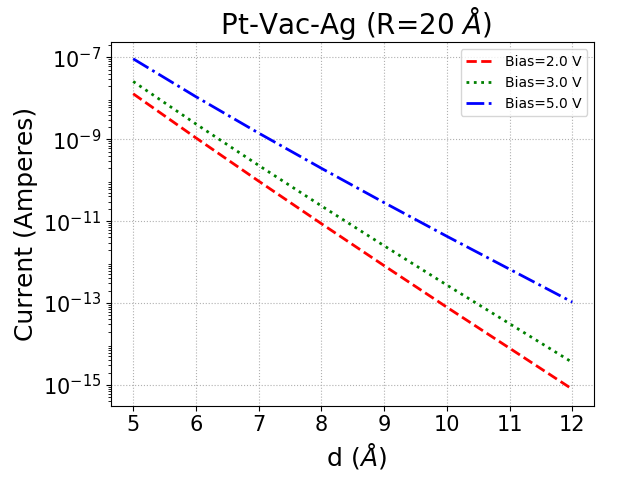}
		\caption{} 
		\label{fig:RussdPtAg}
	\end{subfigure}
	\caption{Plot of Current\, versus\, tip-sample distance for $R=20$ \r{A} and Bias $=2,\,3,\, 5 \, V$ for similar (Left) and dissimilar (Right) electrodes.}
	\label{fig:RussdAlAlPtAg}
\end{figure}
\bigskip

Fig. \ref{fig:AlAlPtAuAgCu} compares the calculated I-V characteristics for Trapezoid and Trapezoid+Image Potential in case of similar (Cu-Vac-Cu) and dissimilar  electrodes (Al-vac-Cu), (Pt-vac-Ag), (Ag-vac-Cu), for tip-sample distance of $\,5 $ \r{A}.
The current is seen to increase rapidly with bias voltage for both Trapezoid and Trapezoid+Image Potential. The currents with image are 3 to 5 orders of magnitude larger than the Trapezoidal potential, which appears to be unrealistic. In contrast, the models described in Chapter 7 suggests that the contribution to image potential by distributed charge is negligible. The latter models are probably more realistic. 

\bigskip

\par Fig. \ref{fig:RussdAlAlPtAg} shows the plots of calculated tunneling currents $I$ versus tip sample distance $d$ in similar (Al-vac-Al) and dissimilar (Pt-Vac-Ag) pairs of electrodes. These plots are for bias voltages of $2$V, $3$V and $5$V, and tip radius of curvature fixed at $R=20$ \r{A}. The current $I$ is found to decrease exponentially with increasing $d$. This  behaviour is also qualitatively reproduced by almost all calculations of tunneling current density including those that use the WKB approximation \cite{simmonsI,simmonsII,SBPZ,hofer2001surface}. 

\bigskip

\begin{figure}[hpt]
	\centering
	\begin{subfigure}{0.49\textwidth}
		\includegraphics[width=3.3in,height=3.0in]{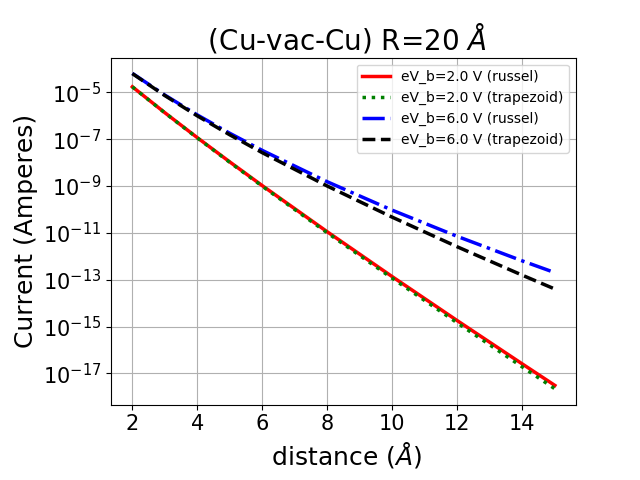}		
		\caption{} 
		\label{fig:RussTrapdCuCu}
	\end{subfigure}
	\hfill
	\begin{subfigure}{0.49\textwidth}	
			\includegraphics[width=3.3in,height=3.0in]{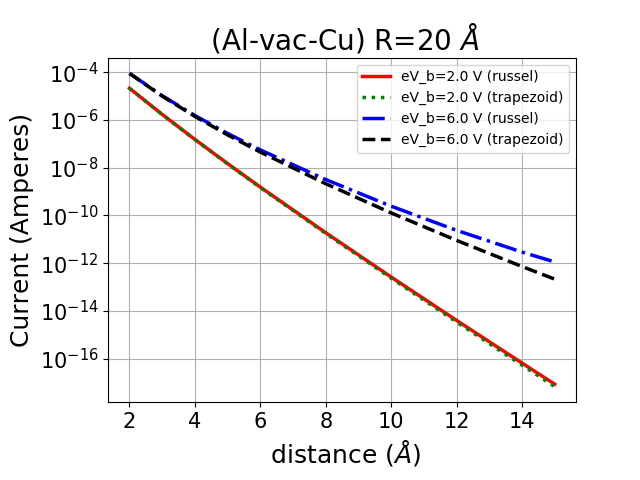}
		\caption{} 
		\label{fig:RussTrapdAlCu}
	\end{subfigure}
	\caption{Plot of Current\, versus\, tip-sample distance for $R=20$ \r{A} and Bias $=2,\,6\, V$ for similar (Left) and dissimilar (Right) electrodes, comparing currents for Russell and Trapezoid potential.}
	\label{fig:RussTrapdCuCuAlCu}
\end{figure}

Fig. \ref{fig:RussTrapdCuCuAlCu} compares the current as a function of tip-sample distance  for Trapezoid and Russel potential in case of similar and dissimilar electrodes, for fixed bias voltage of $2,\,6,\,V$. In both the cases the current decreases exponentially with distance. 

\begin{figure}[hpt]
	\centering
	\begin{subfigure}{0.49\textwidth}
		\includegraphics[width=3.35in,height=3.0in]{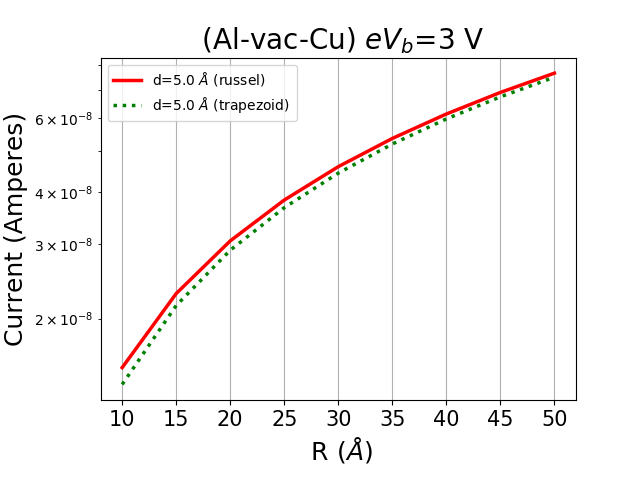}		
		\caption{} 
		\label{fig:IvsRd5}
	\end{subfigure}
	\hfill
	\begin{subfigure}{0.49\textwidth}	
		\includegraphics[width=3.35in,height=3.0in]{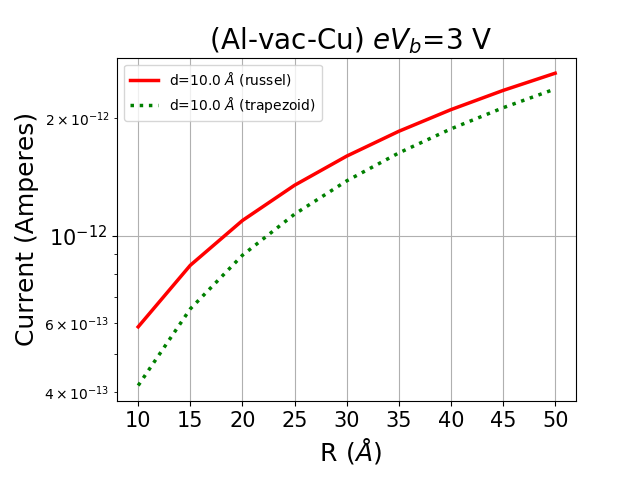}
		\caption{} 
		\label{fig:IvsRd10}
	\end{subfigure}
	\caption{\,Plot \,of \,Current as a function of Radius of curvature for fixed bias of $3\, V$ and for tip-sample distance of $d=5$ \r{A} and $d=10$ \r{A}.}
	\label{fig:IvsRd510}	
\end{figure}

\bigskip

\par For fixed bias voltage $V$ and tip sample distance $d$, increasing the tip radius $R$, leads to a flatter tip, which means that more area in the vicinity of the center lies closer to the sample and the current density for these areas is greatly increased. Since the current $I$ is an integral of the current density over tip area, $I$ is expected to increase with increasing $R$ for all $V$ and $d$. Fig.\ref{fig:Ting}, and \ref{fig:IvsRd510} indeed confirms this behaviour by showing how the $I-V$ curves  correspond to higher currents for larger $R$. The functional dependence of $I$ upon $V$ is seen to be uninfluenced by the values of $R$. 
\begin{figure}[hpt]
	\centering
	\begin{subfigure}{0.49\textwidth}
		\includegraphics[width=3.3in,height=2.9in]{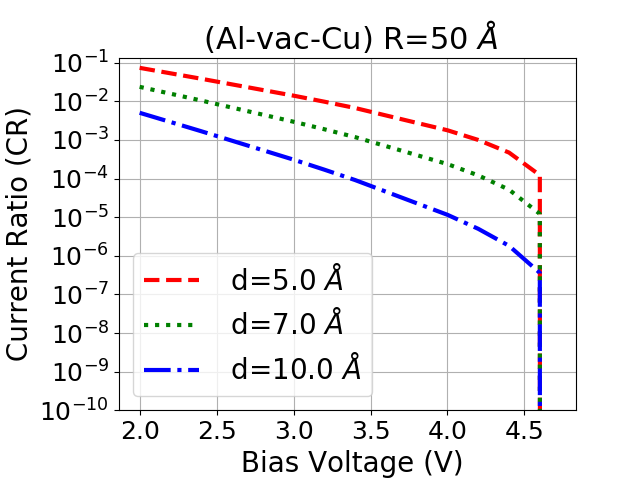}		
		\caption{} 
		\label{fig:CRvsVLog}	
	\end{subfigure}
	\hfill
	\begin{subfigure}{0.49\textwidth}	
		\includegraphics[width=3.3in,height=2.9in]{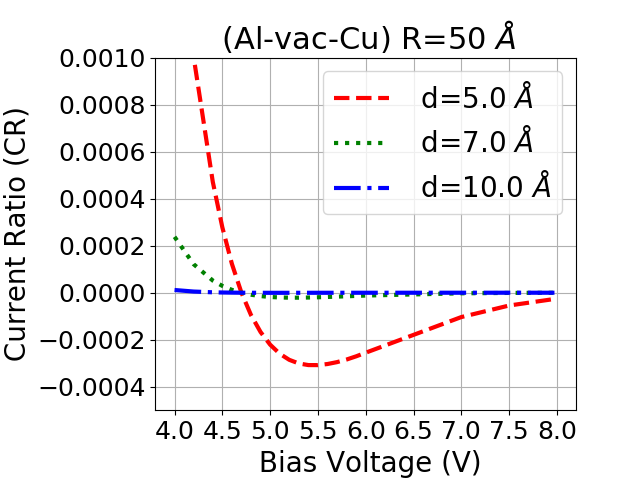}	
		\caption{} 
		\label{fig:CRvsVLin}
	\end{subfigure}
	\	\caption{\,Plot \,of \,Current \,Ratio \,(CR) as a function of Bias Voltage (a) $2\,V$ to $5\, V$ Log plot and (b) $4\, V$ to $8\, V$ Linear Plot for $R=50$ \r{A} and $d=5,7,10$ \r{A}.}
	\label{fig:CR1N2}	
\end{figure}  

\bigskip 
 
\par Pauli Effects have been discussed in some detail in Chapter 3, Figures 3, 4, and 5.  
The effect of the third term on the tunneling currents is studied, by considering the Current Ratio $\text{CR}= 1-(I_\text{Non-Pauli}/I)$, where $I_\text{Non-Pauli}$ is the net tunneling current obtained by using only the Non-Pauli contribution to the Fermi Factor. The total current $I$ is obtained by using all of $\mathcal{F}(E_x)$ including the so called `third term'.
The twin plots in Fig. \ref{fig:CR1N2} plot CR as a function of bias voltage for Al-vac-Cu for which $\eta_1-\eta_2=4.7 \,\,eV$. The value of CR and hence the strength of the Pauli Effect decreases very fast with increasing tip-sample distance. It is found that CR $ \,> 0$ for $eV_b < \eta_1-\eta_2$, for which voltage range, $I > I_\text{Non-Pauli}$. This seems counter intuitive $i.e.$ the Pauli Effect does not always cause Pauli Blocking. For $eV_b>\eta_1-\eta_2$, CR $< \, 0$, implying that $I < I_\text{Non-Pauli}$ which is consistent with the notion of Pauli Blocking. The strength of the Pauli Effect $|CR|$ decreases with increasing bias voltage.

\section{\label{sec:level4}Tunneling Current I For Very Low Bias}
\begin{figure}[hpt]
	\centering
	\includegraphics[width=4.5in,height=3.5in]{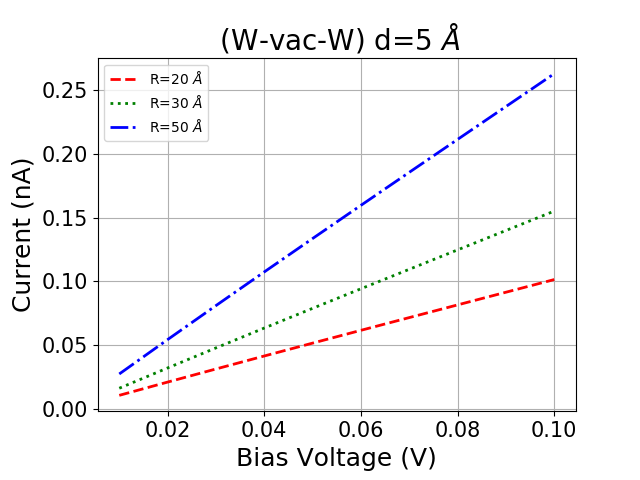}
	\caption{Plot of Current versus Bias Potential $(\le 0.1\, V)$ for $R = 20,\, 30,\, 50$ \r{A} and $ d = 5 $ \r{A} for W-vac-W.} 
	\label{fig:Ting}	
\end{figure}
\par Xie {\it et. al.}\cite{ting} have reported values of tunneling currents in the W-vac-W electrode system for very low bias voltages and for several cross currents $I_\text{cross}$, including for $I_\text{cross}=0$. Here $I_\text{cross}$ is a current made to flow in the surface of the sample. For $I_\text{cross}=0$, their experimental setup corresponds to the tip-vacuum-sample system for which tunneling currents are calculated in this thesis. Tunneling currents of $0.1 \text{nA}$ are reported by Xie {\it et. al.}\cite{ting} for low bias voltages in the range $\,[0 \,\,\text{to} \,70\, mV]$. The calculation of currents for low bias using Green function method is described in Chapter 3, section 3.3. 
Fig. \ref{fig:Ting} plots the calculated tunneling current for (W-vac-W) for very low bias for a fixed tip sample distance $d = 5 $ \r{A} and several tip radii of curvature $(R=20, 30, 50$ \r{A}. These plots show the tunneling current increasing with the bias voltage as expected. The tunneling currents are seen to increase with increasing tip radii. Similar behaviour is also seen for higher bias voltages. Further for $R=30$ \r{A}, a good numerical agreement with the very low bias results for $I_\text{cross}=0$ of Xie {\it et. al.}\cite{ting} is obtained.

\section{\label{sec:level8}Resolving Power (RP)}
\begin{figure}[hpt]
	\centering
	\includegraphics[width=4.5in,height=3.5in]{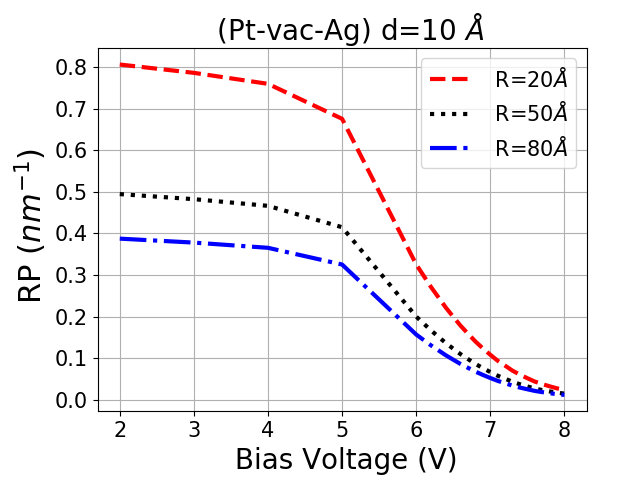}
	\caption{\,\,Plot \,\,of\,\, Resolving Power (RP)\,\, (Measure\, \,of \,\,lateral Resolution) vs Bias Voltage for Radii of curvature \,\,of \,\,the \,\,tip = $20$, $50$, $80$  \r{A} and Tip sample distance $= 10 $ \r{A}.}
	\label{fig:RPvsBias}	
\end{figure}
\par The ratio $\dfrac{J_{Net}(\xi=1)}{I}$ gives the inverse of area $\sigma^{-1}$, where the area $\sigma$  can be regarded as the area of a circle of lateral resolution; {\it i.e.}, sample profile features that lie within this circle would be poorly resolved. The radius $r=\sqrt{\sigma/\pi}$ of this circle would therefore be a good measure of the lateral resolution limit, and can be named as the Lateral Resolution Parameter (LRP). The inverse of the LRP is more relevant for discussing resolving power (RP) and results concerning the behaviour of the RP = $\dfrac{1}{r}$ will be displayed.

\bigskip

\par Fig. \ref{fig:RPvsBias} shows the plots of the RP as a function of the bias voltage ($2\,V$ to $8\, V$) for a fixed tip sample distance $d=10$ \r{A} and for three different tip radii of curvature $viz.$ $R=20,\,50,\,80 $ \r{A}. For low bias voltages the RP is almost constant, and seems very nearly the same for all tip radii. However as the bias voltage increases, the resolution is degraded ($i.e. \,RP$ decreases). This decrease in RP is the fastest for the largest tip radius. 
\begin{figure}[hpt]
	\centering
	\begin{subfigure}{0.49\textwidth}
			\includegraphics[width=3.2in,height=3.0in]{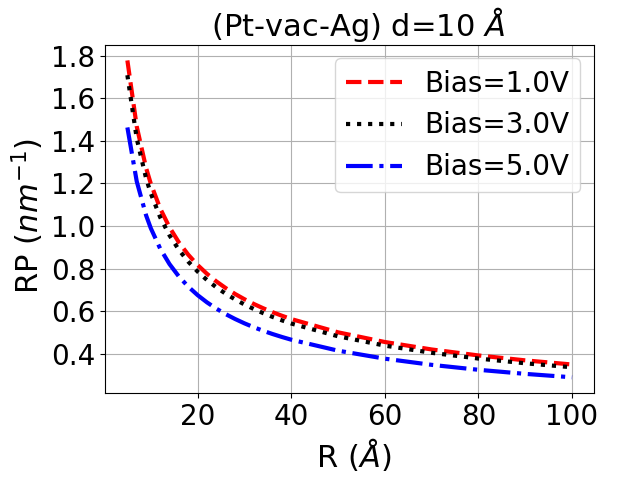}	
		\caption{} 
		\label{fig:RPvsRV5dvar}	
	\end{subfigure}
	\hfill
	\begin{subfigure}{0.49\textwidth}	
		\includegraphics[width=3.2in,height=3.0in]{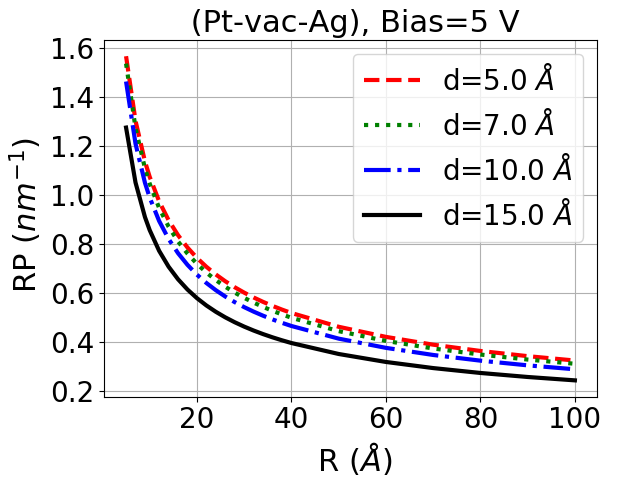}	
		\caption{} 
		\label{fig:RPvsRd10Vvar}	
	\end{subfigure}
	\caption{\,\,Plot of RP (Measure of lateral Resolution) vs R (Radius of curvature of the tip) for (a) Bias Voltage = $5\,V$ and Tip-Sample distances = $5,\,\,7,\,\,10,\,\,15$ \r{A} and (b) Tip-Sample distance = $10 $ \r{A} and Bias Voltage = $1,\,\,3,\,\,5$ V .}
	\label{fig:RPvsR}	
\end{figure} 

\bigskip

\par The twin plots in Fig. \ref{fig:RPvsR} show the dependence of the RP with the tip radius $R$  which is varied from ($5 $ \r{A}, to $100$  \r{A}). Both plots show that sharper the tip ($viz$ lower the value of $R$), the greater is the lateral resolution ($viz$ greater the value of RP). The plot in Fig. \ref{fig:RPvsRV5dvar} shows that for fixed bias the lateral resolution is degraded with increasing tip sample distance. In the plot of Fig. \ref{fig:RPvsRd10Vvar} the tip sample distance is kept constant, and the RP is plotted as a function of $R$ for bias voltages of $1,3,\text{and}\, 5$ volts. This plot shows that the resolution is degraded ($i.e.\, RP$ is smaller) for larger bias voltages and RP decreases faster with tip radius R for larger bias voltages than for smaller bias voltages.

\bigskip

In this Chapter, the results of the calculation are presented via plots of currents as a function of bias, tip-sample distance, and radius of curvature of the tip. The comparison of the currents in case of Trapezoidal and Russell potential as a function of bias, tip-sample distance, for similar and dissimilar electrodes is also presented. Resolving Power (RP) is studied as a function of bias voltage, tip-sample distance, radius of curvature, and it is shown to be degraded for blunter tips, increasing bias voltages, and increasing tip–sample distances. Scope for future work lies in the study of temperature dependence of the tunnel junction characteristics, and in particular when the electrode temperatures are unequal. Many non symmetric configurations and different shaped tips can also be studied by carrying out a detailed Finite Element Calculation. Also the effects of the tunnel currents on the force between tip and sample can also be derived through the use of the calculations in this thesis as well as extending them through the use of Finite Element calculations.

%
%
%
%
%
%
%
%
%
%
%
%
%


    \newpage
    \pagestyle{plain}	    
	\addcontentsline{toc}{chapter}{\textbf{Appendix A-1}}
\renewcommand{\theequation}{A-1.\arabic{equation}}
\setcounter{equation}{0}
\noindent {\Large \textbf{A-1 Schr\"odinger equation for linear potentials and the Airy differential equation}}\\

The Schr\"odinger equation for the trapezoidal potential is 
$$
\Big [\frac{- \hbar^2}{2m} \frac{d^2}{dx^2} + U_{II}(x) \Big ] \Psi_b(x) = E_x \Psi_b(x)
$$
where $U_{II}(x)=(\eta_1+\phi_1)-(\phi_1-\phi_2+eV_b)\dfrac{x}{d} $. Substituting $U_{II}(x)$ in above equation and rearranging 
$$
\Big [\frac{d^2}{dx^2} - \frac{2m}{\hbar^2}  \Big (\eta_1+\phi_1- E_x-(\phi_1-\phi_2+eV_b)\frac{x}{d} \Big ) \Big ] \Psi_b(x) = 0
$$
Defining A and B as 
\begin{equation}
A=\dfrac{2m}{\hbar^2} (\eta_1+\phi_1-E_x) \,, \quad B=\dfrac{2m}{\hbar^2 d} (\phi_1-\phi_2+eV_b) 
\end{equation}	
The Schr\"odinger equation reduces to a following differential equation
\begin{equation}
\frac{d^2 \Psi_b}{dx^2}-(A-Bx)\Psi_b= 0 
\end{equation} 
Put $ h = (A - Bx) B^q \,\,$  so that $dh =  -B^{q+1} dx $.
This implies that $\,\,\dfrac{d}{dh} = -B^{-(q+1)}\dfrac{d}{dx}\,\,$  and  $\,\,\dfrac{d}{dx} = -B^{(q+1)}\dfrac{d}{dh}\,\,$ so that $\dfrac{d^2}{dx^2} = B^{-2(q+1)}\dfrac{d^2}{dh^2} $.
Substituting 
$$
B^{2q+2}\frac{d^2 \Psi_b}{dh^2} - B^{-q} h\Psi_b= 0 
$$
choosing $q = -2/3$, and substituting in above equation, we get Airy differential equation
\begin{equation}
\frac{d^2 \Psi_b}{dh^2} - h\Psi_b= 0
\end{equation}

    \newpage
    \pagestyle{plain}	  
	\addcontentsline{toc}{chapter}{\textbf{Appendix A-2}}
	\renewcommand{\theequation}{A-2.\arabic{equation}}
\renewcommand{\thefigure}{A-2.\arabic{figure}}
\setcounter{equation}{0}
\setcounter{figure}{0}
\noindent {\Large \textbf{A-2 Airy functions and their derivatives, and properties.}}\\

\begin{figure}[h]	
	\begin{subfigure}{0.49\textwidth}
		\includegraphics[width=3.0in,height=2.7in]{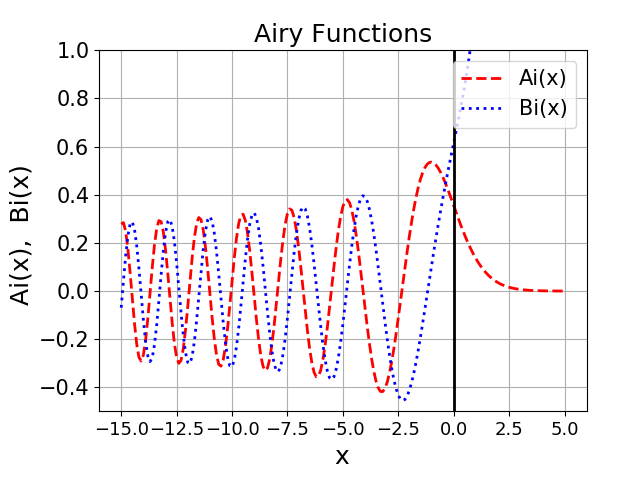}	
		\caption{}
		\label{fig:Airy}
	\end{subfigure}
	\hfill
	\begin{subfigure}{0.49\textwidth}
		\includegraphics[width=3.0in,height=2.7in]{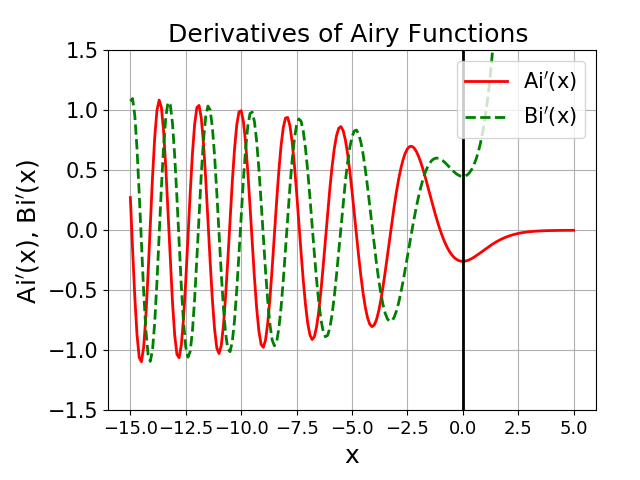}	
		\caption{}
		\label{fig:dAiry}
	\end{subfigure}
	\caption{Variation of (a) Airy function with x (b) derivatives of Airy function with x}
\end{figure}

The Airy functions $Ai(x)$ and $Bi(x)$ are the special functions (pair of linearly independent solutions) of the Airy differential equation	
\begin{equation}
y^{\prime \prime} - x\,y = 0
\end{equation}
The general solution to Airy equation is given by 
$$y(x)=c_1 Ai(x) + c_2 Bi(x)$$
where $c_1$ and $c_2$ are constant coefficients to be determined by the boundary conditions of the problem.
The behaviour of Airy functions can be either oscillatory or exponential, depending on the sign of x. 
For positive values of $x\,$,  $i.e. \, \, x > 0$, the functions $Ai(x)$ and $Bi(x)$ and their derivatives behave monotonically. For negative values of $x\,$,  $i.e. \, \, x < 0$, the functions $Ai(x)$ and $Bi(x)$ and their derivatives are oscillatory functions. The behaviour of $Ai(x)$,and $Bi(x)$ and their derivatives is shown in Figures \ref{fig:Airy} and \ref{fig:dAiry}.

 \bigskip
 
 Both the functions $Ai(x)$ and $Ai^\prime (x)$ are bounded for positive values of $x$. For negative values of $x$, $Ai(x)$ is oscillatory and bounded. However its derivative is oscillatory but its amplitude appears to increase without bound.  $Bi(x)$ on the other hand diverges for large positive $x$ and so does its derivative. For negative values of $x$, $Bi(x)$ is oscillatory and bounded, while and $Bi^\prime (x)$ is oscillatory with increasing amplitudes, The values of the functions $Ai(x),Bi(x)$ and their derivatives are provided as intrinsic functions in Python (a computer programming language). 
The Wroskian of $Ai(x)$, $Bi(x)$ is given by
$$\text{W}[Ai(x),Bi(x)] = [Ai(x) Bi^\prime (x) - Ai^\prime(x) Bi(x)]  = \frac{1}{\pi }$$

    \newpage
    \pagestyle{plain}	 
	\addcontentsline{toc}{chapter}{\textbf{Appendix A-3}}
	\renewcommand{\theequation}{A-3.\arabic{equation}}
\setcounter{equation}{0}
\noindent {\Large \textbf{A-3 The Fermi Factor}}\\

The Fermi Factor is an integral over $E_r$ described in equation (3.18) and is given by is 
\begin{equation}\label{A3eq1}
\mathcal{F}(E_x)=\int\limits_0^\infty dE_r \Bigg [f_1(E)\Big[1 - f_1(E')\Big] \,\dfrac{k_2}{k_1} - \Big[1 - f_1(E)\Big] f_1(E') \, \dfrac{k_1}{k_2} \Bigg] 
\end{equation}
where $E=E_x+E_r$ and $E^\prime = E_x+E_r+eV_b$.

\bigskip

\noindent The integrand of the above equation can be simplified to
\begin{equation}\label{intg}
 \frac{k_2}{k_1} f_1(E)-\frac{k_1}{k_2} f_1(E') + \Big(\frac{k_1}{k_2} - \frac{k_2}{k_1}\Big)f_1(E)f_1(E')
\end{equation}
Define $\quad a = \exp[\beta (E_x-\eta_1)]$ \quad \text{and} \quad $\quad$ $b = \exp[\beta (E_x+eV_b-\eta_1)]$

\bigskip

\noindent The integral over $E_r$ of the first term in the above equation (\ref{intg})
\begin{equation}
\begin{aligned}
I_1 = \int\limits_0^\infty dE_r f_1(E) = & \int\limits_0^\infty dE_r\dfrac{1}{\exp(\beta (E_x + E_r - \eta_1))} \\ = & \int\limits_0^\infty \frac{dE_r}{(1+a\,\exp[\beta E_r])} 
= \dfrac{1}{\beta} \text{ln} \big[ 1+\exp(-\beta (E_x-\eta_1))\big]
\end{aligned}
\end{equation}
Likewise the integral over $E_r$ of the second term in equation (\ref{intg}) is evaluted to 
$$I_2 = \int\limits_0^\infty dE_r f_1(E')= \int\limits_0^\infty \frac{dE_r}{(1+b\,\exp[\beta E_r])} = \dfrac{1}{\beta} \text{ln} \big[ 1+\exp(-\beta (E_x-\eta_1+eV_b))\big]$$

\bigskip

\noindent The third term in equation (\ref{intg}) involves the product $f_1(E)f_1(E')$. In terms of $a$ and $b$ this product is expanded in partial fractions as 
\begin{multline}
f_1(E)f_1(E') = \dfrac{1}{(1+a\,\exp[\beta E_r])(1+b\,\exp[\beta E_r])} \\
= \dfrac{1}{(b-a)}\Big [ \dfrac{b}{1+b\,\exp[\beta E_r]}-\dfrac{a}{1+a\,\exp[\beta E_r]}\Big]
\end{multline}
Hence the integral  of this term over $E_r$ is given by
\begin{equation}
I_3 = \int\limits_0^\infty dE_r f_1(E)f_1(E') = \dfrac{1}{(b-a)}\int\limits_0^\infty dE_r \Big [ \dfrac{b}{1+b\,\exp[\beta E_r]}-\dfrac{a}{1+a\,\exp[\beta E_r]}\Big]  
\end{equation}
$$ = \dfrac{-1}{\beta (b-a)}\Big [b\,\ln(1+\dfrac{1}{b}) - a\,\ln(1+\dfrac{1}{a})  \Big]  $$

\bigskip

\noindent Putting all together in equation (\ref{A3eq1})
\begin{equation}
\mathcal{F}(E_x)= \dfrac{k_2}{k_1}F_1(E_x) - \dfrac{k_1}{k_2}F_1(E_x+eV_b)\, + \frac{\Big [\dfrac{k_1}{k_2}-\dfrac{k_2}{k_1}\Big ]}{(1-e^{-\beta eV_b})}\Big [F_1(E_x+eV_b)-e^{-\beta eV_b}F_1(E_x)\Big] 
\end{equation}
where
\begin{equation}
F_1(E_x)=\dfrac{1}{\beta}\text{Log}[1+e^{-\beta(E_x-\eta_1)}]
\end{equation}
The function $\zeta(E_x)$ used in Chapter 7 is defined to be 
\begin{equation}\label{intg2}
\zeta(E_x) = \int\limits_0^\infty dE_r \Bigg [\frac{k_2}{k_1} f_1(E)+\frac{k_1}{k_2} f_1(E') - \Big(\frac{k_1}{k_2} + \frac{k_2}{k_1}\Big)f_1(E)f_1(E') \Bigg]
\end{equation}
\begin{equation}
 = \dfrac{k_2}{k_1}I_1 + \dfrac{k_1}{k_2}I_2 - I_3 \Big [\dfrac{k_1}{k_2}+\dfrac{k_2}{k_1} \Big]
\end{equation}
Therefore
\begin{equation}
\zeta(E_x) =   \dfrac{1}{\beta}\left(   \dfrac{k_2}{k_1}F_1(E_x) + \dfrac{k_1}{k_2}F_1(E_x+eV_b)\, - \frac{\Big [\dfrac{k_1}{k_2}+\dfrac{k_2}{k_1}\Big ]}{(1-e^{-\beta eV_b})}\Big [F_1(E_x+eV_b)-e^{-\beta eV_b}F_1(E_x)\Big] \right) \\
\end{equation}

	\newpage
	\addcontentsline{toc}{chapter}{\textbf{Bibliography}}
	\bibliographystyle{ieeetr} 
	\bibliography{thesis-Print}

\begin{thebibliography}{10}

\bibitem{binnig5}
G.~Binnig, H.~Rohrer, C.~Gerber, and E.~Weibel, ``Surface studies by scanning
  tunneling microscopy,'' {\em Physical review letters}, vol.~49, no.~1, p.~57,
  1982.

\bibitem{Chen7}
C.~J. Chen, {\em Monographs on the Physics and Chemistry of Materials
  “Introduction to Scanning Tunneling Microscopy”}.
\newblock Oxford University Press, 2008.

\bibitem{Wiesendanger1}
S.~Heinze, M.~Bode, A.~Kubetzka, O.~Pietzsch, X.~Nie, S.~Blugel, and
  R.~Wiesendanger, ``Real-space imaging of two-dimensional antiferromagnetism
  on the atomic scale,'' {\em Science}, vol.~288, no.~5472, pp.~1805--1808,
  2000.

\bibitem{binnig1982tunneling}
G.~Binnig, H.~Rohrer, C.~Gerber, and E.~Weibel, ``Tunneling through a
  controllable vacuum gap,'' {\em Applied Physics Letters}, vol.~40, no.~2,
  pp.~178--180, 1982.

\bibitem{o1998atomic}
S.~O'shea and M.~Welland, ``Atomic force microscopy at solid- liquid
  interfaces,'' {\em Langmuir}, vol.~14, no.~15, pp.~4186--4197, 1998.

\bibitem{horng2012vibration}
T.-L. Horng, ``Vibration responses of atomic force microscope cantilevers,'' in
  {\em Atomic Force Microscopy-Imaging, Measuring and Manipulating Surfaces at
  the Atomic Scale}, IntechOpen, 2012.

\bibitem{ternes2008force}
M.~Ternes, C.~P. Lutz, C.~F. Hirjibehedin, F.~J. Giessibl, and A.~J. Heinrich,
  ``The force needed to move an atom on a surface,'' {\em Science}, vol.~319,
  no.~5866, pp.~1066--1069, 2008.

\bibitem{albrecht1991frequency}
T.~R. Albrecht, P.~Gr{\"u}tter, D.~Horne, and D.~Rugar, ``Frequency modulation
  detection using high-q cantilevers for enhanced force microscope
  sensitivity,'' {\em Journal of applied physics}, vol.~69, no.~2,
  pp.~668--673, 1991.

\bibitem{israelachvili1973van}
J.~Israelachvili and D.~Tabor, ``Van der waals forces: theory and experiment,''
  in {\em Progress in surface and membrane science}, vol.~7, pp.~1--55,
  Elsevier, 1973.

\bibitem{israelachvili2013intersection}
J.~N. Israelachvili, K.~Kristiansen, M.~A. Gebbie, D.~W. Lee, S.~H.
  Donaldson~Jr, S.~Das, M.~V. Rapp, X.~Banquy, M.~Valtiner, and J.~Yu, ``The
  intersection of interfacial forces and electrochemical reactions,'' {\em The
  Journal of Physical Chemistry B}, vol.~117, no.~51, pp.~16369--16387, 2013.

\bibitem{zhang2021intermolecular}
J.~Zhang and H.~Zeng, ``Intermolecular and surface interactions in engineering
  processes,'' {\em Engineering}, vol.~7, no.~1, pp.~63--83, 2021.

\bibitem{hou2022intermolecular}
X.~Hou, J.~Li, Y.~Li, and Y.~Tian, ``Intermolecular and surface forces in
  atomic-scale manufacturing,'' {\em International Journal of Extreme
  Manufacturing}, vol.~4, no.~2, p.~022002, 2022.

\bibitem{hubbard2011america}
A.~Hubbard, ``America journal, 57, 1241--1246. israelachvili, jn, 1985.
  intermolecular and surface forces. new york: Academic, pp. 187--193. khan,
  t., and so, b., 2006. effect of soil dispersion on the pro-cesses of
  infiltration, redistribution and evaporation (available,'' {\em Encyclopedia
  of Agrophysics}, p.~304, 2011.

\bibitem{atkin2009afm}
R.~Atkin, S.~Z. El~Abedin, R.~Hayes, L.~H. Gasparotto, N.~Borisenko, and
  F.~Endres, ``Afm and stm studies on the surface interaction of [bmp] tfsa and
  [emim] tfsa ionic liquids with au (111),'' {\em The Journal of Physical
  Chemistry C}, vol.~113, no.~30, pp.~13266--13272, 2009.

\bibitem{enevoldsen2008detailed}
G.~H. Enevoldsen, H.~P. Pinto, A.~S. Foster, M.~C. Jensen, A.~K{\"u}hnle,
  M.~Reichling, W.~A. Hofer, J.~V. Lauritsen, and F.~Besenbacher, ``Detailed
  scanning probe microscopy tip models determined from simultaneous
  atom-resolved afm and stm studies of the tio 2 (110) surface,'' {\em Physical
  Review B}, vol.~78, no.~4, p.~045416, 2008.

\bibitem{patil}
S.~Patil, A.~V. Kulkarni, and C.~Dharmadhikari, ``Study of the electrostatic
  force between a conducting tip in proximity with a metallic surface: Theory
  and experiment,'' {\em Journal of Applied Physics}, vol.~88, no.~11,
  pp.~6940--6942, 2000.

\bibitem{date}
K.~S. Date, A.~V. Kulkarni, and D.~CV, ``Force on a conducting tip near a
  metallic surface coated with a polarizable dielectric layer: Theory and
  experiment,'' {\em e-Journal of Surface Science and Nanotechnology}, vol.~9,
  pp.~206--209, 2011.

\bibitem{hartmann1999magnetic}
U.~Hartmann, ``Magnetic force microscopy,'' {\em Annual review of materials
  science}, vol.~29, no.~1, pp.~53--87, 1999.

\bibitem{kazakova2019frontiers}
O.~Kazakova, R.~Puttock, C.~Barton, H.~Corte-Le{\'o}n, M.~Jaafar, V.~Neu, and
  A.~Asenjo, ``Frontiers of magnetic force microscopy,'' {\em Journal of
  applied Physics}, vol.~125, no.~6, 2019.

\bibitem{koblischka2003recent}
M.~R. Koblischka and U.~Hartmann, ``Recent advances in magnetic force
  microscopy,'' {\em Ultramicroscopy}, vol.~97, no.~1-4, pp.~103--112, 2003.

\bibitem{wadas1989theoretical}
A.~Wadas and P.~Gr{\"u}tter, ``Theoretical approach to magnetic force
  microscopy,'' {\em Physical Review B}, vol.~39, no.~16, p.~12013, 1989.

\bibitem{ocola}
L.~E.Ocola, {\em "Scanning Thermal Microscopy” Characterization of
  Materials}.
\newblock John Wiley and Sons, 2012.

\bibitem{doi:10.1146/annurev.matsci.37.052506.084342}
W.~Wulfhekel and J.~Kirschner, ``Spin-polarized scanning tunneling microscopy
  of magnetic structures and antiferromagnetic thin films,'' {\em Annual Review
  of Materials Research}, vol.~37, no.~1, pp.~69--91, 2007.

\bibitem{shik2003theoretical}
A.~Shik and H.~E. Ruda, ``Theoretical problems of scanning capacitance
  microscopy,'' {\em Surface science}, vol.~532, pp.~1132--1135, 2003.

\bibitem{ruda2005scanning}
H.~Ruda and A.~Shik, ``Scanning capacitance microscopy of nanostructures,''
  {\em Physical Review B}, vol.~71, no.~7, p.~075316, 2005.

\bibitem{binnig1}
G.~Binnig and H.~Rohrer, ``Scanning tunneling microscopy,'' {\em Surface
  science}, vol.~126, no.~1-3, pp.~236--244, 1983.

\bibitem{binnig3}
G.~Binnig, C.~F. Quate, and C.~Gerber, ``Atomic force microscope,'' {\em
  Physical review letters}, vol.~56, no.~9, p.~930, 1986.

\bibitem{binnig1987scanning}
G.~Binnig and H.~Rohrer, ``Scanning tunneling microscopy—from birth to
  adolescence,'' {\em reviews of modern physics}, vol.~59, no.~3, p.~615, 1987.

\bibitem{wickramasinghe1990scanning}
H.~K. Wickramasinghe, ``Scanning probe microscopy: Current status and future
  trends,'' {\em Journal of Vacuum Science \& Technology A: Vacuum, Surfaces,
  and Films}, vol.~8, no.~1, pp.~363--368, 1990.

\bibitem{wickramasinghe2000progress}
H.~K. Wickramasinghe, ``Progress in scanning probe microscopy,'' {\em Acta
  materialia}, vol.~48, no.~1, pp.~347--358, 2000.

\bibitem{wickramasinghe1989scanned}
H.~K. Wickramasinghe, ``Scanned-probe microscopes,'' {\em Scientific American},
  vol.~261, no.~4, pp.~98--105, 1989.

\bibitem{abraham1988noise}
D.~W. Abraham, C.~Williams, and H.~Wickramasinghe, ``Noise reduction technique
  for scanning tunneling microscopy,'' {\em Applied physics letters}, vol.~53,
  no.~16, pp.~1503--1505, 1988.

\bibitem{wiesendanger1992scanning}
R.~Wiesendanger, H.-J. G{\"u}ntherodt, and W.~Baumeister, {\em Scanning
  tunneling microscopy II: further applications and related scanning
  techniques}, vol.~28.
\newblock Springer, 1992.

\bibitem{baumeister2013scanning}
W.~Baumeister, P.~Gr{\"u}tter, R.~Guckenberger, H.-J. G{\"u}ntherodt,
  T.~Hartmann, H.~Heinzelmann, H.~Knapp, H.~Mamin, E.~Meyer, D.~Pohl, {\em
  et~al.}, {\em Scanning Tunneling Microscopy II: Further Applications and
  Related Scanning Techniques}, vol.~28.
\newblock Springer Science \& Business Media, 2013.

\bibitem{friedbacher1999classification}
G.~Friedbacher and H.~Fuchs, ``Classification of scanning probe microscopies,''
  {\em Pure and applied chemistry}, vol.~71, no.~7, pp.~1337--1357, 1999.

\bibitem{hansma1987scanning}
P.~K. Hansma and J.~Tersoff, ``Scanning tunneling microscopy,'' {\em Journal of
  Applied Physics}, vol.~61, no.~2, pp.~R1--R24, 1987.

\bibitem{chen2021introduction}
C.~J. Chen, {\em Introduction to Scanning Tunneling Microscopy Third Edition},
  vol.~69.
\newblock Oxford University Press, USA, 2021.

\bibitem{holt2004scanning}
K.~B. Holt, A.~J. Bard, Y.~Show, and G.~M. Swain, ``Scanning electrochemical
  microscopy and conductive probe atomic force microscopy studies of
  hydrogen-terminated boron-doped diamond electrodes with different doping
  levels,'' {\em The Journal of Physical Chemistry B}, vol.~108, no.~39,
  pp.~15117--15127, 2004.

\bibitem{weis}
R.~Weisendanger, {\em Scanning Probe Microscopy and Spectroscopy: Methods and
  Applications}.
\newblock Cambridge University Press, 1998.

\bibitem{Wiesendanger}
R.~Wiesendanger, {\em Scanning probe microscopy and spectroscopy: methods and
  applications}.
\newblock Cambridge university press, 1994.

\bibitem{FN}
R.~H. Fowler and L.~Nordheim, ``Electron emission in intense electric fields,''
  {\em Proceedings of the Royal Society of London. Series A, Containing Papers
  of a Mathematical and Physical Character}, vol.~119, no.~781, pp.~173--181,
  1928.

\bibitem{simmonsI}
J.~G. Simmons, ``Generalized formula for the electric tunnel effect between
  similar electrodes separated by a thin insulating film,'' {\em Journal of
  applied physics}, vol.~34, no.~6, pp.~1793--1803, 1963.

\bibitem{hartman}
T.~E. Hartman and J.~S. Chivian, ``Electron tunneling through thin aluminum
  oxide films,'' {\em Physical Review}, vol.~134, no.~4A, p.~A1094, 1964.

\bibitem{dessai2022calculation}
M.~Dessai and A.~V. Kulkarni, ``Calculation of tunneling current across
  trapezoidal potential barrier in a scanning tunneling microscope,'' {\em
  Journal of Applied Physics}, vol.~132, no.~24, 2022.

\bibitem{AS}
M.~Abramowitz, I.~A. Stegun, {\em et~al.}, {\em Handbook of mathematical
  functions}.
\newblock Dover, New York, 1968.

\bibitem{shu2002exactly}
Q.~Shu, Y.~Jiang, S.~Meng, G.~Lin, and W.~Ma, ``Exactly solvable model for
  metal--insulator--metal stepped boundary tunnel junctions,'' {\em Thin solid
  films}, vol.~414, no.~1, pp.~136--142, 2002.

\bibitem{arfken}
G.~Arfken, {\em Mathematical Methods for Physicists}.
\newblock American Press, New York, 1970.

\bibitem{morse}
P.~Morse and H.~Feshbach, {\em Methods of Theoretical Physics}.
\newblock McGraw-Hill Book Company, 1953.

\bibitem{Shu}
Q.~Shu, Y.~Jiang, S.~Meng, G.~Lin, and W.~Ma, ``Exactly solvable model for
  metal--insulator--metal stepped boundary tunnel junctions,'' {\em Thin solid
  films}, vol.~414, no.~1, pp.~136--142, 2002.

\bibitem{Ebeling}
W.~Ebeling, D.~Blaschke, R.~Redmer, H.~Reinholz, and G.~R{\"o}pke, ``The
  influence of pauli blocking effects on the properties of dense hydrogen,''
  {\em Journal of Physics A: Mathematical and Theoretical}, vol.~42, no.~21,
  p.~214033, 2009.

\bibitem{ting}
T.~Xie, M.~Dreyer, D.~Bowen, D.~Hinkel, R.~Butera, C.~Krafft, and I.~Mayergoyz,
  ``On local sensing of spin hall effect in tungsten films by using stm-based
  measurements,'' {\em IEEE Transactions on Nanotechnology}, vol.~17, no.~5,
  pp.~914--919, 2018.

\bibitem{tersoff}
J.~Tersoff and D.~R. Hamann, ``Theory of the scanning tunneling microscope,''
  {\em Physical Review B}, vol.~31, no.~2, p.~805, 1985.

\bibitem{garcia}
N.~Garcia, C.~Ocal, and F.~Flores, ``Model theory for scanning tunneling
  microscopy: application to au (110)(1$\times$ 2),'' {\em Physical review
  letters}, vol.~50, no.~25, p.~2002, 1983.

\bibitem{bardeen}
J.~Bardeen, ``Tunnelling from a many-particle point of view,'' {\em Physical
  review letters}, vol.~6, no.~2, p.~57, 1961.

\bibitem{chen}
C.~J. Chen, ``Role of tip material in scanning tunneling microscopy,'' {\em MRS
  Online Proceedings Library (OPL)}, vol.~159, 1989.

\bibitem{SBPZ}
S.~Banerjee and P.~Zhang, ``A generalized self-consistent model for quantum
  tunneling current in dissimilar metal-insulator-metal junction,'' {\em AIP
  Advances}, vol.~9, no.~8, 2019.

\bibitem{simmonsII}
J.~G. Simmons, ``Electric tunnel effect between dissimilar electrodes separated
  by a thin insulating film,'' {\em Journal of applied physics}, vol.~34,
  no.~9, pp.~2581--2590, 1963.

\bibitem{KH}
K.~Huang, {\em Statistical Mechanics,Page 245 (Equation 11.24)}.
\newblock John Wiley and Sons, 1987.

\bibitem{RScIn}
J.~Zhang, P.~Wang, X.~Zhang, H.~Ji, J.~Luo, H.~Wang, and J.~Wang, ``Systematic
  electrochemical etching of various metal tips for tunneling spectroscopy and
  scanning probe microscopy,'' {\em Review of Scientific Instruments}, vol.~92,
  no.~1, 2021.

\bibitem{SG}
J.~J. S{\'a}enz and R.~Garc{\'\i}a, ``Near field emission scanning tunneling
  microscopy,'' {\em Applied physics letters}, vol.~65, no.~23, pp.~3022--3024,
  1994.

\bibitem{moon}
P.~Moon and D.~E. Spencer, {\em Field Theory Handbook}.
\newblock Springer-Verlag Berlin Heildelberg New York, 1971.

\bibitem{russel}
A.~M. Russell, ``Electron trajectories in a field emission microscope,'' {\em
  Journal of Applied Physics}, vol.~33, no.~3, pp.~970--975, 1962.

\bibitem{CRC75}
S.~Shelbey, {\em CRC Standard Mathematical Tables}.
\newblock CRC Press, ISBN 0-87819-622-6, 1975.

\bibitem{cutler}
P.~Cutler, J.~He, J.~Miller, N.~Miskovsky, B.~Weiss, and T.~Sullivan, ``Theory
  of electron emission in high fields from atomically sharp emitters: Validity
  of the fowler-nordheim equation,'' {\em Progress in surface science},
  vol.~42, no.~1-4, pp.~169--185, 1993.

\bibitem{Huang}
Z.-H. Huang, M.~Weimer, and R.~E. Allen, ``Internal image potential in
  semiconductors: Effect on scanning tunneling microscopy,'' {\em Physical
  Review B}, vol.~48, no.~20, p.~15068, 1993.

\bibitem{hofer2001surface}
W.~Hofer, A.~Fisher, R.~Wolkow, and P.~Gr{\"u}tter, ``Surface relaxations,
  current enhancements, and absolute distances in high resolution scanning
  tunneling microscopy,'' {\em Physical Review Letters}, vol.~87, no.~23,
  p.~236104, 2001.

\end{thebibliography}
	
	
	\newpage
	\addcontentsline{toc}{chapter}{\textbf{Brief Bio-data of the candidate}}
	\chapter*{Brief Bio-data of the candidate}
Ms. Malati Dessai\\
Research Scholar\\
Department of Physics\\
BITS Pilani K K Birla Goa Campus\\
State: Goa, PIN:403 726, Country: India.\\
Permanent Address :\\
V-16, CD Scenic Acres \\
Ambaji Road, Fatorda, Margao - Goa\\
PIN: 403602; Country: India\\
mobile: +91 8007904896\\
e-mail: p20130101@goa.bits-pilani.ac.in\\
e-mail: malatidessai@gmail.com\\
\textbf{Education:}\\
Masters of Science (M.Sc) Physics: Goa University, Goa, India\\
Bachelor of Science (B.Sc) Physics: Dhempe College of Arts and Science, Panaji, Goa, India\\	
~~~~~\textbf{Publication in International Journal}\\
Malati Dessai, Arun V. Kulkarni \textit{Calculation of tunneling current across trapezoidal potential barrier in a scanning tunneling microscope.} Journal of Applied Physics 132, 244901 pages 1-11, (2022)\\
DOI:10.1103/PhysRevC.99.052801\\
\textbf{Teaching:}\\
Assistant Professor\\
Parvatibai Chowgule College of Arts ans Science, Margao, Goa\\
\textbf{Competence with programing language:}\\
FORTRAN 90, Python, Wolfram Mathematica, LATEX.

	\newpage
	\addcontentsline{toc}{chapter}{\textbf{Brief Bio-data of the supervisor}}
	\chapter*{Brief Bio-data of the supervisor}
	
Prof. Arun V Kulkarni is a Professor in the Physics Department, BITS Pilani, KK Birla Goa campus. He obtained his Ph.D. degree  in Sept 1994 from The University of Wyoming, Laramie, WY-82071 USA.  under the supervision of Prof. Glen A Rebka.  His thesis title was :- A calculation of $\dfrac{d^2\sigma}{d \Omega dT_\pi}$  for Inclusive $\pi-^4He$ Double Charge Exchange at $T_\pi \,\leqslant \, 270$ MeV.  His research interest spans Theoretical Computational Physics, Numerical Calculation of E-M fields, Theretical modelling of Scanning Probe Microscopes.  He has finished guiding one student in the area of Scanning Tunneling Microscope. He has guided 3 Masters Thesis, 7 First degree Study Oriented Projects(SOP) and 3 Computer Oriented projects. He was an external examiner of 3 Masters Thesis of students from Goa University. He has 16 publications. He has taught 23 different courses including some in the Math Department, and one in the Chemical Engg. Department. He is currently an IUCAA Associate and is also carrying out research in Gravitational Waves. He has been associated with IAPT (Indian Association of Physics Teachers) and was President of the Goa Regional Council for 2 consecutive terms. He was also the Vice President (Western Zone) of the IAPT(All India). He conducted a 4 - day Joint Workshop for College students called IAPT-UGCP with the Physics Dept.of BITS-Pilani-Goa Campus and IUCAA- Pune in Dec 2019. Was member of the Board of Studies (BOS) on several occassions both in the Physics Departments of Goa University and Chowgule College(Goa). Have been invited to give talks in several Colleges in Goa.
\end{document}